\documentclass[]{aa} 
\usepackage[normalem]{ulem}
\newcommand{\beforeReferee}[1]{}
\newcommand{\afterReferee}[1]{#1}
\usepackage{graphicx}
\usepackage{txfonts}
\usepackage{xcolor}

\usepackage{natbib}
\bibpunct{(}{)}{;}{a}{}{,} 

\newcommand{\ignoreThis}[1]{ }

\providecommand{\dt}[1]{{\tt #1}} 
\newcommand\gaia{Gaia}

\newcommand\gdr[1]{\gaia~DR#1}
\def\GDR1{\gdr{1}}
\def\preDR1{pre-DR1}

\def\bt{$B_{\rm T}$}
\def\vt{$V_{\rm T}$}
\def\btmvt{$B_{\rm T}-V_{\rm T}$}
\def\bmv{$B-V$}
\def\vmi{$V-I$}
\newcommand\bpminrp{\ensuremath{G_\mathrm{BP}-G_\mathrm{RP}}}
\def\bprp{\bpminrp}
\def\gmhp{$G-H{\rm p}$}
\def\ghst{$G_{\rm HST}$}
\def\goggle{$G_{\rm OGLE}$}

\def\Kjk{$K_{\rm J-K}$}
\def\MK{$M_{\rm K}$}
\def\Ks{$K_{\rm s}$}
\def\gmhp{$G-Hp$}
\def\chitwo{$\chi^2~$}

\providecommand{\secref}[1]{Sect.~\ref{#1}}
\providecommand{\tabref}[1]{Table~\ref{#1}}
\providecommand{\figref}[1]{Fig.~\ref{#1}}

\newcommand{\smallImage}[1]{\includegraphics[width=0.49\columnwidth]{#1}}
\newcommand{\columnImage}[1]{\includegraphics[width=0.8\columnwidth]{#1}}

\def\deg{\degr}

\def\arcsec{\,$''$}

\providecommand{\kms}{\,km\,s$^{-1}$}
\providecommand{\masyr}{\,mas\,yr$^{-1}$}
\providecommand{\muas}{\,$\mu$as}

\def\mmag{mmag}
\def\parallax{$\varpi$}

\def\a0{$A_{\rm 0}$}

\def\av{$A_{\rm V}$}

\def\mg{$M_{\rm G}$}

\def\gmag{$G$}
\def\gbp{$G_{\rm BP}$}
\def\grp{$G_{\rm RP}$}

\def\apjl{ApJ}%
\def\apjs{ApJS}%
\def\ao{Appl.~Opt.}%
\def\aap{A\&A}%
\def\ltsim{\ifmmode\stackrel{<}{_{\sim}}\else$\stackrel{<}{_{\sim}}$\fi}

\begin{document} 

   \title{Gaia Data Release 1}

   \subtitle{Catalogue validation 
   }

\author{F.~Arenou\inst{\ref{inst:gepi}}, 
X. Luri\inst{\ref{inst:ieec}}, C.~Babusiaux\inst{\ref{inst:gepi}}, C.~Fabricius\inst{\ref{inst:ieec}}, 
A.~Helmi\inst{\ref{inst:kapteyn}}, A.C.~Robin\inst{\ref{inst:utinam}}, A.~Vallenari\inst{\ref{inst:padova}}, 
S. Blanco-Cuaresma\inst{\ref{inst:Geneva}}, T.~Cantat-Gaudin\inst{\ref{inst:padova}}, K.~Findeisen\inst{\ref{inst:gepi}}, 
C.~Reyl\'e\inst{\ref{inst:utinam}}, L.~Ruiz-Dern\inst{\ref{inst:gepi}}, R.~Sordo\inst{\ref{inst:padova}},
C.~Turon\inst{\ref{inst:gepi}}, N.A.~Walton\inst{\ref{inst:ioa}}, I-C.~Shih\inst{\ref{inst:gepi}}, 
E.~Antiche\inst{\ref{inst:ieec}},
C.~Barache\inst{\ref{inst:SYRTE}}, M.~Barros\inst{\ref{inst:FCUL}}, M. Breddels\inst{\ref{inst:kapteyn}}, 
J.M.~Carrasco\inst{\ref{inst:ieec}}, G. Costigan\inst{\ref{inst:leidenO}}, S. Diakit\'e\inst{\ref{inst:utinam}}, 
L. Eyer\inst{\ref{inst:Geneva}}, F.~Figueras\inst{\ref{inst:ieec}},
L.~Galluccio\inst{\ref{inst:nice}}, J.~Heu\inst{\ref{inst:gepi}}, C.~Jordi\inst{\ref{inst:ieec}}, 
A.~Krone-Martins\inst{\ref{inst:FCUL}}, R.~Lallement\inst{\ref{inst:gepi}}, S.~Lambert\inst{\ref{inst:SYRTE}}, 
N.~Leclerc\inst{\ref{inst:gepi}}, P.M.~Marrese\inst{\ref{inst:roma},\ref{inst:asdc}}, A.~Moitinho\inst{\ref{inst:FCUL}}, 
R.~Mor\inst{\ref{inst:ieec}}, M.~Romero-G\'omez\inst{\ref{inst:ieec}},
P.~Sartoretti\inst{\ref{inst:gepi}}, S.~Soria\inst{\ref{inst:ieec}}, C.~Soubiran\inst{\ref{inst:Bordeaux}}, 
J.~Souchay\inst{\ref{inst:SYRTE}}, J. Veljanoski\inst{\ref{inst:kapteyn}}, H. Ziaeepour\inst{\ref{inst:utinam}},
\afterReferee{G.~Giuffrida}\inst{\ref{inst:asdc}}, \afterReferee{E.~Pancino}\inst{\ref{inst:0024}}, \afterReferee{A.~Bragaglia}\inst{\ref{inst:0043}}
}


\institute{
GEPI, Observatoire de Paris, PSL Research University, CNRS, Univ. Paris Diderot, 
Sorbonne Paris Cit{\'e}, 5 Place Jules Janssen, 92190 Meudon, France\\
\email{Frederic.Arenou@obspm.fr}
\label{inst:gepi}
\and
Institut de Ci\`encies del Cosmos,
Universitat de Barcelona (IEEC-UB), Mart\'i Franqu\`es 1, E-08028 Barcelona, 
Spain
\label{inst:ieec}
\and
Kapteyn Astronomical Institute, University of Groningen, Landleven 12, 9747 AD Groningen, The Netherlands
\label{inst:kapteyn}
\and
Institut UTINAM, CNRS, OSU THETA Franche-Comt\'e Bourgogne, Univ. Bourgogne Franche-Comt\'e, 25000 Besan\c{c}on,
France\label{inst:utinam}
\and
INAF, Osservatorio Astronomico di Padova, Vicolo Osservatorio, Padova, I-35131, Italy
\label{inst:padova}
\and
Observatoire de Gen\`eve, Universit\'e de Gen\`eve, CH-1290 Versoix, Switzerland
\label{inst:Geneva}
\and
Institute of Astronomy, University of Cambridge, Madingley Road, Cambridge CB30HA, United Kingdom
\label{inst:ioa}
\and 
SYRTE, Observatoire de Paris, PSL Research University, CNRS, Sorbonne Universit{\'e}s, UPMC Univ. Paris 06, LNE, 61 avenue de l'Observatoire, 75014 Paris, France
\label{inst:SYRTE}
\and
CENTRA, Universidade de Lisboa, FCUL, Campo Grande, Edif. C8, 1749-016 Lisboa, Portugal
\label{inst:FCUL}
\and
Leiden Observatory, Leiden University, Niels Bohrweg 2, 2333 CA Leiden, The Netherlands
\label{inst:leidenO}
\and
Laboratoire Lagrange, Univ. Nice Sophia-Antipolis, Observatoire de la C\^{o}te d'Azur,CNRS,CS 34229, 06304 Nice cedex, France
\label{inst:nice}
\and
INAF - Osservatorio Astronomico di Roma, Via di Frascati 33, 00078 Monte Porzio Catone (Roma), Italy
\label{inst:roma}
\and
ASI Science Data Center, Via del Politecnico, Roma
\label{inst:asdc}
\and
Laboratoire d'astrophysique de Bordeaux, Univ. de Bordeaux, CNRS, B18N, all{\'e}e Geoffroy Saint-Hilaire, 33615 Pessac, France
\label{inst:Bordeaux}
\and
INAF - Osservatorio Astrofisico di Arcetri, Largo Enrico Fermi 5, I-50125 Firenze, Italy                                                                                                         \label{inst:0024}
\and 
INAF - Osservatorio Astronomico di Bologna, via Ranzani 1, 40127 Bologna,  Italy                                                                                                                 \label{inst:0043}
}

   \date{ }

\abstract
{
Before the publication of the {\gaia} Catalogue, the contents of the first data release have 
undergone multiple dedicated validation tests. 
}
{
These tests aim at analysing in-depth the Catalogue content to detect anomalies, 
individual problems in specific objects or in overall statistical properties, 
either to filter them before the public release, or to describe the different caveats of the 
release for an optimal exploitation of the data.
}
{
Dedicated methods using either Gaia internal data, external catalogues or models have been 
developed for the validation processes. 
They are testing normal stars as well as various populations like open or globular 
clusters, double stars, variable stars, quasars. 
Properties of coverage, accuracy and precision of the data are provided by the numerous 
tests presented here and jointly analysed to assess the data release content.
}
{
This independent validation confirms the quality of the published data, {\GDR1} being the 
most precise all-sky astrometric and photometric catalogue to-date. However, several 
limitations in terms of completeness, astrometric and photometric quality are identified
and described. Figures describing the relevant properties of the release are shown
and the testing activities carried out validating the user interfaces are also described. 
A particular emphasis is made on the statistical use of the data in scientific exploitation.
}
{}

   
   \titlerunning{Gaia Data Release 1 -- Catalogue validation} 
   \authorrunning{F. Arenou et al.}

   \maketitle

%


\section{Introduction}

This paper describes the validation of the first data release 
from the European Space Agency mission {\gaia} \citep{DPACP-18}.
In a historical perspective, {\gaia}, following in the footsteps 
of the great astronomical catalogues since the first by Hipparchus of Nicaea, 
describes the state of the sky at the beginning of the $21^\mathrm{st}$ century.
It is the heir of the massive international astronomical projects, 
initiated in the late $19^\mathrm{th}$ century with the Carte du Ciel \citep{2000A&G....41e..16J}, 
and a direct successor of the ESA Hipparcos mission \citep{1997A&A...323L..49P}.

Despite the precautions taken during the acquisition of the satellite 
observations and when building the data processing system, it is a difficult
task to ensure perfect astrometric, photometric, spectroscopic and 
classification data for a one billion source catalogue built from the intricate 
combination of many data items for each entry. 
However, several actions have been undertaken to ensure the quality of the 
{\gaia} Catalogue through both internal and external data validation processes 
before each release.
The results from the external validation work are described in this paper.

\paragraph{The {\GDR1}:}
There is an exhaustive description of the {\gaia} operations and 
instruments in \citet{DPACP-18}, of the {\gaia} processing in \citet{DPACP-8} 
and the astrometric and photometric pre-processing is also detailed in \citet{DPACP-7}.
For this reason we mention here only what is strictly necessary and
invite the reader to refer to the above papers or to the {\gaia}
documentation for details.

The {\gaia} satellite is slowly spinning and measures the fluxes and observation times 
of all sources crossing the focal plane (their {\gaia} {\it transit}), sending to the 
ground small windows of pixels around the sources. These times correspond to 
one-dimensional, along-scan positions ({\it AL} in what follows) 
which are used in an astrometric global iterative solution process 
\citep[AGIS,][]{DPACP-14} which also needs to simultaneously calibrate the 
instruments and reconstruct the attitude of the satellite. A star crossing
the focal plane is measured on 9 \afterReferee{CCDs} in the astrometric instrument so the number
of observations of a star can be up to 9 times the number of its transits.
On-board resources are able to cope with various stellar densities; however, 
for very dense fields above 400\,000 sources per square degree, 
the brighter sources are preferentially selected.

The photometric instrument is composed of two prisms, a Blue Photometer (BP)
and a Red Photometer (RP). This colour information is not present in the
{\GDR1}, only the $G$-band photometry, derived from the fluxes measured in 
the astrometric instrument being given. The CCD dynamic range does
not allow to observe all sources from the brightest up to $G\sim 21$: sources 
brighter than $G\sim 12$ would be saturated. To avoid this, 
Time Delay Integration (TDI) gates are present
on the CCD and can be activated for bright sources, which in practice reduce 
their integration time (but also complicates their calibration).

Astrometry and photometry are then derived on-ground in independent pipelines, 
which are part of the work developed under the responsibility of the
body in charge of the data processing for the Gaia mission, the 
Gaia Data Processing and Analysis Consortium \citep[DPAC,][]{DPACP-8}.

This first data release 
contains preliminary results based on observations collected during the
first 14 months of mission since the start of nominal operations in July
2014.  At the start of nominal operations of the spacecraft on 25 July
2014, a special scanning law was followed, the Ecliptic Pole Scanning
Law (EPSL).  In EPSL mode, the spin axis of the spacecraft always lies
in the ecliptic plane, such that the field-of-view directions pass the
north and south ecliptic poles on each six-hour spin.  Then followed
the Nominal Scanning Law (NSL) with a precession rate of 5.8 revolutions
per year, starting on 22 August 2014.  As we will notice below, the EPSL
mode left some imprints on the Catalogue content and scientific results.

{\GDR1} contains a total of 1 142 679 769 sources, the astrometric part of {\GDR1} 
being built in two parts: the {\it primary} sources contains positions, parallaxes, 
and mean proper motions for 2\,057\,050 of the stars brighter than about
magnitude $V=11.5$ (about 80\% of these stars).
This data set, the Tycho Gaia Astrometric Solution (TGAS), was obtained through the combination 
of the {\gaia} observations with the positions of the sources obtained by Hipparcos
\citep{1997ESASP1200.....E} when available, or Tycho-2 \citep{2000A&A...355L..27H}.
The second part of {\GDR1}, the {\it secondary} sources, contains the positions and $G$ magnitudes
for 1\,140\,622\,719 sources brighter than about magnitude $G=21$. An annex of variable
stars located around the south ecliptic pole \beforeReferee{could also be published} \afterReferee{is also part of the release} thanks to the 
large number of observations made during the EPSL mode. 

\paragraph{The Catalogue Validation:}
In terms of scientific project, the quality of the released data has been controlled 
by two complementary approaches: the {\it verifications} done internally at each
step of the processing development in order to answer the question: are we building
the Catalogue correctly? and the {\it validations} at the end: is the
final Catalogue correct?

It is fundamental to note that the first step of the validations is logically 
represented by the many tests implemented in the {\gaia} DPAC 
groups before producing their own data, and which are described
in dedicated publications, \citet{DPACP-14} for the astrometry,
\citet{DPACP-11} for the photometry, and \citet{DPACP-15}
for the variability. 

To assess the Catalogue properties and as a final check before publication,
the DPAC deemed useful to implement a second and last step: a validation of the Catalogue
as a whole and actually, this must be stressed, a fully independent validation.

The actual Catalogue validation operations began after data 
from the DPAC groups had been collected and a consolidated Catalogue 
had been built before publication. At this step, no re-run of the data processing
\beforeReferee{being} \afterReferee{was} possible, only the rejection of some stars (if strictly needed)
and some cosmetic changes on the data fields could be done.
After the rejection of problematic stars, 
a process labelled as {\it filtering},
the validation was again performed, and most of the catalogue properties 
described in this paper refer to this post-filtering, published, final {\GDR1} data. 

\bigskip
The organisation of this paper is as follows: 
\secref{sec:description} summarises the data and models
used. Section \ref{sec:wp942:dupes} describes the erroneous or duplicate entries
found and partly removed. The main properties of the {\GDR1} Catalogue are discussed, 
\secref{sec:completeness}, for the sky coverage and completeness,
with a multidimensional analysis in \secref{multidim}, the
astrometric quality of {\GDR1} in \secref{astroqual} and the
photometric properties in \secref{photoqual}. As a conclusion, 
recommendations for data usage are given in \secref{sec:conclusions}.
The validation procedures employed in testing the design and interfaces 
of the archive systems are described in Appendix
together with some illustrations of the statistical properties of the Catalogue.

\section{Data and models}\label{sec:description}

\subsection{Data used}\label{data}

\subsubsection{Gaia data}\label{gaia-data}
Two months before the final go-ahead to publish the {\GDR1} Catalogue, we received the 
official preliminary Catalogue, called {\preDR1} in what follows, which was validated,
then subsequently filtered, as described in \secref{sec:wp942:dupes}, to produce the {\GDR1} Catalogue.
Generally speaking, the validation work has had access to the same fields as
published in {\GDR1} so that any user can reproduce the work indicated below. For 
example we did not have access to any individual transit data or calibration data,
or more generally to the main {\gaia} database, and this
fostered developing methods independent from the work done within the Gaia 
groups producing the data.
A few supplementary fields were however kindly made available for validation
purposes, such as the preliminary {\gbp} and {\grp} magnitudes (in order to study possible 
chromatic effects).

\subsubsection{Simulated Gaia data}\label{simulated-gaia-data}\label{agislab}

In the course of the preparation of the data validation, we also needed 
simulated data, mostly for testing the astrometry of the
TGAS solution. For this purpose we built a simulated catalogue,
called Simu-AGISLab in what follows, which contained astrometric data
for the Tycho-2 stars, on top of which were added simulated TGAS astrometric errors.
Simu-AGISLab used as simulated proper motions the Tycho-2 ones, but they were
``deconvolved'' using the formula indicated in \citet[Eq.~10]{1999ASPC..167...13A}
to avoid a spurious increase of their dispersion with the TGAS astrometric errors added
by the simulation. The simulated parallaxes were a weighted average of
``deconvolved'' Hipparcos parallaxes (for nearby stars) and the photometric parallaxes 
from the \cite{2011yCat.6135....0P} catalogue (for more distant stars). 
The simulated TGAS astrometric errors were produced as described in the 
Tycho-Gaia Astrometric Solution document \citep{2015A&A...574A.115M}, based
on solution algorithms described in \citet[][Sect. 7.2]{2012A&A...538A..78L}.

In addition, global simulations of the Gaia data generated by the 
DPAC group devoted to this purpose
were also used for validation tasks comparing models with data (see 
\secref{models-intro}).


\subsubsection{External data}\label{external-data}
The comparison of {\GDR1} to external catalogues is a tricky task as the Gaia Catalogue is unique in many ways: it combines the angular resolution of \beforeReferee{Hubble}\afterReferee{the Hubble Space Telescope} with a complete survey all over the sky in optical wavelength, down to a \gmag-magnitude $\simeq$ 21, unprecedented astrometric accuracy and all-sky homogeneous photometric data. 

However, the comparison with external catalogues is one way towards a deeper understanding of many of the parameters describing the performance of the Catalogue: overall sky coverage, spatial resolution, catalogue completeness and, of course, precision and accuracy of the different types of data for the various categories of objects observed by Gaia. Besides the Hipparcos and Tycho-2 catalogues, many other catalogues have been used, especially chosen for each of these tests. They are described in each of the relevant subsections.

The cross-match between TGAS and the external catalogues or compilations has been done using directly Tycho-2 or Hipparcos identifiers, either provided in the publications or obtained through SIMBAD queries \citep{2000A&AS..143....9W} using the identifiers given in the original papers. For the full {\GDR1} tests, a positional cross-match has been used.

\subsection{Data integrity and consistency}\label{sec:wp942:consistent}

{\GDR1} is the combined work of hundreds of people divided into dozens of groups working on several complementary yet independent pipelines. 
In addition to testing the data themselves, therefore, we tested the data \emph{representations} to ensure that all catalogue entries were valid and self-consistent. We checked that catalogue values were \emph{finite}, that data were \emph{present} (or missing) when expected, that all fields were in their expected \emph{ranges}, that observation counts agreed with each other, that source identifiers were \emph{unique}, that correlation coefficients formed a \emph{valid} correlation matrix, that fluxes and magnitudes were related as expected, that the positions obtained from the equatorial, ecliptic, and galactic coordinates agreed, and so on. We also confirmed that the {\GDR1} in different data formats indeed contained the same data.

All data integrity issues were fixed before the data release. 
%
%
For TGAS solutions we also checked individual values of proper motions and parallax
looking for e.g.\ \afterReferee{negative parallaxes or} unrealistic tangential velocities. We then 
checked the \emph{uncertainties} of the five astrometric parameters to make sure that
they decreased with the number of observations, or to see if there were Healpix pixels with an
unusually high fraction of large uncertainties. All in all we were particularly
interested in regions on the sky where dubious values occur with higher
frequency than in typical areas, with the aim of excluding if needed such regions from
the release. Although some poorly scanned regions were identified as
problematic, none were finally excluded.

Sources brighter than \afterReferee{about} 12~mag are observed with ``gates'', i.e.\ with
reduced exposure time. We therefore checked that the astrometric standard
uncertainties did not show rapid changes as a function of magnitude.

We found only a few minor issues in the {\GDR1} astrometry as for the data ranges. 
Large values of fields like \dt{astrometric\-\_excess\_noise}\footnote{Roughly speaking, this 
is the noise which should be added to the uncertainty of the observations to obtain a perfect
fit for the astrometric model. The
fields of the Gaia Catalogue are described at {\small\url{https://gaia.esac.esa.int/documentation/GDR1/datamodel/}}} 
and \dt{astrometric\-\_excess\-\_noise\_sig} that statistically were expected for only about a thousand sources are actually present in about 205 million sources, including nearly the entire TGAS sample. These large values reflect the large errors introduced by the preliminary attitude solution for the Gaia spacecraft; a better solution will be used in future releases \citep{DPACP-14} and we expect this problem will be solved. In addition, 4\,288 sources have positions based on only two one-dimensional measurements, providing an astrometric solution with no degrees of freedom. These minimally constrained solutions are expected to go away as more data are collected.

We tested whether sources had enough astrometric measurements 
to allow for a 2- or 5-parameter solution, as appropriate. We then compared the distribution of 
astrometric goodness-of-fit indicators with their expected distributions.

Photometry and astrometry were derived in independent pipelines each of which
could decide to reject or downweight a number of individual observations for a
given source.  We therefore checked if the number of valid observations was
similar in the two pipelines. If more than half of the observations were
rejected, and if the number of valid observations in each pipeline adds up to
less than the total number of observations for the source, there is a problem:
it is not possible to know if the astrometric and
photometric results refer to the same object or e.g.\ to different components
of a binary star. This problem affects less than 9\,000 sources in {\GDR1}
and we expect it to be also solved in future releases\footnote{These stars are not 
flagged, but can be found using \dt{phot\_g\_n\_obs}, \dt{astrometric\_n\_good\_obs\_al}, 
\dt{matched\_observations}}.

\subsection{Galaxy models}\label{models-intro}


Models contain a summary of our present knowledge about the stars in the Milky Way. This knowledge is obviously imperfect and one expects that many of the discrepancies between models and real Gaia data to be due to the models themselves. However, at the level of our current knowledge, if a model performs with a satisfactory accuracy compared to existing data, it can be used for Gaia validation (at the level of this accuracy). This is what we have done in the set of tests based on models. These tests \afterReferee{may} supersede the validation \beforeReferee{with existing}\afterReferee{using external} data in regions of the sky where data are too scarce, or in magnitude ranges where existing data are not accurate enough or incomplete, or in case they do not exist in large portions of sky (such as e.g. parallaxes).

On {\GDR1}, three kinds of tests have been performed: tests on stellar densities, tests on proper motions, and tests on parallaxes. In all tests we analysed the distribution on the sky of the model densities and of the statistical distribution of astrometric parameters (proper motions and parallaxes) and compared them with Gaia data. In order to establish a threshold for test results we compared the model with previous catalogues on portions of sky when available. For this first data release only the Besan\c{c}on Galactic Model~\citep{bgm} has been used for comparisons with Gaia data. 


\section{Erroneous or duplicate entries}\label{sec:wp942:dupes}

The {\preDR1} Catalogue received for validation was subject to several tests
concerning possible erroneous entries. This led to the filtering of a 
significant number of sources (37\,433\,092 sources were removed, 
3.2\% of the input sources).
As this filtering was obviously \beforeReferee{non} \afterReferee{not} perfect (removing actual sources 
while conserving erroneous ones), and had an impact on the Catalogue
content, the rationale, methods used and results are described in this section.

\subsection{Erroneous faint TGAS sources}

\subsubsection{Data before filtering}\label{sec:wp942:photofilter}

As can be seen in \figref{fig:942_tgashist}a, there was a significant number 
of objects (2\,381 sources) in the {\preDR1} version of TGAS 
that had $G\gtrsim 14$~mag,
i.e. clearly fainter than what was expected for Tycho-2. This led to the study
of the $G$ photometry for these stars and, beyond, for the whole catalogue.

A particular concern has been to catch coarse processing errors in the photometry.
For bright sources, the exposure time in each CCD on-board {\gaia} is reduced by activating
special TDI gates on the device as the star image crosses the CCD. This smaller exposure
time is then taken into account when computing the flux. However, in some rare occasions 
the information on gate activation did not
reach the photometric pipeline. The result was artificially low fluxes in that
particular transit, and for reasons beyond the scope of this paper, this
could upset the processing and lead to erroneous $G$ magnitudes. 

We therefore specifically checked if sources appeared much fainter 
in $G$ than in both {\gbp} and {\grp}, the preliminary 
versions of photometry to be published in later releases \citep{DPACP-10}. 
In practice the limit was set
at 3~mag in order not to eliminate diffuse objects with a bright core,
e.g. galaxies, which were expected to be bright in the diaphragm photometry
of {\gbp} and {\grp}; stars with $G-G_{\rm BP} > 3$ and $G-G_{\rm RP} > 3$, 
thus where a problem with $G$ was suspected, were filtered (164\,446 TGAS or 
secondary sources).

While the median number of $G$-band observations per source is 72 in {\GDR1},
it was also found that roughly half of the too faint TGAS sources had fewer 
than 10 CCD observations, and indeed, on the whole catalogue
stars with less than 10 observations clearly behaved incorrectly.
This led to the removal of all sources with less than 10 $G$ observations from {\preDR1} 
(746\,292 TGAS or secondary sources).

\subsubsection{Data after filtering}

\begin{figure}
\centering
\includegraphics[width=0.49\columnwidth]{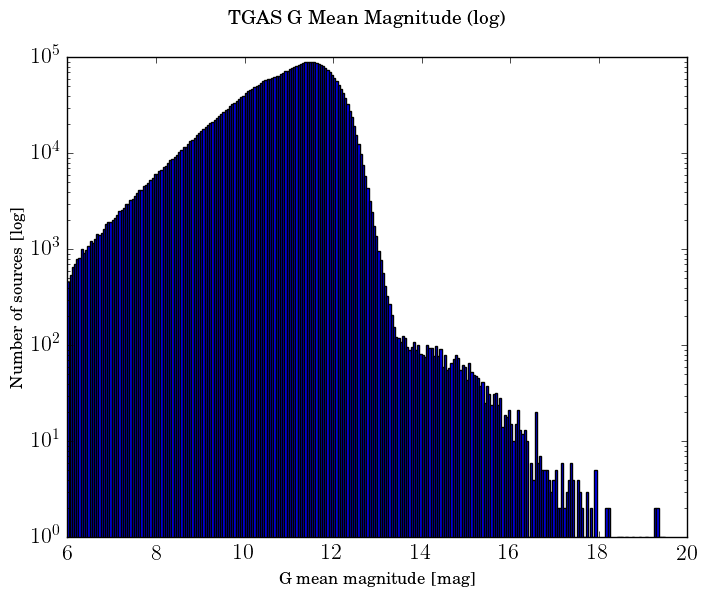}
\includegraphics[width=0.49\columnwidth]{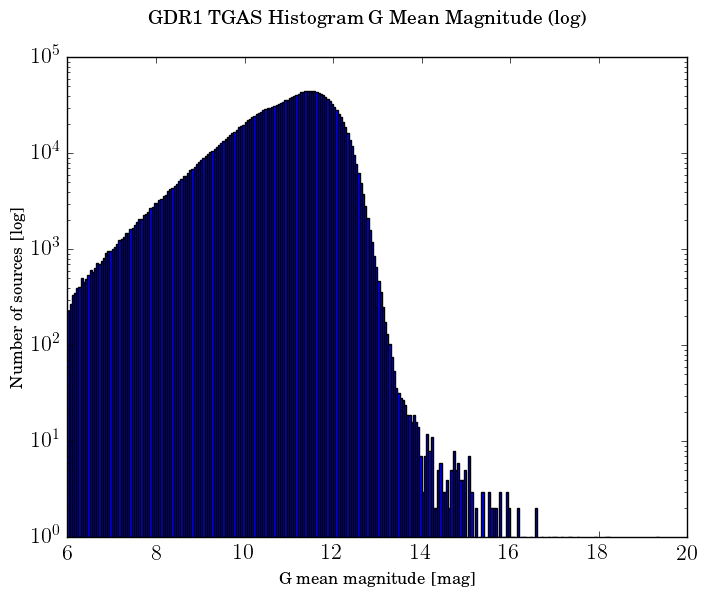}
\caption{Histogram of $G$ magnitudes for TGAS stars (a) before
and (b) after validation filtering.}
\label{fig:942_tgashist}
\end{figure}

Figure~\ref{fig:942_tgashist}b shows the resulting magnitude distribution for TGAS
in {\GDR1}, i.e. after full filtering. There is a remaining tail with 352 sources fainter 
than $G = 13.5$\,mag, and the presence 
of such sources in TGAS calls for an explanation. We have taken a closer look 
at the 60 faintest TGAS stars of which the brightest
has $G = 14.98$\,mag. Of these 60 stars, 25 have a neighbour brighter than $G =
13.5$\,mag and closer than 5\arcsec\ in {\GDR1} suggesting that the wrong star
may have been used in the TGAS solution, which is therefore not valid.  Of the
remaining 35 stars, just over half (18) have from one to four neighbours within
5\arcsec. In these cases we may be dealing with spurious Tycho-2 stars.
Tycho-2 \citep{2000A&A...357..367H} was using an input star list dominated by
photographic catalogues, and a blend of sources may therefore have been seen as
a single bright source. It may then happen that a Tycho-2 solution was derived
from the mixed signal of contaminating sources. We see that as a likely
explanation for most of these cases. For stars that are isolated in {\GDR1},
spurious Tycho-2 stars cannot be excluded, but in at least one case, the faint
Gaia source turns out to be a variable of the R\,CrB type. This star
(HIP\,92207) has $G = 16.57$\,mag in Gaia DR1, but is as bright as $V_T =
10.29$\,mag in Tycho-2. This is in good agreement with available light curves.
It is too early to say if there are more high amplitude variables in the
sample.

\subsection{Duplicate entries}

\subsubsection{{\GDR1} before filtering}\label{sec:duplicates}

Before launch, a catalogue with known optical astrometric and photometric information 
of sources up to magnitude $G=21$ had been built in order to be used as
Initial {\gaia} Source List \citep[IGSL,][]{2014A&A...570A..87S}.

Stars from IGSL may have initially contained duplicates originating from e.g. overlapping plates.
Automatically generated catalogues such as {\GDR1} may also have multiple copies of a source for a variety of reasons, including poor cross-matching of multiple observations, inconsistent handling of close doubles, or other observational or processing problems, beside the duplicates originating from the IGSL. 
To test for duplicate sources we cross-matched the Gaia catalogue against itself, identifying pairs of sources that could not possibly be real doubles, either because they fell within one pixel (59~mas) of each other or because their positions were consistent to within $5\sigma$. Only reference epoch positions were used, with no corrections for high proper motion stars.

It was found that the {\preDR1} Gaia catalogue contained 71~million sources 
with a counterpart within one pixel or $5\sigma$. 
Most appeared in pairs, but some were clustered in groups of up to eight duplicates. 
Up to one third of sources around $G \sim 11 \textrm{ mag}$ were affected, far more 
than at much brighter or much fainter magnitudes. 

For {\GDR1}, we removed all but one source from each group of close matches, selecting the source with the more precise parallax (if present) and breaking ties by the source with \beforeReferee{the} more observations, followed by the better position or photometric error. Because duplicated sources may have compromised astrometry or photometry (e.g., if a source was duplicated because of a cross-matching problem), the surviving sources were marked with the \dt{duplicated\_source} flag in the final catalogue (35\,951\,041 TGAS or secondary sources).

\begin{figure}
\begin{center}
\includegraphics[width=0.8\columnwidth]{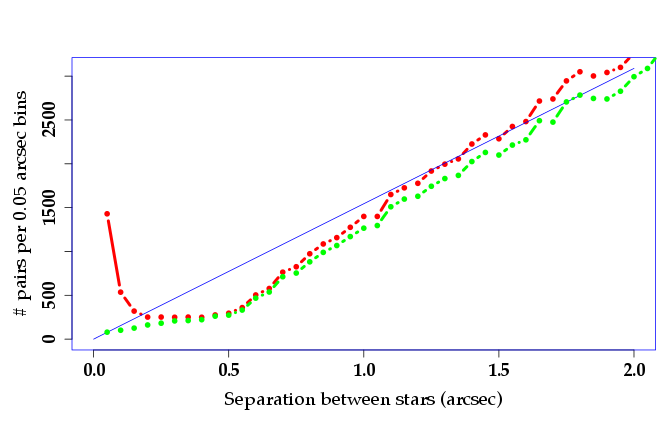}
\caption{\small Number of pairs of sources vs their angular separation in the field 
($l=350$\deg, $b=0$\deg) before (red) and after filtering (green).
The line corresponds to a random distribution up to 10\arcsec\ of the latter.}\label{fig:cu9val_942_int_closedoubles}
\end{center}\end{figure}

\begin{figure}
\begin{center}
\includegraphics[width=0.32\columnwidth]{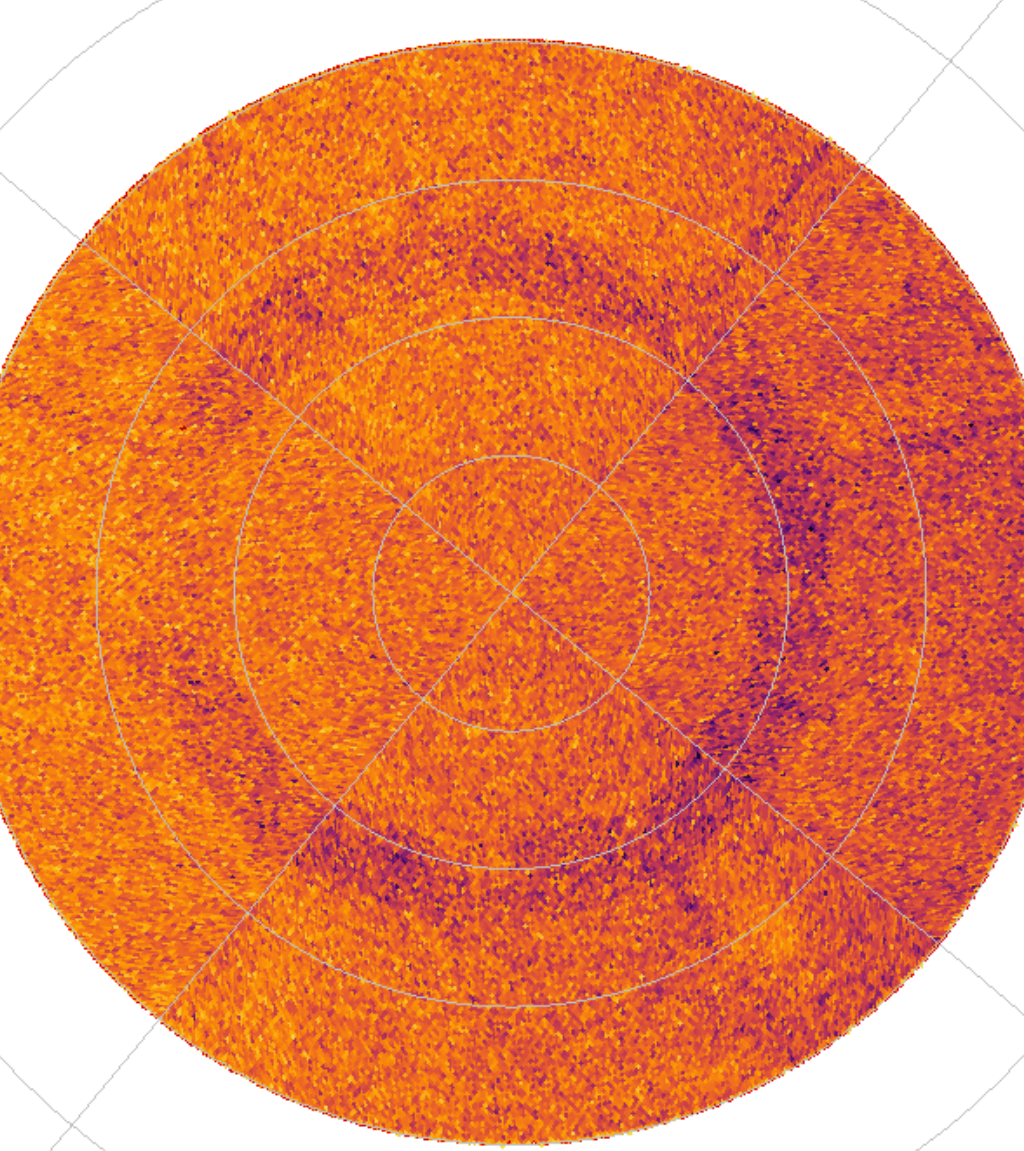}
\includegraphics[width=0.32\columnwidth]{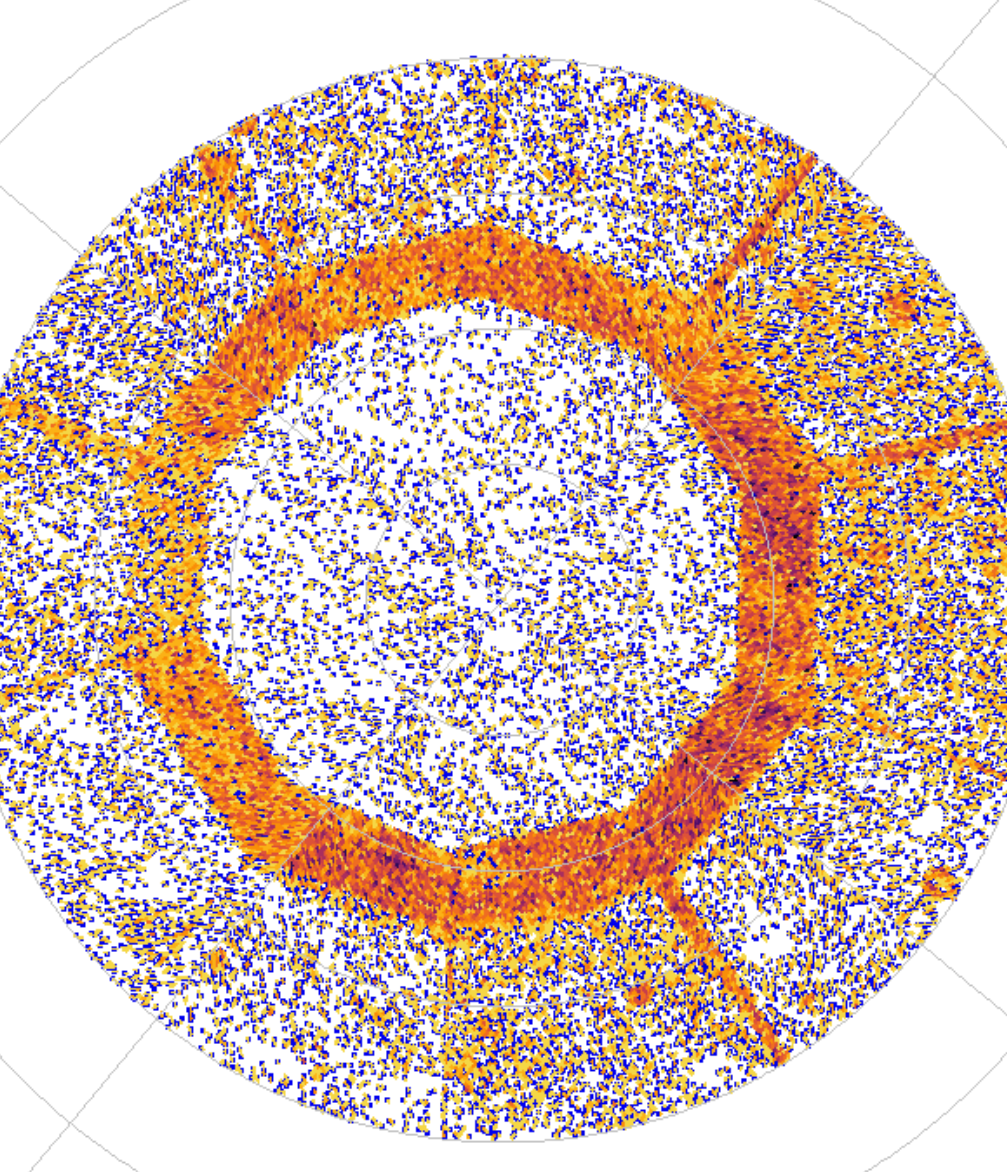}
\includegraphics[width=0.32\columnwidth]{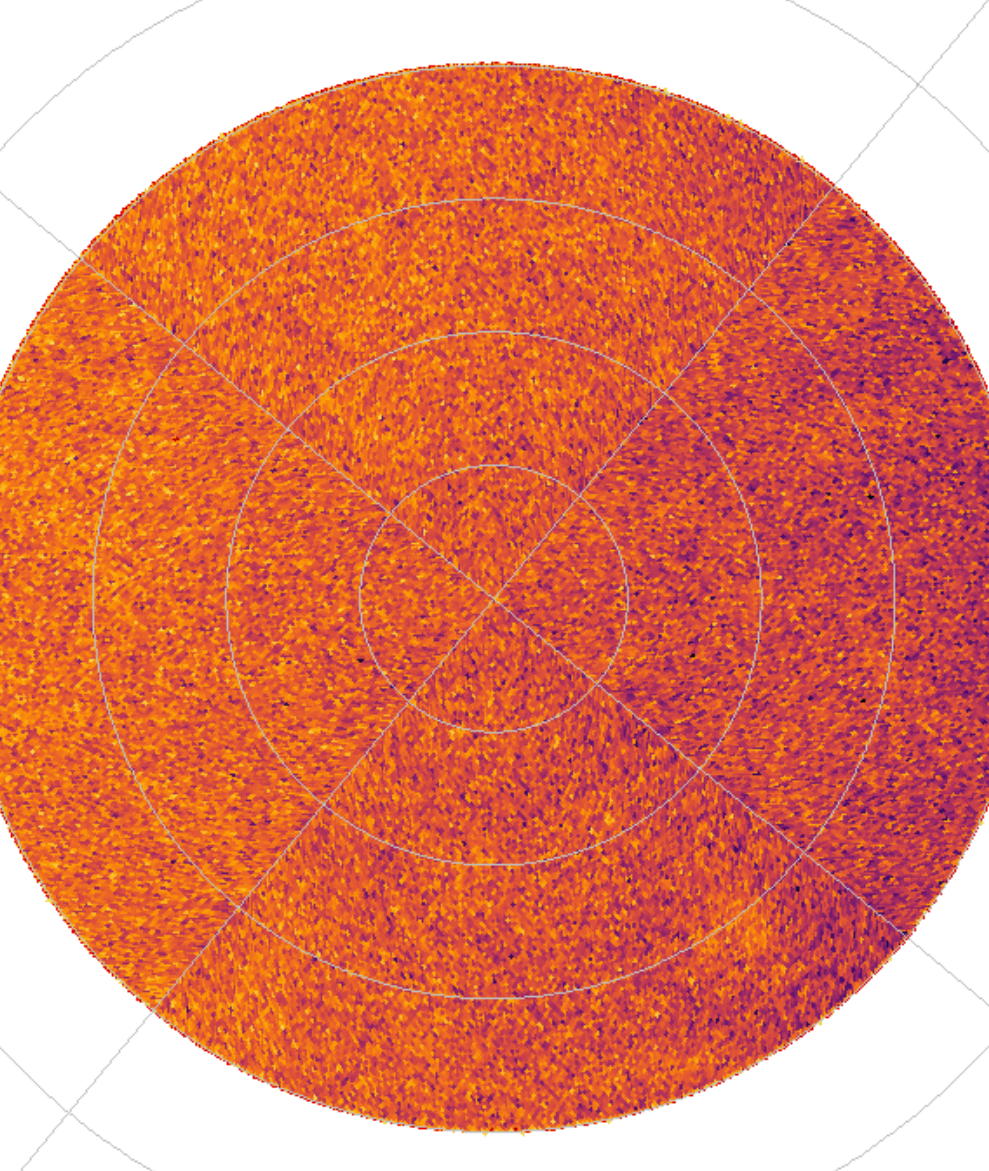}
\caption{\small Effect of duplicate stars in a field of radius 4\deg\ around the South pole: (a) original density 
map in {\preDR1} before validation filtering, (b) duplicates found, (c) after 
duplicates filtering.}\label{fig:cu9val_942_int_southPoleDuplicates}
\end{center}\end{figure}

Two examples of the effect of the filtering of duplicate sources are shown in Figs. \ref{fig:cu9val_942_int_closedoubles} and  \ref{fig:cu9val_942_int_southPoleDuplicates}. The result of the filtering as done for {\GDR1}
is illustrated in Figs. \ref{fig:cu9val_942_int_closedoubles} and \ref{fig:cu9val_942_int_southPoleDuplicates}c. The artefacts in 
Figs. \ref{fig:cu9val_942_int_southPoleDuplicates}a and \ref{fig:cu9val_942_int_southPoleDuplicates}b are
the traces of the overlaps of photographic plates used in some of the surveys from which
the IGSL catalogue was built, causing an excess of duplicate sources in {\GDR1}.

\subsubsection{{\GDR1} after filtering}

Although it is estimated that about 99\% of the duplicates have been removed, 
spurious sources may still remain in {\GDR1}. Formal uncertainties on positions 
of these duplicates may have been underestimated, and the $5\sigma$ criterion 
on positional difference used for rejection may finally not have been large enough. 
This underestimation was suspected the following way: a pair made of one duplicate source and the source
it duplicates actually refers to one single source which dispatched part of its observations
between both (depending on the orientation of the satellite scans). We used this property to
compare the positions and magnitudes in pairs and found that uncertainties were underestimated
by a factor 2 for positions and 4 for magnitudes. While this result cannot be extrapolated
to all normal (not duplicated) stars, this gives at least an upper limit and justifies in any
case the presence of the \dt{duplicated\_source} flag.

A comparison with the Washington Visual Double Star Catalogue \citep[WDS,][]{WDS}
confirms that some duplicates remain, as can be seen with the excess of stars with a near 
zero separation in the bottom left of \figref{fig:wp944_WDS}b.

\beforeReferee{Remaining duplicates may be also present in particular in high density fields; here, however, computing
the probability to get several stars close to each other just by chance (mostly optical doubles)
shows that actual stars may have been removed too so}
\afterReferee{In high density fields, there is a chance to get several stars 
very close to each other by chance only, i.e. optical doubles. 
Trying to remove more duplicates would lead to removing actual stars by mistake.}
The adopted filtering
may \beforeReferee{finally}\afterReferee{actually} have been a reasonable compromise, until the expected improvement
in Gaia DR2.

\section{Sky coverage and completeness of DR1\label{sec:completeness}}

The {\GDR1} release is expected to be incomplete in various ways, full detail of these limitations 
being described in \cite{DPACP-14,DPACP-8}: 
\begin{itemize}
\item {\GDR1} is based on 14 months of data only. As a result, some regions, especially at low ecliptic latitudes, have been poorly observed, both in terms of the number of observations and of the coverage in scanning directions, see for example Fig.~2 of \cite{DPACP-8}. Stars with less than 5 focal plane transits have been filtered out; 
\item stars with a low quality astrometry solution for whatever reason have been filtered out; 
\item bright stars or high proper motions stars may be missing; 
\item faint stars are missing in very dense areas (for stellar densities higher than $\sim$ 400\,000 stars per square degree at $G<20$); 
\item stars with extremely blue or red colours have been filtered out during the photometric calibration. 
\end{itemize}

The tests presented in this section aim at a better characterisation of the object content of DR1, including TGAS,
as for the homogeneity of the sky distribution and the small scale completeness of the Catalogue.
These tests have been performed from different points of view, for various populations 
and using various inputs and methods: using the characteristics of Gaia data only (internal tests), using external data (all sky external catalogues, detailed catalogues of specific samples of stars or of specific regions of the sky), or using Galaxy models.

\subsection{Limiting magnitude\label{sec:faintlimit}}

The completeness of {\GDR1} is the result of a complex interplay between high stellar densities implying a possible overlap of the images on the focal plane, scanning law defining the number of times a region was observed, and data processing. Due to limited telemetry resources, the star images sent to ground followed a decision algorithm which is a complex function of the magnitude. In addition, at the end of the data processing a filtering was applied to discard poor solutions both in the astrometry and in the photometry. As a result, the density distribution over the sky in the final Catalogue is not a simple function of the stellar density, as usually expected. 

\begin{figure}
\centering
\includegraphics[width=0.7\columnwidth]{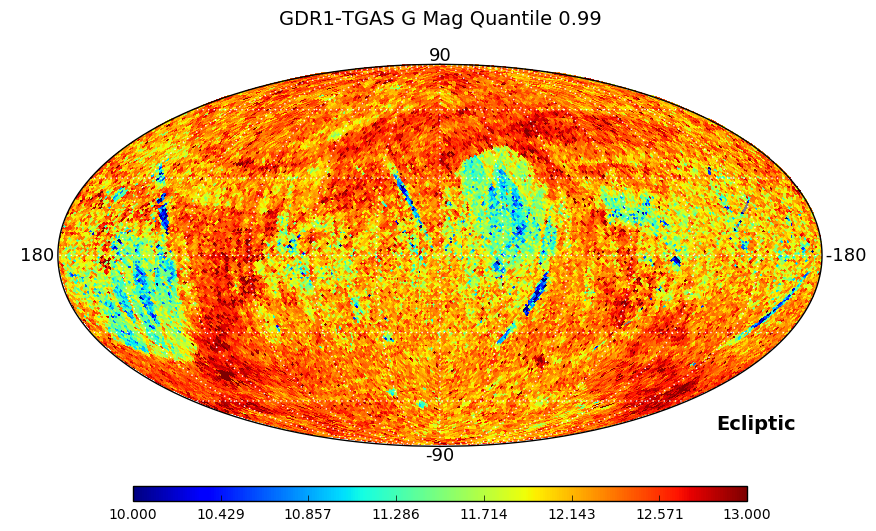}
\includegraphics[width=0.7\columnwidth]{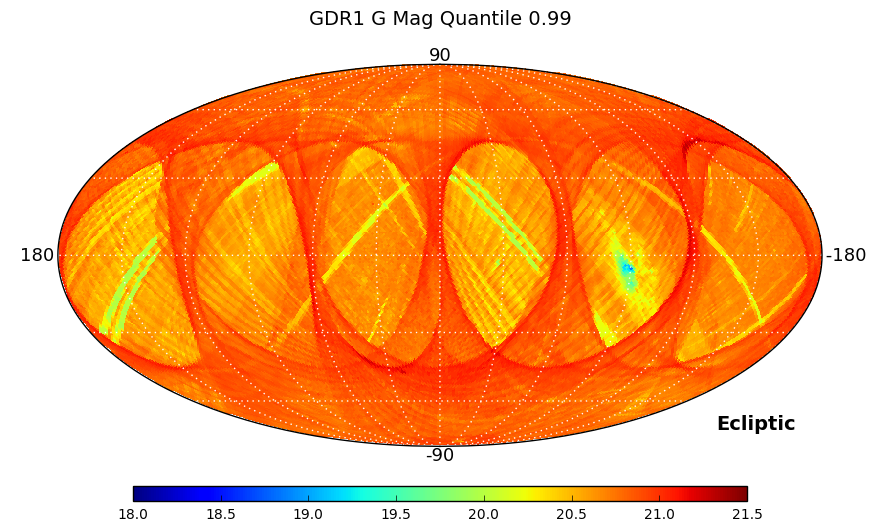}
\caption{Limiting magnitude: 99\% percentile of the $G$ distribution in ecliptic coordinates: 
a) TGAS, b) full Catalogue.}
\label{fig:942_limitmag}
\end{figure}

A first, indirect information about the completeness is brought by the limiting magnitude 
of the Catalogue. Sky variations of the 0.99 quantile of the $G$ magnitude are
shown in \figref{fig:942_limitmag} for TGAS and the whole Catalogue. Concerning the latter,
it appears that Gaia will easily reach at the end of mission \beforeReferee{the} $G>21$  
in a significant fraction of the sky, even if this is still very limited for {\GDR1}; 
it seems however that one magnitude has been 
lost in the under-scanned regions, and two magnitudes in the Baade window. 
The limiting magnitude of TGAS stars \afterReferee{also has an amplitude of two magnitudes} over the sky,
with the brightest regions being also those with some astrometric deficiencies,
as shown below.

\subsection{Overall large scale coverage and completeness} \label{sec:sky-coverage}

\subsubsection{Overall sky coverage and completeness of TGAS}

\begin{figure}
    \begin{center}
        \includegraphics[width=0.8\columnwidth]{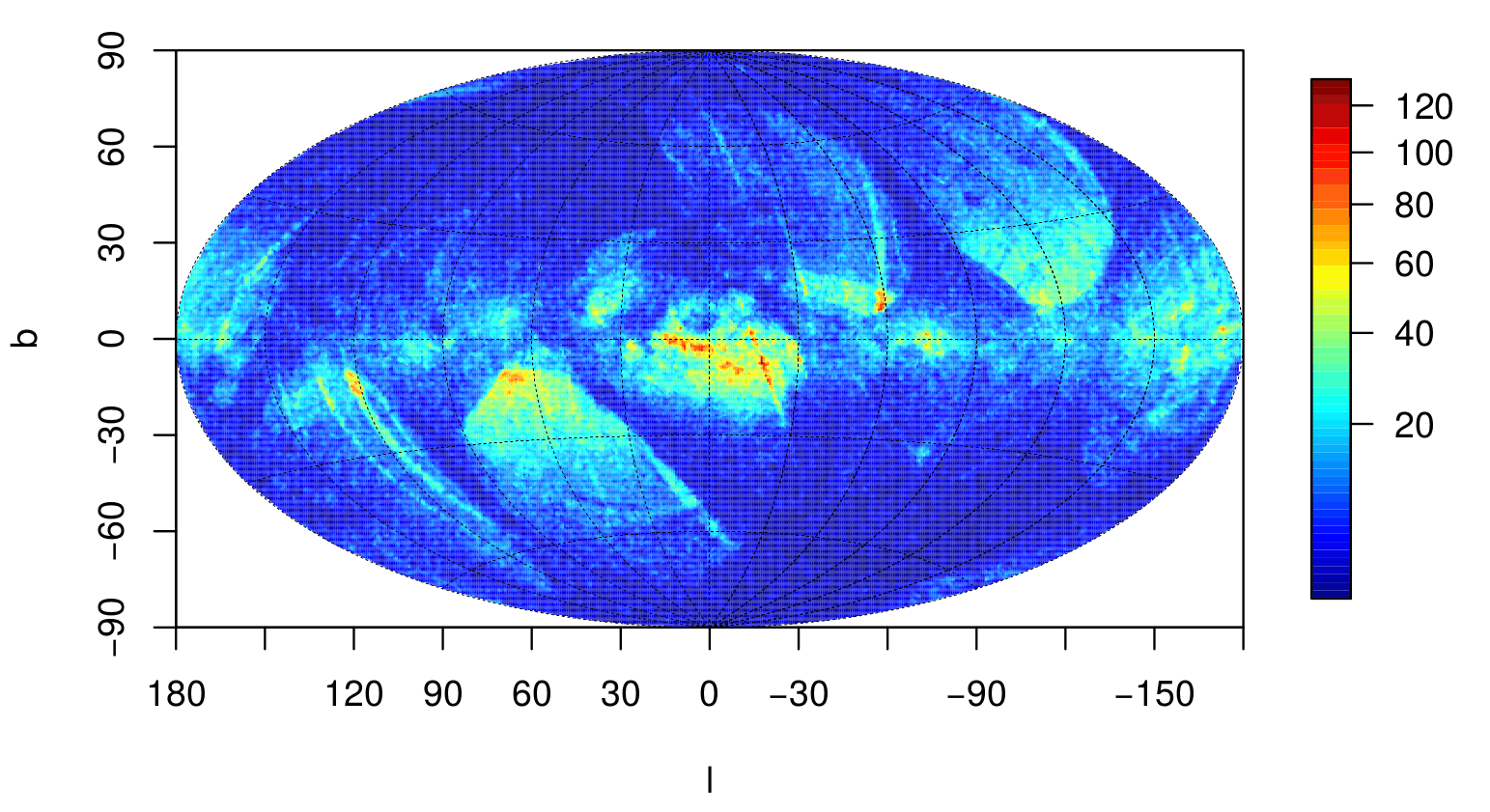}
        \caption[Tycho-2 stars not in TGAS]{Sky distribution of Tycho-2 stars not in TGAS, in galactic coordinates.}
        \label{fig:wp944_tgas_tycho2} 
    \end{center}
\end{figure}

The overall TGAS content has been tested with respect to the Tycho-2 \citep{2000A&A...355L..27H} and Hipparcos Catalogues \citep{1997A&A...323L..49P,1997ESASP1200.....E} 
for detection of possible duplicate entries and characterisation of missing entries. 
TGAS contains 79\% of the Hipparcos and  80\% of the Tycho-2 stars. 
One of the reasons for the missing stars is a bad astrometric 
solution, as all sources with a parallax uncertainty above 1 mas were not kept 
in TGAS (validation tests done on preliminary data had indeed shown several problems
associated to these stars).
The sky distribution of the Tycho-2 sources not present in TGAS is presented \figref{fig:wp944_tgas_tycho2}, showing the impact of the Gaia scanning law (the number of observations and the orientation of the scans being correlated with the solution reliability criteria filters applied for {\GDR1}). 

The detail of the histogram of \figref{fig:942_tgashist} shows that stars fainter than
10.5~mag have suffered a higher loss than average, a likely reason is the occasional 
source duplication described in \secref{sec:wp942:dupes}, 
which affects these magnitudes more.
The loss is clearer for stars brighter than 6\,mag,
partly due to an insufficient number of bright calibration sources for the
broad band photometers, so no colour was available. 
The $G$ magnitude calibration
includes a colour term \citep{DPACP-9}, so a missing colour means that
no $G$-band photometry was produced, and the source did not enter the release.
Stars brighter than about 5, and a fraction of sources fainter than this, 
were also among the sources not kept in TGAS due to the bad quality of their 
astrometric solution.

TGAS completeness has also been tested with respect to high proper motion stars: a selection of 1\,098 high proper motion (HPM) stars has been made with SIMBAD on stars with a Tycho or HIP identifier and a proper motion larger than 0.5\,arcsec\,yr$^{-1}$ (proper motions mainly from Tycho-2 and Hipparcos). 
40\% of this selection is not found in the TGAS solution, in particular bright stars. All stars with a proper motion larger than 3.5 arcsec\,yr$^{-1}$ are absent from TGAS. 
Stars with a proper motion larger than 1~arcsec\,yr$^{-1}$ in TGAS have been confirmed to have a large proper motion in SIMBAD. 

\subsubsection{Overall sky coverage of {\GDR1} from external data.}
The overall sky coverage of {\GDR1} has been tested by comparison with two deeper all sky catalogues:  2MASS \citep{2006AJ....131.1163S} and UCAC4 \citep{2013AJ....145...44Z}. 
The tests performed here use the crossmatch between {\GDR1} and these two catalogues provided to the users in the Gaia Archive \citep{DPACP-17}. The variation over the sky of four key parameters are checked: the number of cross-matched sources, the mean number of neighbours (stars which could have been considered as cross-matched, but for which the cross-match was not as good as for the selected source = the {\it best neighbour}), the number of Gaia stars with the same {\it best neighbour}, and the number of Gaia sources without any match. Finally, a random subset of about 5 million sources has been selected in order to check, if any, the different properties in magnitude, colour, proper motion, goodness of fit, etc... of the above four categories of stars.

\begin{figure*}
    \begin{center}
        \includegraphics[width=0.33\textwidth]{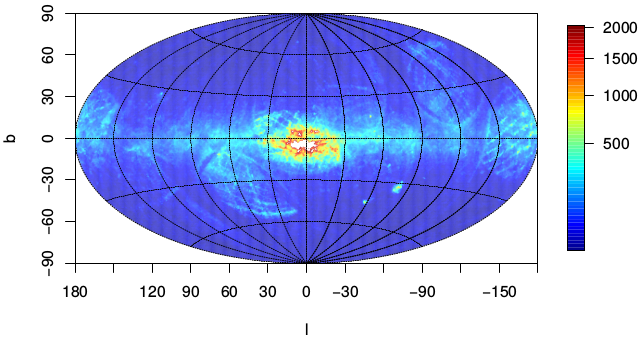}
        \includegraphics[width=0.33\textwidth]{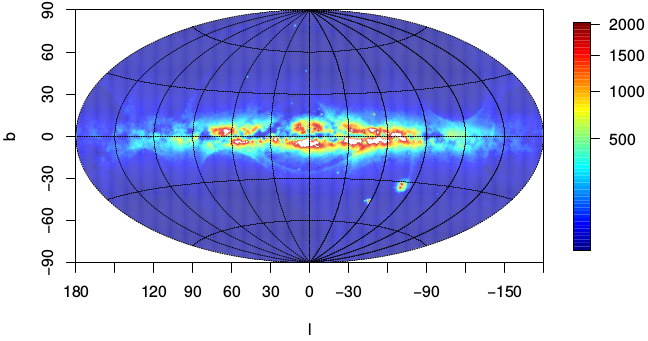}
        \includegraphics[width=0.33\textwidth]{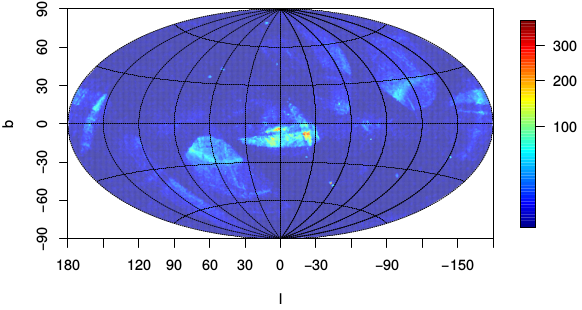}
        \caption[Sky distribution versus UCAC4]{Sky distribution versus UCAC4, in galactic coordinates; a) UCAC4 sources not in Gaia {\GDR1} (5\%); b) UCAC4 sources with multiple matches in \GDR1; c) {\GDR1} sources with \gmag$<14$ not in UCAC4. }
        \label{fig:wp944_ucac4} 
    \end{center}
\end{figure*}

\paragraph{UCAC4.} Only 5\% of the UCAC4 catalogue does not have a match in Gaia DR1. Their sky distribution (\figref{fig:wp944_ucac4}a) shows the footprint of the Gaia scanning law. 
7\% of the UCAC4 sources appear more than once in the cross-match table. 
We will refer to them as multiple-matches, it does not mean that this refer to (or only to) duplicate Gaia entries as discussed in \secref{sec:duplicates}: the Gaia resolution is much better than ground-based instruments so that multiple objects may appear where ground-based catalogues see one object only; those multiple-matches are distributed mainly in high density region, as expected, but their sky distribution also shows the Gaia scanning law footprint (\figref{fig:wp944_ucac4}b). 
258\,605 sources with \gmag$<14$ appear in the Gaia catalogue but not in UCAC4 which is supposed to be complete to about magnitude $R=16$; their sky distribution (\figref{fig:wp944_ucac4}c) follows the Gaia scanning law footprint and recalls the footprint of the Tycho-2 stars not in TGAS (\figref{fig:wp944_tgas_tycho2}). A detailed inspection of those sources indicates that a large portion of them are actually present in the UCAC4 catalogue but that the cross-match could not be done, the positional differences being beyond the astrometric uncertainties. This may be linked to the fact that a large portion of those sources have been measured along uneven scan orientations. 

\begin{figure}
    \begin{center}
        \includegraphics[width=0.33\textwidth]{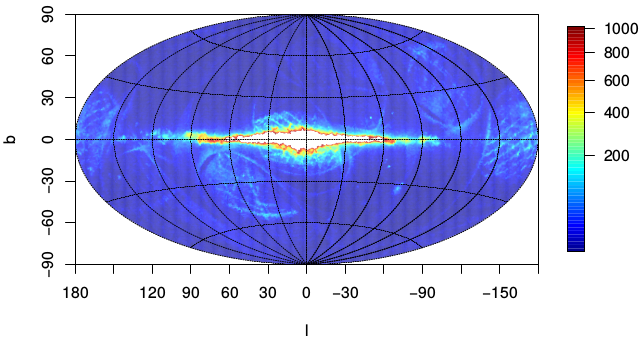}
        \includegraphics[width=0.33\textwidth]{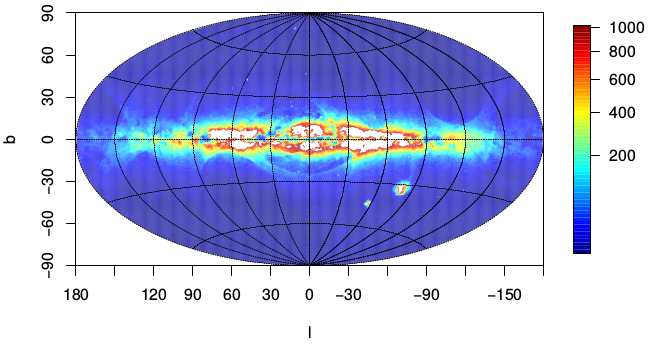}\\
        \caption[Sky distribution versus 2MASS]{Sky distribution versus 2MASS, in galactic coordinates. a) 2MASS sources with J$<$14 not in {\GDR1}; b) 2MASS multiple-matches in {\GDR1}.}
        \label{fig:wp944_2mass} 
    \end{center}
\end{figure}

\paragraph{2MASS.} For this test, we selected 2MASS stars with photometric quality flag AAA and magnitude $J<14$ (this limit corresponds roughly to $V<20$ for \av$<5$). As expected, most of the missing sources are located in high extinction regions along the galactic plane, but some extra features are also apparent showing the Gaia scanning law footprint (\figref{fig:wp944_2mass}a).
The 2MASS multiple-matches have a sky pattern (\figref{fig:wp944_2mass}b) similar to the one observed with UCAC4, with the main concentration being as expected along the dense areas added to a smaller Gaia scanning law footprint. 

\paragraph{Quasars.}
Quasars are essential objects for various reasons and several tests verify that they have been correctly observed by Gaia and identified. \beforeReferee{This}\afterReferee{The} first test compares {\GDR1} quasars with ground-based quasar compilations: GIQC \citep{GIQC}, LQAC3 \citep{2015A&A...583A..75S} and SDSS DR10 \citep{2014A&A...563A..54P} catalogues. It is a check for completeness, duplication and magnitude consistency. 
While the quasars were also affected by the duplicated sources issue (\secref{sec:duplicates}), the filtering seems to have removed them nicely. 
81\% of GIQC, 53\% of LQAC3 and 11\% of SDSS quasars are present in \GDR1, a ratio that reaches 93\% for the LQAC3 sources with a magnitude $B$ brighter than 20. 

\paragraph{Galaxies.} For galaxies, the cross-match has been done with SDSS DR12 \citep{2015ApJS..219...12A} sources with a galaxy spectral classification. The properties of cross-matched galaxies are compared to those of missing galaxies (magnitudes, redshift, axis-ratios and radii). Unfortunately, only $\sim$0.2\% of the SDSS galaxies are present in {\GDR1} due to the different filters applied. Still some large resolved galaxies can have multiple detections associated to them, tracing their shape. 

\subsubsection{Completeness from comparison with a Galaxy model}

\begin{figure*}
\begin{center}
\includegraphics[width=0.24\textwidth]{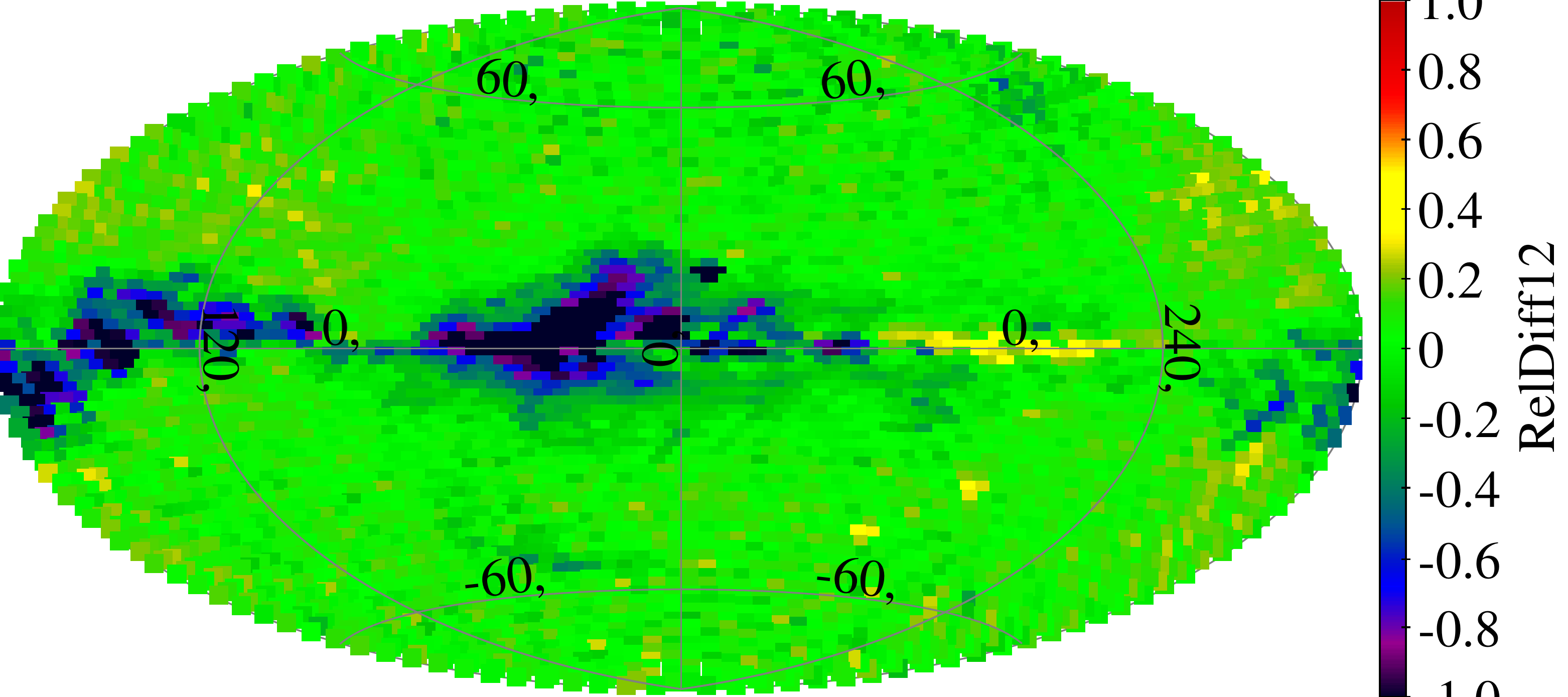}
\includegraphics[width=0.24\textwidth]{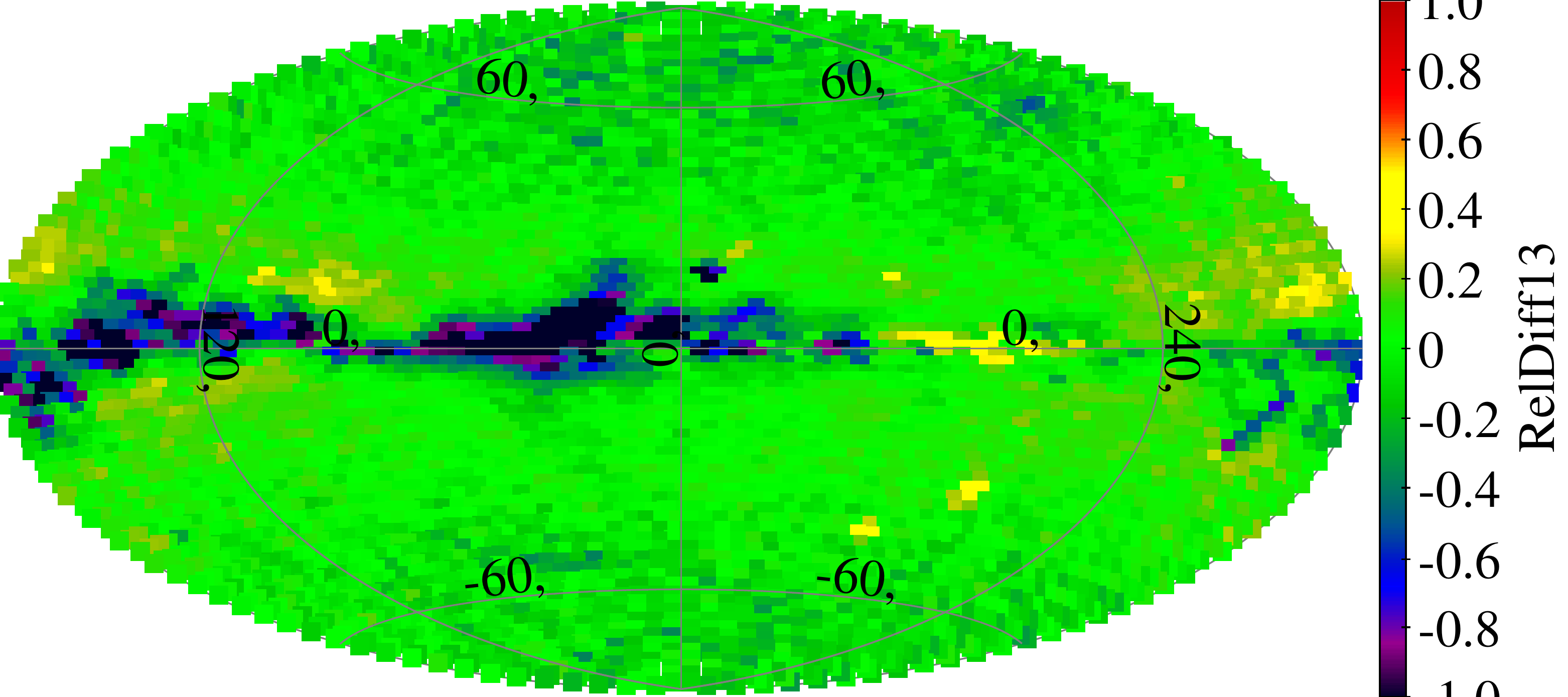}
\includegraphics[width=0.24\textwidth]{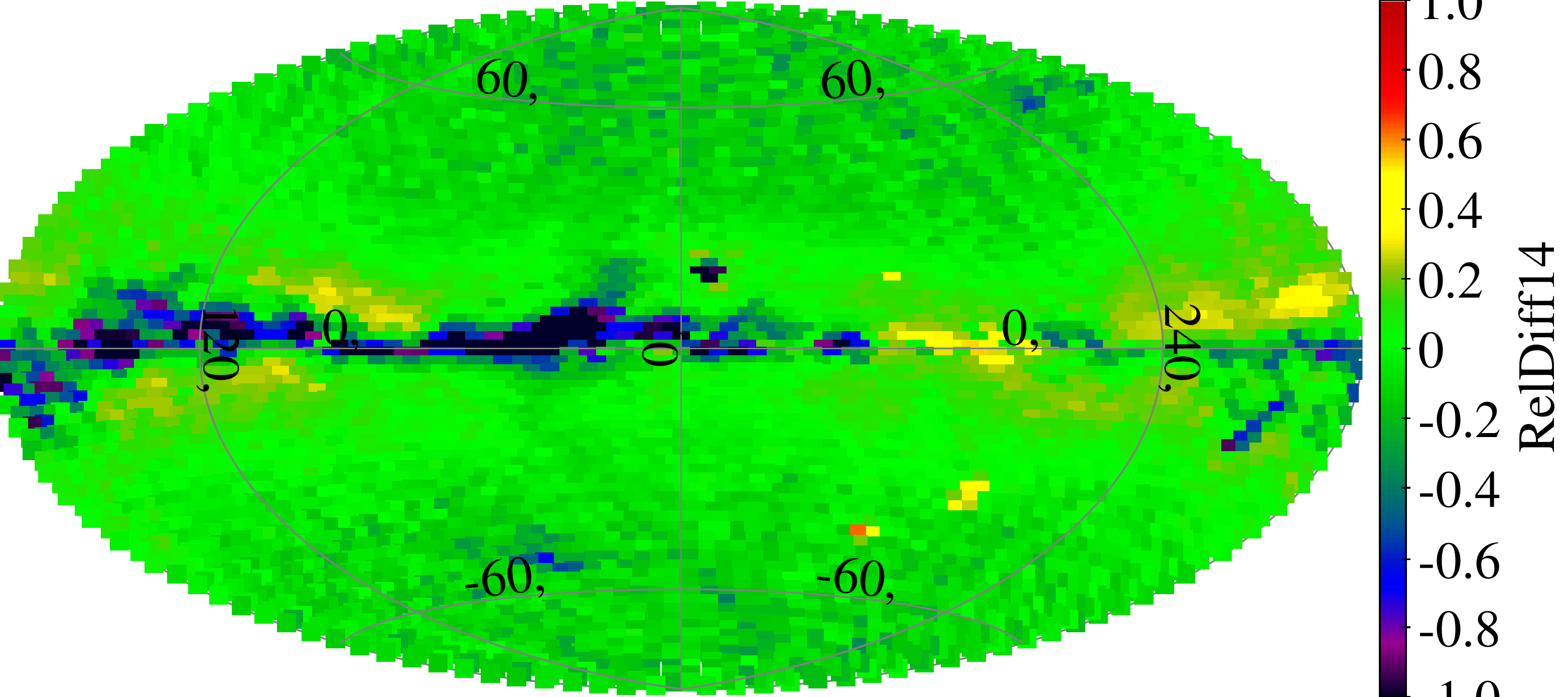}
\includegraphics[width=0.24\textwidth]{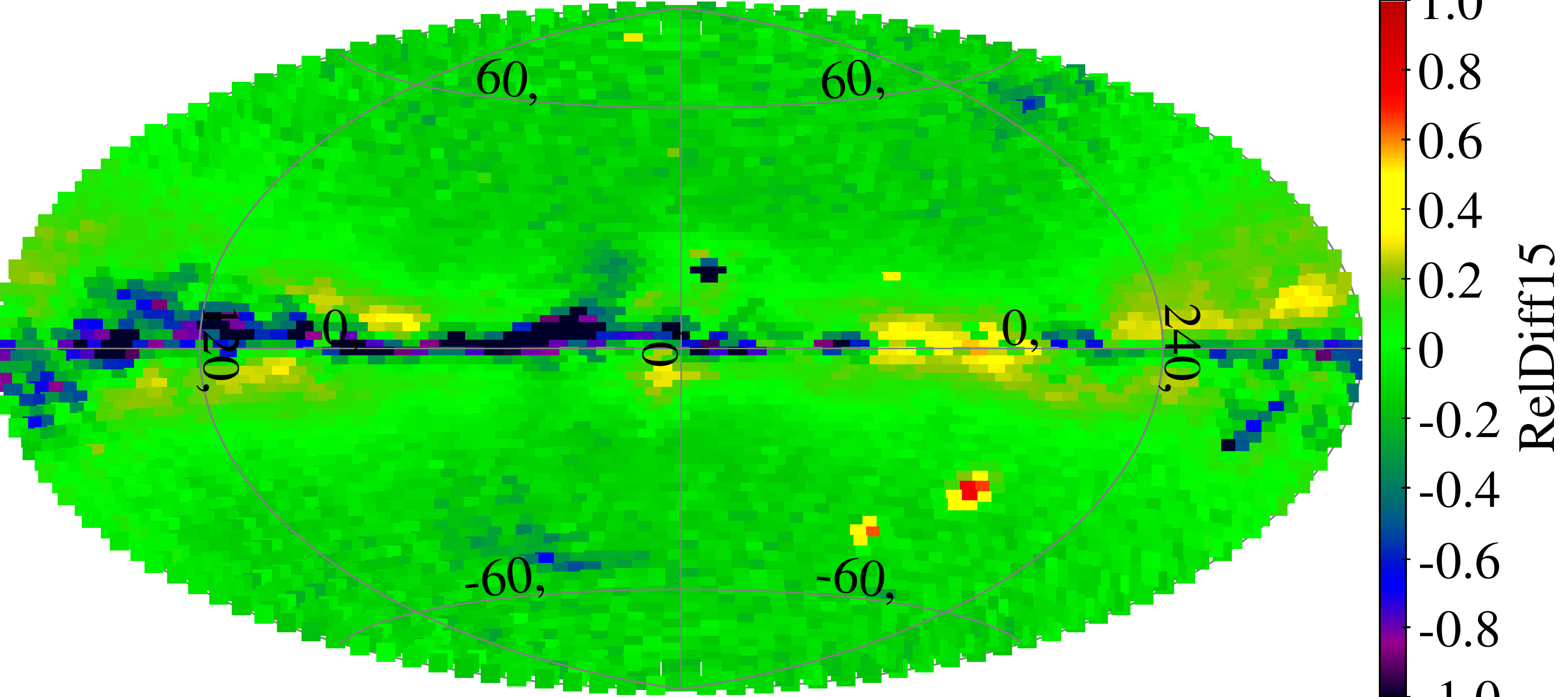}
\includegraphics[width=0.24\textwidth]{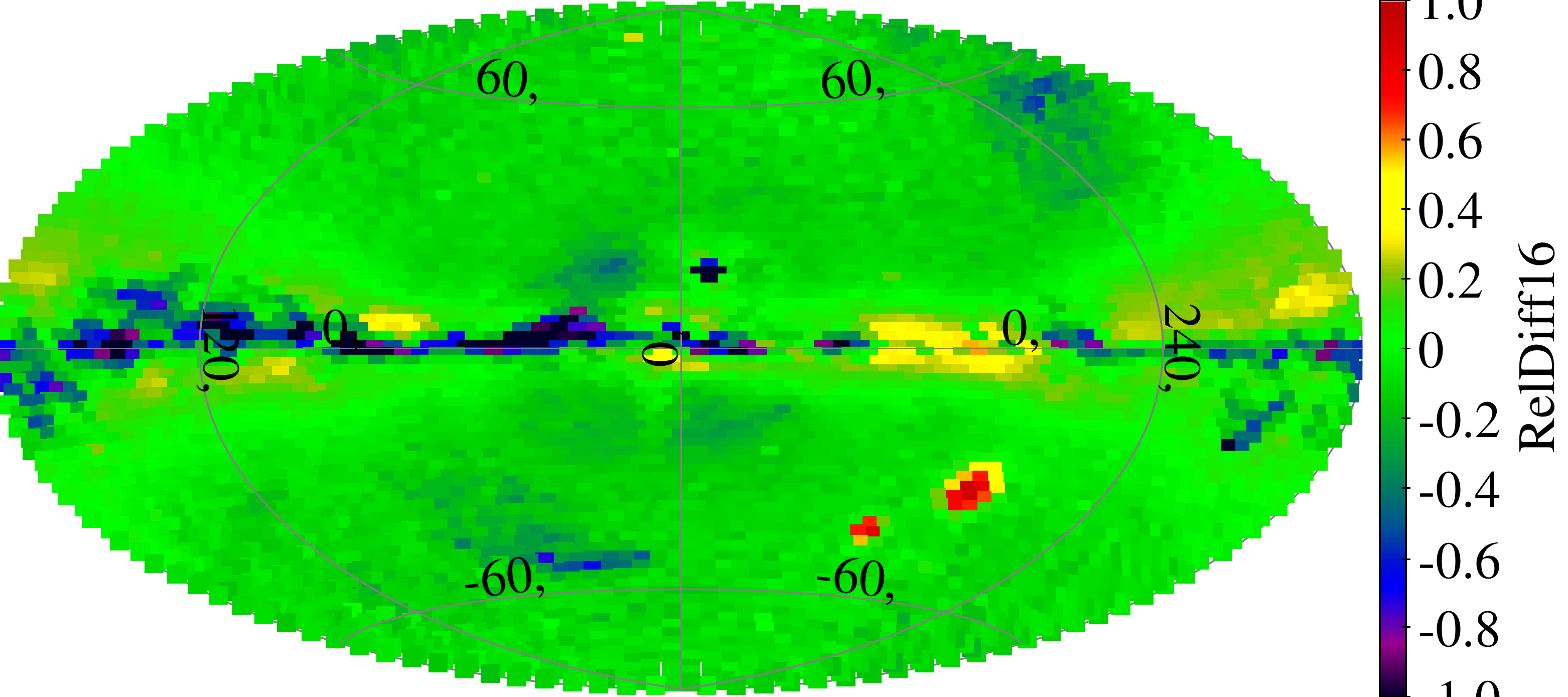}
\includegraphics[width=0.24\textwidth]{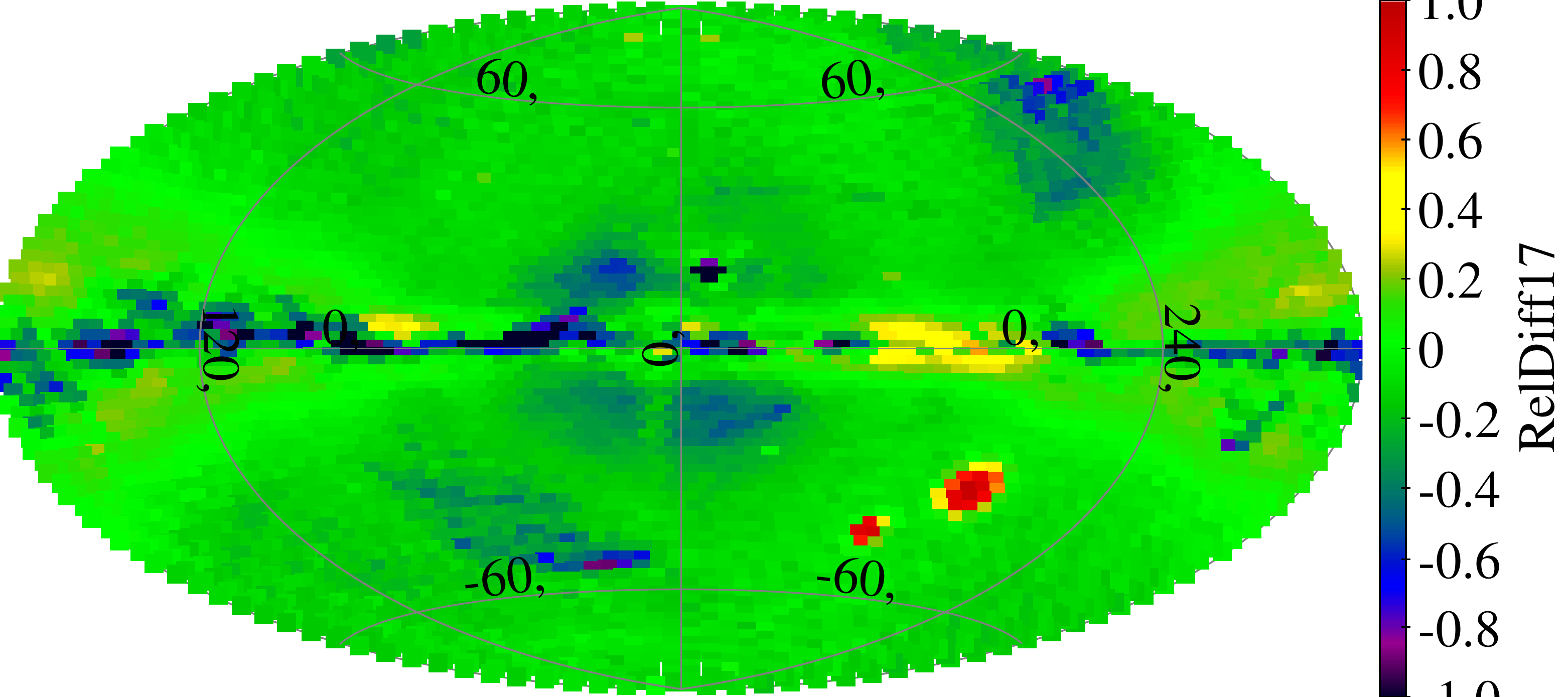}
\includegraphics[width=0.24\textwidth]{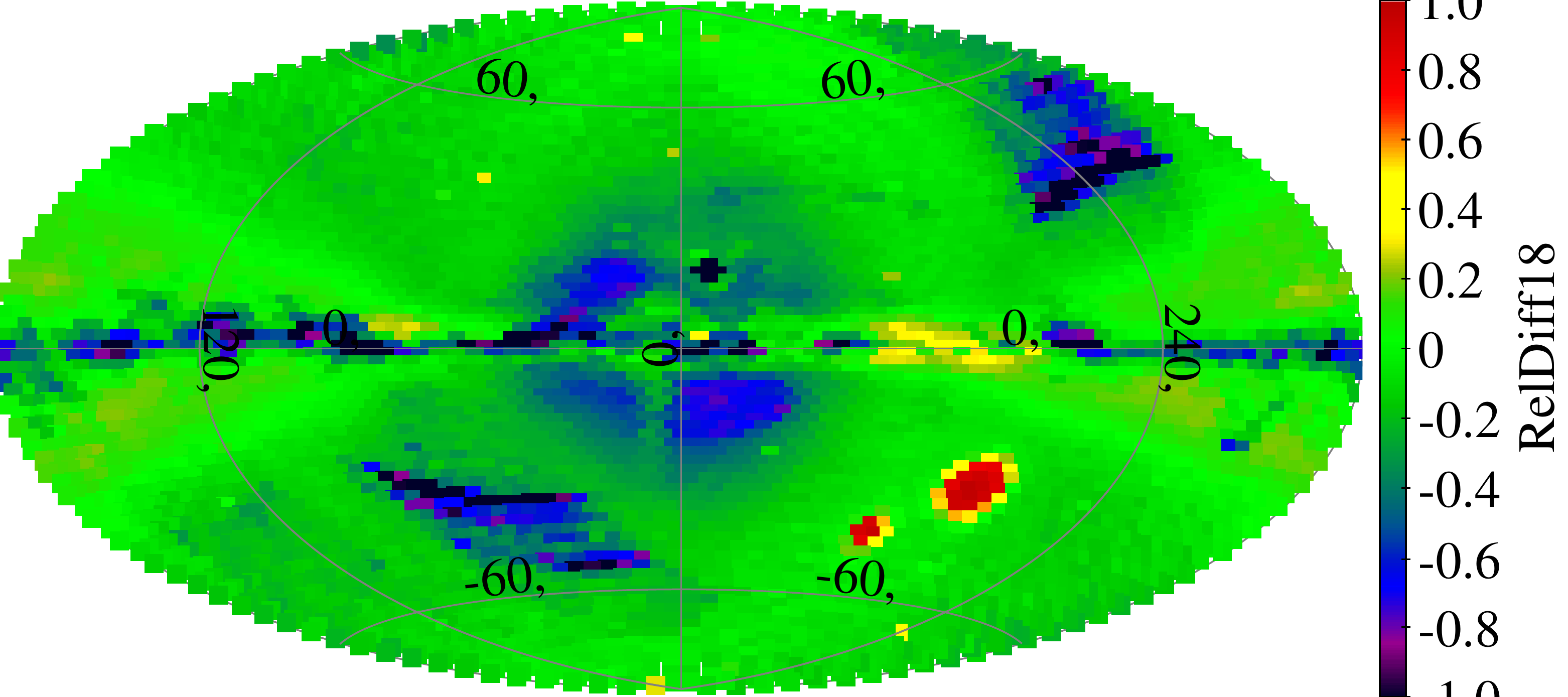}
\includegraphics[width=0.24\textwidth]{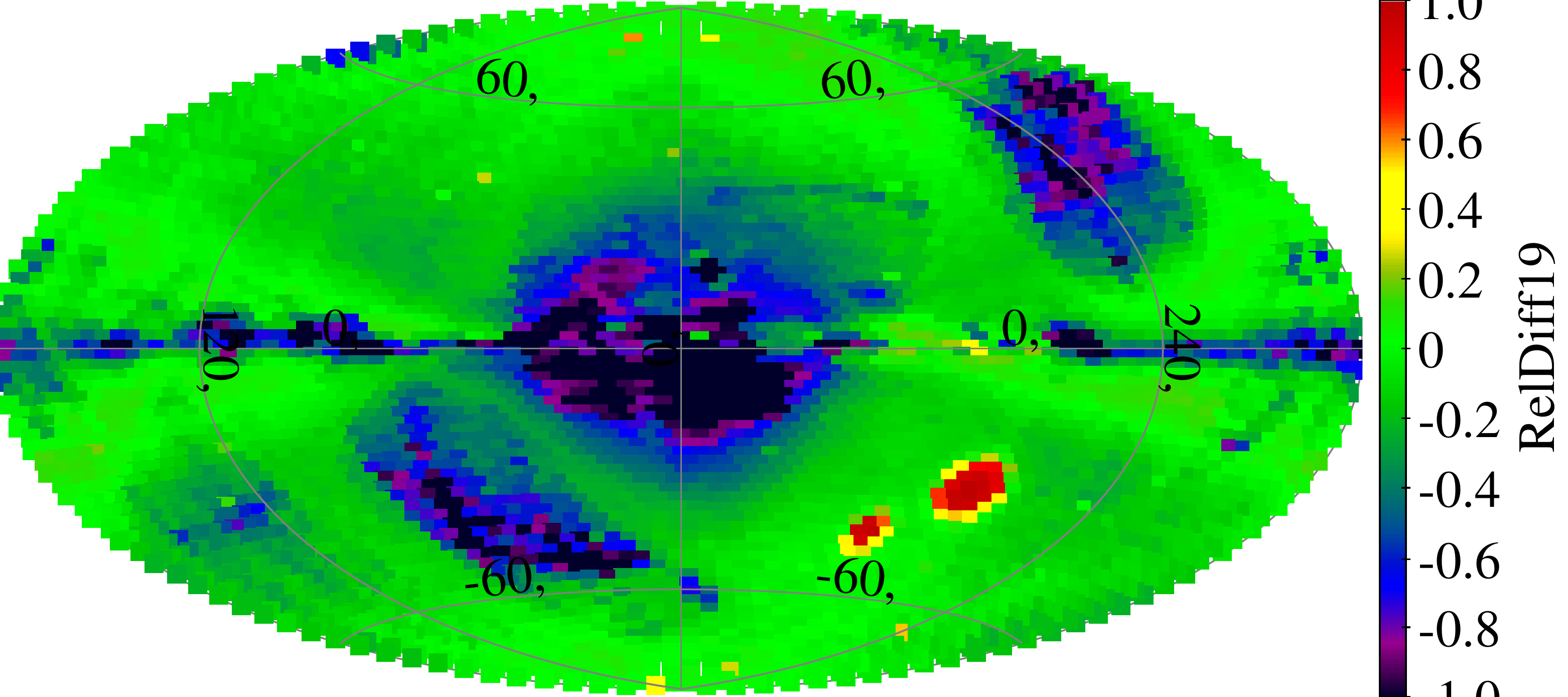}
\caption{Relative star count differences between {\GDR1} and GOG18 simulation in different magnitude bins, 
from $12<G<13$ to $19<G<20$ by step of one magnitude, in galactic coordinates. 
Beside the prominent feature of the Magellanic Clouds (absent from the Galaxy model), 
and inadequacies of the 3D extinction model in the galactic plane, 
the {\gaia} incompleteness around the ecliptic plane due to the scanning law starts 
clearly to appear from $G>16$.}
\label{fig:skymap}
\end{center}
\end{figure*}

\begin{figure*}
\begin{center}
\includegraphics[width=0.25\textwidth,height=0.15\textwidth]{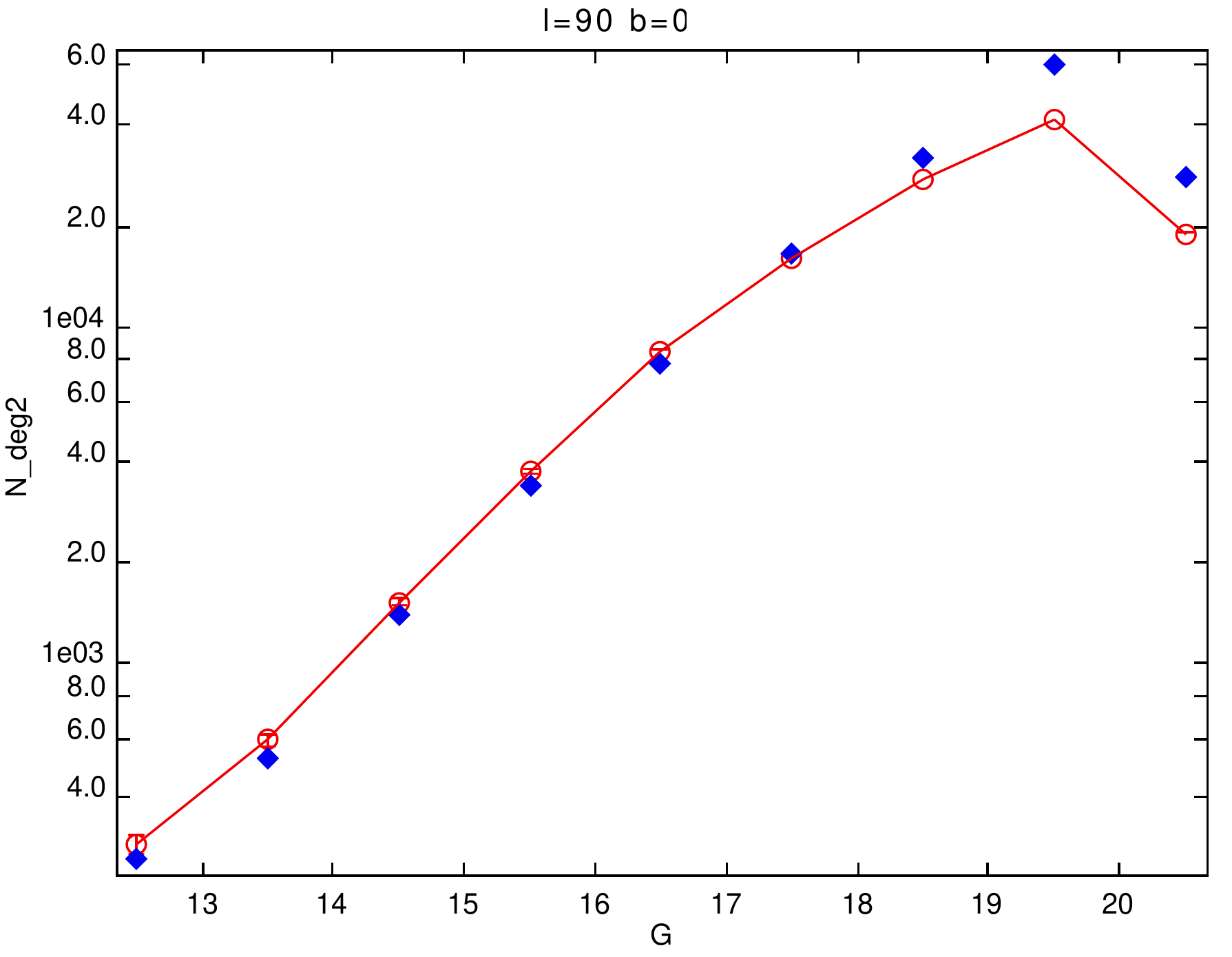}
\includegraphics[width=0.25\textwidth,height=0.15\textwidth]{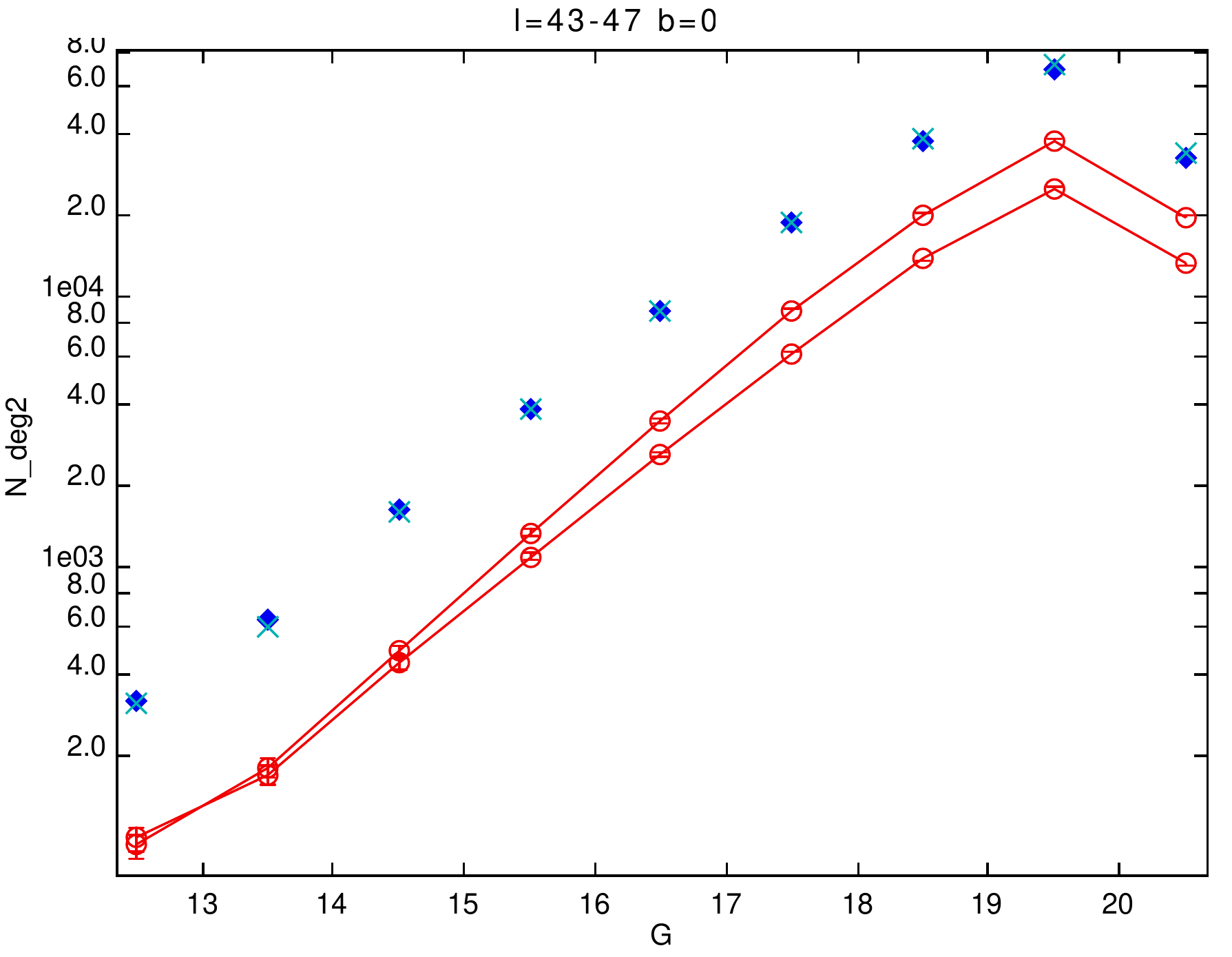}
\includegraphics[width=0.25\textwidth,height=0.15\textwidth]{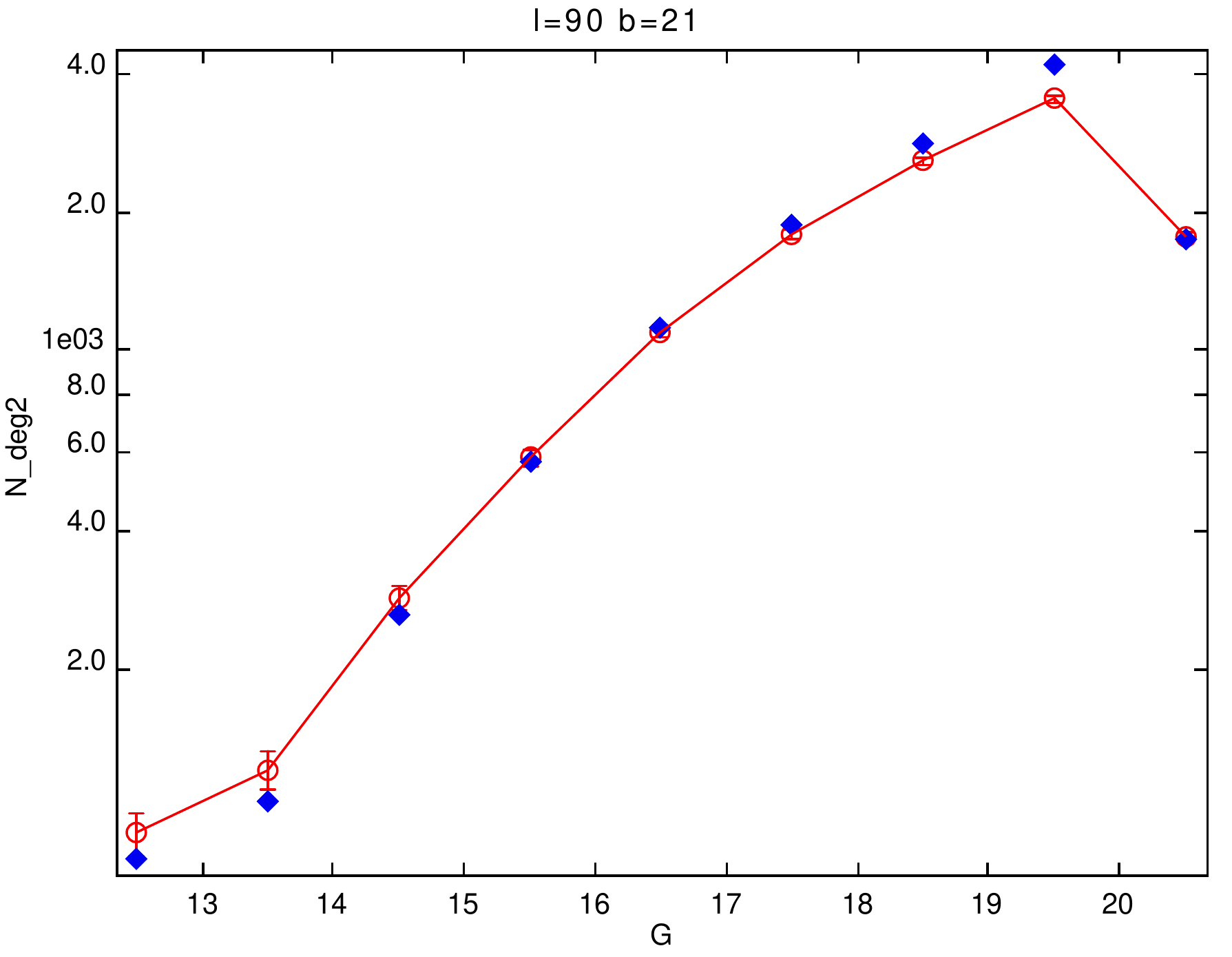}
\includegraphics[width=0.25\textwidth,height=0.15\textwidth]{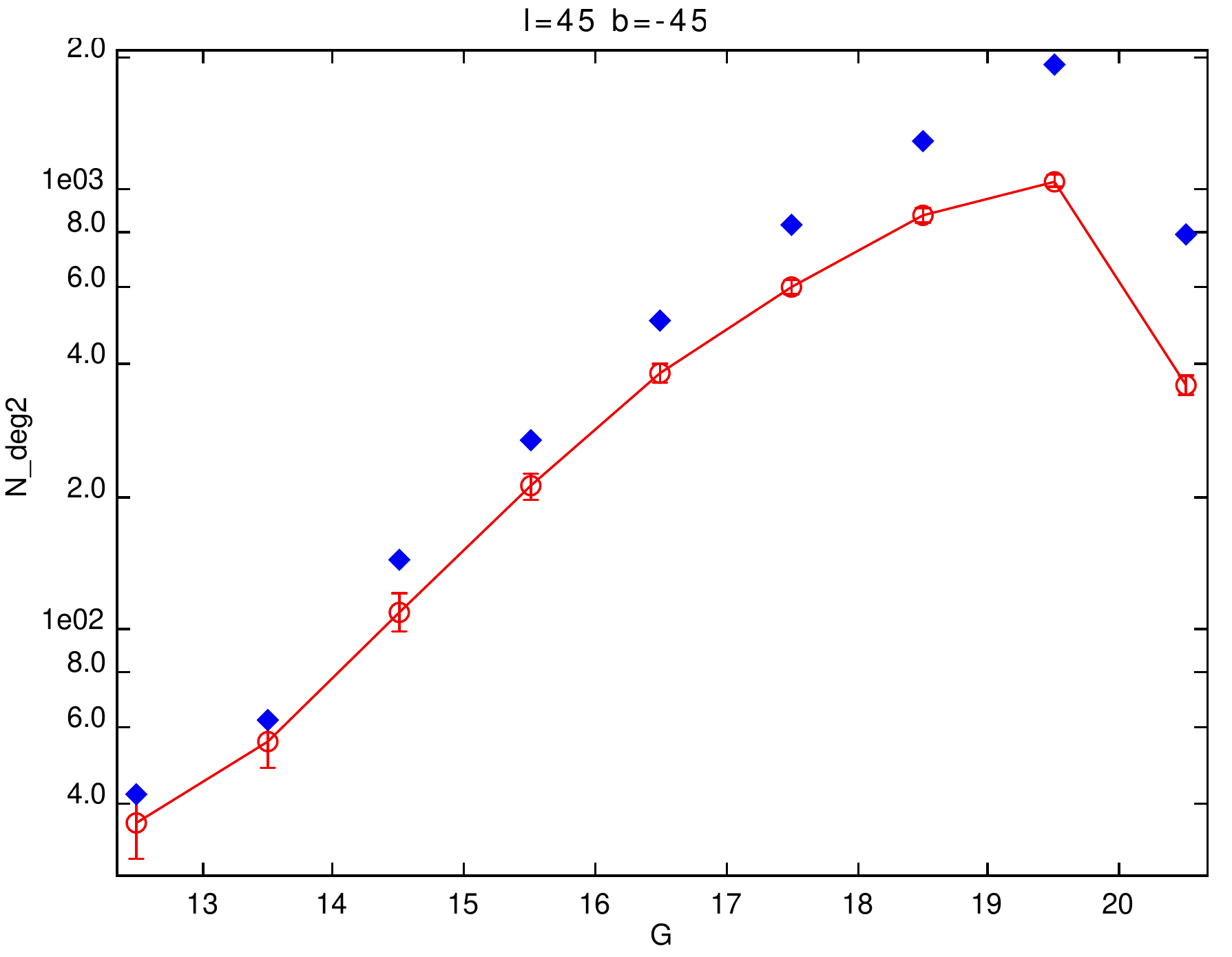}
\includegraphics[width=0.25\textwidth,height=0.15\textwidth]{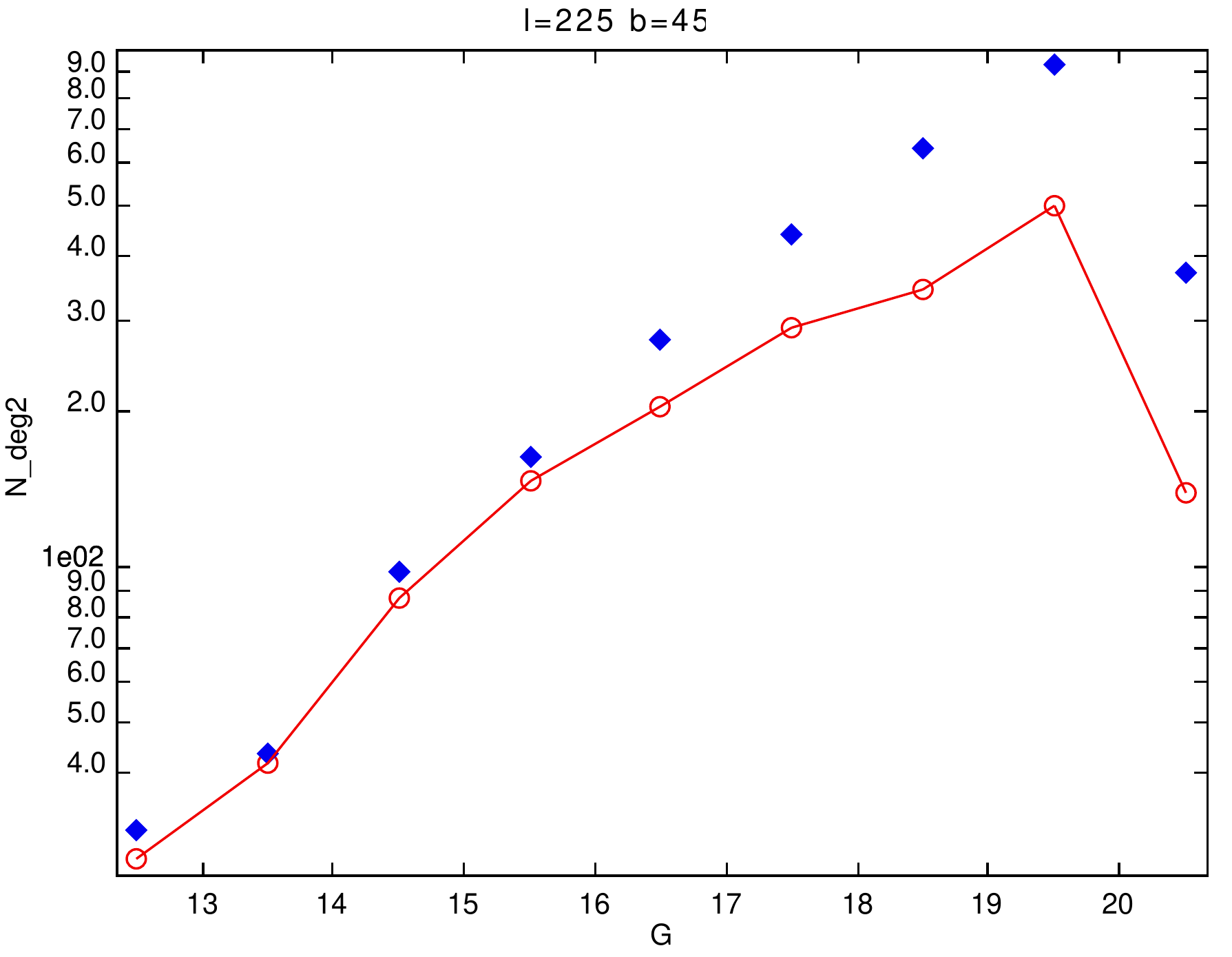}
\includegraphics[width=0.25\textwidth,height=0.15\textwidth]{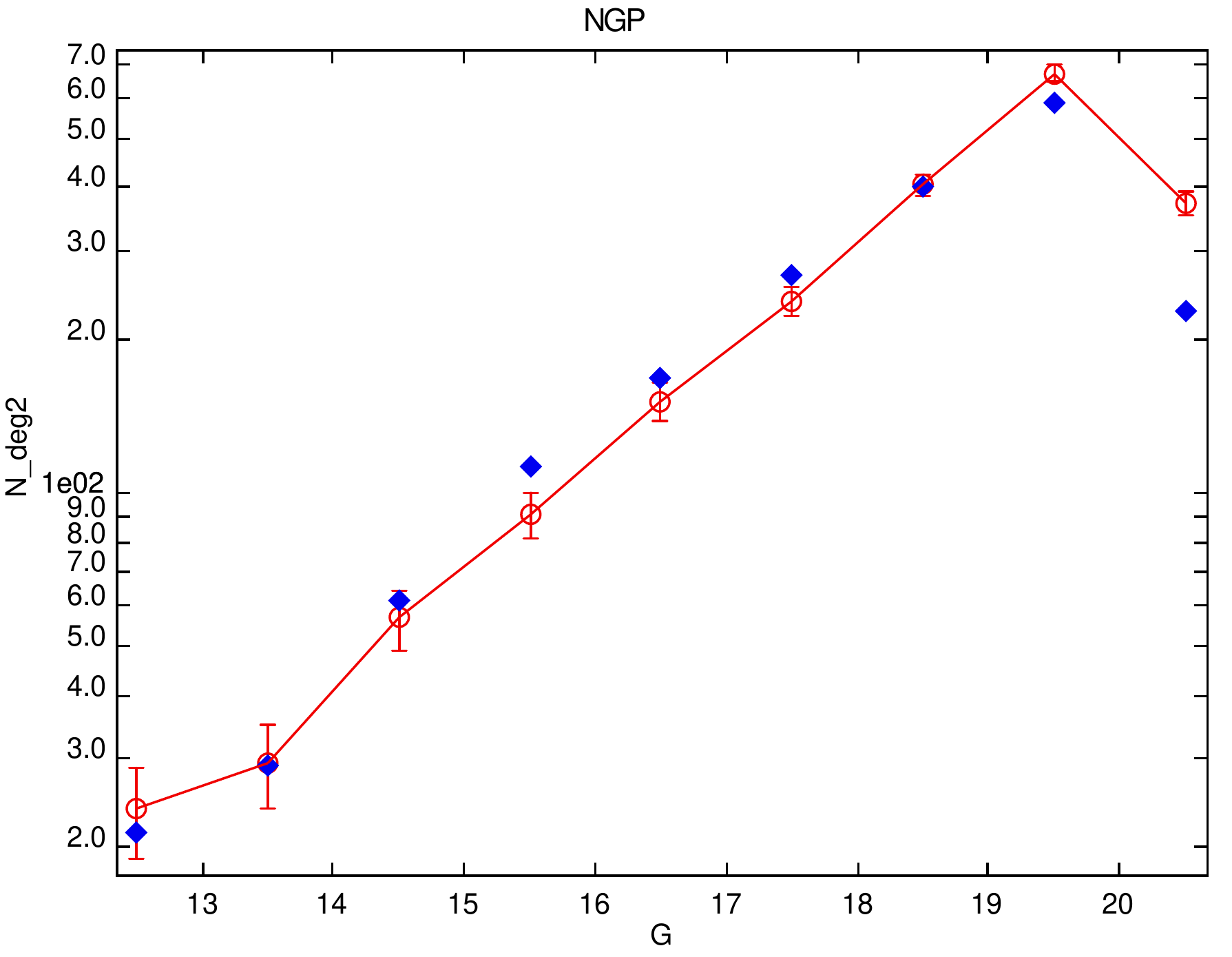}
\caption{Star counts per square degree as a function of magnitude in several directions. Open circles linked with red lines are for {\GDR1} data, filled blue diamonds are simulations from GOG18. Error bars represent the Poisson noise for one square degree field. The bottom row shows regions impacted by the scanning law and the filtering of stars with a low number of observations.}
\label{fig:histograms}
\end{center}
\end{figure*}

Since {\GDR1} only contains $G$ magnitudes and positions, the validation with models consists in the comparison between the distribution of star densities over the sky and a realisation of the Besan\c{c}on Galactic Model \citep[BGM,][]{bgm}, hereafter version 18 of the Gaia Object Generator ~\citep[GOG18, ][]{2014A&A...566A.119L}. The simulation contains 2 billion stars including single stars and multiple systems, and incorporates a model for the expected errors on {\gaia} photometric and astrometric parameters. 

In the validation process, star counts as a function of positions and in magnitude bins have been compared with the model (\figref{fig:skymap}). Systematic differences in Galactic plane fields are mostly due to 3D extinction model problems, but could also be due to other inadequacies of the model (such as local clumps not taken into account in a smooth model). These systematics are seen even in bright magnitude bins. On the other hand, differences at intermediate latitudes in the region of the Magellanic Clouds are not to be considered because these galaxies have not been included in this GOG catalogue. 
There is no other strong difference between data and model that could warn about the quality of the data at magnitudes brighter than 16. However at fainter magnitudes, some regions have significantly less stars than expected from the model. These regions are located specifically around $l=200-250${\deg}, $b=30-60${\deg} and $l=30-80${\deg}, $b=-60;-30${\deg}. At magnitudes fainter than 19, regions all along the ecliptic suffer from this smaller number of sources due to the scanning law and the filtering of objects with a too low number of observations. 
Also at $G>16$ some discrepancies appear in the outer bulge regions, which might be due to incompleteness of the data when the field is crowded (see \secref{sec:underdensity} and \figref{fig:942_holes}).


To estimate in more details the completeness in specific fields, we compared histograms of star counts from {\GDR1} and the GOG18 simulation as a function of magnitude. Figure~\ref{fig:histograms} shows such histograms in some regions of the galactic plane, at intermediate latitudes and at the Galactic poles.  
In the Galactic plane (\figref{fig:histograms}a) the star counts show a drop in the Gaia data at magnitudes brighter than in the model. This could be a priori due to inadequate extinction model or model density laws, or to incompleteness in the Gaia data at faint magnitudes due to undetected or omitted sources. Since the bright magnitude counts are fairly well fitted, the latter hypothesis is most probable. This is also pointed out by comparison with previous catalogues. In the outer Galaxy, GOG18 simulation is probably a too rough model of the Galactic structures, as can be seen in the fields at longitude 180{\deg} 
where the some substructures such as the Monoceros ring or the anticentre overdensity might contribute.
In \figref{fig:histograms}b, the field at longitude 43-47{\deg} and latitude 0{\deg} is for 2 lines of sights, where the model (in blue) gives similar star counts for the two lines while the data (in red) do not. We believe that this is due to varying extinction, which is underestimated in the model for these specific fields.

Over the whole sky, up to magnitude 18, there is a relative difference of a few percent (from less than 3\% at magnitude 12 to 10\% at magnitude 18). Between 18 and 19 the relative difference is 15\%. In the range 19 to 20, the difference is 25\% on the average. 
At high latitudes, and specifically at the Galactic poles, the agreement between the model and the data is also quite good. The regions where the Gaia data seem to suffer from incompleteness are located in the specific regions around $l=225${\deg}, $b=45${\deg} and $l=45${\deg}, $b=-45${\deg}, most probably related to the filtering of sources with a low number of observations. 
The data are however probably complete up to $G=16$ in those regions ($l=225${\deg}, $b=45${\deg}), although the incompleteness could also occur at brighter magnitudes in some areas (at $G=14$ in $l=45${\deg}, $b=-45${\deg}). 

These comparisons show that Gaia data have a distribution over the sky and as a function of magnitude which is close to what is expected from a Galaxy model in most regions of the sky. However it points towards an incompleteness at magnitudes fainter than 16 in some specific areas less observed due to the scanning law, and because sources with a small number of observations have been filtered out. The completeness is also reduced in the Galactic plane due to undetected or omitted sources in crowded regions. This is expected to be solved in future releases where a larger number of observations will be available.

\subsection{Small scale completeness of {\GDR1}} 

\subsubsection{Illustrations of under-observed regions\label{sec:underdensity}}

\begin{figure}
\centering
\includegraphics[width=0.7\columnwidth]{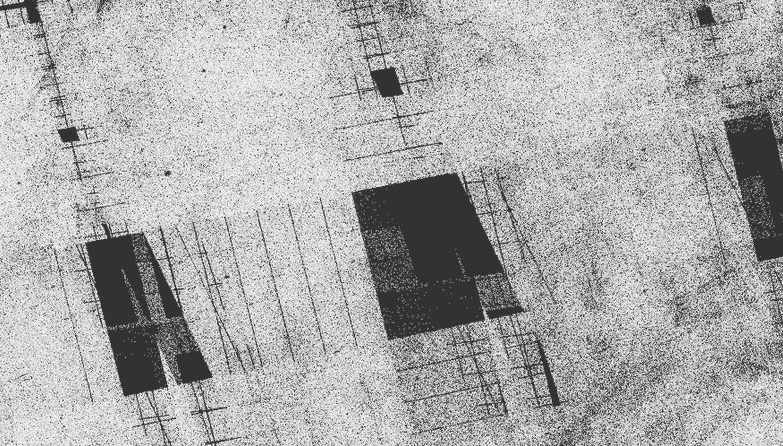}
\includegraphics[width=0.7\columnwidth]{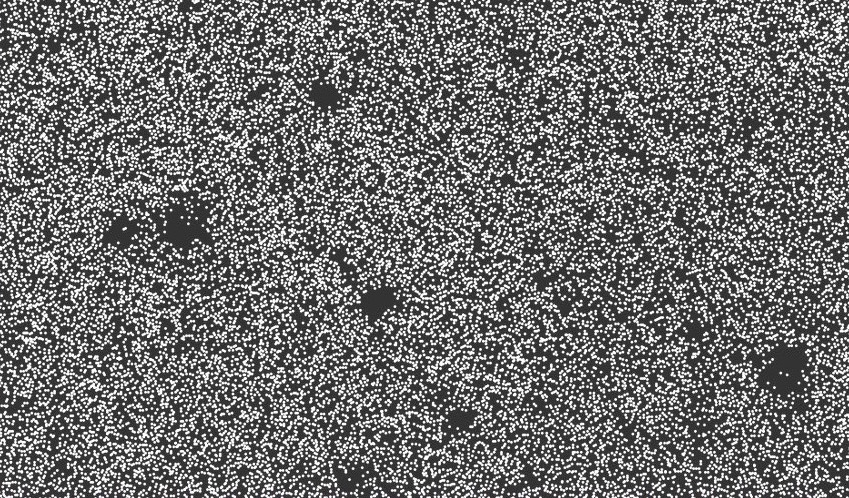}
\caption{Regions with under-densities in DR1: a) under-scanned field near $l=354${\deg}, $b=-3${\deg}, 
size $\sim 3$ square degrees;
b) holes created by lack of on-board resources in another dense field near $l=330${\deg}, $b=3${\deg}, size $\sim 200$ square arcmin.}
\label{fig:942_holes}
\end{figure}

Empty regions due to the threshold on the number of observations are 
illustrated in \figref{fig:942_holes}a near the galactic center; regions under-scanned 
like these ones are not frequent and have a
limited area, below 0.1 square degree \citep[see also][Sect. 6.2]{DPACP-8}. The field shown in \figref{fig:942_holes}b
near the bulge suffered from limited on-board resources, which created holes
in the sky coverage, as shown also for globular clusters in \figref{fig:947_patchy_6GCs}.

\subsubsection{Tests with respect to external catalogues\label{sec:wp944_smallscalecompleteness}}
The small scale completeness of {\GDR1} and its variation with the sky stellar density has been tested in comparison with two catalogues: \beforeReferee{the} Version 1 of the Hubble Space Telescope (HST) Source Catalogue \citep[HSC,][]{2016AJ....151..134W} and a selection of fields observed by OGLE \citep{2008AcA....58...69U}. 

\paragraph{Hubble Source Catalogue.} The HSC is a very non-uniform catalogue based on deep pencil-beam HST observations made using a wide variety of instruments (Wide Field Planetary Camera 2 (WFPC2), Wide Field Camera 3 (WFC3) and the Wide Field Channel of the Advanced Camera for Surveys (ACS) and observing modes. The spatial resolution of Gaia is comparable to that of Hubble and the HSC is therefore an excellent tool to test the completeness of {\GDR1} on specific samples of stars. To check the completeness as a function of \gmag, we computed an approximate \gmag-band magnitude from HST F555W and F814W magnitudes (\ghst) using theoretical colour-colour relations derived following the procedure of \cite{2010A&A...523A..48J}. 

The first test was made in a crowded field of one degree radius around Baade's Window. Nearly 13\,000 stars were considered, observed in both the F555W and F814W HST filters with either WFPC2 or WFC3. 

The second test was made on samples of stars observed with one of the three HST cameras, using the red filter F814W and either F555W or F606W. Sources were selected following the recommendations of \cite{2016AJ....151..134W} to reduce the number of artefacts.
Moreover, only stars with an absolute astrometric correction flag in HST set to yes have been selected, leading to a typical absolute astrometric accuracy of about 0.1\arcsec. The size of the resulting samples varies from 1600 stars for ACS-F555W to nearly 120\,000 stars for ACS-F606W, going through 15-23\,000 stars for the four other samples. The completeness of Gaia observations for these samples, position differences and colour-colour relations have been tested. 

\begin{figure}
    \begin{center}
        \includegraphics[width=0.49\columnwidth]{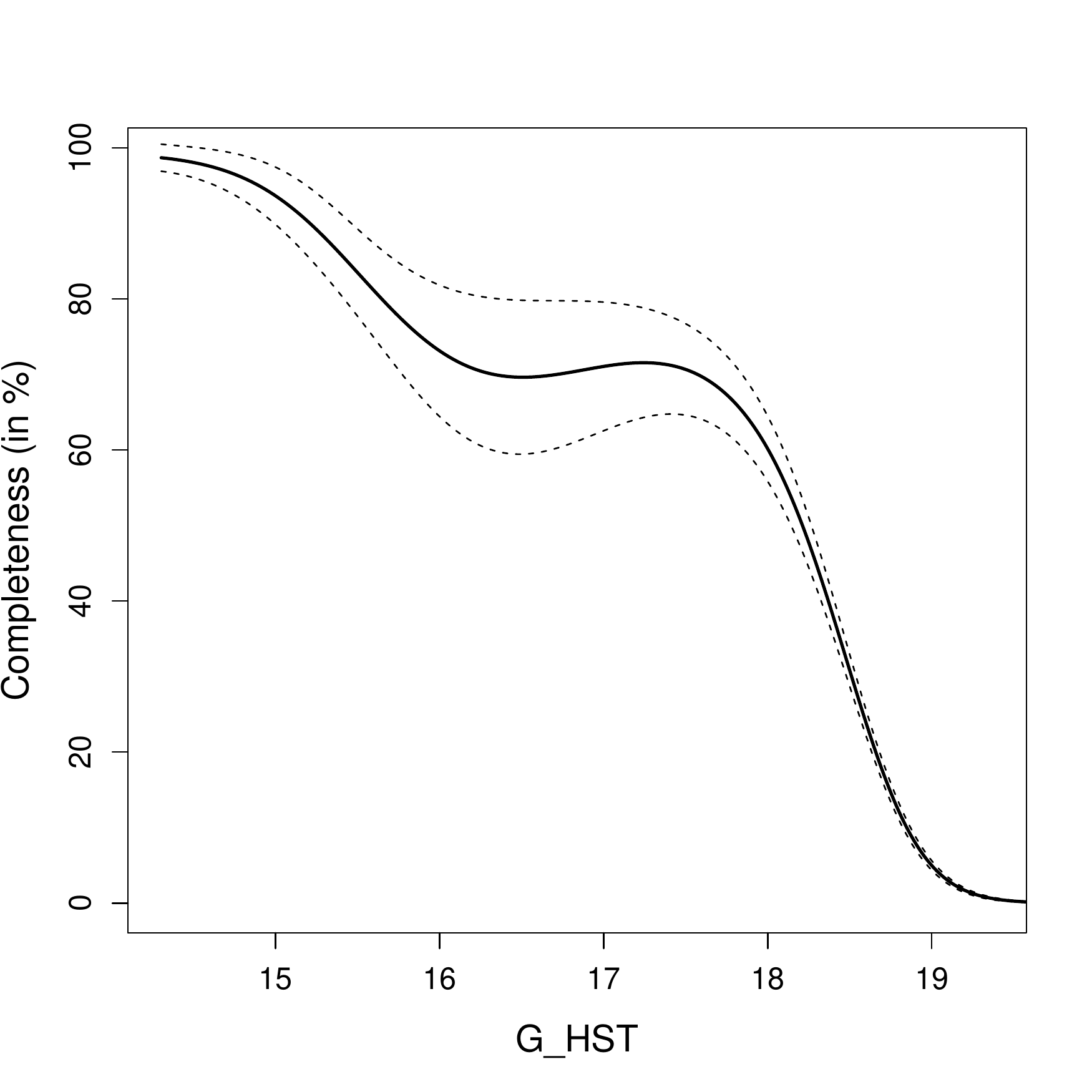}
        \includegraphics[width=0.49\columnwidth]{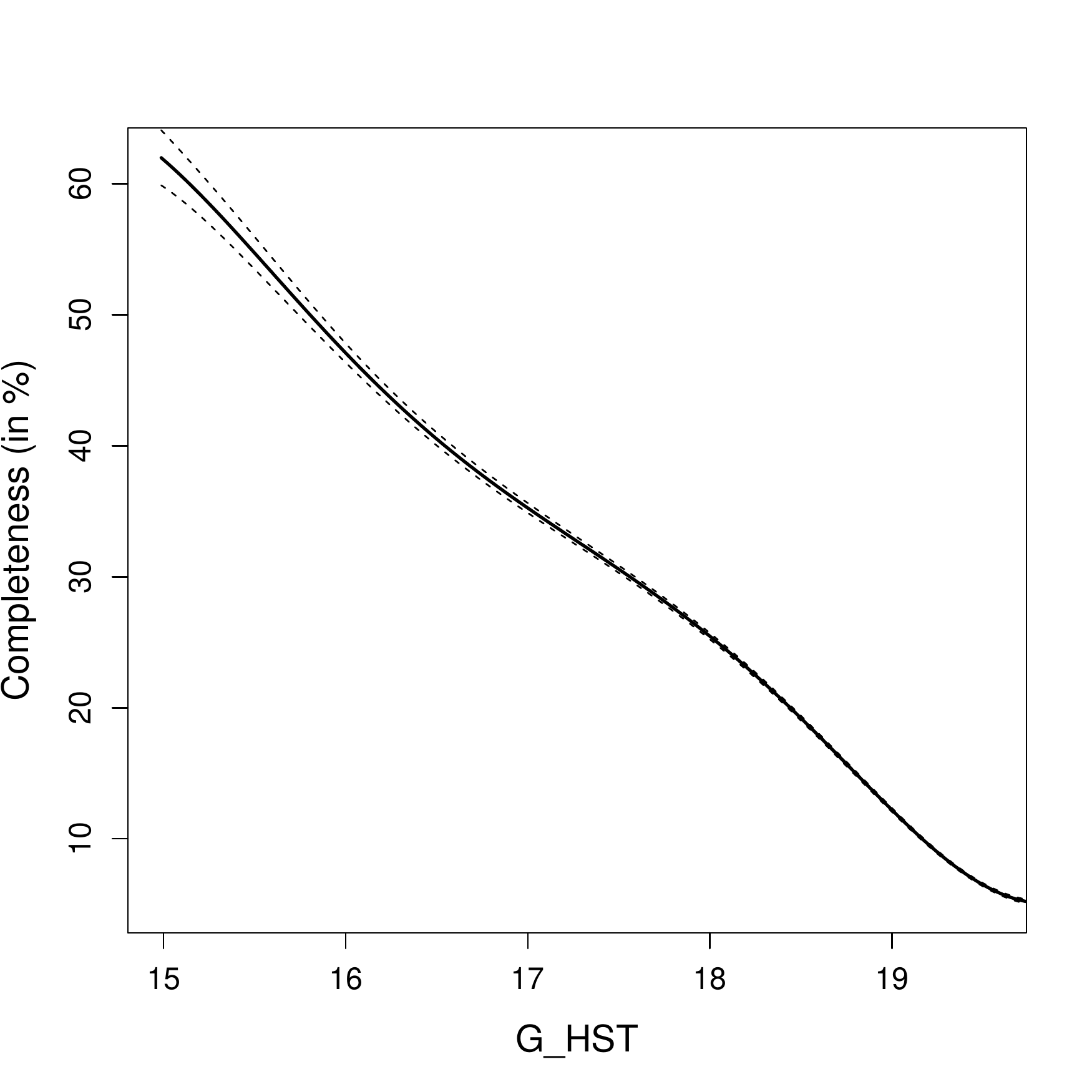}
        \caption[{\GDR1} completeness versus HST]{{\GDR1} completeness (in \%) versus the Hubble Source Catalogue as a function of \ghst\ magnitude. The dotted lines correspond to the 1$\sigma$ confidence interval; a) in Baade's Window ($l=1${\deg}, $b=-4${\deg}); b) for all-sky HSC sources observed with the ACS and the F606W filter. }
        \label{fig:cu9val_wp944_HSTcompl} 
    \end{center}
\end{figure}

The completeness results of both tests are presented in \figref{fig:cu9val_wp944_HSTcompl}. In Baade's Window, the completeness follows the expectations for DR1: in this very dense area, on-board limitations lead to a brighter effective magnitude limit. The ``all-sky'' result (using here 128\,000 ACS stars with F606W$<20$~mag)
is at first sight more surprising, but in fact bright source observations with HST are quite rare and are done mainly in very dense areas (which need the HST resolution) such as globular clusters, which also suffer from Gaia on-board limitations.  We further checked this interpretation by using individual HST observations and images around a few positions : the test made for a low density area around the dwarf spheroidal galaxy Leo II \citep{2011ApJ...741..100L} leads to a completeness at magnitude 20 of nearly 100\%, while a test for a high density area around the globular cluster NGC 7078 \citep{2014ApJ...797..115B} leads to a completeness worse than the one presented \figref{fig:cu9val_wp944_HSTcompl}.

\begin{figure*}
\centering
\includegraphics[width=0.6\textwidth]{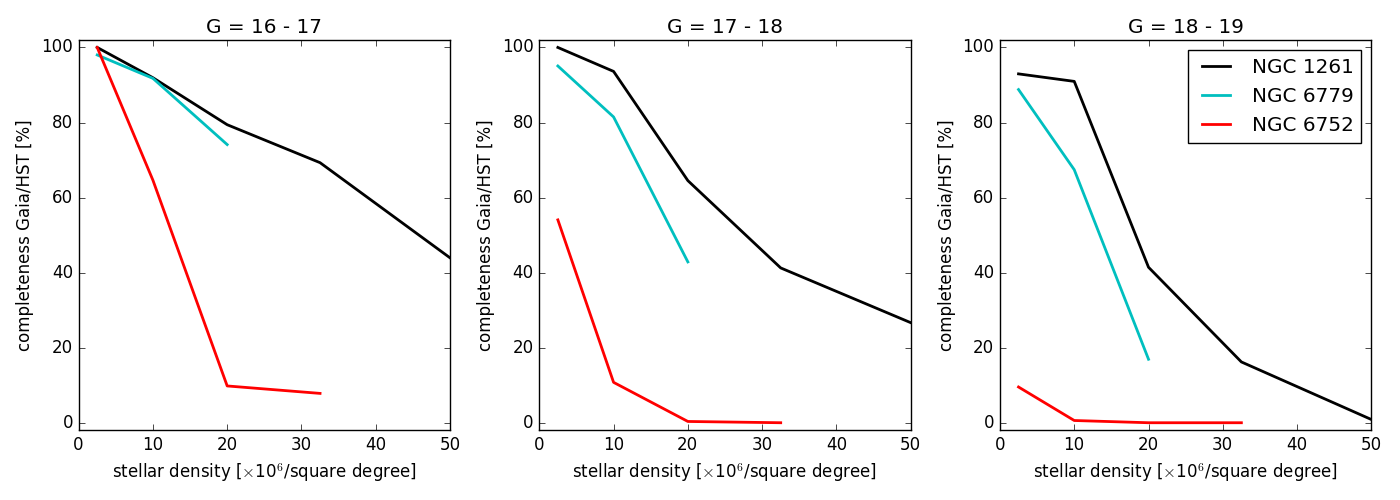}
\caption{Completeness against density in the field of three chosen GCs, in different magnitude ranges. Fields such as NGC\,1261 have a median of 220 observations, allowing for a much better completeness in the denser regions than NGC\,6752 (40 observations). } \label{fig:947_completeness_3GCs}
\end{figure*}

\begin{figure*}
\centering
\includegraphics[width=0.75\textwidth]{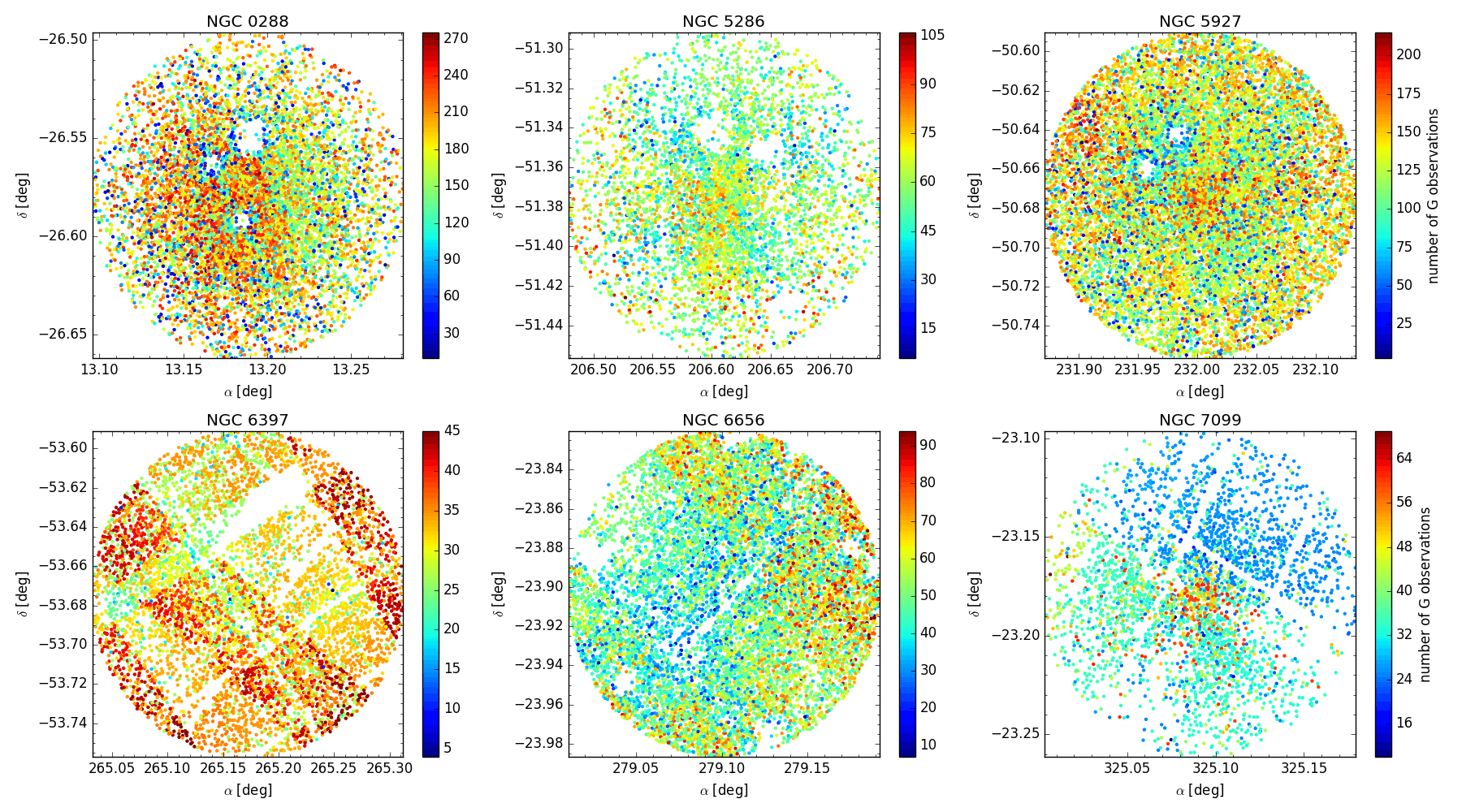}
\caption{Stellar distribution for six chosen GCs, colour-coded by number of $G$ observation for each star. \textit{Top row:} examples of holes caused by limited on-board resources or bright stars. \textit{Bottom row:} in some regions patterns are visible corresponding to stripes where no stars had a sufficient number of observations. } \label{fig:947_patchy_6GCs}
\end{figure*}

\begin{figure}
\centering
\includegraphics[width=0.8\columnwidth]{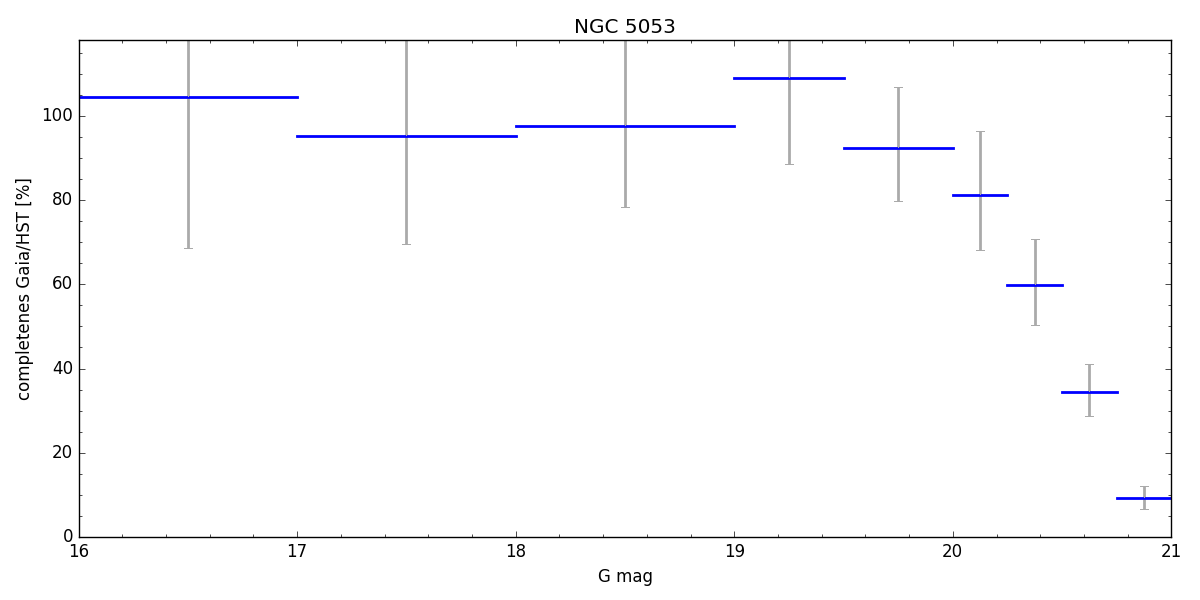}
\caption{Completeness of Gaia relative to HST in the area around NGC\,5053 featuring stellar densities under 1 million per square degree.} \label{fig:947_completeness_5053}
\end{figure}

\paragraph{HST observations of Globular Clusters.} We run detailed completeness tests within globular clusters using HST data specifically reduced for the study of those crowded fields. We used 26 globular clusters for which HST photometry is available from the archive of \citet[][see Table \ref{chap:cu9val_T1}]{2007AJ....133.1658S}. 
The data for all GCs were acquired with the ACS and contain magnitudes in the bands F606W and F814W. The observations cover fields of 3\,arcmin$\times$3\,arcmin size.  For M4 (NGC~6121), data by  \citet{2013AN....334.1062B}, and \citet{2015MNRAS.454.2621M} 
 taken in the HST project GO-12911 in WFC3/UVIS filters were used. For this test, the photometric transformations HST bands to Gaia $G$-band were adjusted for each cluster to fit a sample of bright stars in order to avoid issues due to variations in metallicity and extinction.

High quality relative positions and relative proper motions are available for these clusters. When artificial star experiments were available in the original HST catalogue (GCs marked  with * in Table \ref{chap:cu9val_T1}), the completeness of HST data has been evaluated by comparing the number of input and recovered artificial stars in each spatial bin. We find the completeness of the HST data to be well above 90\% and close to 100\% in all cases for  stars brighter than $V=21$, but for the very crowded cluster NGC5139 (OmegaCen).
The GCs are chosen to present different level of crowding down to $G\sim 22$. In general, HST data cover the inner core of the clusters, where the stellar densities are above $10^6$ stars per square degree in almost all regions (above 30 million in many cases, and up to 110 million stars per square degree in the core of NGC~104/47~Tuc). In a few cases, lower densities are reached in the external regions. We therefore expect Gaia to be very severely incomplete in most of the regions studied in this test.
The HST magnitudes were converted to Gaia $G$ magnitudes using the same transformations as previously between $G$ and F814W, F606W but on the Vega photometric system. 

\begin{table}
\centering\footnotesize
\caption{GCs used in the completeness test. 
Asterisks denote the ones with artificial star experiments available 
in the original HST catalogue.\label{chap:cu9val_T1}}
\begin{tabular}{lrr}
\hline\hline
\multicolumn{1}{c}{cluster} &\multicolumn{1}{c}{$\alpha$ (J2000)} &\multicolumn{1}{c}{$\delta$ (J2000)} \\
\hline
  LYN07 & 242.7619 & -55.315\\
  NGC~104* & 6.0219 & -72.0804\\
  NGC~288 & 13.1886 & -26.5791\\
  NGC~1261 & 48.0633 & -55.2161\\
  NGC~1851 & 78.5267 & -40.0462\\
  NGC~2298 & 102.2465 & -36.0045\\
  NGC~4147 & 182.5259 & 18.5433\\
  NGC~5053 & 199.1128 & 17.6981\\
  NGC~5139* & 201.6912 & -47.476\\
  NGC~5272 & 205.5475 & 28.3754\\
  NGC~5286 & 206.6103 & -51.3735\\
  NGC~5466 & 211.364 & 28.5342\\
  NGC~5927 & 232.002 & -50.6733\\
  NGC~5986 & 236.5144 & -37.7866\\
  NGC~6121* & 245.8974 & -26.5255\\
  NGC~6205 & 250.4237 & 36.4602\\
  NGC~6366 & 261.9349 & -5.0763\\
  NGC~6397* & 265.1725 & -53.6742\\
  NGC~6656* & 279.1013 & -23.9034\\
  NGC~6752* & 287.7157 & -59.9857\\
  NGC~6779 & 289.1483 & 30.1845\\
  NGC~6809* & 294.998 & -30.9621\\
  NGC~6838* & 298.4425 & 18.7785\\
  NGC~7099 & 325.0919 & -23.1789\\
  PAL~01 & 53.3424 & 79.5809\\
  PAL~02 & 71.5245 & 31.3809\\
\hline
\end{tabular}
\end{table}

For each GC, the total density of stars in square bins of 0.008\,deg = 0.5\,arcmin was evaluated, then in each bin we counted the number of stars present in the HST photometry and in the {\GDR1}, by slice in magnitude.

The completeness of {\GDR1} is shown in \figref{fig:947_completeness_3GCs} for three clusters, as a function of the stellar density observed in the HST data. Different crowded regions present different degrees of completeness, depending on the number of observations in that region. In addition, holes are found around bright stars (typically for $G<11-12$ mag), and entire stripes are missing, as illustrated in \figref{fig:947_patchy_6GCs}. 



In less crowded regions, such as in the field around NGC\,5053 where stellar densities are under 1 million per square degree, the completeness is very high, as shown in \figref{fig:947_completeness_5053}.

\begin{figure*}
    \begin{center}
        \includegraphics[height=0.5\columnwidth]{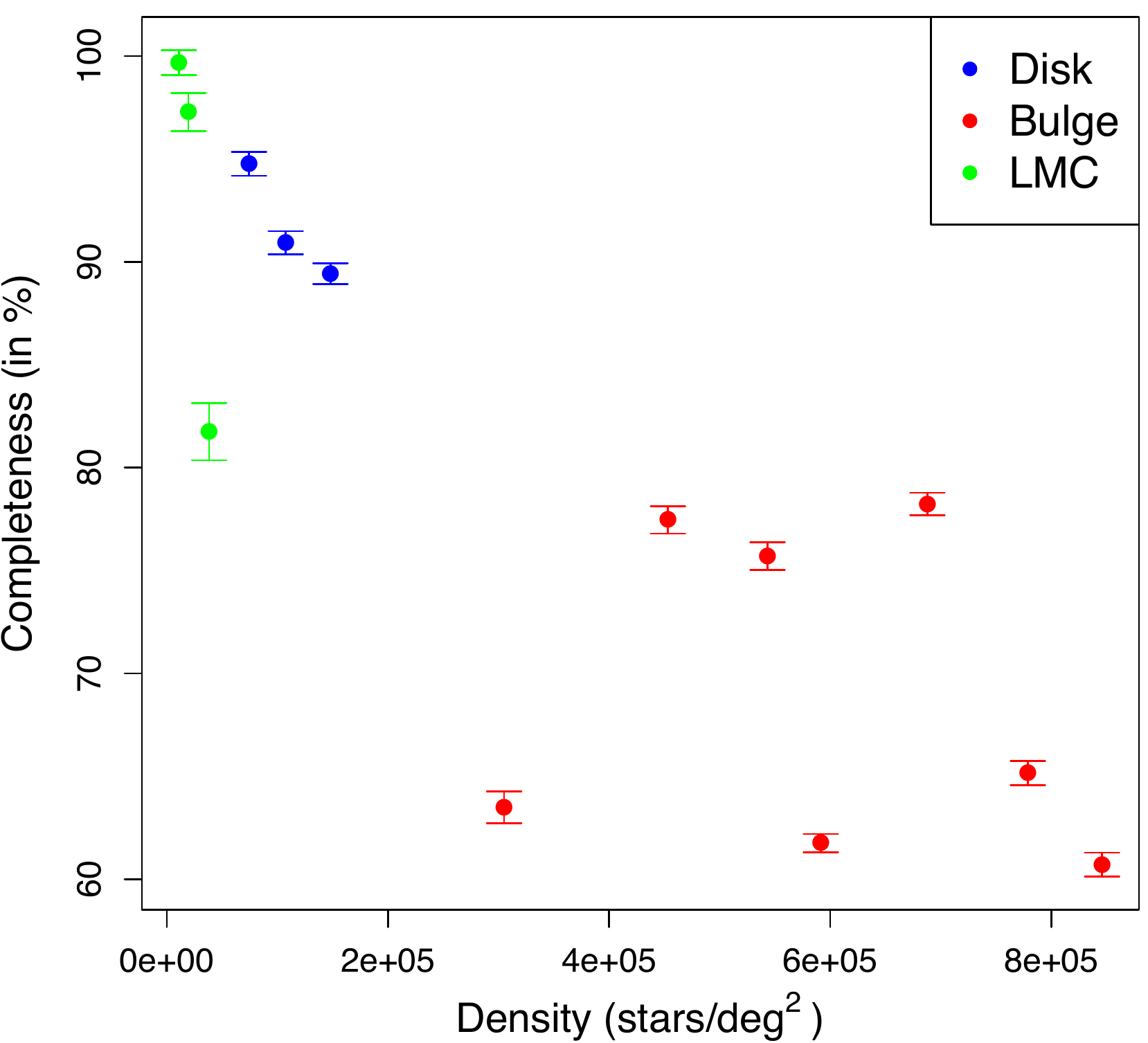}
        \hspace{0.15\columnwidth}
        \includegraphics[height=0.5\columnwidth]{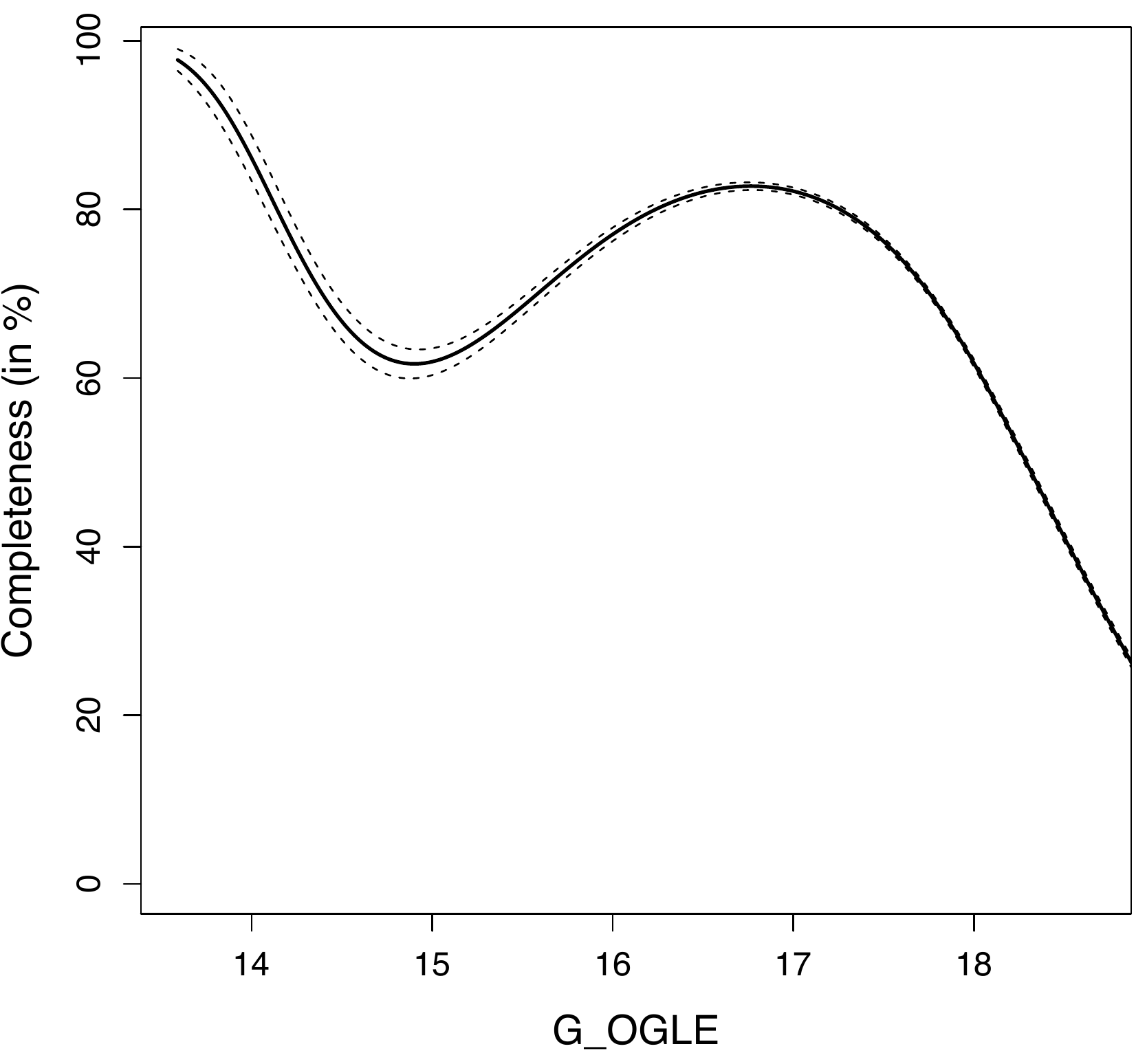}
        \includegraphics[height=0.49\columnwidth]{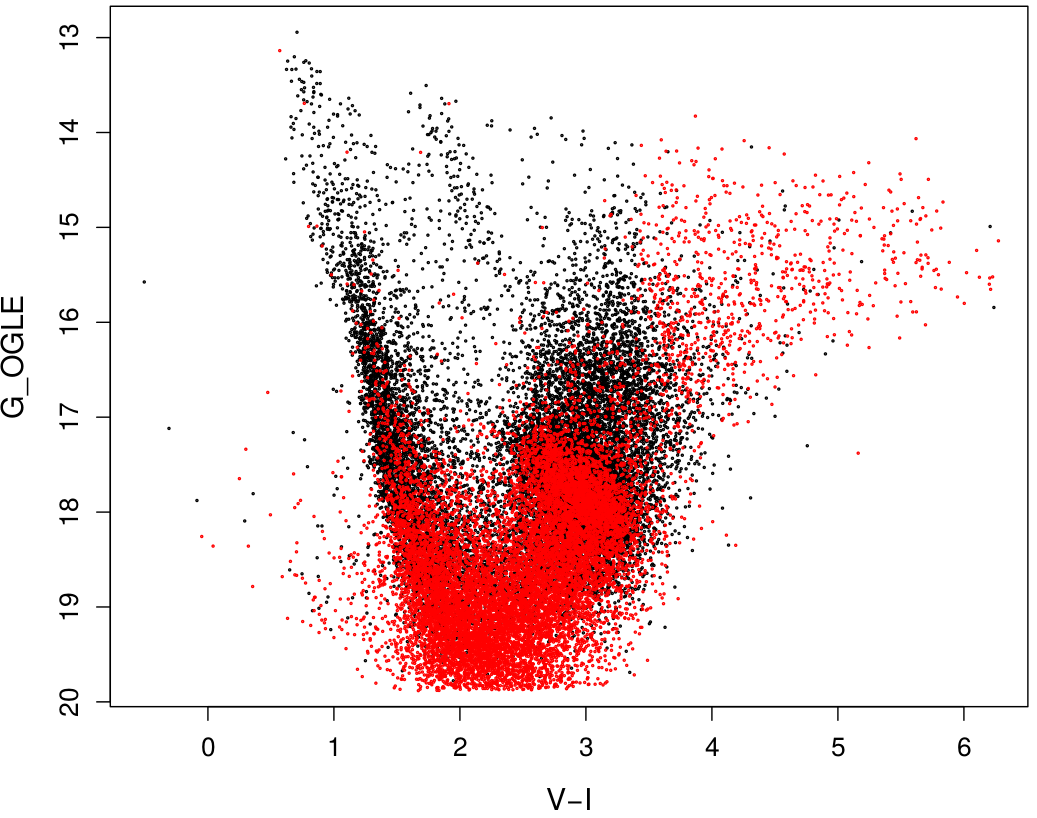}
        \caption[Completeness versus OGLE]{{\GDR1} completeness versus some OGLE Catalogues. a) Completeness at \gmag=18 of some OGLE fields as a function of the measured density at \gmag=20; b) Completeness in OGLE Bulge field blg100 ($l=-0.3${\deg}, $b=-1.55${\deg}), density: 970\,000 stars/deg$^2$; c) associated color-magnitude diagram, stars in red being missing in {\GDR1}. }
        \label{fig:cu9val_wp944_Oglecompl} 
    \end{center}
\end{figure*}

\paragraph{OGLE catalogues.} To further test the variation of the completeness with sky density, we looked at the completeness versus OGLE data using a few fields in the OGLE-III Disk \citep{2010AcA....60..295S}, OGLE-III Bulge \citep{2011AcA....61...83S} and OGLE-IV LMC \citep{2012AcA....62..219S} surveys. A \gmag-band magnitude was computed from OGLE $V$ and $I$ magnitudes (\goggle) using an empirical relation derived from the matched Gaia/OGLE sources (two relations were derived, one for OGLE-III and one for OGLE-IV due to their different filters). The stellar densities were estimated from the OGLE data themselves, therefore they are certainly slightly under-estimated. As can be seen in \figref{fig:cu9val_wp944_Oglecompl}, the completeness is not only dependent on the sky density, but also on the sky position, linked to the Gaia scanning law, as we saw above. In the bulge fields, the completeness may show a drop around \gmag =15 (as seen in \figref{fig:cu9val_wp944_Oglecompl}b, confirming the feature of \figref{fig:cu9val_wp944_HSTcompl}a). This is due to the fact that the reddest stars have not been kept in {\GDR1} (because of filtering at calibration level) and those missing stars correspond to the reddened red giant branch of the bulge (\figref{fig:cu9val_wp944_Oglecompl}c).

\subsection{Completeness and angular resolution}

Although there are no doubts about the excellent, spatial angular resolution of  
{\gaia}\footnote{e.g. Pluto and Charon could easily be separated with a 0.36" along-scan 
separation, see {\scriptsize\url{http://www.cosmos.esa.int/web/gaia/iow_20160121}}}, the {\it effective} angular
separation in {\GDR1} can be questioned, e.g. due to possible cross-match problems.

\subsubsection{Distribution of the distances between pairs of sources \label{sec:neighbours}}

A simple way of checking the angular resolution of a catalogue is to look
at the distribution of the distances between pairs of sources. For a random
star field with $\rho$ stars per unit area, a ring of radius $r$, centred on
a given star, will contain $\rho 2 \pi r \Delta r$ stars, where $\Delta r$ is
the width of the ring. For a sample of $N$ stars, we will have
$N \rho \pi r \Delta r$ unique pairs at that separation.

\begin{figure}
\centering
\includegraphics[width=0.8\columnwidth]{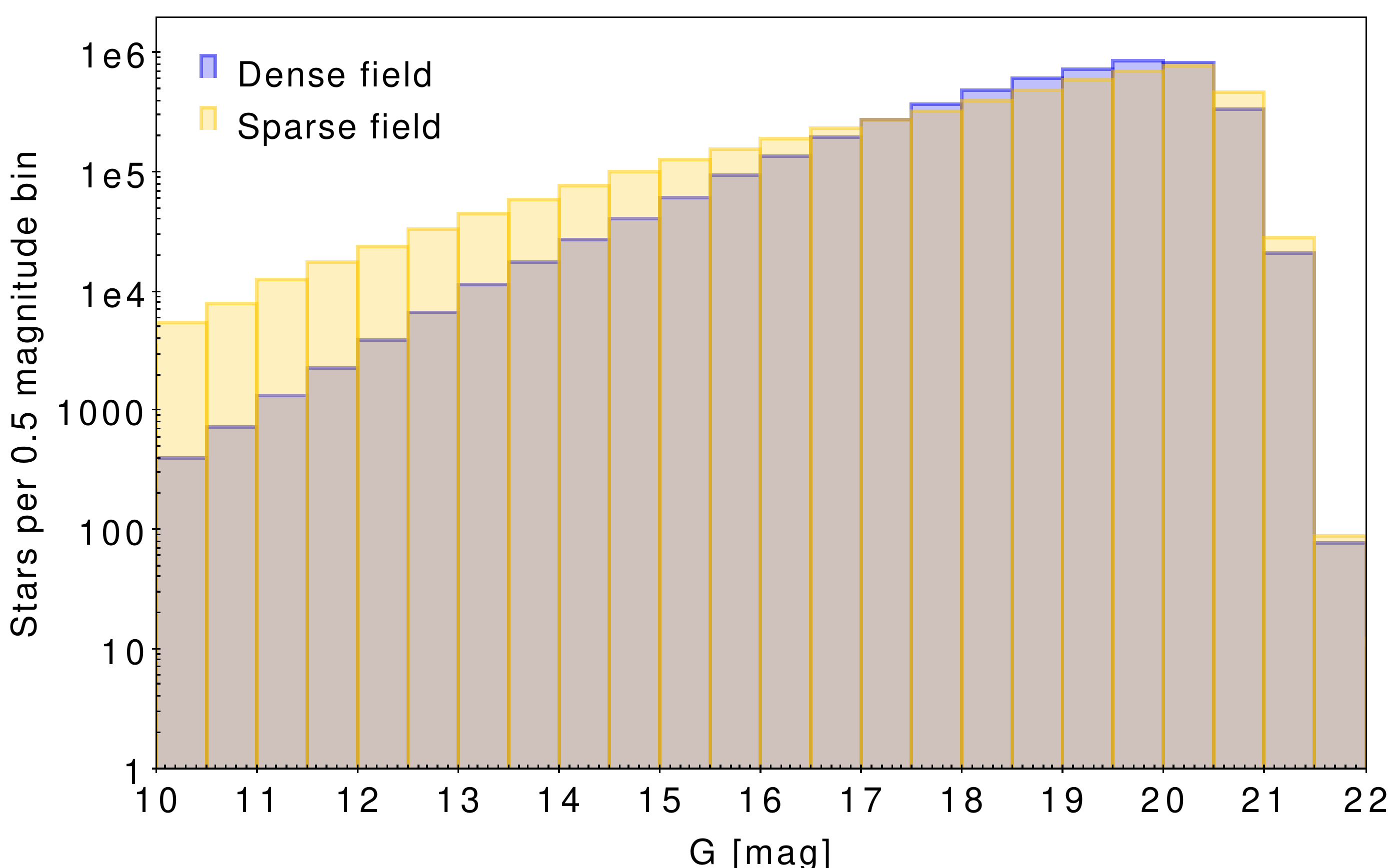}
\caption{$G$ magnitude histograms for a dense field ($l=330$\deg, $b=-4$\deg, $\rho=2$\deg) 
and a sparse field ($l=260$\deg, $b=-60$\deg, $\rho=15$\deg). The sparse field has
been scaled as to give about the same number of sources as the dense field.
\label{fig:942_ghist}}
\end{figure}

We have looked at two fields, a dense field of radius 2\degr\ centred at $(l,b)
= (330\degr, -4\degr)$ with 400\,000 stars per square degree and a sparse 
field of radius 15\deg ($l=260$\deg, $b=-60$\deg) with 2\,900
stars per square degree, scaled to produce the same number of sources. 
Figure~\ref{fig:942_ghist} shows the distribution of $G$ magnitudes in these two fields.
The difference of slopes comes from the fact that the dense field may integrate  
disk stars on a larger distance, with extinction not that large at $b=-4$\degr,
whereas the sparse field at higher latitude quickly leaves the disk and
integrates the thick disk, less dense. 

\begin{figure}
\centering
\includegraphics[width=0.49\columnwidth,height=0.4\columnwidth]{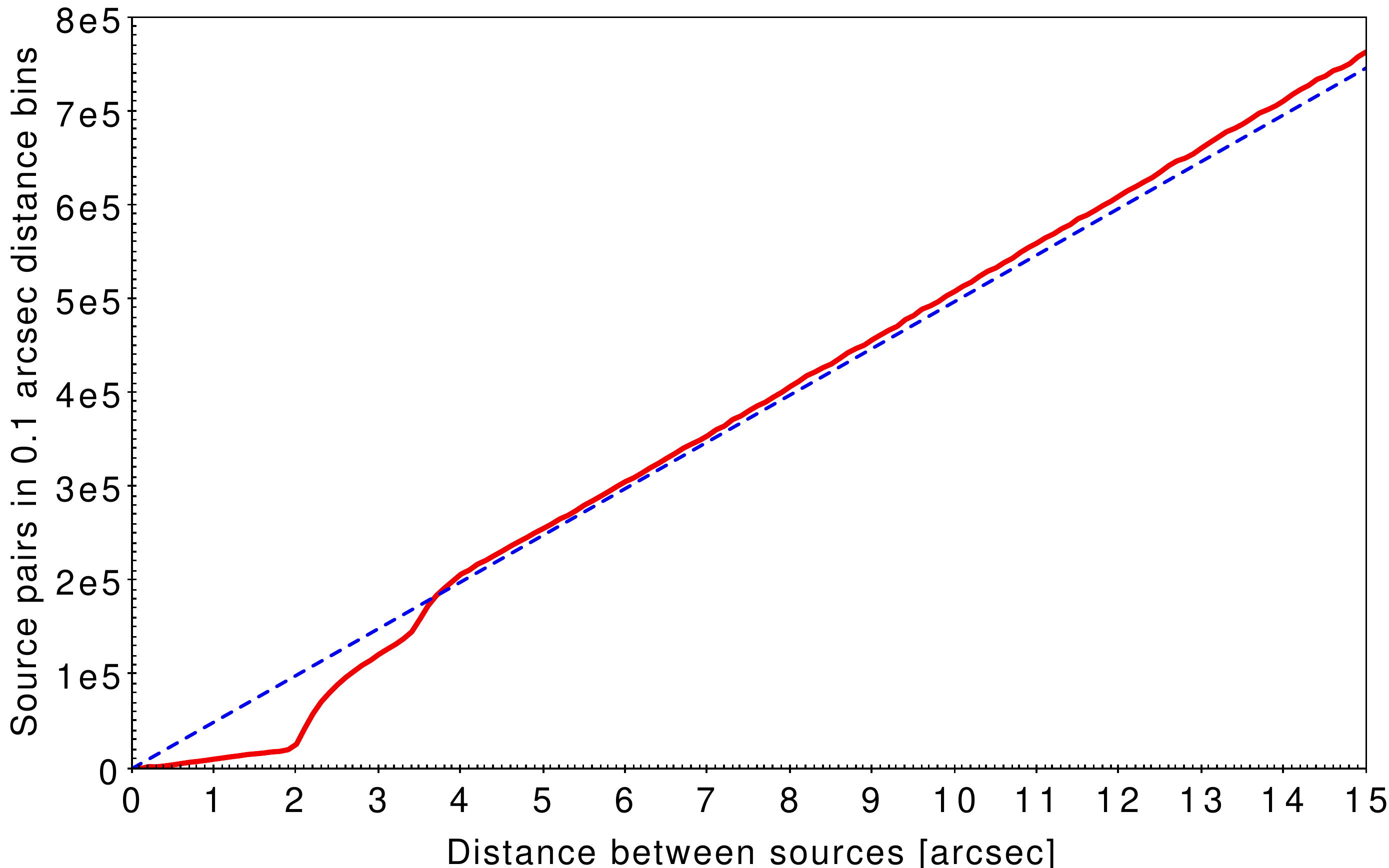}
\includegraphics[width=0.49\columnwidth,height=0.4\columnwidth]{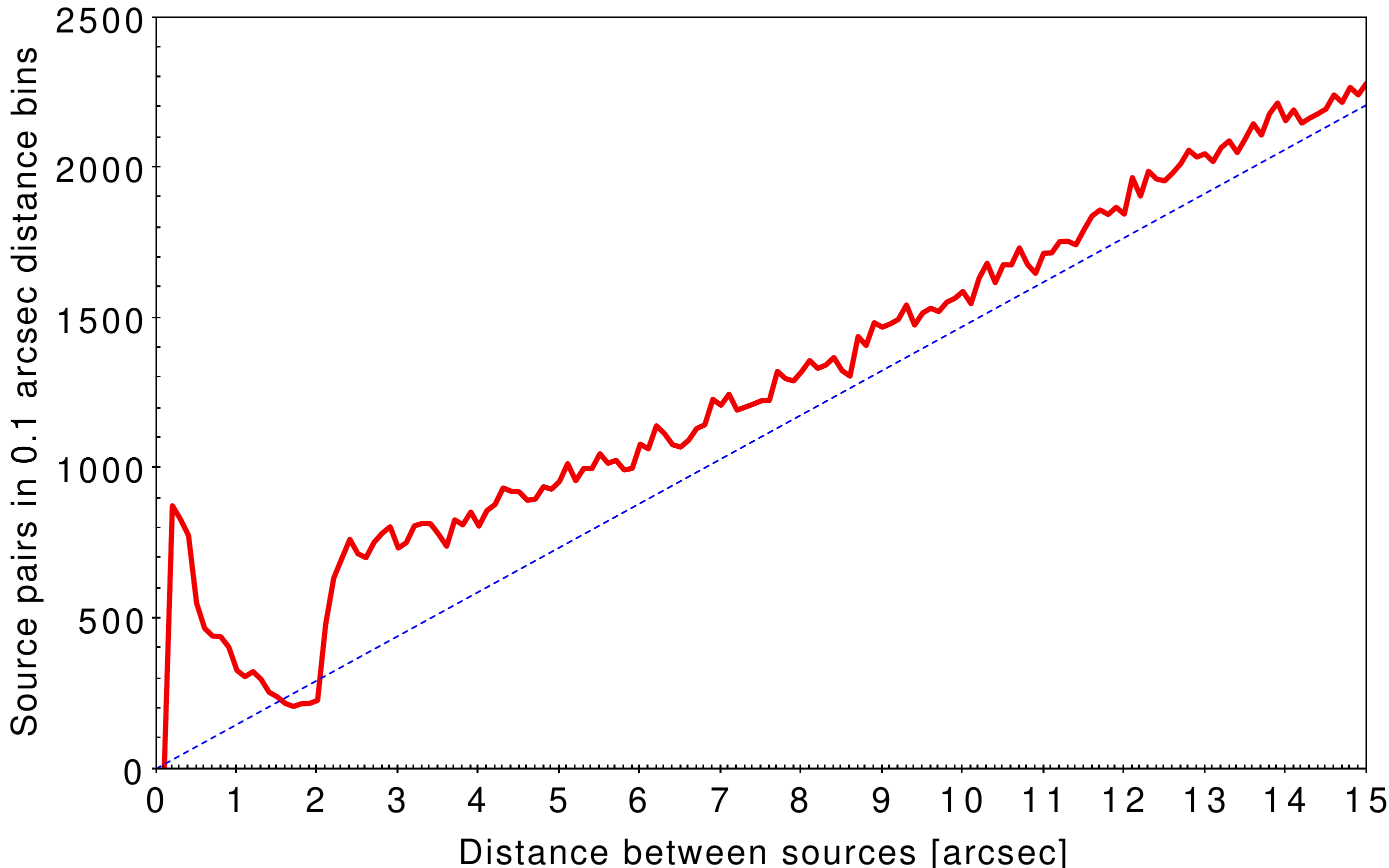}
\caption{Distribution of source to source distances in {\GDR1} for a dense
($l=330$\deg, $b=-4$\deg, $\rho=2$\deg, left) 
and sparse ($l=260$\deg, $b=-60$\deg, $\rho=15$\deg, right) star field. 
The dashed lines show the
relation corresponding to a random distribution of the sources.
\label{fig:942_pairs}}
\end{figure}

The resulting distributions of distance between sources are shown in
\figref{fig:942_pairs}. For the dense field (left) the distribution is
close to random for separations above 4\arcsec, but drops for smaller
separations with a sharp drop at 2\arcsec. In the shallow field, which is much 
larger and not as uniform, the sharp drop between 2\arcsec\ and 2\farcs5 is also
seen, but not the drop at 3\farcs5. In order to improve the uniformity of the sparse
field, three small areas around galaxies and clusters were left out when deriving 
the distribution.

\begin{figure}
\centering
\includegraphics[width=0.49\columnwidth,height=0.4\columnwidth]{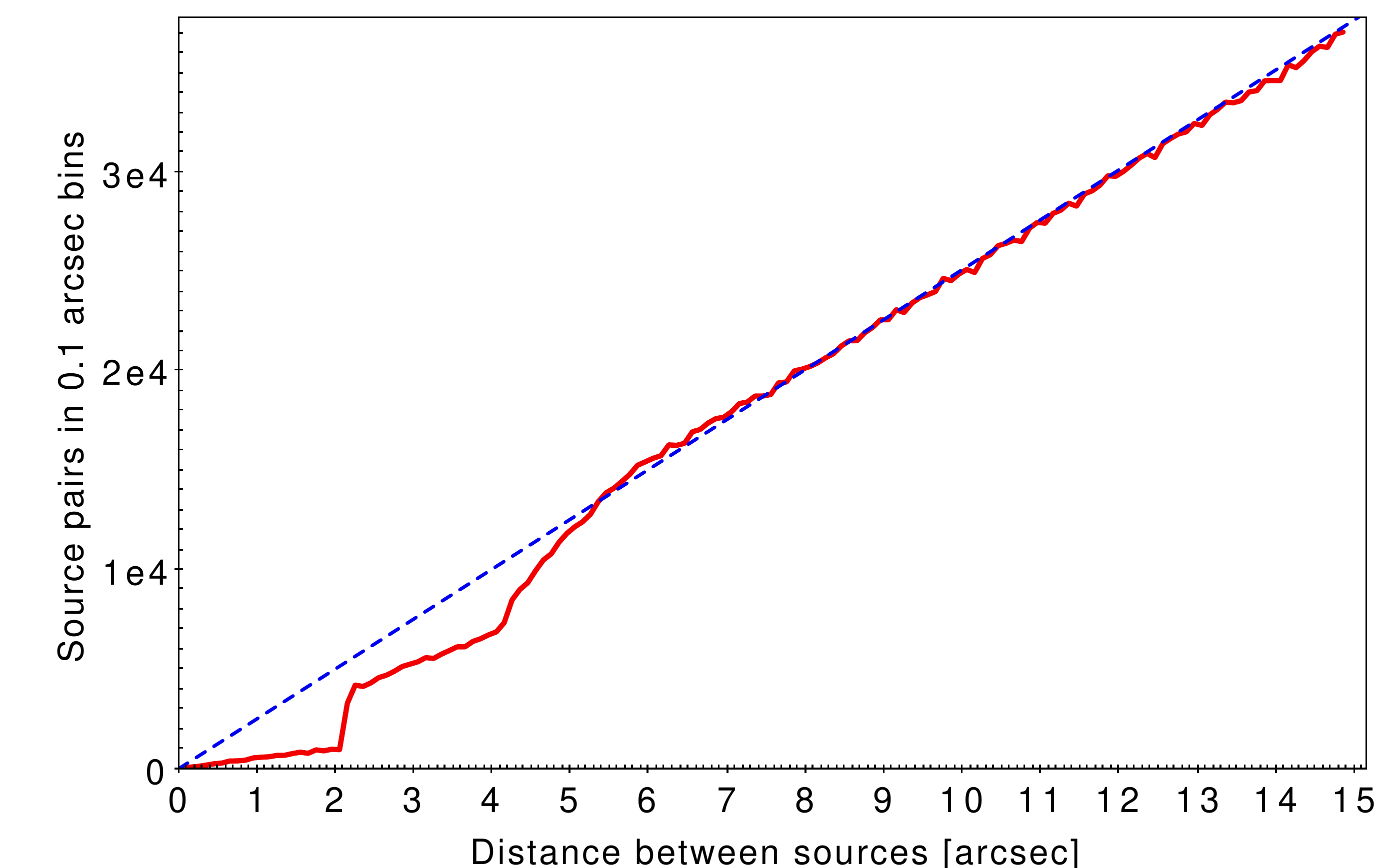}
\includegraphics[width=0.49\columnwidth,height=0.4\columnwidth]{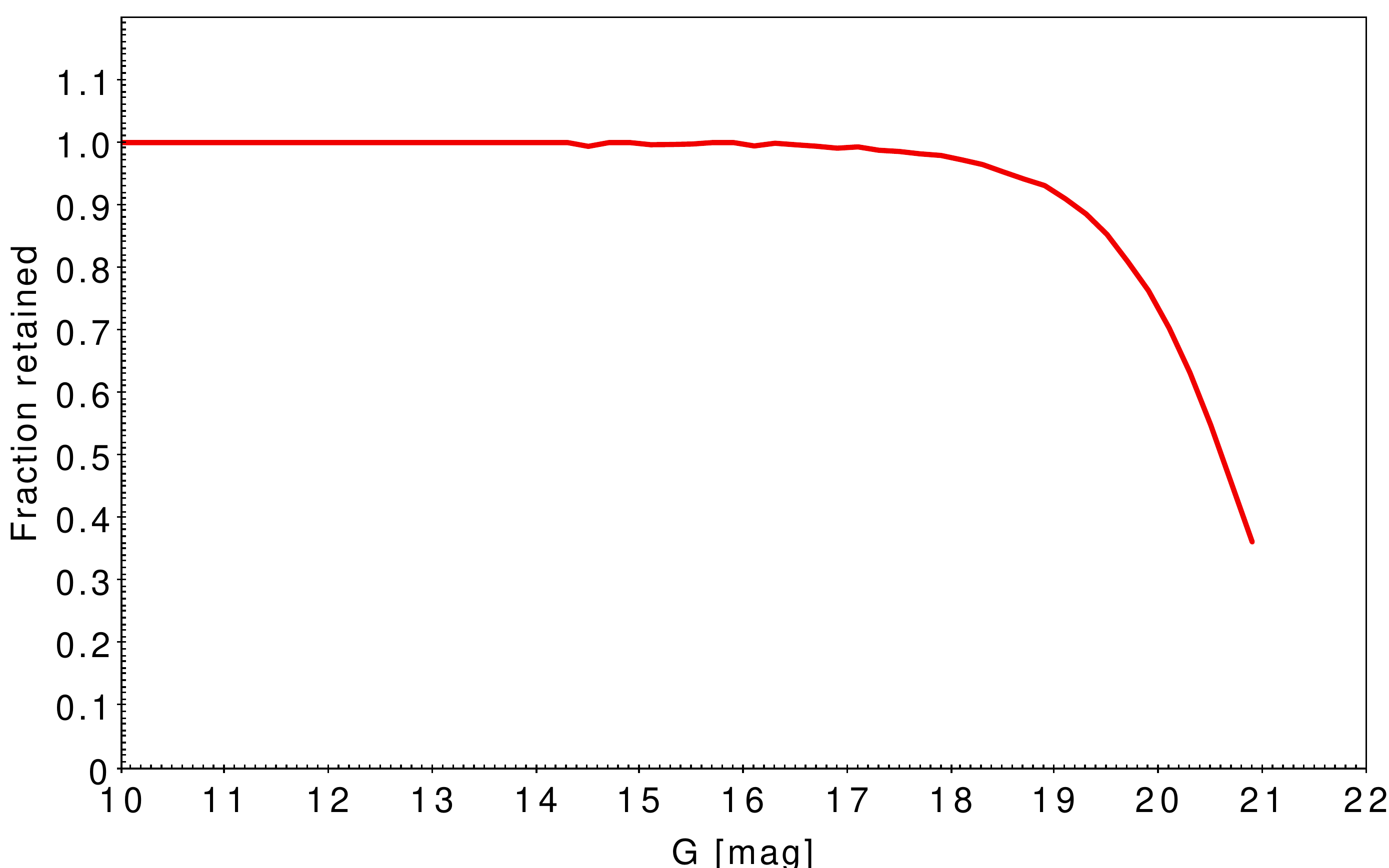}
\caption{Simulation of the distribution of source to source distances in 
a dense, random field (left) after applying selection criteria similar to
Gaia DR1. The fraction retained is shown in the right panel.
The field has a true source density of 500\,000 stars per square degree,
but only 322\,000 remain after applying the selection criteria.
\label{fig:942_simul}}
\end{figure}

To better understand these results, we made a simple simulation of a dense,
random field, starting with 500\,000~stars in a square degree. We then removed
sources which had very poor chances of ever getting a clean photometric
observation. The photometric windows are quite large, 2\farcs1 in the across
scan direction and a diagonal size of 4\farcs1. If a source had either a
significantly brighter neighbour within 2\farcs1 or at least two such
neighbours between 2\farcs1 and 4\farcs1, it was removed. We took neighbours 
brighter by more than 0.2~mag. The criterion of two bright
neighbours is very simplistic and is taken to represent the cases where a star
is unlikely to ever get a clean photometric observation, irrespective of the
scanning direction.
Figure \ref{fig:942_simul}a shows the resulting distribution, which reproduces
many of the same characteristics seen in the real data (separations below 4\arcsec) shown in
\figref{fig:942_pairs}a.

We can therefore expect that the population of pairs closer than 2\arcsec\
consists of sources of similar brightness, where in a given transit either
source had a fair chance of being detected as the brighter and therefore got a
full observation window instead of the truncated window assigned to the fainter
detection in case of overlapping windows. 
For a brief description of the on-board conflict resolution see e.g.
\citet[][Sect.\ 2]{DPACP-7}.
There is of course still the risk, that a few of the closest pairs are in
reality two catalogue instances of the same source (duplicates) as discussed in
\secref{sec:duplicates}. 

We can now further understand the drop between 2\arcsec\ and 4\arcsec\ as being
due to conflicts between the photometric windows for the sources. This drop is
not present in the sparse field, where the chance of having two disturbing
sources in the right distance range is much smaller than in a dense field.

An important lesson from the simulation is illustrated in the second panel
of \figref{fig:942_simul}. Of the original 500\,000 stars in the simulation
only 322\,000 (64\%) survived the selection criteria described above. 
This has a significant impact on the fainter couple of magnitudes.

Below 2\arcsec\ separation, the dense field shows the expected small fraction of
field stars of similar magnitude. However, the sparse field shows a peak below
half an arcsecond, suggesting a high frequency of binaries in that area. We
looked in more detail at the 73 pairs brighter than 12\,mag to see if the Tycho 
Double Star Catalogue \citep[TDSC,][]{2002A&A...384..180F} could confirm 
the duplicity. Of the 65 pairs found in Tycho-2, 
47 are listed as doubles in TDSC, while 7 may be doubles missing in TDSC,
and 11 are possibly duplicated {\gaia} sources. 
This small test thus indicates that the majority of the {\GDR1} doubles are
actual double stars.


\subsubsection{Tests of the angular resolution using the WDS}\label{ssec:angular-resolution-944}

The spatial resolution of the Gaia catalogue has also been tested using the Washington Visual Double Star Catalogue \citep[WDS,][]{WDS}. A selection was made of sources composed of only 2 components, with the magnitudes for both the primary and the secondary brighter than 20 mag and a separation smaller than 10\arcsec. Sources had also to have been observed at least twice with differences between the two observed separations smaller than 2\arcsec and magnitude differences had to be smaller than 3 magnitude, and must not have a note indicating an approximate position (!), a dubious double (X), uncertain identification (I) nor photometry from a blue ($B$) or near-IR band ($K$). 
The resulting selection contains 43\,580 systems. The completeness of {\GDR1} versus the observation of these systems shows the performance of Gaia detection and observation of double systems as a function of the separation and magnitude difference between the components. 

\begin{figure}
    \begin{center}
        \includegraphics[width=0.48\columnwidth]{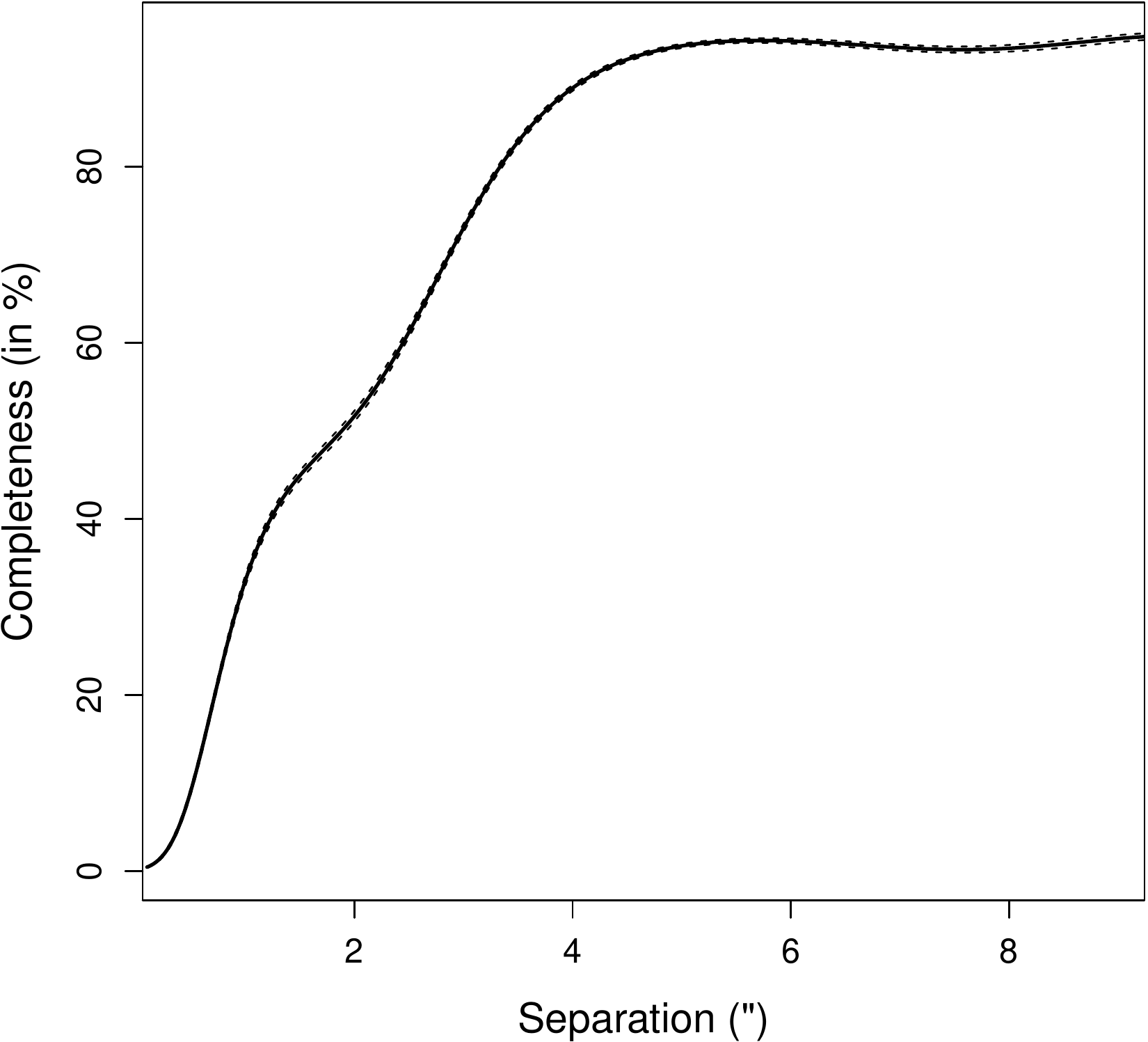}
        \hspace{0.05\columnwidth}
        \includegraphics[width=0.42\columnwidth]{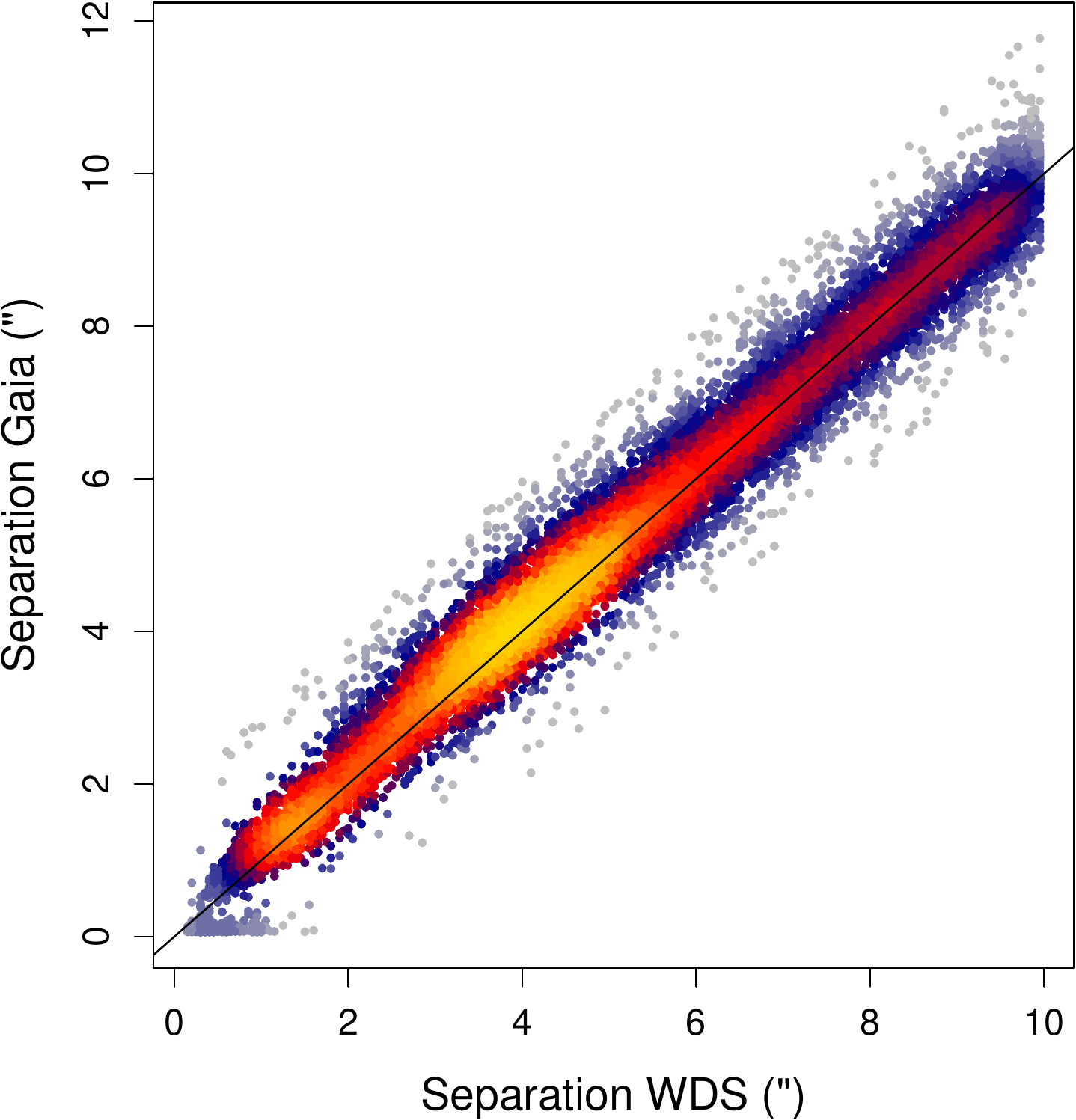}
        \caption[Visual double stars completeness versus WDS]{\beforeReferee{Visual binaries completeness}\afterReferee{Completeness of double stars} versus WDS; a) Completeness as a function of the \beforeReferee{visual binary separation}\afterReferee{separation between components}; 
        b) Separation between the components found with Gaia vs WDS separation (arcsecond).}
        \label{fig:wp944_WDS} 
    \end{center}
\end{figure}

The results are illustrated by a plot of completeness versus separation \beforeReferee{and 2D plots of completeness versus separation and magnitude difference, }presented in \figref{fig:wp944_WDS}a. As discussed in previous section, the angular resolution of {\GDR1} degrades rapidly below 4\arcsec. Although the filtering of {\preDR1} removed most of the duplicated sources, 
the excess of points with a very small Gaia separation and a WDS separation below about 1\arcsec\ 
in \figref{fig:wp944_WDS}b shows that a few duplicates ($\sim$0.5\% of the WDS sample) may still be present.

\subsection{Summary of the Catalogue completeness}\label{sec:summary-content}

A large filtering has been done on the main {\gaia} database to avoid spurious 
stars, for example a minimum of 5 focal plane transits for a star to be published 
in {\GDR1}. Due to the scanning law, and the resulting varying number of observations, 
the consequence is that some sky regions have a poor coverage, or are, locally, 
not covered at all. On the positive side, the filtering has succeeded to avoid 
spurious stars or ghosts which could be produced in the surroundings of bright stars,
or at least our statistical tests did not detect special features due to false
detections.

The limiting magnitude is therefore very inhomogeneous over the sky,
and the completeness as a function of magnitude is as well inhomogeneous:
starting from $G=16$ some sky zones appear clearly incomplete. 
Dense areas are, as expected, more affected due to the window and gate conflicts and the lack of on-board resources \citep{DPACP-18}.
High extinction regions also suffers from an increased, colour dependent, 
completeness issue due to the removal of the very red sources by the photometric pipeline \citep{DPACP-12}.

Duplicate sources which have been one of the main problems of {\preDR1} have mostly
been removed, although not completely, and their effect on the astrometric
or photometric properties of a fraction of bright star is probably still present.

Due to the preliminary nature of this data release the effective
angular resolution of the {\GDR1} data (not the angular resolution of
the Gaia instrument itself which is as expected) is also degraded, with
a deficit of close doubles.  In sparse regions, however, the spatial
capabilities of Gaia may already overcome the ground-based ones.  

As for TGAS, a significant fraction (20\%) of Tycho-2 stars are not present,
also due to the scanning coverage and to calibration problems, in particular at 
the bright end. A large fraction of high proper motion stars are missing, 
and those redder and fainter. 

It thus appears that {\GDR1} is not complete in any sense (magnitude,
colour, volume, resolution, proper motion, duplicity, etc.), so that any
statistical analysis should be careful to produce unbiased results.

The current completeness is however not representative of the future Gaia capabilities. 
That this will be corrected at the next data release triggers another
warning for the users preparing star lists: the \dt{source\_id} list present in DR2 
(and further releases) may be partly different from {\GDR1}. 
On one hand the gains to expect on the 
cross-matching performances (at small angular separations) and the larger number 
of transits (i.e. less stars with not enough observations to be published)
imply that many more stars will be present in DR2. On the other hand, a significant
number of \dt{source\_id} may disappear, caused by both splitting and merging sources.

\section{Multidimensional analysis}\label{multidim}

\subsection{Description of statistical methods}\label{sec:stat}

To understand whether the statistical properties of the {\GDR1} dataset
are consistent with expectations, we compared the distribution of the data  
(and in particular their degree of clustering) to suitable simulations 
for all two-dimensional subspaces. In the case of
TGAS, the comparison data is the simulation designated as
``Simu-AGISLab-CS-DM18.3cor'' (\secref{agislab}), while for {\GDR1} it is GOG18. 

To this end, we use the Kullback-Leibler divergence (KLD):
\begin{equation}
\label{eqn:KLD_wp945}
p_\mathrm{KLD} = -\int {\rm d}^2 x p({\bf x}) \log p({\bf x})/q({\bf x})
\end{equation}
where ${\bf x}$ is a (sub)space of observables, $p({\bf x})$ is the
distribution of the observables in the dataset, and $q({\bf x})$ is
some comparison distribution. When $q({\bf x}) = \Pi_i p_i(x_i)$,
i.e. the product of the marginalized 1D distribution of each of the
observables, the KLD gives the mutual information. This expression
shows that the mutual information is sensitive to clustering or
correlations in the dataset, with a high degree leading to large values
while in their absence $p_\mathrm{KLD}$ would be zero.

We thus computed $p_\mathrm{KLD}$ for more than 300 subspaces for the
data, as well as for the simulations. In both cases, we used a range
for the observables defined by the data after 3-$\sigma$ clipping the
top and bottom regions. Since the simulated and the observed data can
have different distributions without this necessarily implying a
problem in the data, we prefered to work with the relative mutual information rankings. If
the structure is similar in data and simulations, we expect the
rankings to cluster around the one-to-one line, while if a subspace
shows very different rankings this would imply very
different distributions. Such a subspace (or observable) is flagged
for further inspection. This is important since the number of
subspaces is very large.

The comparison to the simulations is sensitive to global issues
(across the whole sky), while there could potentially be systematic
problems in the data restricted to small localized regions of the
sky. Therefore, we also compared the values of the mutual information
obtained for different regions of the sky (e.g. symmetric with respect
to the Galactic plane) and with similar number of observations.

\subsection{Results from the KLD statistical methods}\label{sec:res_stat}

\subsubsection{TGAS and comparison to AGISLab simulations}

\begin{figure}
\begin{center}
\includegraphics[width=0.8\columnwidth]{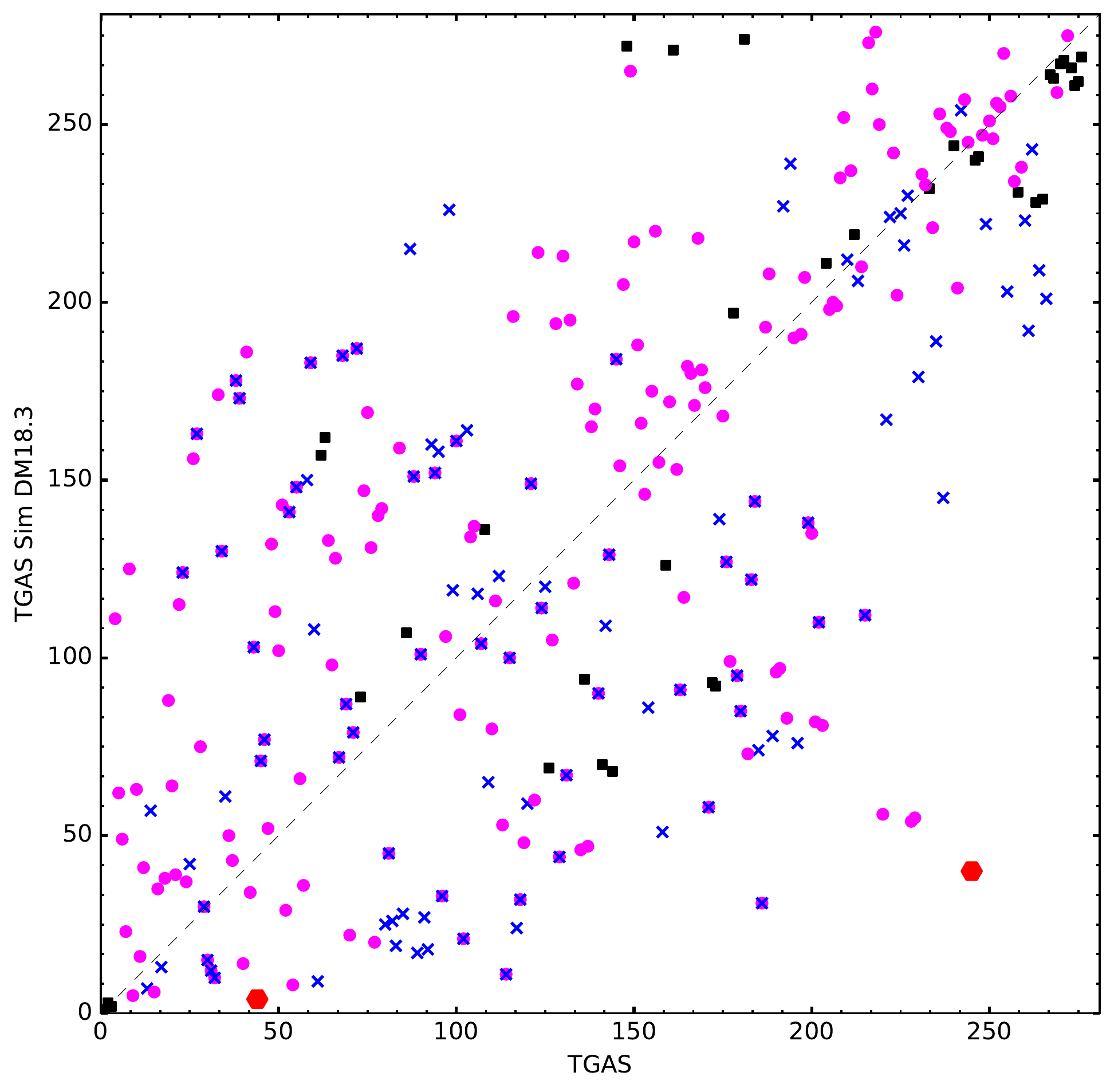}
\caption{Ranking of two-dimensional subspaces according to their mutual
information in the TGAS data (x-axis) vs. the simulation (y-axis). The black squares correspond to subspaces formed only from observables, while the blue crosses are those containing an uncertainty, and the 
magenta circles contain a correlation parameter. The red hexagons correspond to the subspaces shown in \figref{fig:subspace-sims-example}.
}\label{fig:rank_wp945}
\end{center}
\end{figure}

\begin{figure}
\begin{center}
\includegraphics[width=0.8\columnwidth]{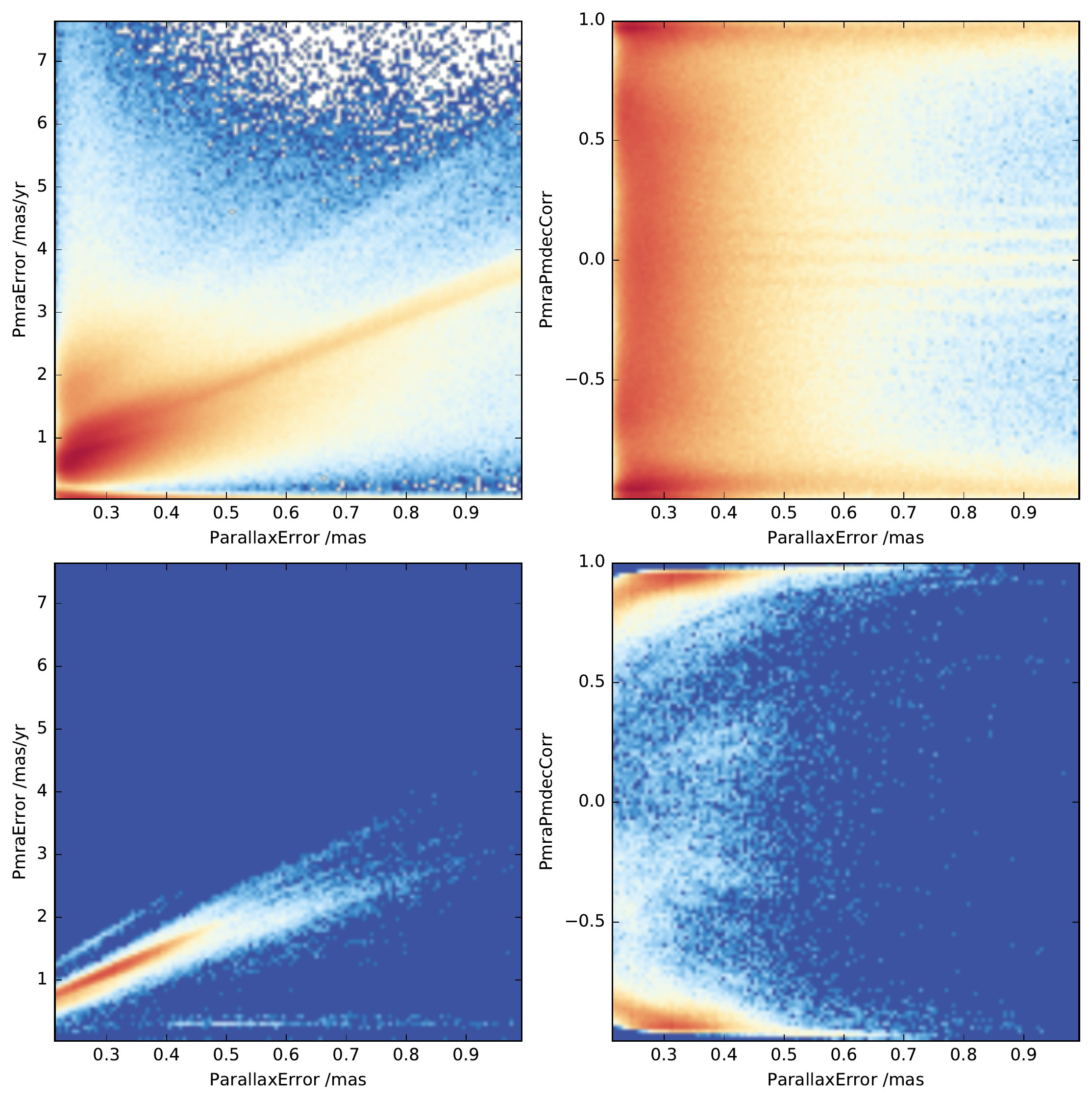}
\caption{Examples of the subspaces showing a strong deviation from the 1:1 expected relation shown in \figref{fig:rank_wp945}, particularly in the astrometric errors (left) and correlations (right) in TGAS (top) compared to those in the simulations (bottom).}\label{fig:subspace-sims-example}
\end{center}
\end{figure}

Figure \ref{fig:rank_wp945} shows the mutual information ranking of
the two-dimensional subspaces from the TGAS data versus the ranking of
the same subspaces in the AGISLab simulation. Most subspaces with
direct observables (e.g. \dt{ra}, \dt{dec}, etc., black points) show
very similar distributions in the data and in the simulations, as
evidenced by their closeness to the 1:1 line.  Subspaces associated to
errors (blue crosses) and to correlations between observables/errors
(magenta circles), tend to deviate more in general. Examples of the
distributions found for some of the subspaces deviating more strongly
(red hexagons in \figref{fig:rank_wp945}) are given in
\figref{fig:subspace-sims-example}.

\subsubsection{TGAS comparison in different sky regions}

\begin{figure}
\begin{center}
\includegraphics[height=0.49\columnwidth, width=0.53\columnwidth]{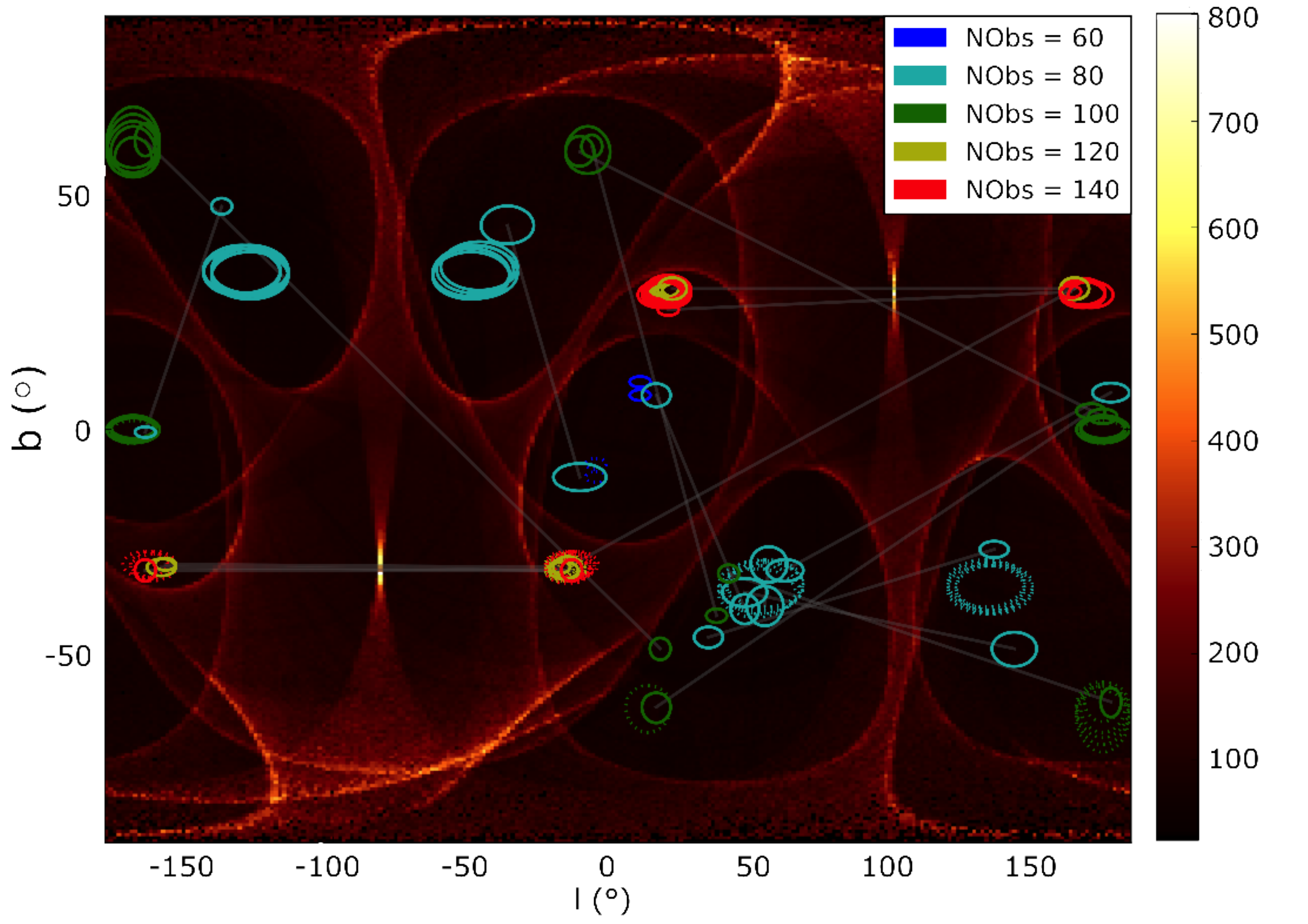}
\includegraphics[height=0.49\columnwidth, width=0.46\columnwidth]{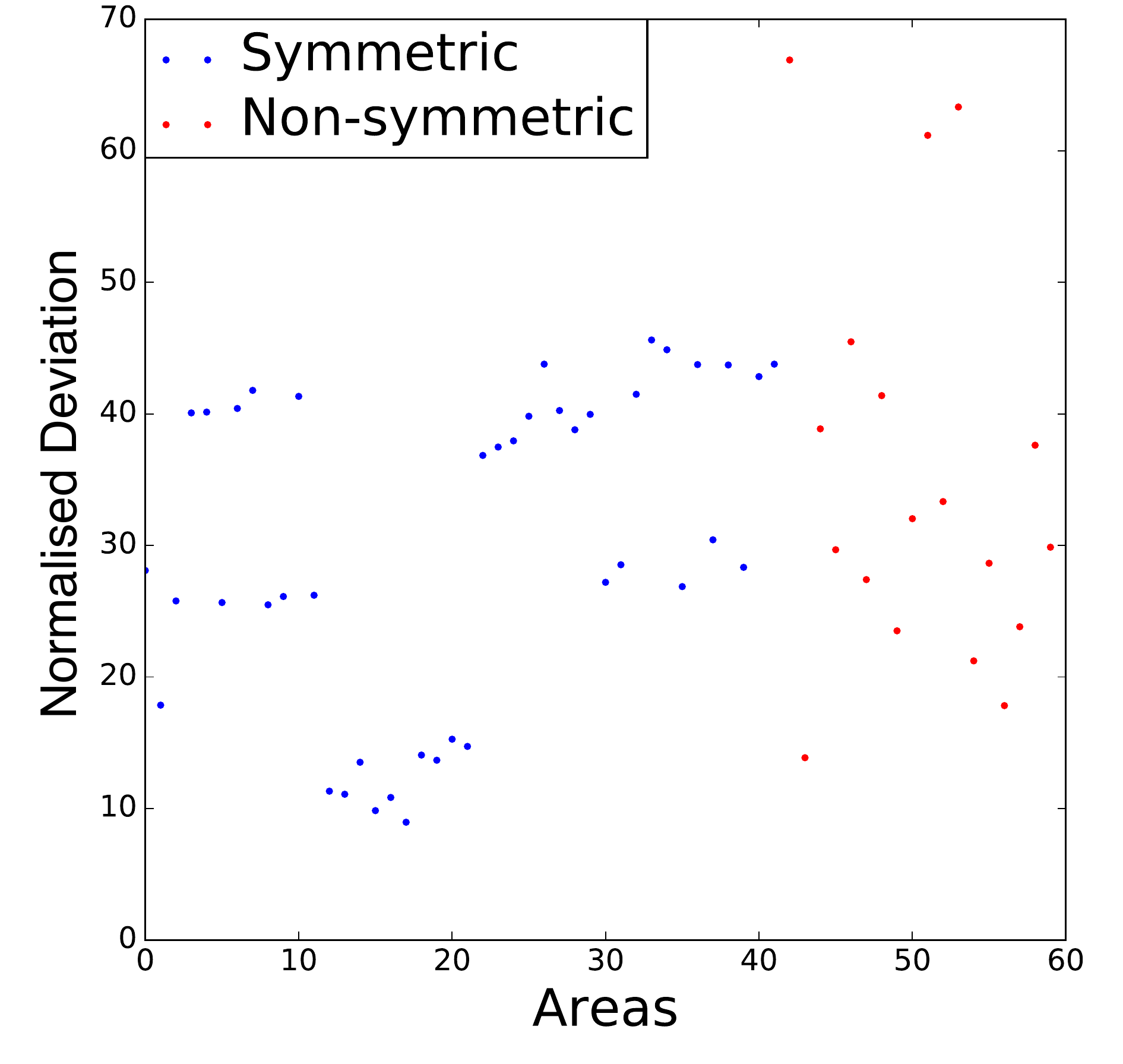}
\caption{Left: Distribution of regions for which the mutual information has been computed, where the inset indicates the number 
of observations inside the regions. The regions are circles in $l-\sin b$ space, with the positive $b$ region in solid and its symmetric counterpart in dashed. Regions that are compared and which are not symmetric are connected by a grey line. Right: average deviation of the mutual information between a region and its counterpart, in (red) blue  for (non) symmetric counterparts.}\label{fig:tgas_sky}
\end{center}
\end{figure}

Naively, one might expect regions with similar number of observations
to have similar distributions of errors, and if symmetric with respect
to the Galactic plane or centre, perhaps also in the distribution of several
of the observables. To check for the presence of \beforeReferee{systematicities} \afterReferee{systematics} in
the data, we selected 60 regions with a similar \dt{astrometric\_n\_obs\_al} 
(in the range 60 to 140), of which (20) 40 have a (non-)symmetric 
counterpart.  The left panel of \figref{fig:tgas_sky}
shows their distribution in Galactic coordinates. For these regions we
have computed the mutual information and compared the values to their
counterpart. The normalised deviation from the naively
expected 1:1 line is plotted in the right panel of 
\figref{fig:tgas_sky}, and is defined as $ \sum_i |p_{i,KLD} -
p_{i,KLD}^*|/[0.5*(p_{i,KLD} + p_{i,KLD}^*)]$, where $i$ runs through
the various subspaces and $p$ and $p*$ are the mutual information for
the region and its counterpart. Blue and red points correspond to comparisons between symmetric and 
non-symmetric regions respectively. This plot shows that non-symmetric
regions sometimes have \beforeReferee{more} different distributions. 
By dividing the normalised deviation (whose median value is $\sim 30$) 
by the number of subspaces (780 for TGAS) we obtain an estimate
of the average deviation per region. In this way we found that on
average there are 4\% differences in the mutual information between
different regions. Comparison to the results of AGISLab simulations
does not reveal pairs of regions whose mutual information appear to be
very different for specific subspaces.

\subsubsection{{\GDR1} comparison to GOG simulations}

\begin{figure}
\begin{center}
\includegraphics[width=0.8\columnwidth]{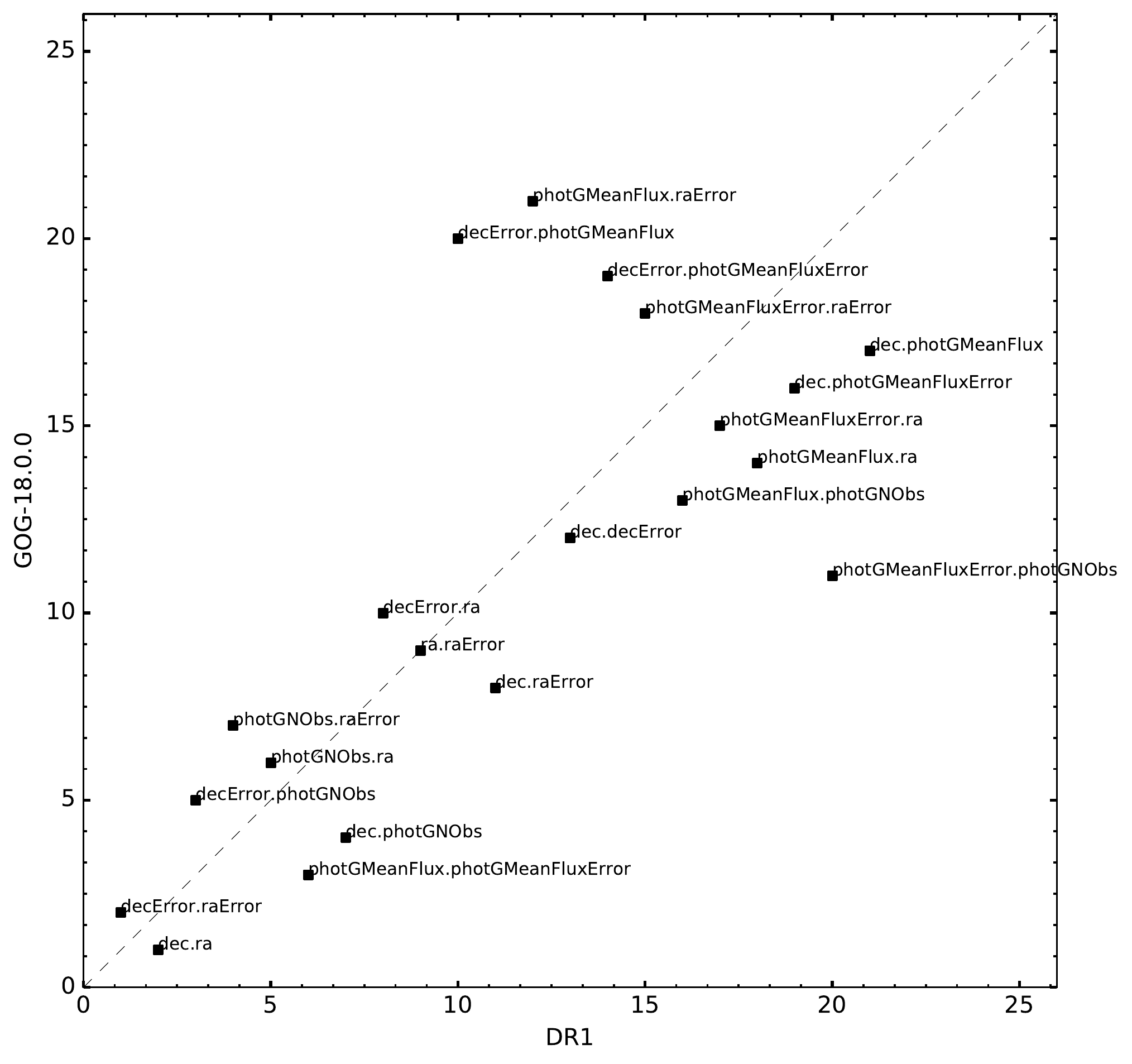}
\caption{Ranking of two-dimensional subspaces according to their mutual
information in the {\GDR1} data (x-axis) vs. the GOG simulation (y-axis).}\label{fig:rank_dr1_wp945}
\end{center}
\end{figure}

In \figref{fig:rank_dr1_wp945} we show the rankings obtained for the
observables and their errors in the full {\GDR1} Catalogue. Because of the smaller number of
observables, only 21 subspaces exist. The relation of
the mutual information in data and simulations is very close to the
1:1 line, implying similar distributions and hence a good
understanding of the data as far as this global statistic can
test. The observables showing the greater deviations are those related
to uncertainties, and this can be understood from the fact that GOG18 models 
the uncertainties expected at the end of mission, rather than those obtained after 14 months of
observations. 

\section{Astrometric quality of {\GDR1}}\label{astroqual}

For the majority of the sources included in {\GDR1}, the 1\,140\,622\,719 {\it secondary sources}, the only available astrometric parameter is the position. 
For the 2\,057\,050 {\it primary sources}, the TGAS subset, the complete set of astrometric parameters is available: position, trigonometric parallax and proper motion. 
As a consequence, most tests concerning astrometry have been devoted to TGAS validation and only \secref{sec:DR1pos} deals with tests on secondary sources astrometry.

We study in \secref{sec:accuracy-plx} the accuracy of the TGAS parallaxes, and in \secref{sec:precision-plx} their precision. 
In both cases, we discuss first the estimation done using internal ({\gaia} only) data, then with external data.
Table \ref{tab:cu9val_wp944_summaryplx} gives a summary of the difference between the TGAS parallaxes and those from external catalogues that are presented in this section. 

\begin{table*}
\caption{Summary of the comparison between the TGAS parallaxes and the external catalogues. The number of outliers (at 5$\sigma$) versus the total number of stars is presented. The parallax difference ($\varpi_G-\varpi_E$, in mas) and the extra uncertainty (in mas) that needs to be quadratically added or subtracted to the data to adjust the residuals are indicated in red [green] when they are [not] significant (p-value limit: 0.01). A global estimate of the parallax offset as given by the weighted average of these various tests is $-0.036\pm 0.002$ mas, very similar to the estimate found using quasars, and the median of the extra dispersion is $-0.14\pm 0.08$ mas.
\label{tab:cu9val_wp944_summaryplx}}
\centering
\begin{tabular}{lccc} 
\hline\hline
{\bf Catalogue} & {\bf Outliers} & {\bf \parallax~difference} & {\bf \parallax~extra dispersion} \\
 \hline
 Hipparcos & 0.09\% & \textcolor{red}{$-0.094 \pm 0.004$} & \textcolor{red}{$0.58\afterReferee{0} \pm 0.005$} \\
  \hline
 VLBI & 0 / 9 & \textcolor{green}{$0.083 \pm 0.12$} & - \\
 HST & 2 / 19 & \textcolor{green}{$-0.11 \pm 0.19$} & \textcolor{green}{$0.6 \pm 0.2$} \\
 RECONS & 0 / 13 & \textcolor{green}{$-1.04 \pm 0.58$} & \textcolor{green}{$-0.9 \pm 0.5$} \\
 VLBI \& HST \& RECONS & 2 / 41 &  \textcolor{green}{$-0.08 \pm 0.12$} & \textcolor{green}{$0.42 \pm 0.13$} \\
 \hline
 Cepheids & 0 / 207 & \textcolor{green}{$-0.014 \pm 0.014$} & \textcolor{red}{$-0.18 \pm 0.01$} \\
 RRLyrae & 0 / 130 & \textcolor{red}{$-0.07 \pm 0.02$} & \afterReferee{\textcolor{red}{$-0.16 \pm 0.02$}} \\
 Cepheids \& RRLyrae & 0 / 337 & \textcolor{red}{$-0.034 \pm 0.012$} & \textcolor{red}{$-0.17 \pm 0.01$} \\
 \hline
 RAVE & 47 / 5144 & \textcolor{red}{$0.07\afterReferee{0} \pm 0.005$} & \textcolor{green}{$-0.06 \pm 0.02$} \\
 APOGEE & 0 / 2505 & \textcolor{red}{$-0.06\afterReferee{0} \pm 0.006$} & \textcolor{red}{$-0.12 \pm 0.01$} \\
 LAMOST & 6 / 317 & \textcolor{green}{$-0.01 \pm 0.02$} & \textcolor{red}{$-0.17 \pm 0.02$} \\
 PASTEL & 1 / 218 & \textcolor{green}{$0.05 \pm 0.02$} & \textcolor{green}{$0.1\afterReferee{0} \pm 0.05$} \\
 APOKASC & 0 / 969 & \textcolor{red}{$-0.07\afterReferee{0} \pm 0.009$} & \textcolor{red}{$-0.15 \pm 0.01$} \\
 \hline
 LMC & 2 / 142 & \textcolor{red}{$0.11 \pm 0.02$} &  \textcolor{green}{$-0.14 \pm 0.03$} \\
 SMC & 0 / 58 & \textcolor{red}{$-0.12 \pm 0.05$} &  \textcolor{green}{$-0.09 \pm 0.09$} \\
  \hline
 ICRF2 QSO auxiliary solution & 1 / 2060 & \textcolor{red}{$-0.046 \pm 0.01\afterReferee{0}$} & \textcolor{red}{$-0.17 \pm 0.01$} \\
 \hline
\end{tabular}
\end{table*}

\subsection{TGAS Parallax accuracy}\label{sec:accuracy-plx}

\subsubsection{Parallax accuracy using quasars}\label{sec:QSO-plx}
In the course of the AGIS astrometric solution, about 135\,000 quasars were included
and solved for parallax and positions, with proper motions being constrained with 
a prior near zero{\masyr} \citep[][Sect. 4.2]{2016A&A...586A..26M,DPACP-14} and made available
for validation (and are not part of {\GDR1}).
As the true parallax for quasars can be considered as null, the study of these parallaxes gives a direct 
information on the properties of the parallax errors. Unfortunately, the available
quasars cover part of the sky only, and in particular they can give little insight
inside the galactic plane. 

The median zero-point of the quasar parallaxes is significantly non-zero: $-0.04\afterReferee{0}\pm 0.003$ mas. 
This is close to the value for the ICRF2 QSO subsample, see \tabref{tab:cu9val_wp944_summaryplx},
and corroborated by other all sky external comparisons in this table and discussed in more details below,
and this is what we adopt as average {\GDR1} parallax zero-point. 

We selected random sky regions with 2{\degr} radius, keeping only those possessing at least 20 quasars,
and computed median parallaxes in these regions. 
The map of the median parallaxes in these regions is represented 
\figref{fig:cu9val_942_QSOKsky}. \afterReferee{Outside of the galactic plane
where the lack of objects (see \figref{fig:cu9val_942_QSOrhoplx}) brings little information, 
there are} large scale spatial effects with characteristic 
amplitude of about 0.3 mas (significant at $2\sigma$). In a few (exceptional) small regions, the 
parallax bias may even reach the mas level. 

\begin{figure}
\begin{center}
\columnImage{{figures-942/TGAS_01.00-ecl-qso-reg-MedPlx-r2}.png}
\caption{\afterReferee{Median parallaxes of quasars in 2{\deg} radius regions (mas), ecliptic coordinates. 
There is little insight in the galactic plane, due to the lack of objects.}
\beforeReferee{Due to the lack of objects (see \figref{fig:cu9val_942_QSOrhoplx}), the galactic plane 
bring few information.} Outside of it, local systematics
with about 0.3 mas characteristic amplitude can be seen.}\label{fig:cu9val_942_QSOKsky}
\end{center}
\end{figure}

\begin{figure}
\begin{center}
\includegraphics[width=0.49\columnwidth, height=0.45\columnwidth]{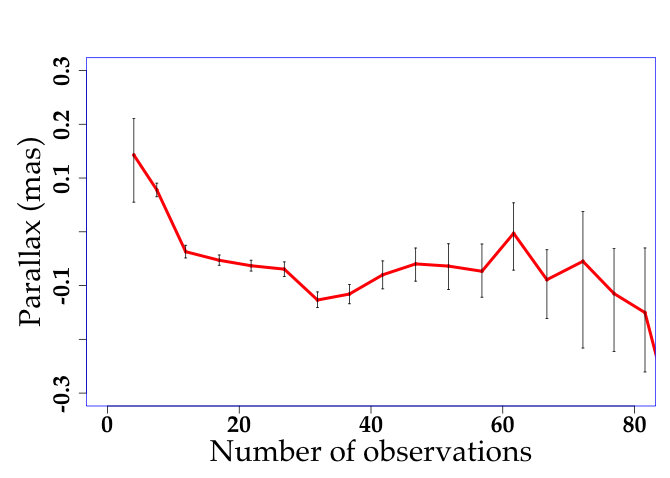}
\includegraphics[width=0.49\columnwidth, height=0.45\columnwidth]{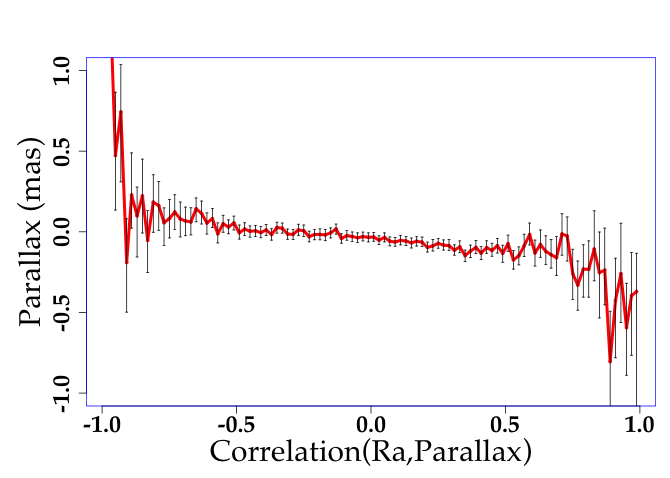}
\caption{Median quasar parallaxes (mas) vs number of observations (left)
and vs correlation between right ascension and parallax (right).}\label{fig:cu9val_942_QSOrhoplxsky}
\end{center}
\end{figure}

\begin{figure}
\begin{center}
\includegraphics[width=0.49\columnwidth, height=0.35\columnwidth]{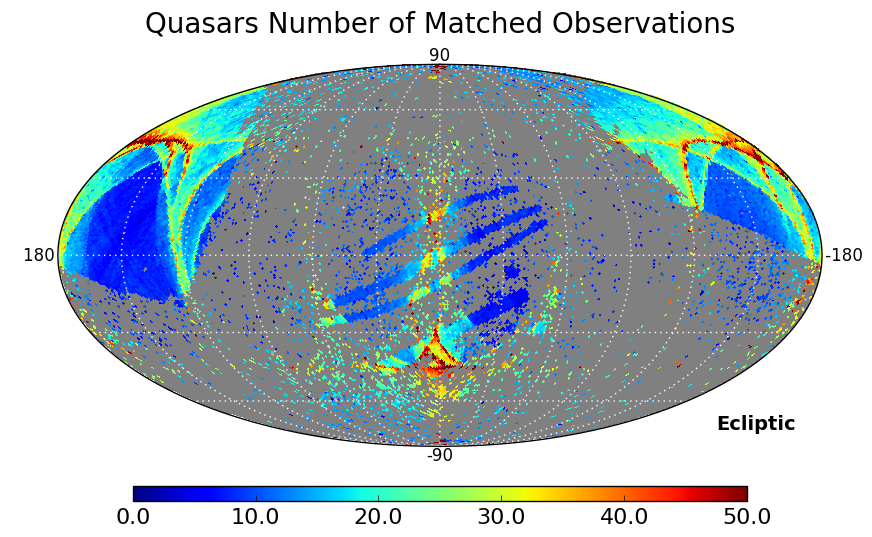}
\includegraphics[width=0.49\columnwidth, height=0.35\columnwidth]{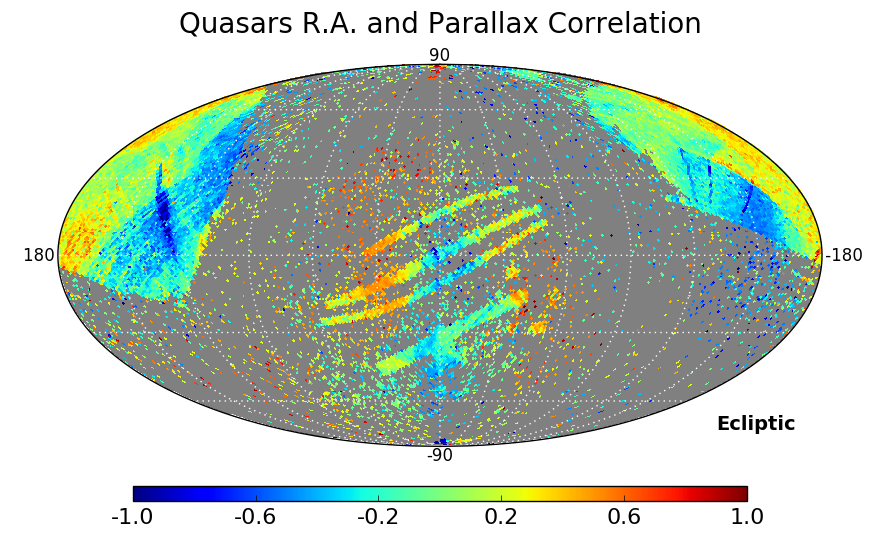}
\caption{Healpix map in ecliptic coordinates of the number of quasar observations (left)
and of the correlation between right ascension and parallax (right).}\label{fig:cu9val_942_QSOrhoplx}
\end{center}
\end{figure}

\begin{figure}
\begin{center}
\includegraphics[width=0.49\columnwidth, height=0.45\columnwidth]{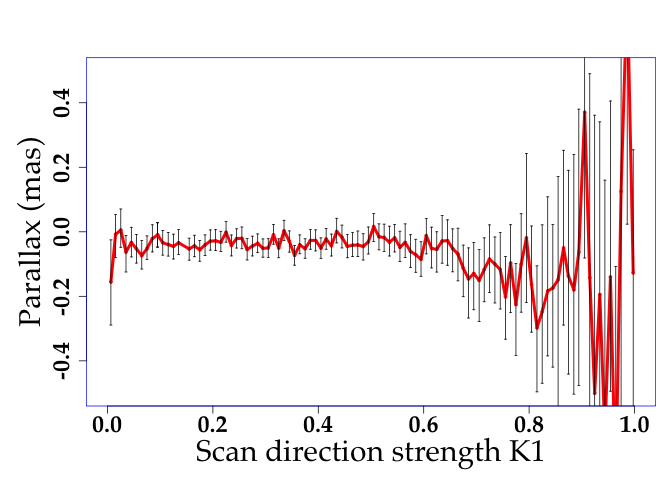}
\includegraphics[width=0.49\columnwidth, height=0.45\columnwidth]{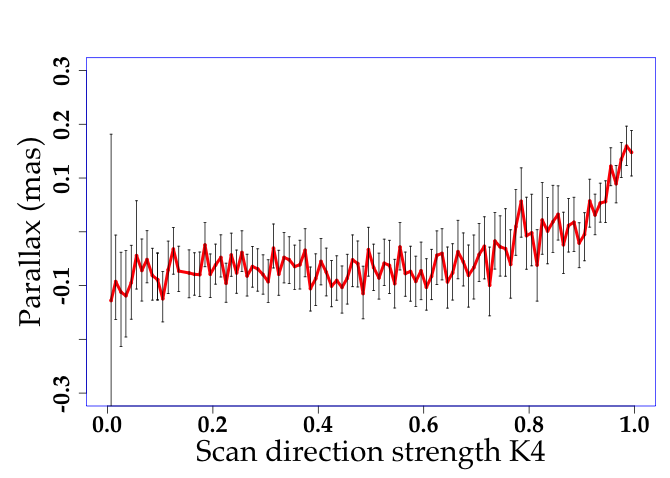}
\caption{Median quasar parallaxes (mas) vs 
scan direction strength K1 (left) and vs K4 (right).}\label{fig:cu9val_942_QSOK1K4}
\end{center}
\end{figure}

The bias variations are directly related to the number of 
measurements (\figref{fig:cu9val_942_QSOrhoplxsky}a, \ref{fig:cu9val_942_QSOrhoplx}a), 
and consequently to the standard uncertainties,
with also a 0.3 mas amplitude. Parallax biases look also related to the  
correlations between right ascension and parallax
(\figref{fig:cu9val_942_QSOrhoplxsky}b, \ref{fig:cu9val_942_QSOrhoplx}b).
In \figref{fig:cu9val_942_QSOKsky} and \figref{fig:cu9val_942_QSOrhoplx}, the regions along 
$\lambda\sim 0$ and 180{\deg} (ecliptic pole scanning law) appear clearly. 

As for the origins of these systematics, possible along-scan measurements 
problems, if \dt{scan\_direction\_strength\_k1}\footnote{The 
``scan direction strength'' fields in the Catalogue quantify the 
distribution of AL scan directions across the source and \dt{scan\_direction\_strength\_k1} 
is the degree of concentration when the sense of direction is taken into account; as for
\dt{scan\_direction\_strength\_k4}, a value near 1 indicates that the scans are concentrated in two
nearly orthogonal directions.} 
is a proxy for this, may be part of the reason (\figref{fig:cu9val_942_QSOK1K4}a), 
with some contribution from possible chromaticity problems. 
The \dt{scan\_direction\_strength\_k4}, associated to small 
numbers of observations,  \afterReferee{also} looks \beforeReferee{also} contributing (\figref{fig:cu9val_942_QSOK1K4}b) with 
here again a 0.3 mas amplitude. 

\afterReferee{It is important to stress that the map illustrating spatial variations of the 
parallax bias of the quasars, \figref{fig:cu9val_942_QSOKsky}, cannot be used to ``correct''
the parallaxes. The quasars are faint, and the TGAS parallaxes, which were 
obtained with a different astrometric solution, may
suffer from supplementary effects due to their bright magnitudes.}

\subsubsection{Parallax accuracy tested with very distant stars}\label{sec:LMCSMC}
The zero point of the parallaxes and their precision can also be tested directly by using stars in TGAS (or quasars, see previous subsection) distant enough so that their measured parallaxes can be considered as null according to the catalogue's expected precision. The normalized parallax distribution of those sources should follow a standard normal distribution. For TGAS we have been looking for stars with $\varpi<0.1$~mas. This limit has been chosen to be consistent with TGAS precision (estimated to be of the order of a few tenths of mas). For \GDR1, only the Magellanic Clouds contain enough confirmed members in TGAS for this test. 

 \paragraph{LMC/SMC.}
 A catalogue containing 250 LMC and 79 SMC Tycho-2 stars has been compiled from the literature: Hipparcos 
\citep[Annex 4 of][]{1992ESASP1136.....T}, \cite{1989ESASP1111B.191P}, \cite{2008AcA....58..163S}, \cite{2009AJ....138.1003B}, \cite{2009ApJS..184..172G}, \cite{2012ApJ...749..177N} for the LMC; Hipparcos \citep[Annex 4 of][]{1992ESASP1136.....T}, \cite{1989ESASP1111B.191P}, \cite{2010AcA....60...91S}, \cite{2004MNRAS.353..601E}, \cite{2010AJ....140..416B}, \cite{2010ApJ...719.1784N} for the SMC. 
 For the 46 Hipparcos stars included, the Hipparcos and Simbad information has been confirmed to be fully consistent with LMC/SMC membership.  
 
A mean parallax of 0.11$\pm$0.02~mas has been found for the LMC and -0.12$\pm$0.05~mas for the SMC with a small over-estimation (by 0.14~mas) of the uncertainties. None of these values is consistent with the all-sky zero-point and this indicates local variations of the parallax zero point across the sky, confirming the spatial variations found \secref{sec:QSO-plx}. Further filtering of the sources has been done by comparing the parallaxes and proper motions of the stars with the mean values of the clouds (taken from SIMBAD) through a $\chi^2$ test. Using a limit p-value of 0.01 on this $\chi^2$ test removes 20\% of the LMC stars (3\% of the SMC). The remaining stars still show a significant parallax bias although reduced as expected. A correlation of the parallax residual with magnitude is observed in all cases (with a larger residual for the brighter stars). This dependency with magnitude and the surprisingly large number of outliers indicated by the $\chi^2$ test are similar to the Hipparcos $\chi^2$ test results (Section \ref{sec:wp944_astrom}), suggesting that a filtering based on the covariance matrix is actually hiding Gaia related issues rather than LMC/SMC membership issues. 

\subsubsection{Parallax accuracy tested with distant stars}\label{sec:distantstars}

An estimation of the parallax accuracy can also be obtained with stars distant enough so that their estimated distance through period-luminosity relation or spectrophotometry is known with a precision better than $\sigma_{\varpi_\mathrm{E}}<0.1$~mas, i.e. much more precise than the TGAS parallaxes. A maximum likelihood method \citep[improved from][Sect. 4]{1995A&A...304...52A} has been implemented to estimate the offset and extra-dispersion that should be \beforeReferee{take}\afterReferee{taken} into account for the Gaia parallaxes to be consistent with these external distance estimates. 

Two catalogues have been tested using the period-luminosity relation: 

\paragraph{Cepheids.} The catalogue of  \cite{2012ApJ...747...50N} has been used. It provides distance modulus for the Cepheids using the Wesenheit function. The error on the distance modulus has been estimated by adding quadratically the dispersion around the Wesenheit function, the uncertainty on the distance modulus of the LMC used to calibrate this relation, the $I$-magnitude error and the overall dispersion seen by \citet{2012ApJ...747...50N} when comparing their distance modulus to other methods (0.2~mag). The latter was needed in order the get distance moduli consistent with the Hipparcos parallaxes. 
  The catalogue contains 233 Tycho-2 stars with $\sigma_{\varpi_\mathrm{E}}<0.1$~mas. 
  
\paragraph{RRLyrae.} For TGAS we used the catalogue of \cite{2005A&A...442..381M}. We computed the distance modulus using the magnitude independent of extinction \Kjk = $K - \frac{A_K}{A_J - A_K} (J-K)$. The extinction coefficients were computed applying the \cite{FitzpatrickMassa07} extinction curve on the \citet{CastelliKurucz03} SEDs. \MK\ was derived from the period-luminosity relation of \cite{2015ApJ...807..127M} \afterReferee{(assuming a mean metallicity of -1.0~dex with a dispersion of 0.2)} and the colours were derived from \cite{2004ApJS..154..633C} transformed in the 2MASS system using the transformations of \cite{2001AJ....121.2851C}. The catalogue contains 150 Tycho-2 stars with $\sigma_{\varpi_\mathrm{E}}<0.1$~mas.
  
A parallax offset of $-0.034 \pm 0.012$~mas and a small overestimation of the standard uncertainty are significative when the Cepheids and the RR Lyrae samples are combined (Table~\ref{tab:cu9val_wp944_summaryplx}). \\

\begin{figure}
\begin{center}
\includegraphics[width=0.7\columnwidth]{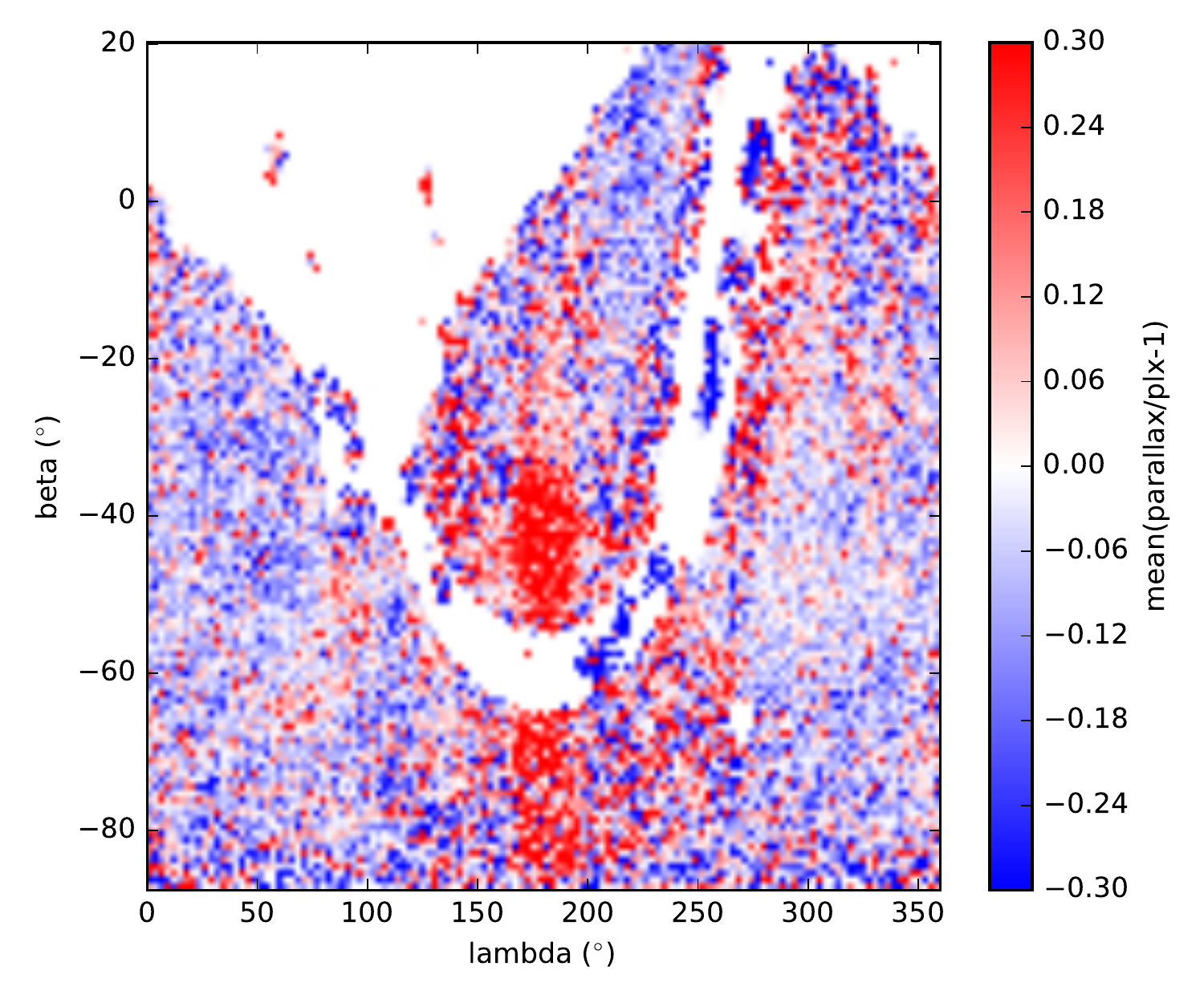}
\caption{Distribution of $\varpi_\mathrm{TGAS}/\varpi_\mathrm{RAVE} - 1$ for $\sim 200\,000$ stars matched in the
RAVE catalogue to the TGAS solution. Stars along EPSL, $\lambda \sim 180${\deg}, 
appear to have a systematically overestimated parallax by
up to $\sim 0.3$~mas, with stars with G magnitudes in the range $10 - 11.5$
and colour $1.4 \le ${\bprp}$ \le 1.8$ being the most strongly affected.}
\label{fig:rave2tgas-wp945}
\end{center}
\end{figure}

For the following catalogues, spectrophotometric distance moduli have been collected or computed. 
\paragraph{RAVE} \citep{2013AJ....146..134K} with distances from \citet{Binney2014MNRAS.437..351B}. 
It contains 6850 Tycho-2 stars with $\sigma_{\varpi_\mathrm{E}}<0.1$~mas. A comparison with Hipparcos has shown the presence of 24\% of outliers, mainly due to dwarf/giant mis-classifications. 
Strong outliers are also seen in the comparison with TGAS but they represent only 1\% of the sample. A global parallax offset of 0.07\afterReferee{0}$\pm$0.005~mas is seen with a strong variation with sky position (with 0.3~mas amplitude). 
This is the only catalogue, together with the LMC, that present a significant positive parallax bias (Table \ref{tab:cu9val_wp944_summaryplx}). 
To further study the presence of systematic effects in localized regions on the sky that could affect the RAVE results, another test has been made using this time all the $192\,655$ stars in common between TGAS and RAVE.  
Thanks to their extended sky coverage, we could identify a
systematic difference in the parallaxes in the region with ecliptic
coordinates $\lambda \sim 180${\deg}, as shown in
\figref{fig:rave2tgas-wp945}. The amplitude of this effect is of
order $\sim 0.3$~mas and affects more strongly the fainter and redder
TGAS stars. It appears that this effect is directly correlated with the 
number of observations along-scan (\dt{astrometric\_n\_obs\_al} parameter) 
and the ecliptic scanning law followed early in the mission\afterReferee{, 
and is consistent with the spatial biases found with quasars at \secref{sec:accuracy-plx}}.

\paragraph{APOGEE DR12} \citep{2015AJ....150..148H}. Distance moduli were computed using a Bayesian method on the Padova isochrones \citep[CMD 2.7]{Bressan12} and using the magnitude independent of extinction \Kjk. The prior on the mass distribution used the IMF of \cite{Chabrier01} while the prior on age was chosen flat. Stars too far from the isochrones were rejected using the $\chi^2_{0.99}$ criterion. It led to 3100 Tycho stars with $\sigma_{\varpi_\mathrm{E}}<0.1$~mas. 
  A global parallax difference of $-0.06\afterReferee{0}\pm 0.006$~mas was found, with a strong variation with magnitude, the brighter the larger the difference. 
  
\paragraph{LAMOST DR1} \citep{LamostDR1}. Same method as for APOGEE. It leads to 451 stars with $\sigma_{\varpi_\mathrm{E}}<0.1$~mas. 
  No significant parallax difference was detected with this sample. 
 
\paragraph{PASTEL} \citep{2016A&A...591A.118S}. Same method as for APOGEE. It leads to 917 Tycho stars with $\sigma_{\varpi_\mathrm{E}}<0.1$~mas. 
  No significant parallax difference was found except for the blue stars ($J$-\Ks$<$0.3), with a difference up to 0.3~mas, most probably linked to the spectro-photometric distance determination that has been less tested on those young massive stars and is more dependent on the age prior. Therefore only stars with $J$-\Ks$>$0.3 are used in the summary \tabref{tab:cu9val_wp944_summaryplx}. 
 
\paragraph{APOKASC} using the distances provided by \cite{2014MNRAS.445.2758R} derived using both Kepler asteroseismologic and APOGEE spectroscopic parameters. It contains 984 Tycho sources with $\sigma_{\varpi_\mathrm{E}}<0.1$~mas. The median $\sigma_{\varpi_\mathrm{E}}$ of this catalogue is 0.02~mas. 
  A global parallax difference of $-0.07\afterReferee{0}\pm 0.009$~mas is seen, with a strong variation with magnitude, similar to what was found with the APOGEE results. Both use the Padova isochrones, have the Kepler region and its spectroscopic parameters in common, but the distance modulus for APOGEE has been computed by us and the APOKASC has a precision on its distance modulus much increased thanks to the usage of the asteroseismology parameters. The variation of the parallax difference with magnitude could come from a feature of the stellar evolution models. Both the APOKASC and APOGEE catalogues present a correlation between magnitude and colour, but in the APOKASC the brighter stars are bluer than the fainter stars (due to the extinction effect on the red clump population) while in APOGEE it is the opposite (due to the more evolved giants being redder); one therefore does not expect the colour to be able to explain the systematics we see in magnitude. 

All those tests with TGAS show significant variations with sky position but with global parallax differences lower than 0.3~mas. 
These tests also show a small correlation with colour ($<$0.2~mas), but not all in the same direction nor with the same amplitude, indicating an expected bias linked to survey parameter correlations and/or stellar isochrones/priors.

\subsubsection{Parallax accuracy tested using distant clusters}\label{sssec:cu9val_ocpar}

This test aims at assessing the internal consistency of parallaxes within a cluster, and checking the parallaxes against photometric distances in order to verify the zero-point of parallaxes. 

Sky coordinates, ages, extinctions and distances have been obtained for all clusters listed in the \citet{2014A&A...564A..79D} database \citep{1995ASSL..203..127M}. Making use of theoretical isochrones \citep{Bressan12}, we retained 488 clusters with an age/distance/extinction combination allowing them to contain stars reaching magnitude $V=11.5$ (the magnitude at which Tycho-2 becomes strongly incomplete). 

All stars within a radius corresponding to a distance of 3\,pc from the center of the cluster were searched, which means that the angular size of the queried field depends on the cluster distance.
Stars were selected based on their identifier in the Tycho-2 catalogue, avoiding \beforeReferee{binary}\afterReferee{double} stars flagged in \citet{2002A&A...384..180F}. When available, a preliminary knowledge of cluster membership was used, but the final cluster membership was determined from the TGAS data itself. The method used was that of \citet{1999A&A...345..471R}, which makes use of proper motions and parallaxes. 

We limited the statistics to clusters more distant than 1\,000\,pc 
so that the uncertainty of the photometric parallaxes is mostly better than the \beforeReferee{one}\afterReferee{uncertainty} of the {\GDR1} parallaxes.
 For every cluster, we computed the average difference $\Delta\mathbf{P}$ between the measured parallax of each star and the reference value (or \textit{photometric} parallax) $\varpi_\mathrm{ref}$ normalised by the uncertainty. 
 In order to compute those values, we need to take into account the  uncertainties on the parallaxes (i.e. $\sigma_\mathrm{\varpi,ref}$ on the reference value and $\sigma_{\varpi}$ on TGAS parallaxes) and the correlation among parameters of nearby stars. We note $S$=diag($\sigma_{i}$) the diagonal matrix made with the standard errors $\sigma_{i}$:

\begin{equation}
S=
\begin{pmatrix}
  \sigma_{\varpi,1} & 0 & ... & 0 \\
  0 & \sigma_{\varpi,2} & ... & 0 \\
  ... & ... & ... & ... \\
  0 & 0 & ... & \sigma_{\varpi,n} \\
 \end{pmatrix}
\end{equation}

\noindent
and we note $\mathbf{C}$ the correlation matrix, where $C_{ij}$ is the correlation coefficient between the parallaxes of star $i$ and star $j$, constructed as in \citet{2010IAUS..261..320H}. The matrix $\mathbf{\Sigma}=\mathbf{SCS}$ is the covariance matrix of $\mathbf{P}$. Noting $\mathbf{D}$ the design matrix $n$-vector (1,1,...,1), we can compute the mean parallax $\varpi= \sigma^2_{\varpi} (\mathbf{D}^T \mathbf{\Sigma}^{-1} \Delta\mathbf{P}) $ with $\sigma^2_{\varpi}=(\mathbf{D}^T \mathbf{\Sigma}^{-1} \mathbf{D})^{-1}$ the square of its standard error.

\begin{figure}
\centering
\includegraphics[width=0.8\columnwidth,height=0.5\columnwidth]{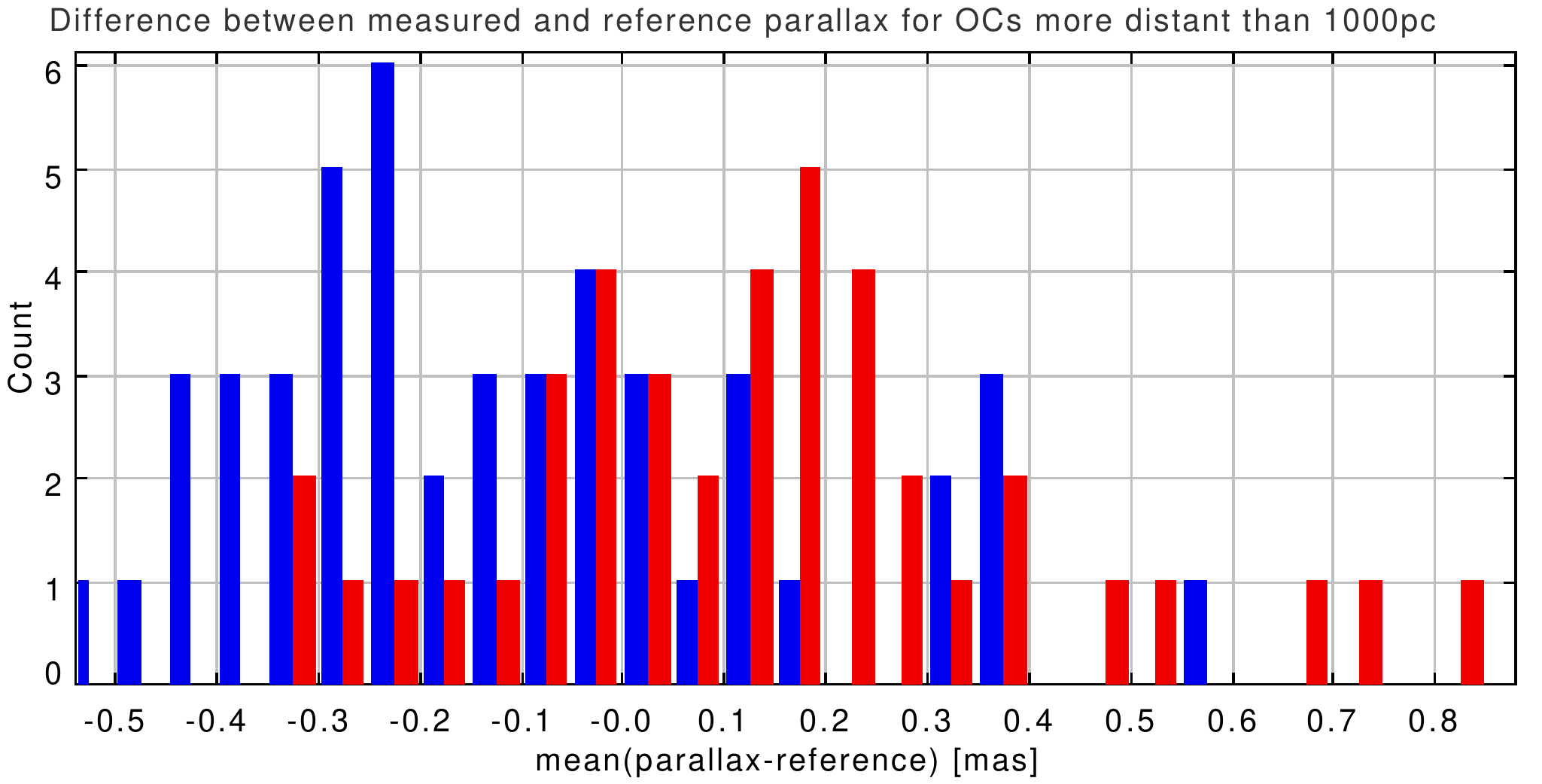}
\caption{Distribution of the differences between the mean TGAS parallaxes and the one from photometric distance for the distant open clusters. 
\afterReferee{Red and blue labels are attributed to the clusters defined \figref{fig:WP947_VAL_010_030_Ocs_dst1000pc}}.} \label{fig:cu9val_947_hist_manual_pmean}
\end{figure}

\begin{figure}
 \begin{center}
\includegraphics[width=0.8\columnwidth,height=0.5\columnwidth]{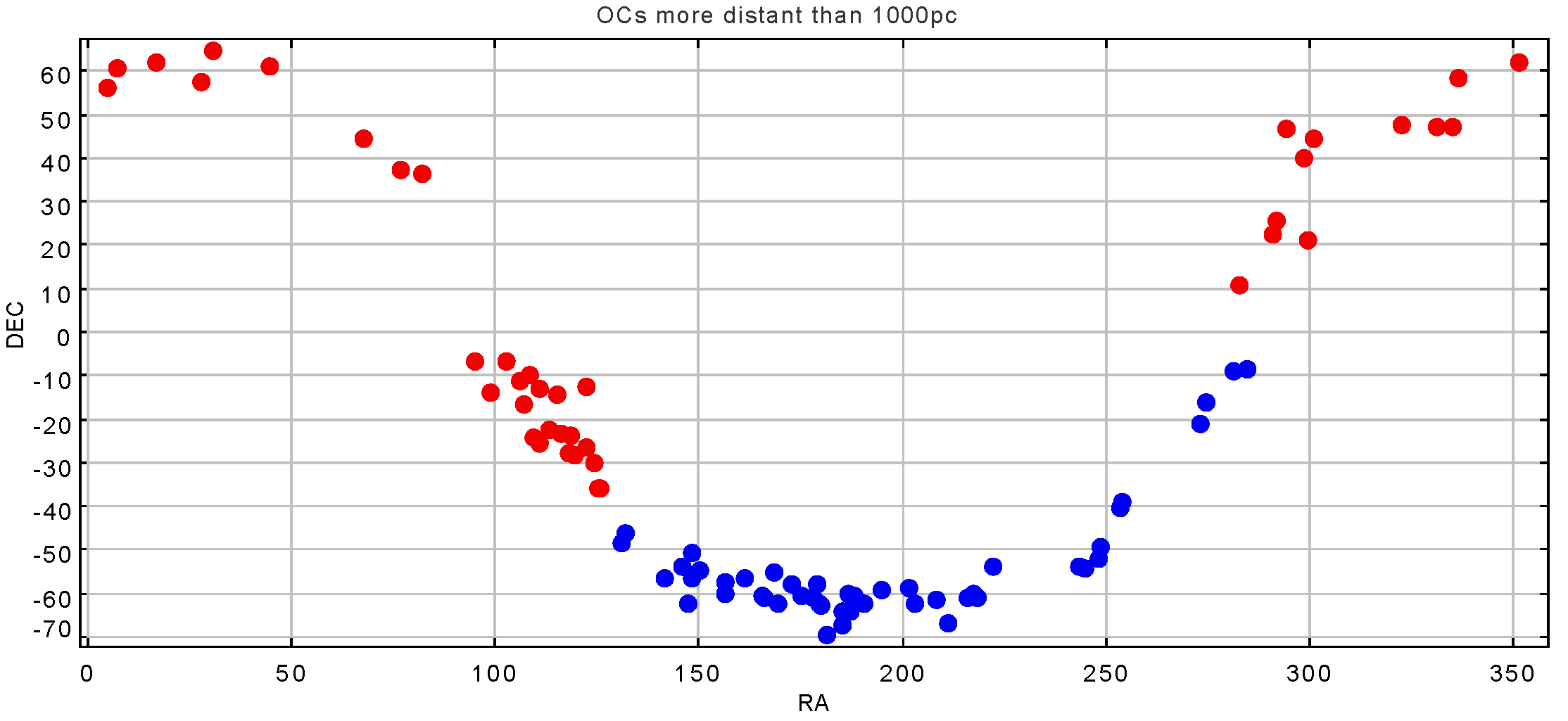}
\end{center}
\caption{Sky distribution of open clusters more distant than 1\,000\,pc. The blue group appears to contain objects with underestimated parallaxes, while the red group contains overestimated parallaxes (\figref{fig:cu9val_947_hist_manual_pmean}).}\label{fig:WP947_VAL_010_030_Ocs_dst1000pc}
\end{figure}

Once an average difference to the reference value ($\overline{\Delta\varpi}$) and associated error ($\sigma_{\overline{\Delta\varpi}}$) was established for each cluster, we studied the global distribution of $\Delta_\mathrm{off}$=$\overline{\Delta\varpi} / \sqrt{ \sigma^2_{\overline{\Delta\varpi}} + \sigma^2_{\varpi,ref}}$  which tells us by how many standard errors the average measured parallax differs from the reference parallax. In the absence of systematics, this distribution is expected to be centred on zero, with a dispersion of one sigma. A mean value differing from zero would indicate a global offset. Conservatively, we considered that all photometric distances listed in the \citet{2014A&A...564A..79D} database are affected by uncertainties of 20\%.  No significant global parallax offset was found, but an apparent systematic error varying with sky position (see \figref{fig:cu9val_947_hist_manual_pmean}). Most clusters with overestimated parallaxes appeared to be located in the Galactic regions with $l<200^{\circ}$ (towards the Galactic anticentre), while most of the underestimated parallaxes were at $l<200^{\circ}$ (see \figref{fig:WP947_VAL_010_030_Ocs_dst1000pc}). The parallax offsets were $-0.16\pm 0.04$ mas for $l>200^{\circ}$ and $+0.13\pm 0.04$ mas for $l<200^{\circ}$.

We investigated the possibility that this effect could be caused by uncertainties in the automatic membership procedure applied. We manually inspected the results of the membership determination and discarded a certain number of clusters for which the cluster membership could not be securely established. The final statistics were computed for a sample of 38 distant clusters with secure membership determinations.
The median value of differences to the reference values for these 38 clusters is $+0.004\pm 0.02$ mas, confirming no obvious global parallax offset. Splitting the sample into two groups ($l>200^{\circ}$ and $l<200^{\circ}$), we find respectively a median of $-0.02\pm 0.032$ mas for the $l>200^{\circ}$ sample, and $+0.044\pm0.027$ mas for the $l<200^{\circ}$ sample, which does not show a significant difference. 

Unfortunately, the low number of tracers available in this experiment did not allow us to draw a map of the bias by averaging values in coordinate space. The slight variation in zero-point between the $l>200^{\circ}$ and $l<200^{\circ}$ groups can then be interpreted either as random variations caused by the uncertainties on the reference values, or as local variations of the parallax zero point (of the order of a few tenths of mas on a scale of several degrees).

\subsection{TGAS astrometric precision}\label{sec:precision-plx}

\subsubsection{Internal estimation of the parallax uncertainty.}\label{sec:wp942:parallaxdeconv}

The quasar analysis in \secref{sec:QSO-plx} \beforeReferee{also} allowed to study the parallax 
dispersion. It was found that the robust unit-weight error (the ratio of the observed
dispersion over the standard uncertainty) decreased with magnitude 
from $\sim 1$ down to about 0.8 at $G=20$. 
It would however be difficult to extrapolate this overestimation of the
uncertainties to the much brighter TGAS sources, so this question was studied differently.

The measured TGAS parallax distribution, at least its small and negative tail, can be used to estimate the parallax uncertainties without referring to the formal uncertainty, following the \afterReferee{deconvolution} procedure of \citet{1995A&A...304...61L}. The procedure models the observed distribution as the convolution of a nonparametric true parallax distribution (subject only to the constraint that all true parallaxes are positive) with a Gaussian error kernel. The Gaussian width parameter that gives the best fit to the observed distribution has been adopted as the parallax uncertainty of the sample.

As noted by \citeauthor{1995A&A...304...61L}, the estimated parallax uncertainty is usually biased, and the process of solving for the true parallax distribution, which resembles Lucy-Richardson deconvolution, suffers from overfitting as the number of iterations increases. Both effects need to be controlled. As the parallax distribution of the TGAS sample differs from that of the Hipparcos sample explored in \citet{1995A&A...304...61L}, we performed simulations to determine the bias correction factor and number of iterations to use for TGAS data. The Simu-AGISLab simulated data (\secref{agislab}) were randomly sampled with new errors to produce a realistic data set large enough for testing. We used cross-validation to test the predictive accuracy of the debiased estimates, including uncertainties in the bias correction factor.

Unlike \citeauthor{1995A&A...304...61L}, we found that 2-3 iterations gave much more accurate results than a few dozen, regardless of the sample being studied; the reasons for this discrepancy are not yet clear. We fit arbitrary (nonlinear) functions to the bias correction factor and the accuracy of the debiased parallax uncertainty, enabling prediction of the bias correction to $\sim 8\%$ and of the accuracy of the final parallax uncertainty (i.e., the uncertainty on the uncertainties) to $\sim 20\%$.
We also found that simulation runs with small ($N \sim 100$) or precise ($\sigma_\varpi \sim 0.1 \textrm{ mas}$) data sets behaved very differently from the trends seen for larger or less precise data; presumably the sharp changes at high precisions are related to the parallax distribution assumed for the TGAS catalogue.

\begin{figure}[tbhp]
\centering
\includegraphics[width=0.49\columnwidth]{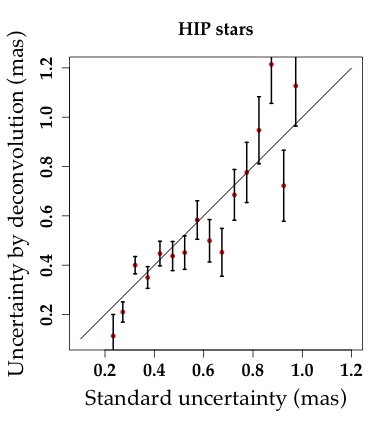}
\includegraphics[width=0.49\columnwidth]{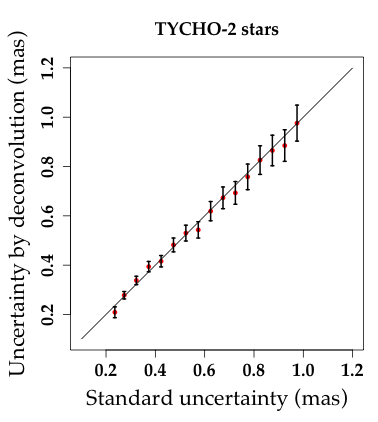}
\caption{Best-fit uncertainties from deconvolution of parallaxes versus standard uncertainties for TGAS Hipparcos stars (left) and for Tycho-2 (non-Hipparcos) stars (right), with bisector represented. Error bars include all sources of uncertainty, including bias correction.}\label{fig:wp942:parallaxdeconv}
\end{figure}

When modelling the observed parallax distribution, we first corrected all parallaxes for the $-0.04$~mas bias found \secref{sec:QSO-plx}, though analysis with and without the correction gave indistinguishable results. We analysed the TGAS data in bins of standard uncertainty of 0.05 mas width, and separately for each type of astrometric solution, in case each group had different error properties.

We show in \figref{fig:wp942:parallaxdeconv} the results of modelling the TGAS 
parallaxes dispersion compared to the standard uncertainties. 
As can be seen the TGAS standard uncertainties $\sigma_\varpi$ on parallaxes 
appear accurate. More quantitatively, a weighted fit for Hipparcos stars is 
$(0.980{\scriptstyle\pm 0.135})\sigma_\varpi - 0.003{\scriptstyle\pm 0.062}$,
while for Tycho-2 stars it is 
$(0.973{\scriptstyle\pm 0.024})\sigma_\varpi +  0.011{\scriptstyle\pm 0.011}$,
both being consistent with a unit-weight error = 1. 
Assuming a unit-weight error = 1, and fitting only for an extra dispersion 
(quadratically added) gives $-0.19\pm 0.02$ mas for Hipparcos and 
$-0.11\pm 0.01$ mas for Tycho-2. This is consistent with the median value obtained 
with external estimates, \tabref{tab:cu9val_wp944_summaryplx}, and it shows
that the standard uncertainties appear (except probably for the most precise 
parallaxes) slightly pessimistic. 


\subsubsection{Comparison with external astrometric data}\label{sec:wp944_astrom}

The comparison of Gaia results with external astrometric data is not straightforward as Gaia will provide the most accurate and the most numerous astrometric data ever produced, at least in the optical domain. However the consistency between Gaia data and carefully selected external astrometric data might be important in order to detect any statistical misbehaviour in one or the other source of data, including Gaia.

Only positions from the Hipparcos or Tycho-2 catalogues have been used as priors in TGAS. The parallaxes and/or the proper motions have not been used, so this ensures that the comparison with TGAS parallaxes and proper motions is meaningful, as they are independent from those of Hipparcos and Tycho-2. Note that another independent comparison with those catalogues is presented in Annex C of \cite{DPACP-14}.
For the Hipparcos and Tycho-2 proper motion tests, the global rotation between the reference frames of Hipparcos and TGAS derived in \cite{DPACP-14} has been applied. A possible (residual-)rotation has been checked. 
For each catalogue, the distribution of the normalized residuals (Gaia-External) of each parameter $R_\mathcal{N}$, e.g. for the parallax $R_\mathcal{N}$=(\parallax$_G$-\parallax$_{E}$)/$({\sigma_{\text{\parallax}_G}^2+\sigma_{\text{\parallax}_{E}}^2})^{1/2}$, has been checked to be consistent with a normal distribution, and correlations of those residuals with magnitude, colour and sky position have been checked too. 

A $\chi^2$ test has been also performed on combined parameters $X$ ($X$ being the positions, or the proper motions, or the parallaxes and proper motions) using the full covariance matrix of both the external ($\Sigma_E$) and the Gaia ($\Sigma_G$) catalogues to compute the normalized residuals $R_{\chi}=(X_G-X_E)^T (\Sigma_G+\Sigma_E)^{-1} (X_G-X_E)$  and their distribution has been tested to follow a chi-squared distribution with $n$ degrees of freedom, $n$ being the number of parameters tested (e.g. 2 for {\GDR1} positions, 2 for TGAS proper motions and 3 for TGAS parallaxes and proper-motions). Similarly to the one dimensional case, correlations with magnitude and colour, and sky distribution of those residuals have also been tested. 

In all the tests, we used a p-value limit of 0.01 (e.g. we indicate that we find a bias, extra variance or a correlation with a confidence level higher than 99\%). For the normalized residuals using individual parameter ($R_\mathcal{N}$), this level corresponds to $\vert R_\mathcal{N} \vert>2.6$, while for the $\chi^2$ residuals on 2 components this level corresponds to $R_{\chi}>9.21$.


For the validation of TGAS, the following astrometric catalogues have been considered:
\paragraph{Hipparcos new reduction.} 
A selection of {\it well behaved} Hipparcos stars has been done using the 5-parameter solution type with a good astrometric solution (goodness of fit $\vert F2 \vert<5$), and without any binary flag indicated in the literature, mainly from WDS \citep{WDS}, CCDM \citep[Catalogue of the Components of Double and Multiple Stars,][]{2000A&A...363..991D} and SB9 \citep[9th Catalogue of Spectroscopic Binary Orbits,][]{2004A&A...424..727P}\beforeReferee{)}. Stars also included in Tycho-2 \beforeReferee{are}\afterReferee{were} kept only if the proper motions from Hipparcos \beforeReferee{are}\afterReferee{were} consistent with those of Tycho-2 (rejection p-value: 0.001). The resulting sample includes 93\,802 well behaved stars, against which both the parallaxes and proper motions of TGAS have been tested. 

\begin{figure}
    \begin{center}
        \includegraphics[width=0.8\columnwidth]{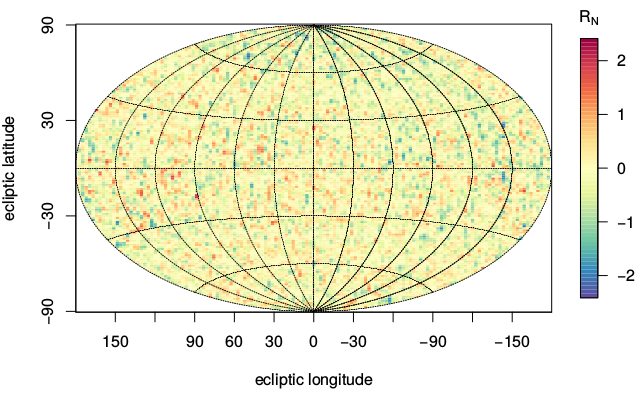}
        \caption[TGAS versus Hipparcos Parallaxes]{Sky variation of the normalized residuals $R_\mathcal{N}$ of the TGAS versus Hipparcos parallaxes in ecliptic coordinates. Although correlation with sky position is significant, no sky region indicate a normalized residual larger than 2.6.}
        \label{fig:cu9val_wp944_hipplx} 
    \end{center}
\end{figure}

A global parallax zero point difference between {\gaia} and Hipparcos of $-0.094\pm 0.004$~mas \beforeReferee{is}\afterReferee{was} found\footnote{If we 
assume, as shown \secref{sec:QSO-plx}, a $-0.04\pm 0.003$~mas zero-point for {\GDR1}, 
an estimate of the Hipparcos zero-point (new reduction) would then be $+0.054\pm 0.005$~mas. This
would also be the zero-point of the first Hipparcos reduction as the average parallax difference between both reductions
is about 0. This value is then marginally consistent with the estimation done two decades ago 
\cite[$-0.02\pm 0.06$~mas,][]{1995A&A...304...52A} with preliminary Hipparcos data, and to what was estimated with the published data, $-0.05\pm 0.05$~mas \cite[][Vol III, Chap. 20.]{1997ESASP1200.....E}}.
The under-estimation of the standard uncertainties for both parallax and proper motions  is significative (extra dispersion of 0.6 mas). Small variations of the parallax and proper motion residuals is seen with sky position  (\figref{fig:cu9val_wp944_hipplx}) and magnitude (smaller than 0.1 mas, most probably due to the gates).

\begin{figure}
    \begin{center}
        \includegraphics[width=0.8\columnwidth]{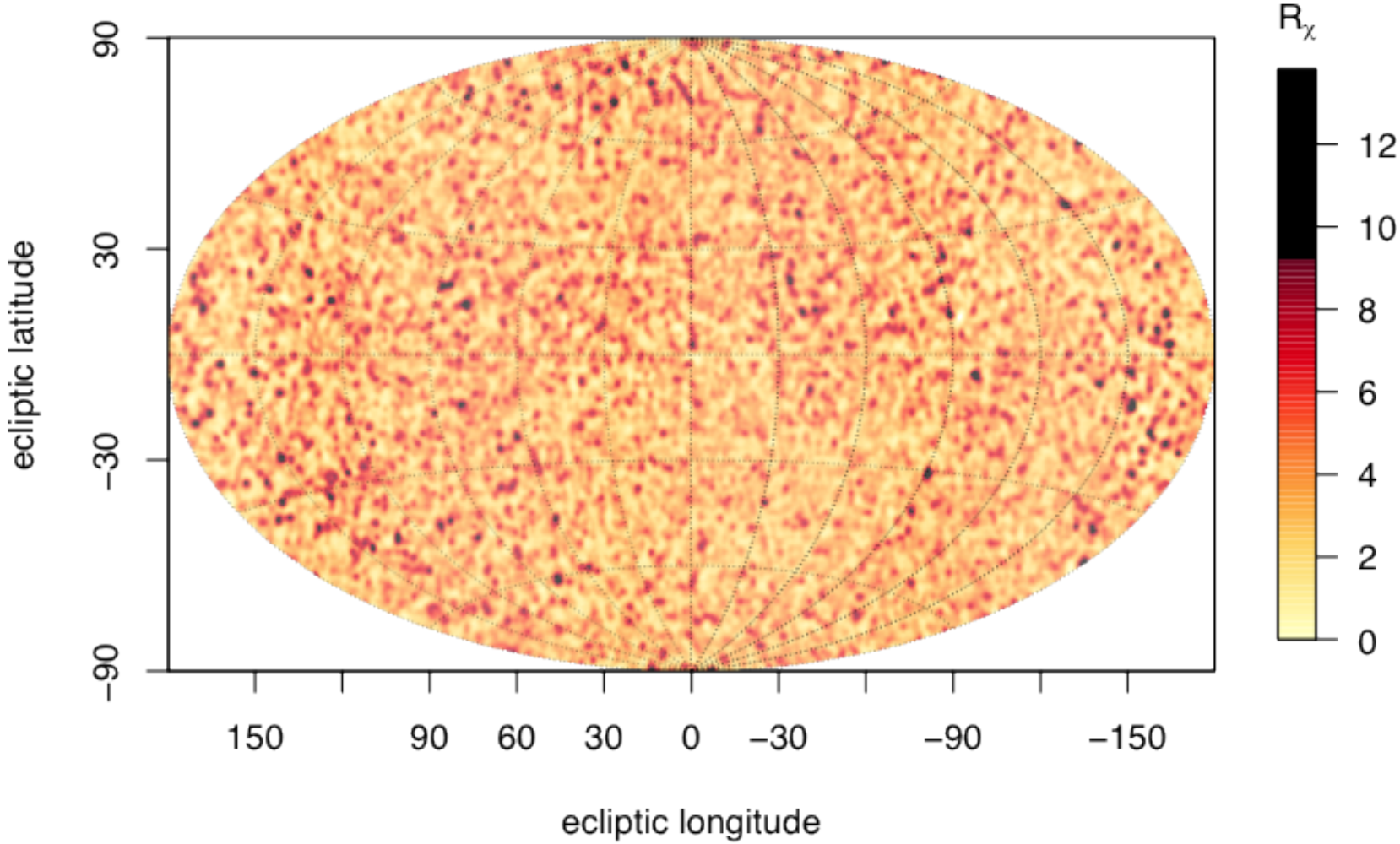}
        \includegraphics[width=6cm]{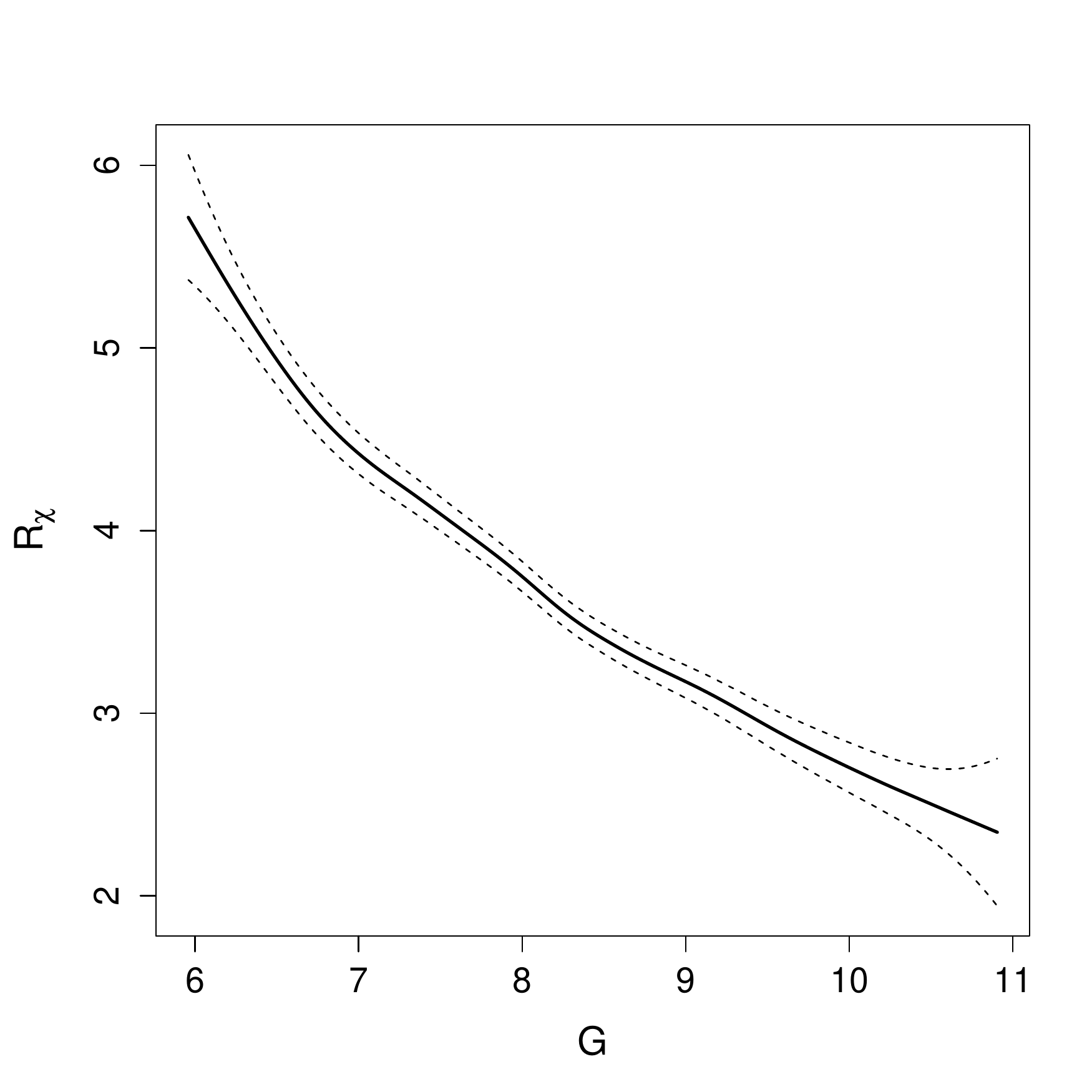}
        \caption[TGAS versus Hipparcos astrometry]{TGAS proper motions versus Hipparcos $\chi^2$ test: the residuals $R_\chi$ should follow a $\chi^2$ of 2 degrees of freedom. Sky regions with a significant residual ($R_\chi>9.21$) are highlighted in black.  a) Sky variation in ecliptic coordinates. b) correlation with the \gmag~magnitude; the dotted lines correspond to the 1$\sigma$ confidence interval. }
        \label{fig:cu9val_wp944_hipPMchi2} 
    \end{center}
\end{figure}

The $\chi^2$ test with Hipparcos, using either parallax and proper motions or proper motions only, shows stronger variations across the sky (\figref{fig:cu9val_wp944_hipPMchi2}a), with areas showing a mean residual $R_\chi$ over 9.21 (the p-value 0.01 limit)  while the residuals of parallax or proper motions components individually stay below the p-value limit ($\vert R_\mathcal{N} \vert<2.6$). 11\% of the sources have a $\chi^2$ p-value$<$0.01, e.g. 11 times more than expected.
Moreover a strong correlation between $R_\chi$ and \gmag~magnitude is observed  (\figref{fig:cu9val_wp944_hipPMchi2}b). This behaviour of $R_\chi$ is also seen with the quasar positions (\secref{sec:DR1pos}), indicating potential issues with the covariance matrix. Those could be due to extra correlations introduced by the attitude and calibration models not taken into account in the provided covariance matrix \citep{2012A&A...543A..14H}.

\paragraph{Hipparcos and Tycho-2 stars with inconsistent proper motions.}
The second sample includes the 1574 stars previously eliminated because of the inconsistency between Hipparcos and Tycho-2 proper motions. A specific test has been done on those stars: most of them are expected to be long period binaries not detected in Hipparcos, and for which the longer time baseline of Tycho-2 could have provided a more accurate value. 

The TGAS solution also has a long time baseline thanks to its Hipparcos/Tycho input position. It has been therefore tested that the TGAS solution for those stars is globally closer to the Tycho-2 solution than to the Hipparcos solution. 
This is indeed the case with 7\% of those TGAS sources being outliers versus the Tycho-2 solution while 50\% are outliers versus the Hipparcos solution. 

\paragraph{Tycho-2.} 
Only Tycho-2 stars with a normal astrometric treatment (no double star with Tycho-2 separate entries, no close known or suspected double star with {\it photocentre treatment}) have been used in this test. Due to the different priors used for the Hipparcos and Tycho-2 stars (Hipparcos positions at the Hipparcos mean epoch, J1991.25, for Hipparcos stars; Tycho-2 positions at the effective Tycho-2 observation epoch, taken to be the mean of the $\alpha$ and $\delta$ epochs, for Tycho-2 stars) in the TGAS solution, the test has been done once for the Tycho-2 sources not in Hipparcos and once for the Tycho-2 sources in the {\it well behaved} Hipparcos sub-sample described above.  

\begin{figure}
    \begin{center}
        \includegraphics[width=0.8\columnwidth]{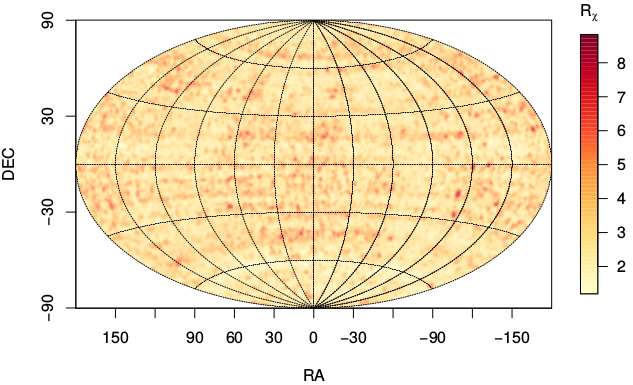}
        \caption[TGAS versus Tycho-2 proper motions]{Sky variation of the residuals $R_\chi$ of the TGAS versus Tycho-2 parallaxes in equatorial coordinates. }
        \label{fig:cu9val_wp944_tycPM} 
    \end{center}
\end{figure}

For the Tycho-2 sources in the Hipparcos well-behaved sub-sample an under-estimation of the standard uncertainties is seen (extra dispersion of 0.6\masyr\ similar to what is found with the Hipparcos sample) and a correlation with magnitude and colour is found with an amplitude smaller than 0.1 mas (the residuals increasing with magnitude and colour). 
 For the Tycho-2 sources not in Hipparcos, a strong variation of the residuals is seen with sky position (\figref{fig:cu9val_wp944_tycPM}) with features parallel to the equatorial declinations which corresponds to the zones of the Astrographic Catalogue used to derive the Tycho-2 proper motions. A very large extra dispersion of 1.8\masyr\ is also observed. We most probably see here the defaults of the Tycho-2 proper motions. A rotation smaller than 0.2\masyr\ is also observed. 
 
\paragraph{VLBI compilation.} VLBI data have mainly been obtained from the USA VLBA, the Japanese VERA and the European EVN: 90 proper motions and 44 parallaxes (including respectively 70 and 30 stars in Tycho-2). Over the years, with increasing baseline length and better calibration of the ionospheric and tropospheric delays, astrometric accuracy using VLBI at centimetre wavelengths is approaching ${\sim}$10~$\mu$as for parallaxes and ${\sim}$1~$\mu$as yr$^{-1}$ for proper motions \citep[][and reference therein]{2014ARA&A..52..339R}. For proper motions, only those with a mean epoch $>$ 2000 were considered as calibration techniques improved drastically at that epoch, especially with new detailed maps of ionospheric delay. The compilation covers all stellar sources for which trigonometric parallaxes and proper motions have been obtained from VLBI astrometry \citep[as quoted in the review of][]{2014ARA&A..52..339R}, but also stars with only proper motions obtained from VLBI positions \citep{2007AJ....133..906B}, and VLBI proper motions of X-ray binaries with an estimation of distance obtained by other means \citep{2014PASA...31...16M}.  

36 stars of this compilation are present in TGAS, including 9 with parallax information. All the tests associated to this catalogue pass (parallax and proper motion bias, variance, correlations), with the exception of the full covariance matrix \chitwo test which indicates that half the stars with both parallax and proper-motion information available (assuming no correlation for the VLBI parameters as this information is rarely available) have a \chitwo p-value larger than 0.01. 

\paragraph{HST compilation.} The Fine Guidance Sensors (FGS) on Hubble Space Telescope have produced high accuracy trigonometric parallaxes of astrophysically interesting objects such as Cepheids, RR-Lyrae, novae, cataclysmic variables or cluster members  \citep{2015IAUGA..2257159B, 2007AJ....133.1810B}. The FGS field of view is small and the parallaxes of target stars have been measured with respect to reference stars which have their own parallaxes estimated by spectro-photomometric measurements. The correction to absolute leads to a median error of absolute parallaxes announced to be 0.2 mas. The present compilation covers 69 stars with parallaxes (including 43 in Tycho-2) and, for about a third of them, proper motions, published up to end 2015. 

19 stars of this compilation are present in TGAS, passing all the tests. 

\paragraph{RECONS.} The REsearch Consortium On Nearby Stars (RECONS, www.recons.org) has built a database of all systems estimated to be closer than 25~pc (parallaxes greater than 40 mas with errors smaller than 10 mas). We have used the database as published on 1st April 2015 \citep{2015IAUGA..2253773H}, leading to 348 stars (including 27 in Tycho-2) with trigonometric parallaxes.

13 stars of this compilation are present in TGAS, passing all the tests.

\begin{table*}
\caption{Comparison in the LMC and SMC of the correlations between astrometric
parameters: the median of the standard correlations given in the Catalogue appear consistent with the empirical ones
computed with the astrometric residuals.}\label{tab:test_correl}
\centering\small
\begin{tabular}{lcccccc} 
\hline\hline
 && \multicolumn{2}{c}{LMC} && \multicolumn{2}{c}{SMC} \\
correlation &~~~& predicted & observed &~~~& predicted & observed \\
\hline
$\rho(\varpi,\mu_{\alpha *})$ && +0.7\afterReferee{47} $\pm$ 0.013 & +0.77\afterReferee{4} $\pm$ 0.063  && -0.2\afterReferee{03} $\pm$ 0.023 & -0.3\afterReferee{56} $\pm$ 0.139 \\
$\rho(\varpi,\mu_{\delta})$ && -0.68\afterReferee{0} $\pm$ 0.012 & -0.42\afterReferee{4} $\pm$ 0.09\afterReferee{0} && -0.8\afterReferee{01} $\pm$ 0.005 & -0.6\afterReferee{75} $\pm$ 0.11\afterReferee{0} \\
$\rho(\mu_{\alpha *},\mu_{\delta})$ && -0.31\afterReferee{1} $\pm$ 0.033 & -0.22\afterReferee{0} $\pm$ 0.097 && -0.1\afterReferee{17} $\pm$ 0.028 & -0.17\afterReferee{2} $\pm$ 0.147 \\
\hline
\end{tabular}
\end{table*}

\subsubsection{Validation of the astrometric correlations.}\label{sec:extcorrel}
As shown above and stressed in \secref{sec:effect-correlations}, the correlation between
astrometric parameters should not be neglected when computing covariance matrices.
After having tested the formal uncertainties above, checking whether these correlations
are accurate is also needed.

It is usually difficult to compute these correlations but there are at least two
different local areas, the LMC and SMC, where average proper motions and parallaxes
are already known to a sufficient precision. The astrometric errors can thus be computed
from the residuals between Gaia proper motions and parallaxes and the external estimation.
We used the Tycho-2 stars only (not the Hipparcos ones) as the internal dispersion 
of the proper motions can be neglected compared to the astrometric uncertainties 
($\sim 1$\masyr) in the former case, not in the latter.

Using the star list indicated in \secref{sec:LMCSMC} restricted to Tycho-2 stars, we rejected 
all sources having in absolute value one of these residuals 3 times larger than their formal
uncertainty, to avoid any contamination by field stars.  

In each Magellanic Cloud, we then computed the medians of the formal correlations 
as given in the Catalogue, and we estimated the actual ones computing the
empirical correlation coefficients between residuals. As shown in \tabref{tab:test_correl},
the various estimations are consistent with the predictions at a p-value=0.01.
The expected internal variations of the proper motions inside the Clouds also explain
the large dispersion.

Although this test has been done on two regions only, it is reasonable to consider 
that the correlations between astrometric parameters, as given in the Catalogue, 
are statistically reliable.

There is however an important caveat. We did not discuss explicitly in this paper the 
angular correlations between stars which we know exist (see \figref{fig:cu9val_942_QSOKsky}).
In principle, this section should have compared the full observed covariance-matrix
(of all stars $\times$ 5 astrometric parameters) to the predicted one, but it is much too
difficult to predict the correlations between stars for now. It is thus possible that
the local comparison done here shows an agreement while a whole sky comparison would 
disagree.

\subsubsection{Comparisons with proper motion from distant open clusters}\label{sssec:cu9val_ocpm}
The aim of this test was two-fold: assessing the internal consistency of proper motions within stellar clusters, and looking for biases and systematics by testing the proper motions zero-point against literature values.

Following the open cluster selection described in \secref{sssec:cu9val_ocpar}, we computed the difference between the proper motion of each star and the reference value for its cluster listed in the MWSC catalogue \citep{2013A&A...558A..53K} and in \citet{2014A&A...564A..79D}. This procedure is designed to take into account possible small-scale correlations between parameters.
For each cluster, we obtained a mean value $\Delta$ of this difference, and its associated error $\sigma$. We flagged the objects for which the difference to the reference value is too large to be explained by the nominal uncertainties, as well as those with discrepant small or large internal dispersions. The test also looks for trends in proper motions against magnitude and colour.

A global zero-point test was performed from the $\Delta$ values obtained for individual clusters, restraining the sample to objects distant enough so that their internal dispersion in proper motions is negligible compared to the uncertainty on the proper motion of individual stars. The expected all-sky average of this quantity should be zero if no bias is present. A clustering test allows us to verify if outliers are randomly distributed, or if they cluster in problematic areas in the sky.

We retained 20 clusters that are sufficiently distant and present secure membership for more than 10 stars.
Scaling the difference $\Delta$ according to the total uncertainty (standard uncertainties listed in TGAS and uncertainty on the literature value), we found no significant differences in proper motions. In units of uncertainty, the all-sky zero-point of $\mu_{\alpha *}$ is $+0.04\pm0.21$, and for $\mu_{\delta}$: $+0.12\pm0.26$. We also found that outliers appear homogeneously distributed across the sky.

\subsubsection{Specific tests on known double and multiple systems} 

In addition to the above general tests, a specific test has also been done on known double and multiple systems from the Hipparcos new reduction (HIP2) and the TDSC in order to detect any possible bias between single and non-single stars. For non-Hipparcos systems, we use the component designation given in the TDSC, m\_TDSC, to distinguish between primary components (A or Aa), unresolved systems (AB), and secondary components (all other entries in TDSC). For Hipparcos systems, four categories with increasing periods were distinguished: stochastic solutions (short period, solution type Sn = 1 modulo 10 in HIP2), acceleration stars with 7- or 9-parameter solution (intermediate period, Sn = 7 or 9 modulo 10 in HIP2), secondary component (long period, separation $\rho>0$ as provided in the original Hipparcos catalogue), other double stars (the remaining non single stars). 
The characteristics of those Hipparcos and Tycho systems were compared to those of the {\it well behaved} Hipparcos sample described in \secref{sec:wp944_astrom}, adding the extra criterion of passing the $\chi^2$ test comparing the parallax and proper motion between Hipparcos and TGAS. Of course, within these ``{\it single star}'' samples, many unknown unresolved binaries may hide.

A difference in behaviour between those different subsets with respect to the {\it single star} samples was looked for, using various parameters: the parallax and proper motion residuals (TGAS-external), and the TGAS errors, goodness of fit and excess noise (source modeling errors). Mainly acceleration solutions are expected to show large discrepancies between their proper motions in TGAS and those from Hipparcos or TDSC. 
Another source of discrepancy may be the fictitious difference created by the comparison of TGAS and Hipparcos proper motions for close systems for which only the photocentre was observed by Hipparcos. 
For example, it was found that the excess noise, which is about 0.5 mas on the average except for very bright stars
(\secref{sec:quality_indic}) did not exhibit significant differences between
single, primaries and secondaries; on the contrary unresolved systems had significantly 
degraded solutions with about 1.2 mas excess noise on the average in the $7\lesssim G\lesssim 12$ mag range. 

Several other tests have also been done on secondary components, checking whether the separation or position angle 
with respect to the primary component had no adverse effect. In the past, during the validation 
of early preliminary Gaia data, it had been found that proper motions of many secondaries 
below 2\arcsec\ separation had a large discrepancy (up to a 80{\masyr} amplitude) compared to TDSC. 
Noting that 2\arcsec\ divided by the time span between Hipparcos
and Gaia (2015-1991) gives about 80{\masyr}, it was deduced that the cross-matching of some close double
stars had been deficient: \afterReferee{most probably} the wrong first epoch position had been used for the Tycho-Gaia astrometric
solution (TGAS), e.g. the Tycho position for the A component was associated to the observations of the B component
because it was closer to it, depending on the position angle of the system, and {\it vice versa}. 

\begin{figure}
    \begin{center}
        \includegraphics[width=0.49\columnwidth]{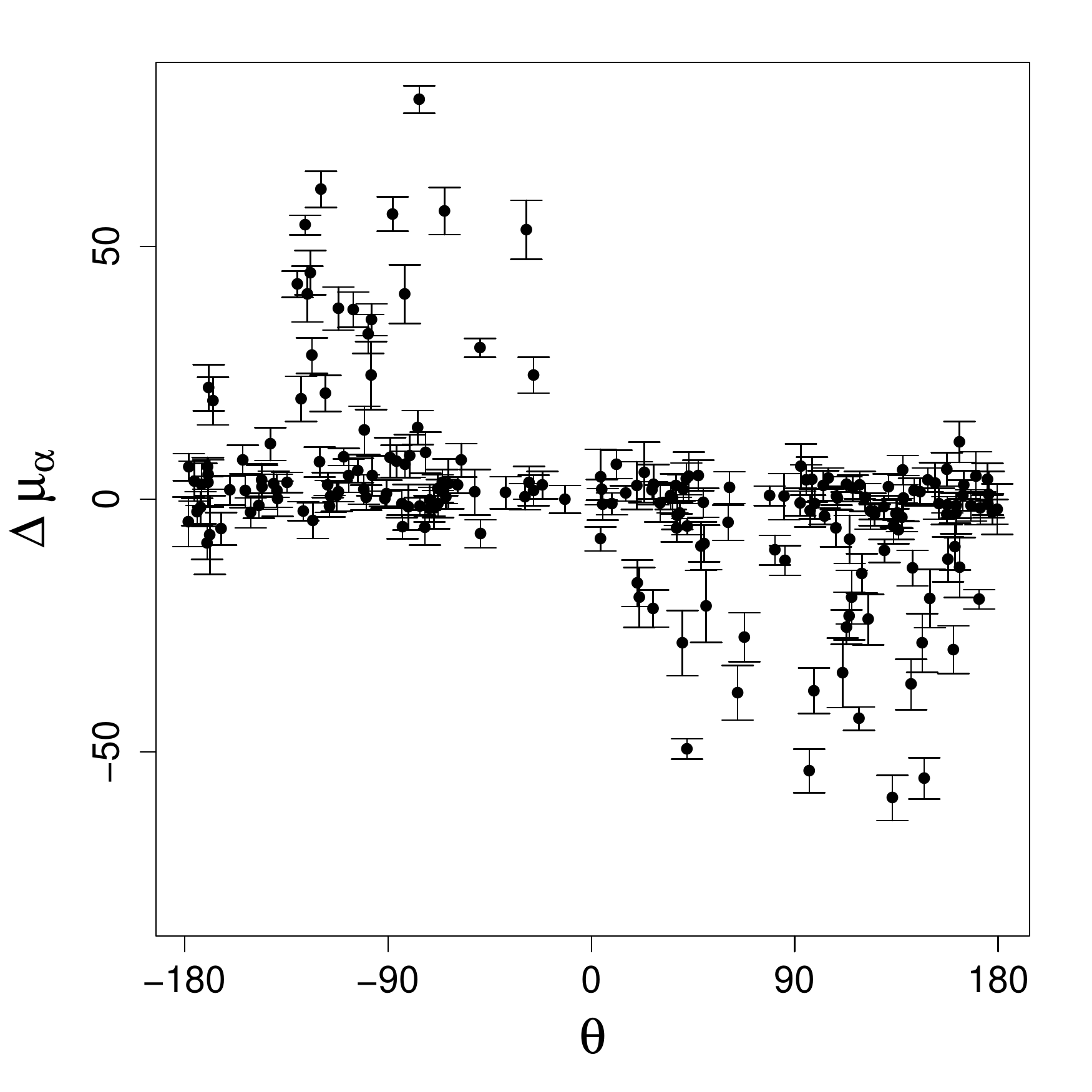}
        \includegraphics[width=0.49\columnwidth]{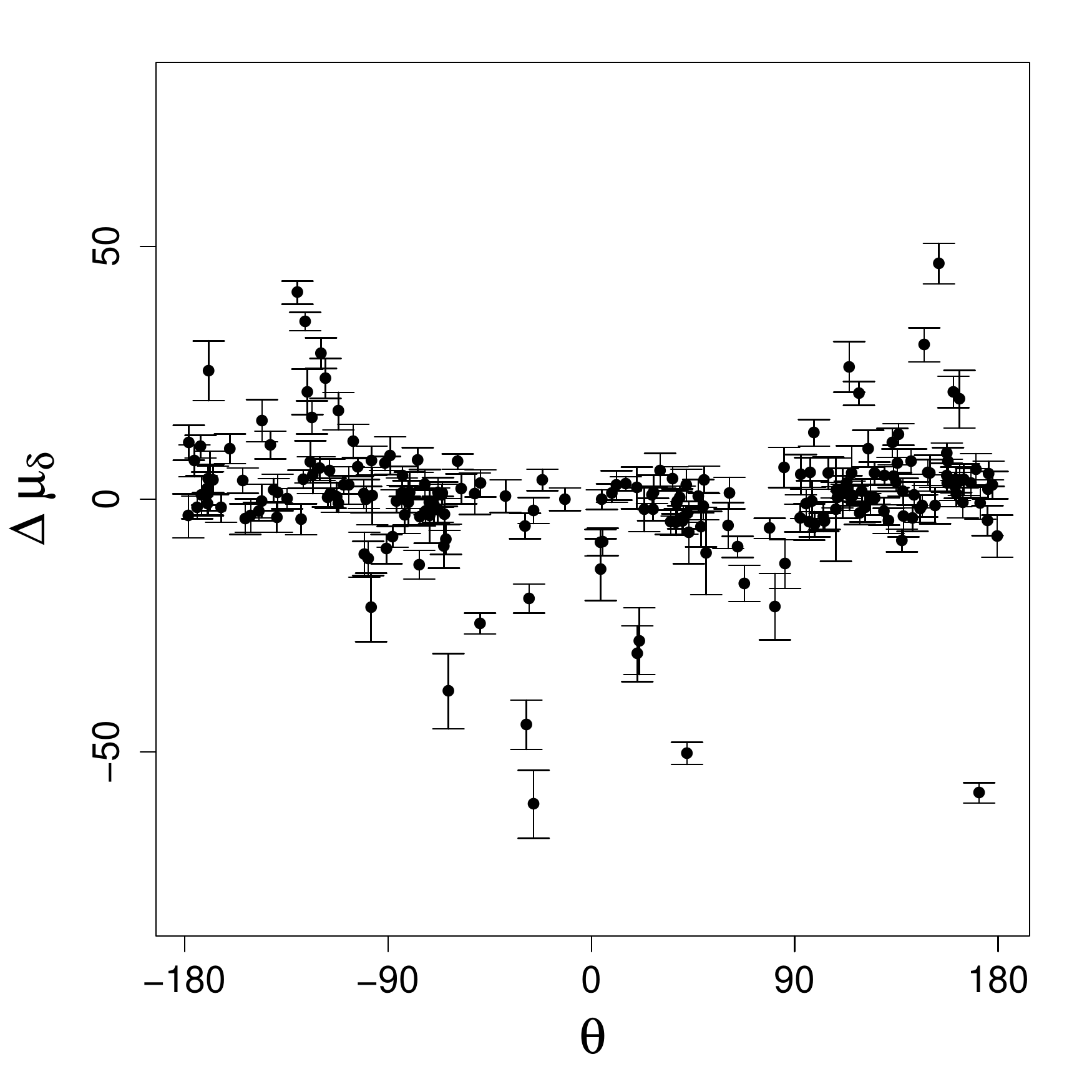}
        \caption{Difference between TGAS and TDSC proper motions ({\masyr}) as a function of position angle $\theta$ (deg) 
        for secondary components of multiple systems with $\rho<2$\arcsec; $\mu_{\alpha *}$ (left) 
        and $\mu_{\delta}$ (right).}
        \label{fig:cu9val_wp944_muTDSC} 
    \end{center}
\end{figure}

\beforeReferee{Since this preliminary solution, stars with a parallax uncertainty 
larger than 1 mas had no more been considered for TGAS and } 
\afterReferee{Unlike the preliminary Gaia data, the TGAS solution disregarded
stars with a parallax uncertainty larger than 1 mas, which}
received a 2 parameter astrometric solution instead. However, for close 
double stars which remain in TGAS, and as can be seen in
\figref{fig:cu9val_wp944_muTDSC}, there are still several pairs mis-identified,
and it is unclear whether the mis-identification comes from Gaia or Tycho in the
first place.
Using this figure, it should be easy for the user to detect and reject the
bad astrometric solutions for pairs (both components) depending on a) separation 
below 2\arcsec, b) position angle in the bad range, c) proper motions differences 
above uncertainties and possibly d) large excess noise.

\subsection{TGAS validation from the comparison with Galaxy models}

Two Besan\c{c}on Galactic Model simulations have been run for TGAS validation, using slightly modified models, both in density laws and kinematics, in order to verify the dependency of the model inputs to the validation. Both simulations were done with the model described in \cite{2014A&A...564A.102C} where the evolutionary scheme has been updated, as well as the IMF, SFR and evolutionary tracks. Moreover, the thick disc and halo populations have been updated, following \cite{2014A&A...569A..13R}, with new density laws. Concerning the kinematics, we used alternatively the standard model kinematics \citep{bgm}, hereafter BGMBTG2, and a revised kinematics from an analysis of RAVE survey (Robin et al, in prep), hereafter called BGMBTG4. BGMBTG2 and BGMBTG4 also differ by several model parameters such as the extinction model and thin disc scale length. 

The use of two different models allows to evaluate what is due to acceptable model variations in the parallax and proper motion distributions. Model parameters are described in Mor et al. 2015 (internal Gaia documentation GAIA-C9-TN-UB-RMC-001). The simulations contain binary systems where the second component is merged with the primary when the separation is smaller than 0.8 arcsec, the estimated resolution of the Tycho-2 catalogue. We also introduced the uncertainties expected in TGAS after 6 months of Gaia observations, following the recipes published in September 2014 after commissioning phase\footnote{\scriptsize\url{http://www.cosmos.esa.int/web/gaia/science-performance}}.

The validation was done by comparing the proper motion and parallax distributions in TGAS catalogue to simulated ones. The sky \beforeReferee{is}\afterReferee{was} divided in healpix rings with healpixsize 20, giving a solid angle of 8.5943 square degree in each bin, and 4800 bins in total. Then bins were grouped in rings of equal galactic latitudes in order to compare the values between latitude rings. Finally, we consider 5 latitude intervals (-90 to -70\deg, -70 to -20\deg, -20 to 20\deg, 20 to 70\deg, and 70 to 90\deg) in order to analyse the characteristics of the distributions in the plane, at intermediate latitudes, and at the poles separately. For each region of the sky considered, we compared the mean and standard deviation between the model and the data for the parallax and proper motion distributions. 

\subsubsection{Parallaxes}

\begin{figure*}
\begin{center}
\includegraphics[width=3.5cm, height=4cm]{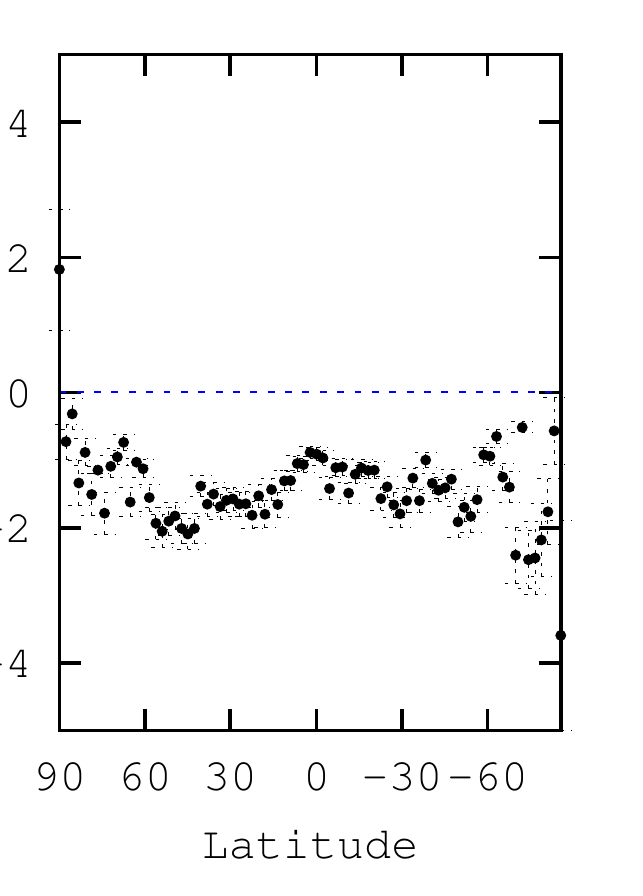}
\includegraphics[width=3.5cm, height=4cm]{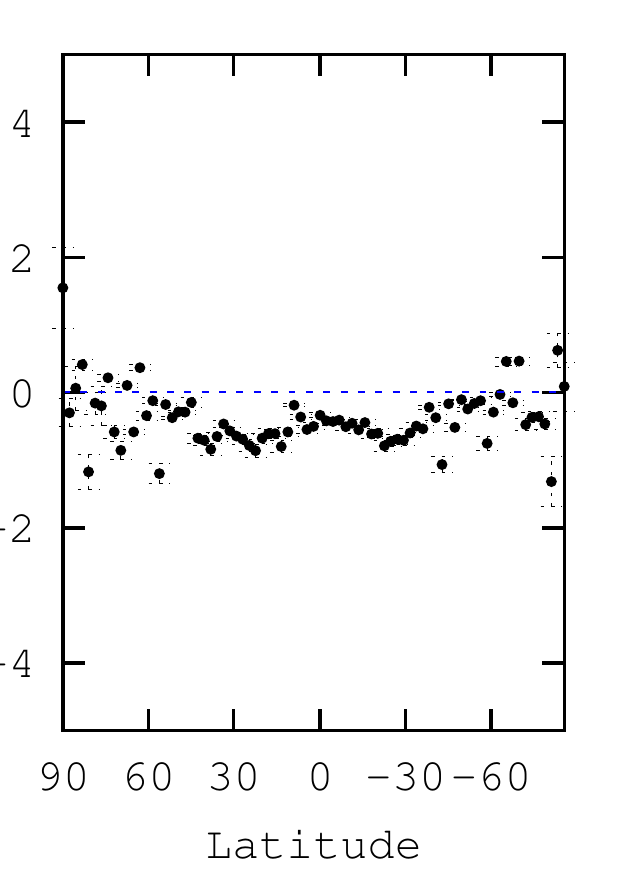}
\includegraphics[width=3.5cm, height=4cm]{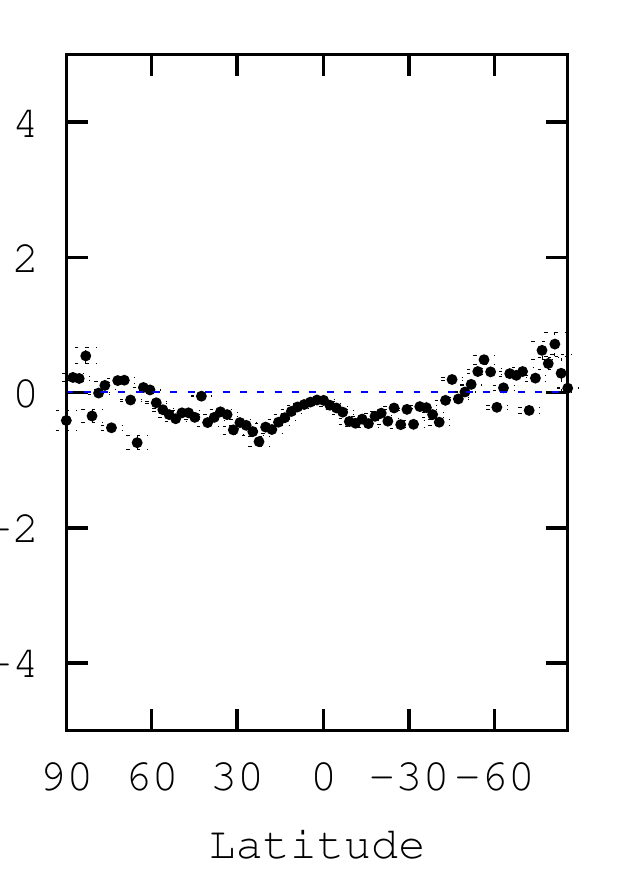}
\includegraphics[width=3.5cm, height=4cm]{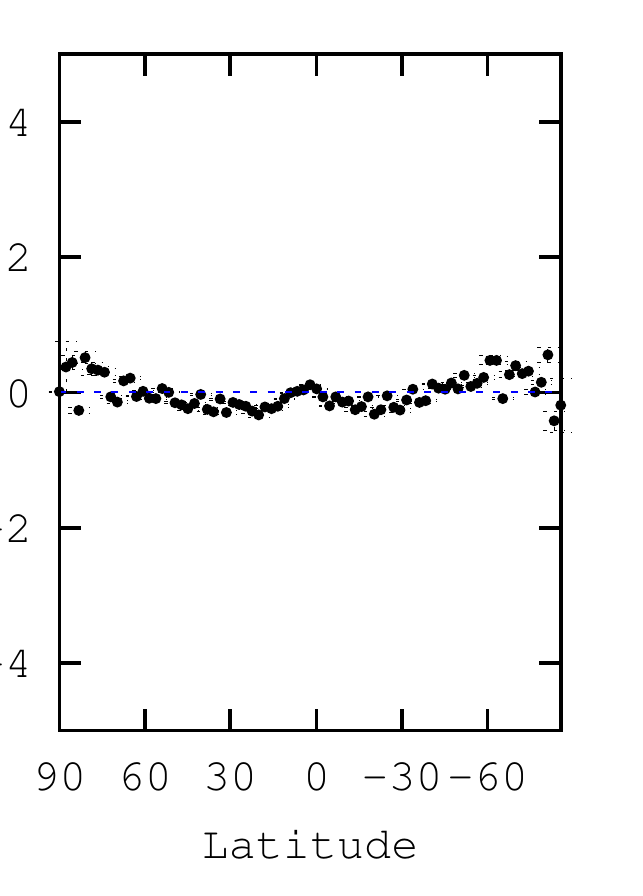}
\includegraphics[width=3.5cm, height=4cm]{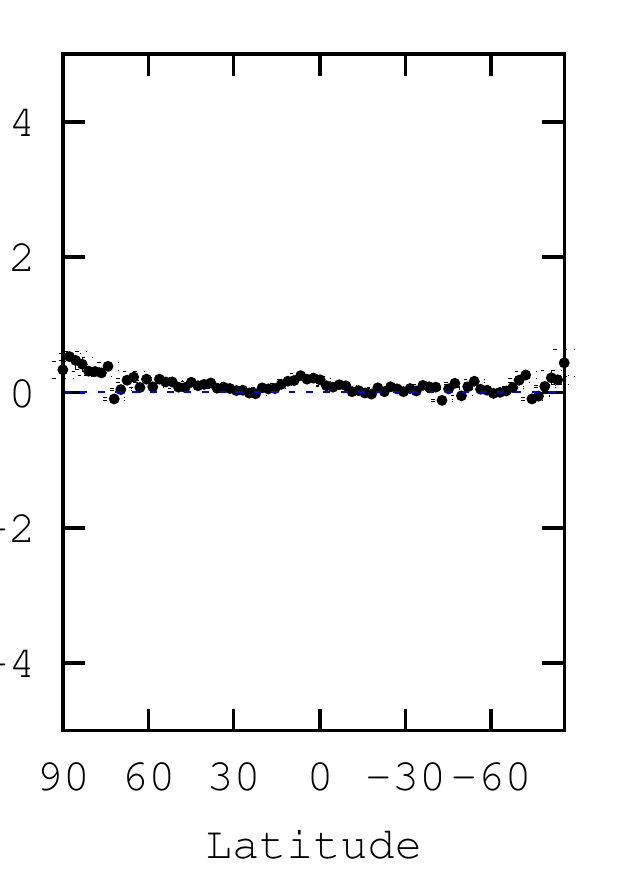}
\caption{Mean difference in parallax in mas between BGMBTG2 model simulation and TGAS data, in different rings of latitude, for five magnitude bins in \vt\ from left to right, from 9-9.5 (left) to 11-11.5 (right).}
\label{fig:Mean_parallax}
\end{center}
\end{figure*}

Figure \ref{fig:Mean_parallax} shows the mean parallax differences between the BGMBTG2 simulation and TGAS data, as a function of latitude rings. Each panel corresponds to a magnitude interval of 0.5 mag width, starting at \vt=9. 

From these comparisons we notice that, for bright stars, the mean parallax differences seem to suffer from a slight zero point offset, which also depends slightly on Galactic latitude. 
The systematic shift between models and TGAS data is of the order or less than 1 mas depending on the region of the sky, 
but it is unclear whether this originates from the data or the model.

In the standard deviation in parallax, the comparison with models shows a good agreement. The dominant factor in the simulation of the parallax standard deviation is the error model assumed to simulate the errors added in the BGM simulations. The good agreement implies that the dependency of the parallax errors on magnitude and latitude is in agreement with the expectations.

\subsubsection{Proper motions}

\begin{figure*}
\begin{center}
\includegraphics[width=3.5cm, height=4cm]{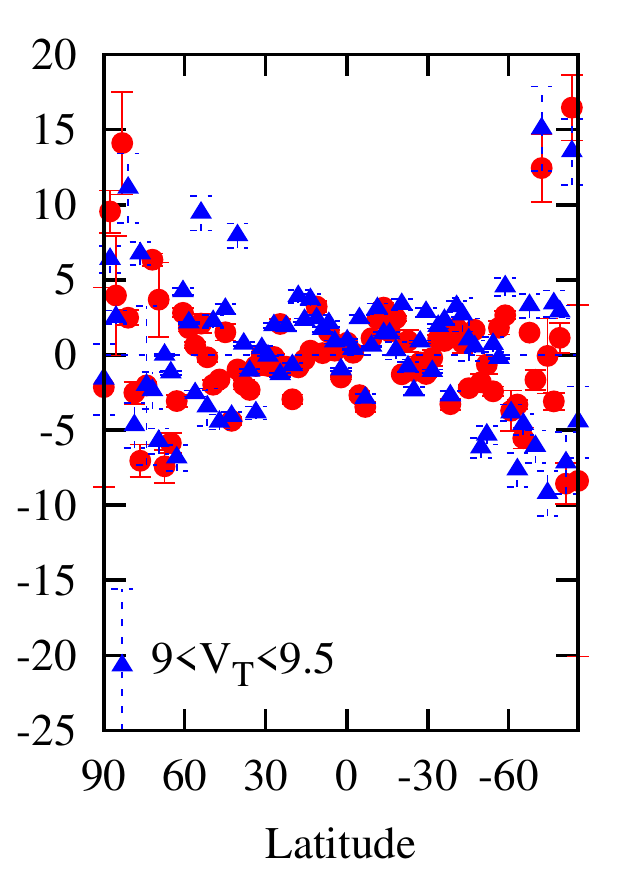}
\includegraphics[width=3.5cm, height=4cm]{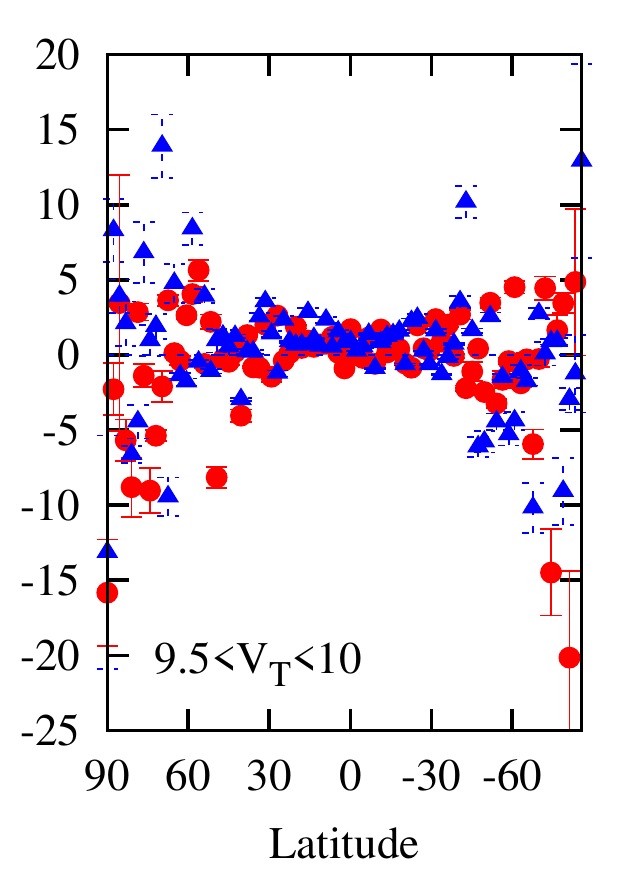}
\includegraphics[width=3.5cm, height=4cm]{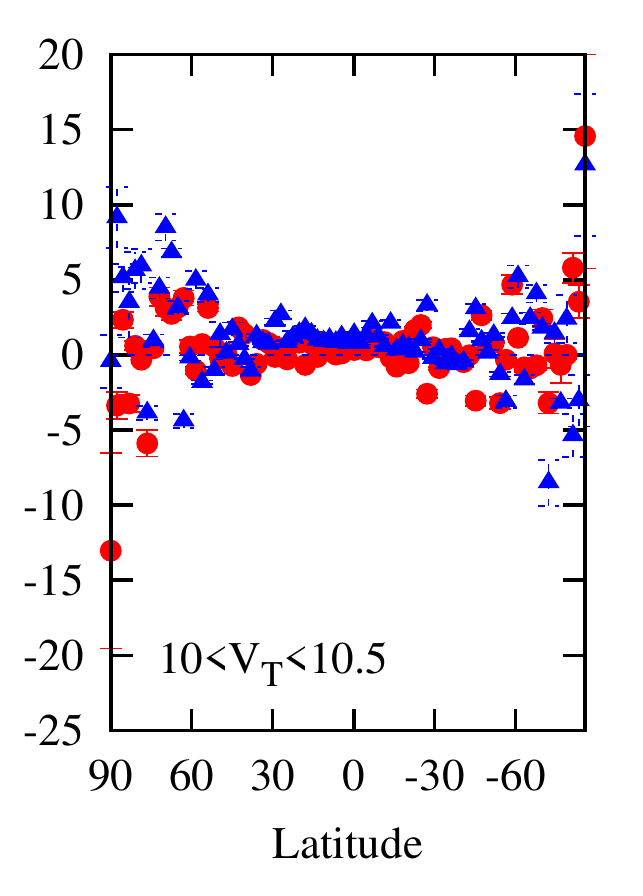}
\includegraphics[width=3.5cm, height=4cm]{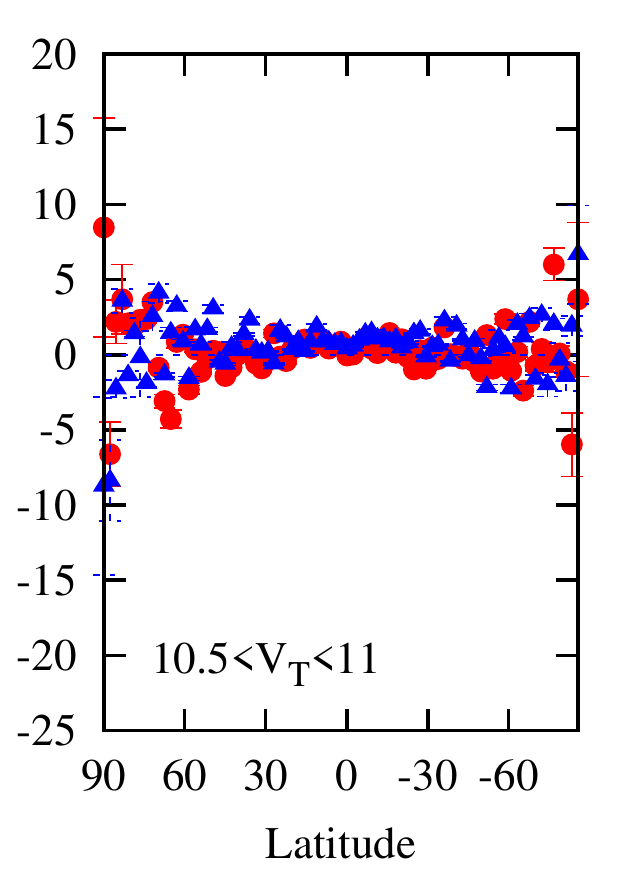}
\includegraphics[width=3.5cm, height=4cm]{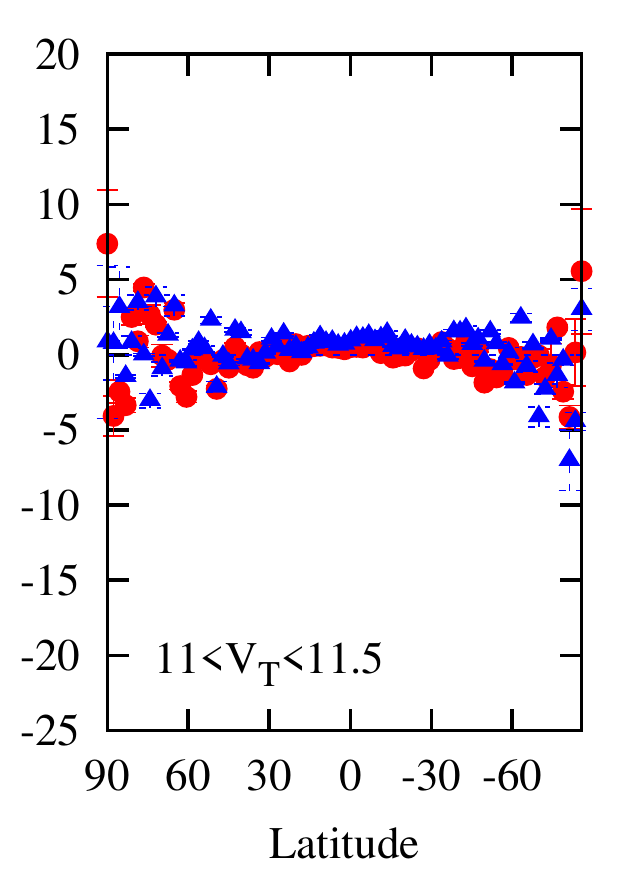}
\caption{Difference in mean proper motion along Galactic longitude ($\mu_{l *}$) between TGAS data and two models: BGMBTG2 (red), BGMBTG4 (blue), in different magnitude intervals, between $V_T$=9 (left) to $V_T$=11.5 (right) by steps of 0.5 magnitude.  }
\label{fig:Mean_properMotion_l}
\end{center}
\end{figure*}

Figure \ref{fig:Mean_properMotion_l} shows the differences in the mean proper motion along Galactic longitude ($\mu_{l *}$) between the BGMBTG2 and BGMBTG4 simulations and TGAS data, as a function of latitude healpix rings. Each panel corresponds to a magnitude interval of 0.5 mag width, starting at \vt=9. Both models show similar difference distributions with the data. 

\begin{figure*}
\begin{center}
\includegraphics[width=3.5cm, height=4cm]{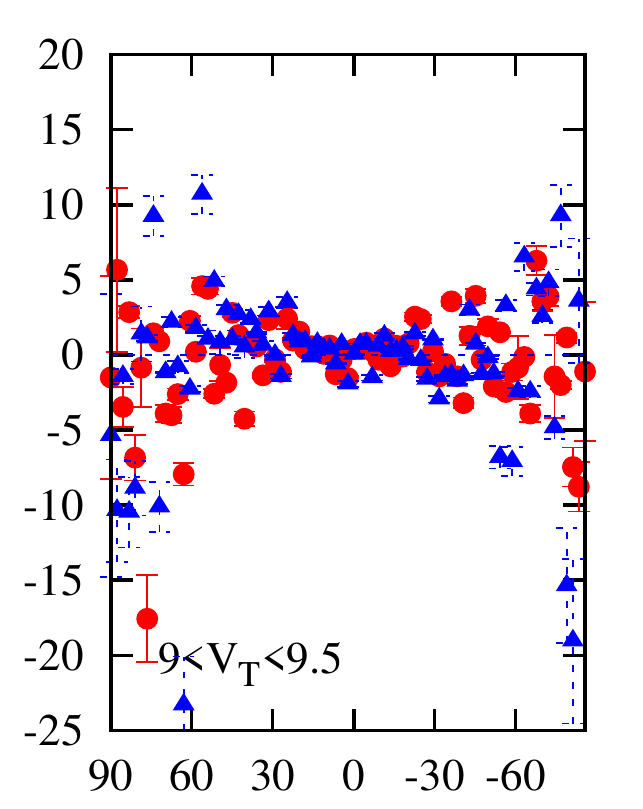}
\includegraphics[width=3.5cm, height=4cm]{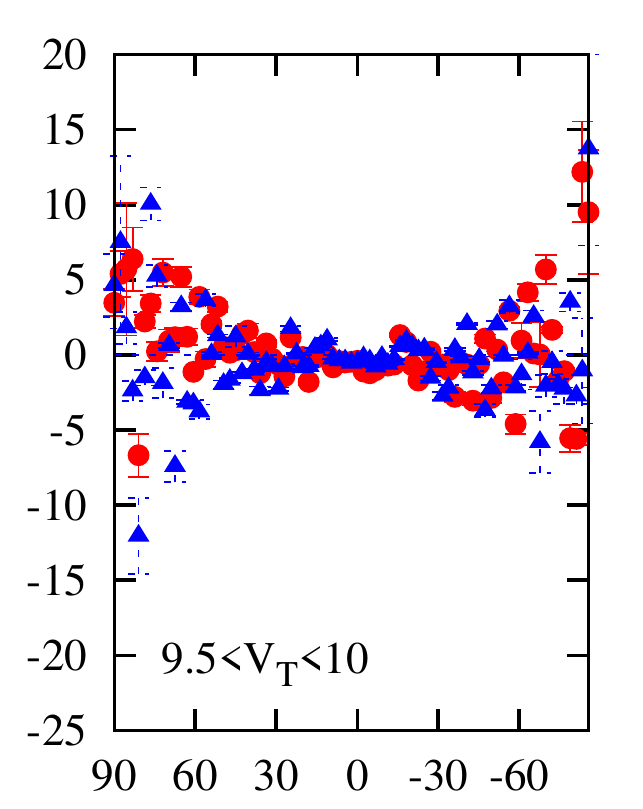}
\includegraphics[width=3.5cm, height=4cm]{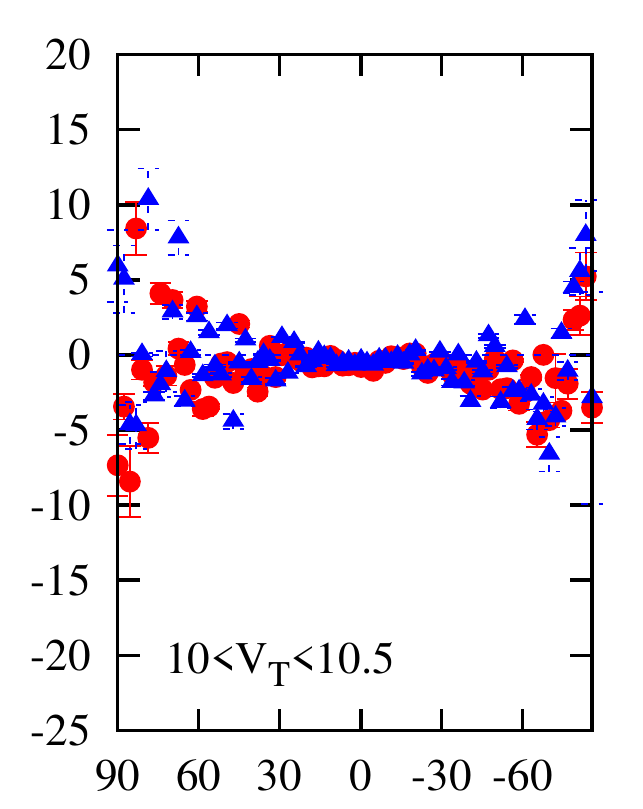}
\includegraphics[width=3.5cm, height=4cm]{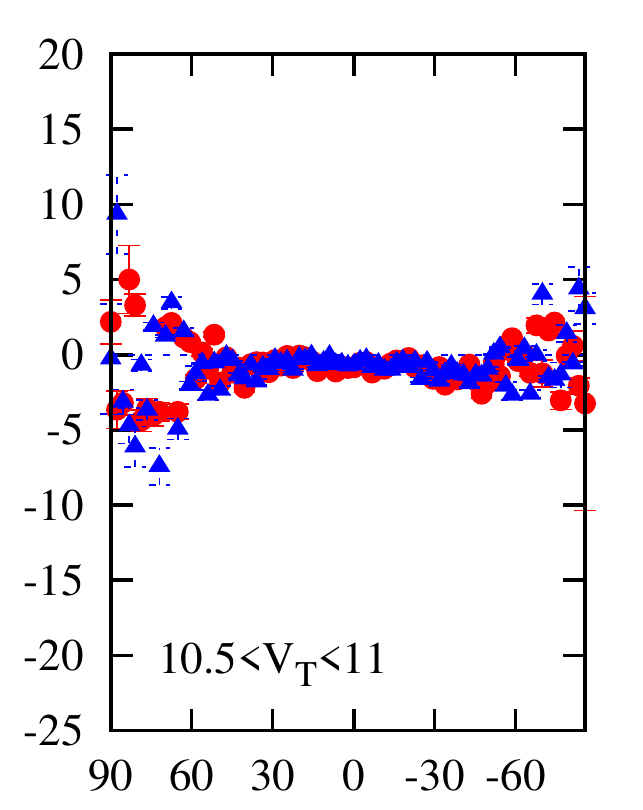}
\includegraphics[width=3.5cm, height=4cm]{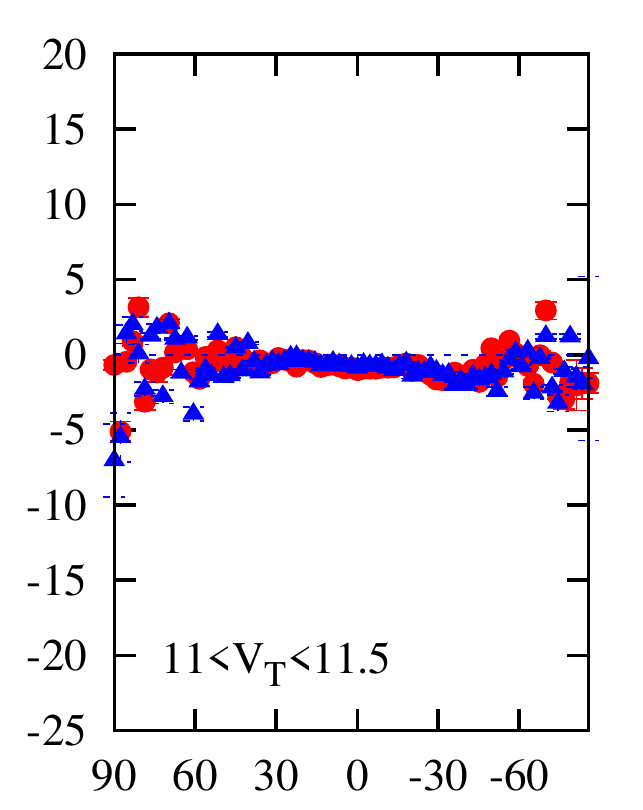}
\caption{Difference in mean proper motion along Galactic latitude ($\mu_b$) between TGAS data and two models: BGMBTG2 (red), BGMBTG4 (blue), in different magnitude intervals, between $V_T$=9 (left) to $V_T$=11.5 (right) by steps of 0.5 magnitude.}
\label{fig:Mean_properMotion_b}
\end{center}
\end{figure*}

Figure \ref{fig:Mean_properMotion_b} shows the differences in the mean proper motion along Galactic latitude ($\mu_b$). The zero point differences between models and data are at the level of the differences between the two models at bright magnitudes. However systematic differences appear in the faintest magnitude bins 
which again can be attributed either to the model or
related to large correlated errors in some regions of the ecliptic plane due to the scanning law. Notice also the higher noise level at the Galactic poles due to the smaller number of sources. 

\subsection{{\GDR1} positions and reference frame}\label{sec:DR1pos}

For the billion+ sources of {\GDR1}, the only astrometric parameters available are the two components of the position. The astrometry of the secondary DR1 dataset has been compared with the following catalogues:
 \paragraph{URAT1 star positions} \citep{urat1}. URAT1 is a catalogue containing stellar positions of 228\,276\,482 stars down to $R$=18.5, at epochs ranging from 2012.3 to 2014.6 with typical standard errors of 10--30 mas. Only stars distant enough to have a proper motion smaller than 100\masyr\ even assuming a tangential velocity of 500\kms\ were used. The Gaia-ESO and LAMOST surveys have been used to estimate the spectrophotometric distances of those stars (see method in \secref{sec:distantstars}), leading respectively to samples of 5\,384 and 136\,234 stars. The cross-match between DR1, including TGAS, with URAT1 was done by position, with multiple detections within 0.2\arcsec~removed. 

Correlations with magnitude, colours, sky positions are seen, but overall this effect stays within an amplitude of 30~mas.

 \paragraph{ICRF2 QSO positions} \citep{2015AJ....150...58F}. The second realisation of the International Celestial Reference Frame (ICRF2) contains very precise positions of 3\,414 compact radio astronomical objects. The positional noise floor is announced to be  of about 40\muas ~and the directional stability of the frame axes of about 10\muas. A least-square method using the covariance matrix of both catalogues allows to estimate the rotation and dipolar deformation between the ICRF2 and the Gaia reference frames. Correlations of differences between {\GDR1} and ICRF2 positions with other parameters such as magnitude and colours were tested, following the same methods as described above for stars. 

The test has been done both on the auxiliary quasar solution and on the main {\GDR1} secondary solution, with the same conclusions so that only the numbers corresponding to {\GDR1} are provided below \citep[note that the priors used in their astrometric reduction are different,][]{DPACP-14,DPACP-26}.  
2\,292 ICRF2 quasars are found in {\GDR1} within a 0.1\arcsec\ radius. As expected by construction \citep{DPACP-14}, no rotation versus the ICRF2 is found, but a deformation (glide) is detected, lower than 0.2~mas. It should be noted that this deformation is not significant anymore if the cross-match radius is increased from 0.1 to 0.5\arcsec\ which adds 15 sources. The residuals of the position differences normalized using the covariance matrix of both {\GDR1} and the ICRF2 $R_\chi$ show a too large number of outliers (10\% with a p-value $<0.01$, i.e. 10 times more than expected) and $R_\chi$ is correlated both with the magnitude and with the number of observations. This behaviour of $R_\chi$ is the same as the one observed in the comparison with Hipparcos (\secref{sec:wp944_astrom}). 
   

More anecdotally, four known quasars were included in the Hipparcos and Tycho-2 catalogues (HIP  60936 = 3C273, TYC 9365-284-1, TYC 259-212-1, TYC 3017-939-1). Only the first and the last ones are present in TGAS. 3C273 has an astrometry consistent with null parallax and proper motion, but this is not the case for the Tycho-2 AGN, TYC 3017-939-1 ($R_\chi$=25.3).

\subsection{Quality indicators of the astrometric solution}\label{sec:quality_indic}

\begin{figure}[]
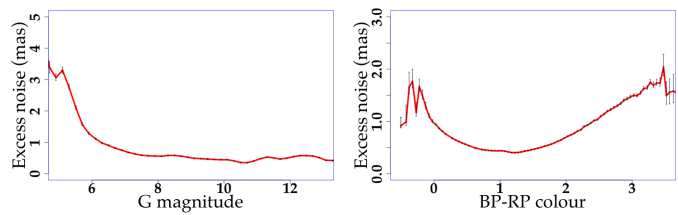

\begin{center}
\smallImage{{figures-942/PhotGMeanMag-AstrometricExcessNoise}.png}
\smallImage{{figures-942/BP-RP-AstrometricExcessNoise}.png}
\caption{Astrometric excess noise (mas) smoothed as a function of $G$ magnitude (left) 
and {\gbp-\grp} (right).}\label{fig:cu9val_942_excessnoise}
\end{center}
\end{figure}

As mentioned before, the {\GDR1} astrometric solution applied only a single star
model to all stars; resolved doubles with small magnitude difference or astrometric 
binaries with noticeable orbital or acceleration motion are thus susceptible
to lead to a bad astrometric fit. Second, as described too, the adopted PSFs are not yet
optimal for all stars (and probably not for very blue or very red stars),
\beforeReferee{the attitude}\afterReferee{and the modelling of the satellite attitude}
can still be improved together with the geometric or CCD calibrations. There is
no {\em stricto sensu} goodness of fit metrics in the catalogue, as they would actually 
never be good given the caveat above. However there are both 
\dt{astrometric\_n\_bad\_obs\_al}, \dt{astrometric\_n\_\-bad\_\-obs\_ac} and 
\dt{astrometric\_excess\_noise} and its significance \dt{astrometric\_\-excess\_\-noise\_sig}.
Beside a median floor at about 0.5 mas due to attitude, etc, 
the \dt{astrometric\_excess\_noise} appears, 
as expected, sensitive to calibration problems for bright stars and extreme colours 
(\figref{fig:cu9val_942_excessnoise}). Outside these cases, and outside some regions
(see corresponding Figure in \secref{sec:statistics}), a star with a larger
and significant excess noise is a candidate to being non-single. It is thus
suggested to take advantage of these fields for the selection of ``cleaner'' samples.

\subsection{Summary of the astrometric validation}\label{sec:summary-astro}

{\GDR1} is the most precise all sky astrometric survey since Hipparcos. 
And indeed, the quoted parallax precision in the catalogue appears correctly 
estimated, or slightly pessimistic only, as found by error deconvolution 
and when compared to external catalogues.
The only exception is the comparison with Hipparcos, which then points to some
underestimation of the errors in Hipparcos itself.

However, the preliminary character of the astrometric solution, and in particular
problems related to imperfect attitude or instrument modelling reveal systematic 
errors of the same order as the random errors.
A global negative parallax zero point (about -0.04 mas) is consistently 
found with many independent estimation methods (quasars, period-luminosity candles,
spectro/astero/photometric parallaxes).
This zero-point is however a consequence of large scale spatial variations
related to the scanning law that may reach at least a 0.3 mas amplitude 
(i.e. comparable to the median precision of stars in the catalogue). 
This is also consistently shown independently with quasars, LMC, SMC or RAVE data. 
In extreme cases, larger local biases may be expected.
Correlation with magnitude is also found towards the bright end.

For the scientific exploitation, the consequence of these systematics is that 
local parallax averages cannot be more precise than about 0.3 mas. 
Any study should take into account that any catalogue parallax is 
$\varpi \pm\sigma_\varpi \mathrm{~(rand.)} \pm 0.3 \mathrm{~(syst.)}$
And because the correlations between parallaxes and the other astrometric parameters 
is frequently very large, systematics must be present as well on the other 
astrometric parameters.

Another consequence of the presence of astrometric systematics is that all  
luminosity or kinematical calibrations must ensure that the 
\beforeReferee{stars are well}\afterReferee{star samples are evenly} distributed, which is in itself another issue, as completeness is difficult 
to ensure, cf. \secref{sec:summary-content}. 

Concerning proper motions, significant differences with Tycho-2 have been found 
which clearly originate from this catalogue, although some correlations 
with Gaia-only parameters may marginally also be interpreted as originating from 
Gaia, but this can only be to a much lesser extent. In particular, several  
components of close double systems have wrong astrometric solutions, due to incorrect 
cross-matches.

In TGAS as well as for the whole catalogue, the astrometric deficiencies
look related to bright stars and small number of observations.
There is no doubt that these problems will be resolved in the next Gaia data 
releases.

\section{Photometric quality of DR1}\label{photoqual}

The photometric quality of {\GDR1}, accuracy and precision, 
has been tested using both internal methods 
(using Gaia photometry only) and by comparisons to external catalogues.

\subsection{Internal test of the photometric accuracy}

Using the \gbp/\grp~photometry, a way was found to check internally the variation of the \gmag~magnitude zero point with magnitude, that we will also check below with the external catalogues. 
One should keep in mind however that the {\gaia} photometric data are correlated due to the calibration procedures. 

We randomly selected sources at high galactic latitude ($\vert b \vert>50$\deg) with photometric quoted uncertainties in \gmag, \gbp~and \grp~$<0.02$~mag and a minimum of 10 observations in each band. We re-sampled this selection to have a uniform distribution in magnitude. An empirical robust spline regression was derived which models the global (\gbp-\grp)/(\gmag-\grp) colour relation
and we computed the residuals of the observed \gmag-\grp\ minus the \gmag-\grp $=f$(\gbp-\grp) spline.

\begin{figure}
\begin{center}
\includegraphics[width=0.7\columnwidth]{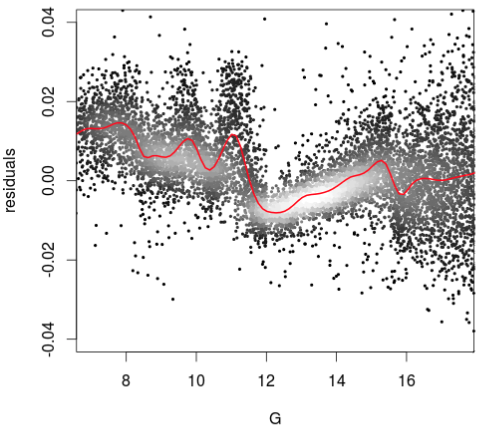} 
\end{center}
\caption[Gaia \gmag\ versus \gbp\ and \grp~photometry]{Gaia \gmag~versus \gbp\ and \grp~photometry. Residuals of \gmag-\grp\ from a global \gmag-\grp=$f$(\gbp-\grp) spline as a function of \gmag~magnitude. The red line is a smoothed spline fit. The sample contains 10\,000 stars with a uniform distribution in magnitude, therefore the lighter grey scale indicates less dispersion in the residuals. 
}
\label{fig:wp944_bprpphot}
\end{figure}

The variation of these residuals with magnitude (\figref{fig:wp944_bprpphot}) is consistent with what we observed in the comparison with external catalogues below. First, the variations at bright magnitudes ($G<$12) are most probably linked to the different gate effects and saturation issues. 
Second, the window size changes on-board at \gmag~magnitudes 13 and 16. In very preliminary data, this induced  a strong jump at \gmag=13, seen and corrected in the calibration process of the DPAC photometric group \citep{DPACP-9}. In {\GDR1} the jump at \gmag=13 seems nicely corrected but a small jump at \gmag=16 is still visible. The increase in the residual dispersion seen in \figref{fig:wp944_bprpphot} at faint magnitudes is linked to the reduced precision of \gbp/\grp. 

\subsection{Internal test of the photometric precision}

\def\gbr{$G_\mathrm{BR}$}
\beforeReferee{With only one band and its quoted precision published, validating the photometric 
precision without external comparisons looks also difficult. A method designated as
``internally homogeneous photometric samples'' was however devised to provide
at least some upper limits on the photometric precision.}

\beforeReferee{The basic principle is to compare in samples the observed dispersion of the $G$
magnitude of stars to the one which is predicted by their standard uncertainty 
(i.e. the quoted precision $\sigma_G$ in the Catalogue). 
There is obviously no way of obtaining an unbiased value about $G$ magnitude
by selecting samples based on the $G$ magnitude itself, so the principle 
is to make a selection of samples using other magnitudes, {\gbp} and {\grp}. 
In practice, we use instead what we will call {\gbr}=({\gbp}+{\grp})$/2$ for two reasons: first
this makes a magnitude closer to $G$, second the precision on {\gbr} is better 
than on {\gbp} or {\grp} alone. }

\beforeReferee{Samples were thus selected around some {\gbr} values, with an extremely
thin width around this value (about one {\mmag} width
for the brightest stars), using stars with a very similar colour, choosing e.g.
$0.78 < $ {\bprp} $ < 0.80$ as it provides samples large enough 
(other colour ranges have been used without significant changes).}

\beforeReferee{In such internally homogeneous samples, one can assume that
the $G$ magnitude has some central value with a very small dispersion 
due to the differences between stars of varying type, distance, extinction
(or variability),
and also due to the $G$ measurement errors, which is what we are trying to estimate.
Obviously, the stars in these samples have to be selected with a {\gbr} 
with a precision better or comparable to the small bin width chosen, which 
in turn strongly reduces the sample sizes. We also removed all sources with
an astrometric excess noise significance above 200 which would
be an indication of duplicity or data reduction problems.}

\beforeReferee{The central $G$ value in such subsamples remains unknown (although
it could be modelled by photometric transformations, though probably not to
the required precision), so we subtract the median value in order to keep
the $G$ residuals only, centred on 0. Then, for a given {\gbr} central
value, all the residuals obtained in all subsamples of one {\mmag} 
bin size are collected for each {\mmag} from {\gbr}-0.5 mag to {\gbr}+0.5 mag
and then added to build one single sample to study. One such resulting sample, 
around $BR=13$, is shown in \figref{fig:942-internal-photometry-hist-disp}a. The
resulting dispersion appears indeed small enough to allow the study of the effect
of the measurement errors at the {\mmag} level. }


\beforeReferee{The variance around 0 is the sum of some intrinsic variance
(due to the {\gbr} bin width and {\gbr} measurement errors, the small variations
between stars, variability, multiplicity, etc) 
and of the variance due to the measurement errors estimated by the $\sigma_G$ standard 
uncertainty. The observed  dispersion 
(robustly estimated by a MAD - Median Absolute Deviation) 
is clearly varying as expected with the $G$ standard uncertainty
(\figref{fig:942-internal-photometry-bins-precision}).}

\beforeReferee{So, a robust weighted fit has been done by cutting in 
equal size bins of standard uncertainty, estimating in each bin the variance by 
the square of the MAD and fitting the linear model:
observed variance = intrinsic variance + unit-weight variance $\times$ 
standard uncertainty squared. What is of interest here is the unit-weight 
error (square root of the unit-weight variance), representing 
the ratio of the ``external'' errors over the ``internal'' ones. 
The intrinsic variance also estimated through this fit, and which is 
of no interest here, has been subtracted in \figref{fig:942-internal-photometry-bins-precision}
to compare only the variance due to the measurement errors to the published $G$ precisions.}

\beforeReferee{The same estimation of the unit-weight error  
has been done for several values of magnitude, and for various colours,
with consistent results: an upper limit of the underestimation 
of the standard uncertainty of the $G$ magnitude
in the Catalogue as a function of magnitude in the
[10-15] magnitude range shown is a factor 1.7 at magnitude 13 
(\figref{fig:942-internal-photometry-hist-disp}b). 
There are no results for stars brighter than 10
(the number of stars being much too small to permit this
method) nor for the faintest stars (as there is a 
too large range of standard uncertainties
and to few {\gbr} precise enough to build uncontaminated bins
for the analysis).}

\beforeReferee{There is however no reason that a simple straight line 
would be valid for the unit-weight on the whole standard 
uncertainty range.
Splitting instead this range in four different bins
(though not trying to get continuity of the results between bins),
\figref{fig:942-internal-photometry-bins-precision}, 
suggests that the uncertainties might be underestimated for
the most precise stars
($\sigma_G\in(0.3,1.5]$ mmag: unit-weight error=$1.59\pm 0.12$, 
\figref{fig:942-internal-photometry-bins-precision}a), 
slightly overestimated in the medium uncertainty range 
($\sigma_G\in(1.5,4]$: uwe=$0.51\pm 0.09$ and
$\sigma_G\in(4,10]$: uwe=$0.47\pm 0.17$), 
and correctly estimated for the least precise
magnitudes
($\sigma_G\in(10,30]$: uwe=$1\pm 0.2$, 
\figref{fig:942-internal-photometry-bins-precision}b).
For the most precise $G$, one cannot exclude
however that the available {\gbr} was not precise enough to 
avoid an unmodelled dispersion, so that this method would 
provide an upper limit only on the unit-weight error.}

\afterReferee{With only one band and its quoted precision published, validating the photometric 
precision without external comparisons is difficult. 
We made experiments using {\gbp} and {\grp} in order to check that the
observed variance vary as expected with the quoted precision $\sigma_G$ 
in the Catalogue, viz. 
observed variance = intrinsic variance + unit-weight variance $\times$ 
standard uncertainty squared. For most stars, there was no indication 
that their standard uncertainties were underestimated.

However,} there are about 12 million stars with $G$ standard uncertainties 
better than 0.5 mmag, which are thus difficult to check. 
There are however indications that some of the best precisions may be too optimistic:
the 53 most precise stars (having $\sigma_G< 0.1$ mmag) have a median value
of about 80 observations while the 1000 most precise have about 500 observations 
as median value. While the latter may explain a good precision, the
former cannot, as they would otherwise beat the Poisson noise (note that
a significant fraction of DR1 sources have standard uncertainties below 
Poisson noise). 
The most precise photometry may thus contain a mix of stars with
a large number of observations (\beforeReferee{OK}\afterReferee{as expected}) 
and of stars with very small apparent scatter, 
by chance or due to correlations, and these uncertainties 
should thus not be taken at face value.

\subsection{Photometric accuracy and precision from external catalogues\label{sect:wp944photometry}}

The following tests compare the photometry of Gaia DR1, including TGAS, with external photometry. We check here the distribution of a mixed colour index, Gaia magnitude minus the external catalogue magnitude, versus an external catalogue colour. An empirical robust spline regression was derived which models the global colour-colour relation. The residuals from this model were then analysed as a function of magnitude, colour and sky position.

\paragraph{HST CALSPEC standard stars} \citep{Bohlin07}. The HST CALSPEC standard spectrophotometric database\footnote{\scriptsize\url{http://www.stsci.edu/hst/observatory/crds/calspec.html}} has been used to compute theoretical \gmag-magnitudes by convolving their spectra with the nominal Gaia passband using the pre-launch nominal passband. As this passband has not yet been adapted to the real Gaia response, expected photometric differences are observed, reaching a difference of up to 0.1~mag  at \bmv=1.2. This confirms that the pre-launch filter should not be used blindly by the community working on {\GDR1} data. Instead colour-colour transformations between Gaia and other photometric systems, available in {\GDR1} documentation, should be used. An updated passband will be provided with DR2. 


\paragraph{$BVRI$ photometric standard stars}  \citep{Landolt92}. 397 stars, mostly within the magnitude range $11.5<V<16.0$ and in the colour range $-0.3<B-V<2.3$, with photometric scatter $<0.02$ mag have been selected for this test. 
The observed dispersion around the colour-colour relation is larger than the quoted errors. This can be explained by an intrinsic stellar variability or by an under-estimation of the errors in one or both catalogues. 

\begin{figure}
 \begin{center}
\includegraphics[width=0.65\columnwidth,height=0.5\columnwidth]{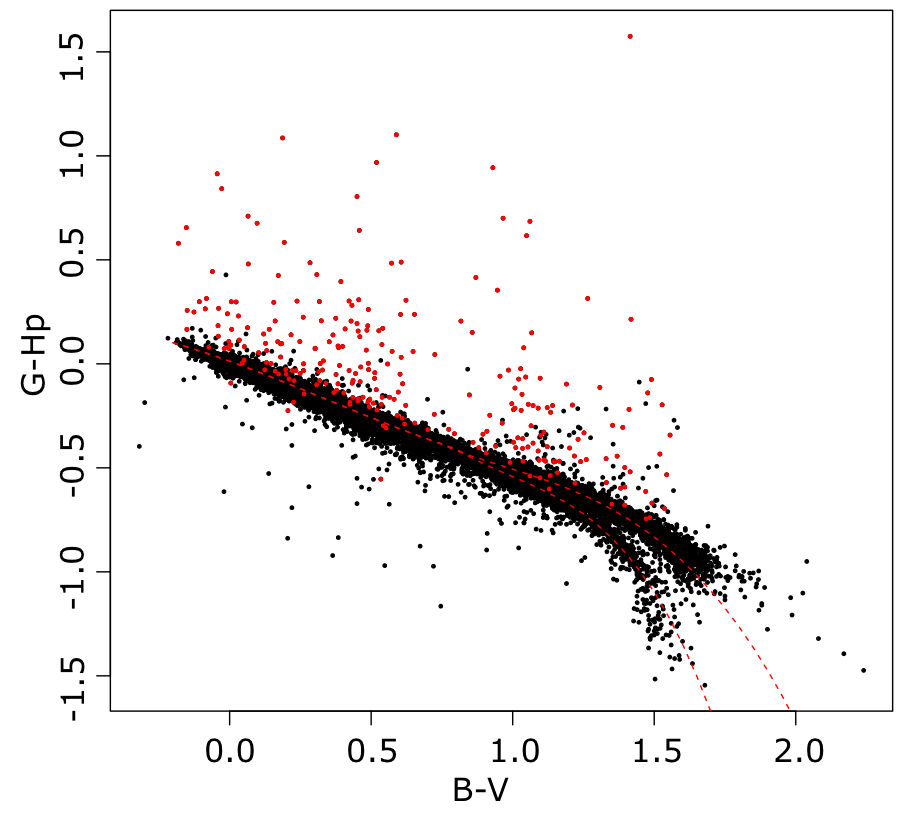} \\
\vspace{0.5cm}
\includegraphics[width=0.65\columnwidth,height=0.5\columnwidth]{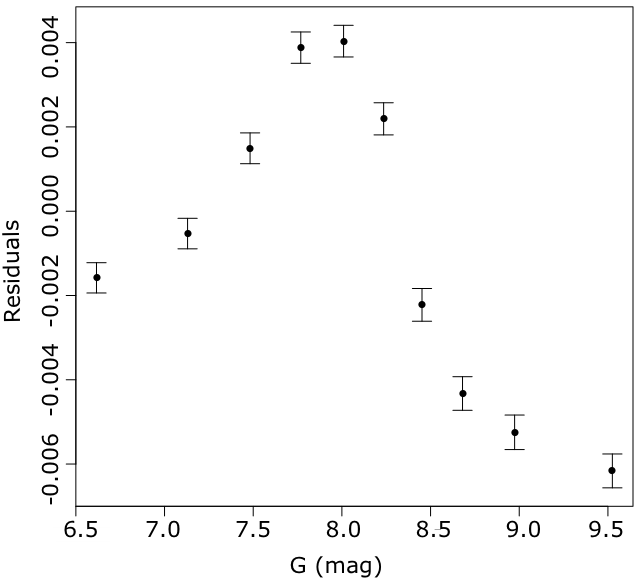} 
\end{center}
\caption[Gaia versus Hipparcos photometry]{{\GDR1} versus Hipparcos photometry. Top: Colour-colour relation \gmhp~as a function of \bmv; in red, stars with $\sigma_G>0.01$~mag; the red dotted lines are the colour-colour polynomial relation provided in the release documentation for dwarfs and giants. 
Bottom: \gmhp\ residuals from a global \gmhp\ $=f(V-I)$ spline relation as a function of \gmag~magnitude for low extinction stars only.
}
\label{fig:wp944_hipphot}
\end{figure}

\paragraph{Hipparcos photometry.} The sample of the {\it well behaved} Hipparcos stars (i.e. excluding known or suspected binaries, see \secref{sec:wp944_astrom}) has been used here with extra filters to exclude variable stars (variability flag VA=0) and restrict the sample to stars with good Hipparcos photometry ($\sigma_{Hp}<0.01$~mag and $\sigma_{B-V}<0.02$ or $\sigma_{V-I}<0.03$~mag). Although the {\preDR1} filtering removed the strongest outliers, a number of outliers are still present in the colour-colour relations, but a large fraction of them can be filtered out using their photometric errors, as illustrated in \figref{fig:wp944_hipphot}a where red dots are stars with $\sigma_G>0.01$~mag.

We have further selected a subset of the Hipparcos stars with low extinction ($A_V<0.05$~mag) using the 3D extinction map of \cite{Puspitarini14} or, when the star reaches the limit of the map, the 2D map of \cite{Schlegel98}. This selection ensures a clean colour-colour spline relation \gmhp\ vs \vmi. The residuals versus this global relation show a strong variation with magnitude (\figref{fig:wp944_hipphot}b), with an amplitude up to 0.01~mag\afterReferee{. Such a systematic is }ten times larger than the uncertainties quoted for \gmag~at magnitude 8. This is most likely due saturation effects near gate changes or residual calibration errors linked to this. 

\begin{figure}
 \begin{center}
\includegraphics[width=0.65\columnwidth,height=0.5\columnwidth]{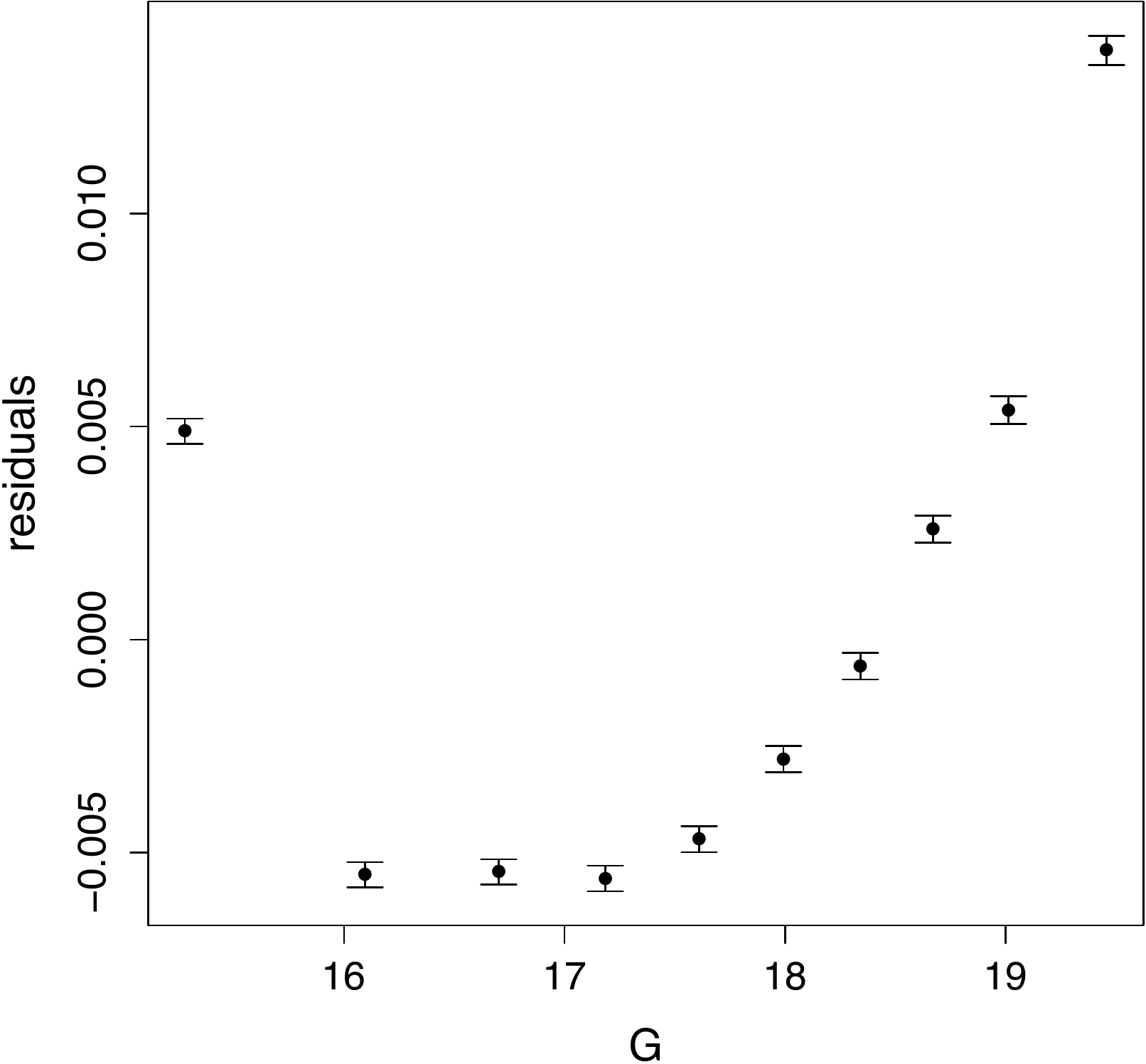} 
\end{center}
\caption[Gaia versus SDSS photometry]{{\GDR1} versus SDSS photometry: 
$G-r$ residuals from a global $G-r=f(g-i)$ spline relation as a function of \gmag~magnitude.
}
\label{fig:wp944_sdssphot}
\end{figure}

\paragraph{SDSS photometry.} \afterReferee{Here} we used \beforeReferee{here} the tertiary standard stars of \cite{Betoule13} calibrated to the HST-CALSPEC spectrophotometric standards with a precision of about 0.4\% in {\it griz}. It covers four CFHT Deep fields and the SDSS strip 82. While the CFHT fields are in low extinction regions, for the SDSS strip only areas with a maximum $E(B-V)<0.03$ according to the \cite{Schlegel98} map are selected.
The residuals versus the global colour-colour spline relation (\figref{fig:wp944_sdssphot}) show a strong increase of the residuals at the faint end in all SDSS and CFHT fields, with an amplitude larger than the quoted uncertainties, of the order of 0.01 at \gmag=20. An increase of the bias at $\sim$16~mag is also seen in the SDSS field (the SNLS is too faint to probe this magnitude) which could be due to window class change but also to saturation in the SDSS data. 
We checked that the increase at the faint end is not due to the random errors alone (as the ordinate is correlated with the abscissa in \figref{fig:wp944_sdssphot}) by checking that this increase was visible using also all the SDSS magnitudes, in particular with $z$ that is fully independent from the other magnitudes used for the residual computation. Note that we did similar checks for all the other external catalogues.

A confirmation of this global behaviour has been obtained with the OGLE data which were used for the completeness tests (\secref{sec:wp944_smallscalecompleteness}). To avoid potential zero point issues, we used data from a single CCD at a time. The large extinction of those fields lead to a less well defined colour-colour relation but the increase of the residuals with magnitude is nevertheless also seen in the OGLE data, confirming the $> 0.02$ mag zero point variation with magnitude of the Gaia photometry at its faint end. 

\paragraph{Tycho-2 photometry.} We selected only stars with photometric errors in \bt~and \vt~$<0.05$~mag and at high galactic latitude ($\vert b \vert>40$\deg) to have a low extinction. To obtain clean colour-colour relations, 
the sample has been roughly separated between dwarfs and giants with a colour cut at \btmvt~=0.9~mag and an absolute magnitude cut at \mg=4.5, taking into account the parallax error at 1~$\sigma$. The residuals show a variation with \gmag~magnitude, confirming the increase seen at $G\sim 8$ with Hipparcos and suggesting an increase at $G\sim 11$ as well. 


\paragraph{2MASS photometry.} The comparison with 2MASS is more difficult due to a sharp feature at $J-$\Ks$\sim 0.8$ for the red dwarfs and the unavailability of parallaxes in {\GDR1}. To remove the red dwarf feature, we selected only stars with $J-$\Ks$<0.7$. As for Tycho-2 we selected only stars with photometric errors in $J$ and \Ks$<0.05$~mag and at high galactic latitude ($\vert b \vert>40$\deg). The residuals also show an important variation with \gmag~magnitude. 

\paragraph{} All the tests above also show a correlation of the \gmag~residuals with Gaia \gbp-\grp~which has not been studied in detail as this colour is not part of {\GDR1}, but this variation does not exceed $\sim$0.01~mag.
Those tests also show a significant correlation between the photometric residuals and the astrometric excess noise which measures the disagreement with the astrometric model. This is expected as the astrometry and the photometry share the same PSF model
and the same windows, possibly contaminated by a neighbour.


\subsection{Testing $G$ photometry using clusters}\label{sssec:cu9val_ocphot}
 
To test the photometric accuracy and precision of {\GDR1} against published photometry of stellar clusters, we made use of a sample of high photometric quality by \cite{2008ApJS..176..262T}. These authors provided high precision photometry in $V$ band (a few  mmag), for 5 open clusters: Hyades, Praesepe, Coma Ber, NGC752 and M67. The photometry in this catalogue is highly homogeneous, both in data reduction and in zero point for all the clusters. In addition, we used M4 HST photometry by \cite{2014MNRAS.442.2381N} in $F606W$ band, where repeated observations allowed to reach a few mmag precision (for the relevant magnitudes, $F606W<$21).

\medskip
For all clusters, the same procedure was adopted, namely:
\begin{itemize}
\item the reference catalogue was checked, removing variable and multiple stars. Variability information was taken from  SIMBAD. Multiplicity information was taken from the Hipparcos catalogue. For the Hyades, we used also \cite{2016A&A...585A...7K} catalogue to remove multiple stars. In the case of M4, variability information is taken from \cite{2014MNRAS.442.2381N}. After this selection, the total number of stars is 40 in M4 (down to $G\sim 14$), and 232 in the open cluster sample.
\item We extracted for each source the {\gaia} data. For the open cluster sample stars, the cross match is straightforward, being all bright stars observed in the Hipparcos catalogue. For M4, at fainter magnitudes and with a high level of crowding, a more sophisticated cross match procedure was followed taking into account proper motions (from L. Bedin, priv. comm.).
\item The difference between $G$ magnitude and a reference magnitude does depend on the apparent colour, \beforeReferee{thus} \afterReferee{and consequently it depends} both on temperature and extinction. In the case of open clusters, to gain statistics yet working with homogeneous extinction levels, we grouped the 5 clusters according to the extinction level \citep[from][]{2008AJ....136.1388T, 2007AJ....133..370T, 2007AJ....134..934T, 2006AJ....132.2453T}. The 3 groups are: Coma and Hyades, ($E(B-V)<0.01$); Praesepe ($E(B-V) \sim 0.1$); M67 and NGC752
($E(B-V) = 0.1$ -- 0.14).  
\item For each group of OCs and for M4, we derived separately the relation between $G$ magnitude and the reference magnitude against colour, using a low order spline.
\item We analysed the residuals of this function against the apparent G magnitude.
\end{itemize}

\begin{figure}
\centering
\includegraphics[width=0.9\columnwidth]{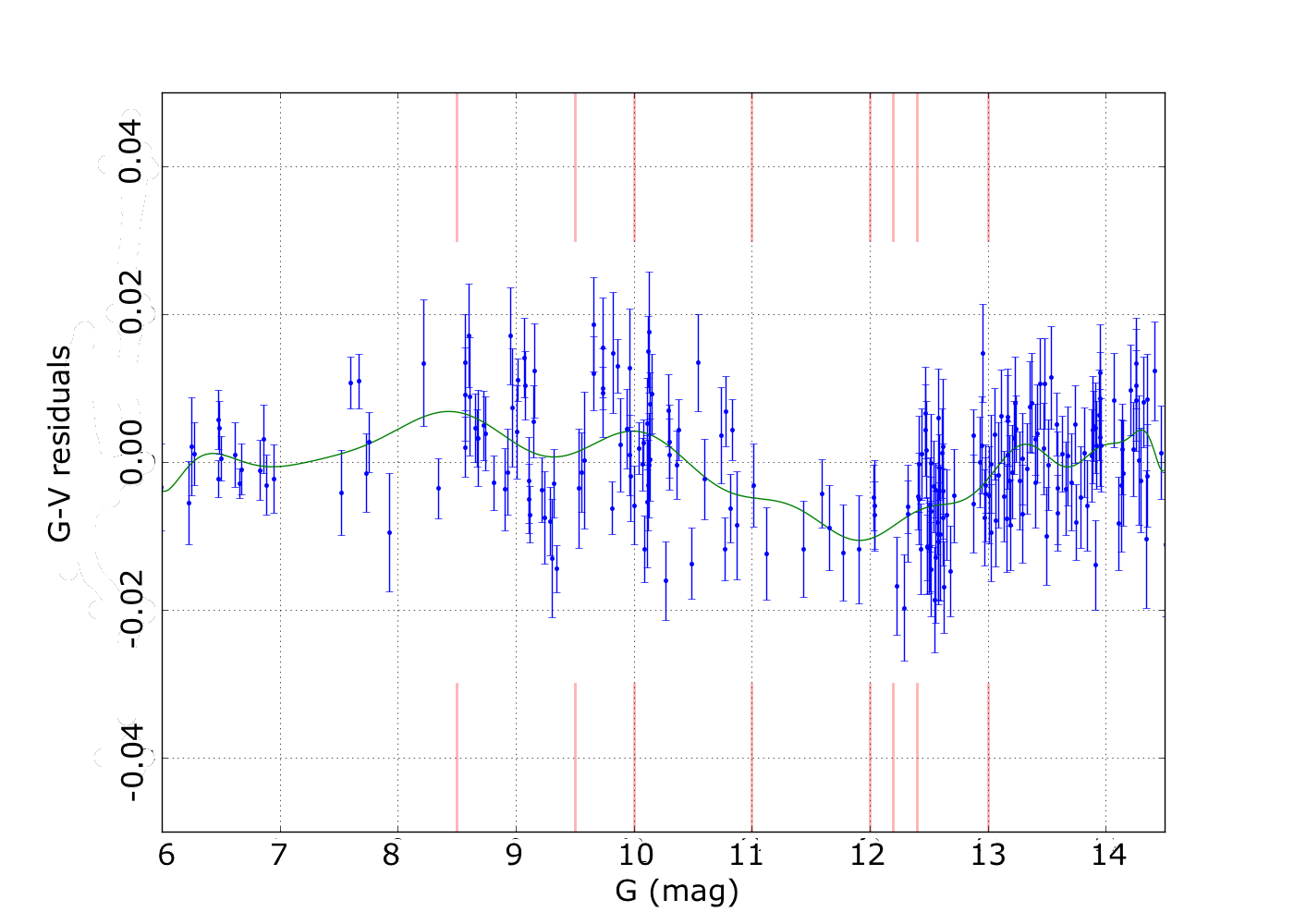}
\caption[$G-V$ vs magnitude for open clusters]{Residuals of the difference $G-V$ against a low order spline, as a function of  the magnitude\afterReferee{, for 5 different clusters}. The $V$ magnitude is from the \cite{2008ApJS..176..262T} catalogue\beforeReferee{, for 5 different clusters}. Red lines marks the gates position in magnitude. The green curve is a high-order spline fit to the data.} \label{fig:cu9val_W947_T08phot_test}
\end{figure}

\begin{figure}
\centering
\includegraphics[width=0.9\columnwidth]{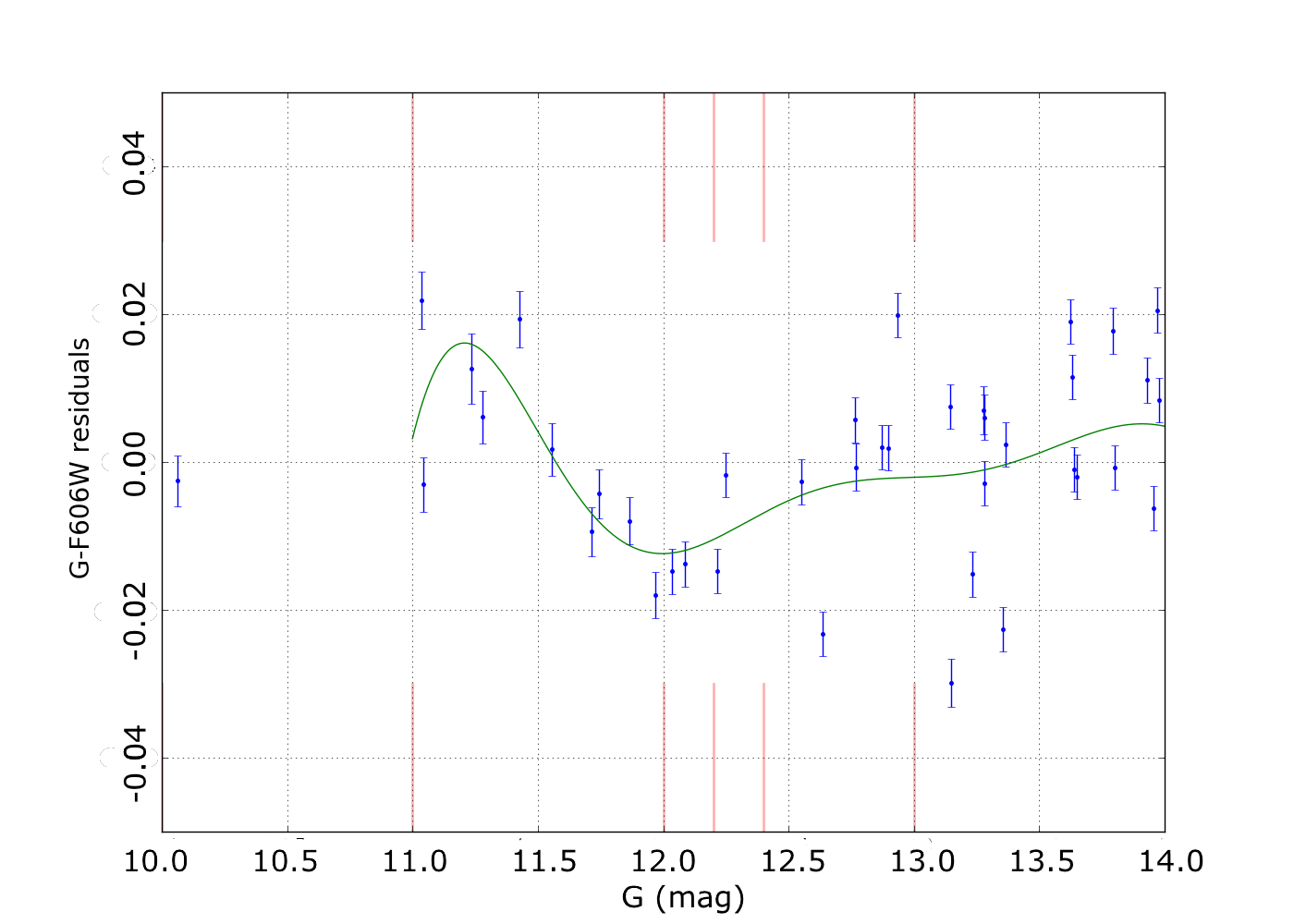}
\caption[HST $F606W$ photometry for M4]{Same as in \figref{fig:cu9val_W947_T08phot_test} but for M4 using $G$ minus HST $F606W$ photometry by \cite{2014MNRAS.442.2381N}.} \label{fig:cu9val_W947_M4phot_test}
\end{figure}

We show the residuals in \figref{fig:cu9val_W947_T08phot_test}, for the 5 open clusters together and in \figref{fig:cu9val_W947_M4phot_test} for M4. In both figures, we fitted a high order spline. The residuals clearly show systematics at a 10 mmag level related to the presence of gates, as discussed in \secref{sect:wp944photometry} using a comparison with large external catalogues.


\subsection{Photometry for variable stars}\label{sec:variables}

Gaia is particularly interesting for stellar variability studies since it provides a remarkable time-domain survey, which is going to help to better characterise already known variables and even detect new ones. {\GDR1} includes light curves for a selection of Cepheids and RR~Lyrae stars as described in \cite{DPACP-15,DPACP-13}. Several tests were developed to validate the data compared to ground-based surveys.

Additionally, objects with intrinsic or extrinsic variability may also affect the Gaia data analysis \citep{2000ASPC..210..482E}. For instance
the instrument and/or the data processing can also introduce false variability that might be interpreted as real. This aspect has been taken into consideration to implement a set of tests which verify that no significant statistical biases are present in {\GDR1}.

\subsubsection{Testing variable stars light curves.}\label{chap:cu9val_var_lc}

We compared the dataset of Cepheids and RR~Lyrae stars included in the {\GDR1} against the OGLE IV SEP catalogue \citep{2012AcA....62..219S}. We found that reported {\GDR1} periods, average $G$ magnitudes and amplitudes are in agreement with the external catalogue and no particular outlier was found. OGLE also classifies stars depending on their variability, no particular disagreement was found with {\GDR1} classification.

\begin{figure}
\centering
\includegraphics[width=0.7\columnwidth]{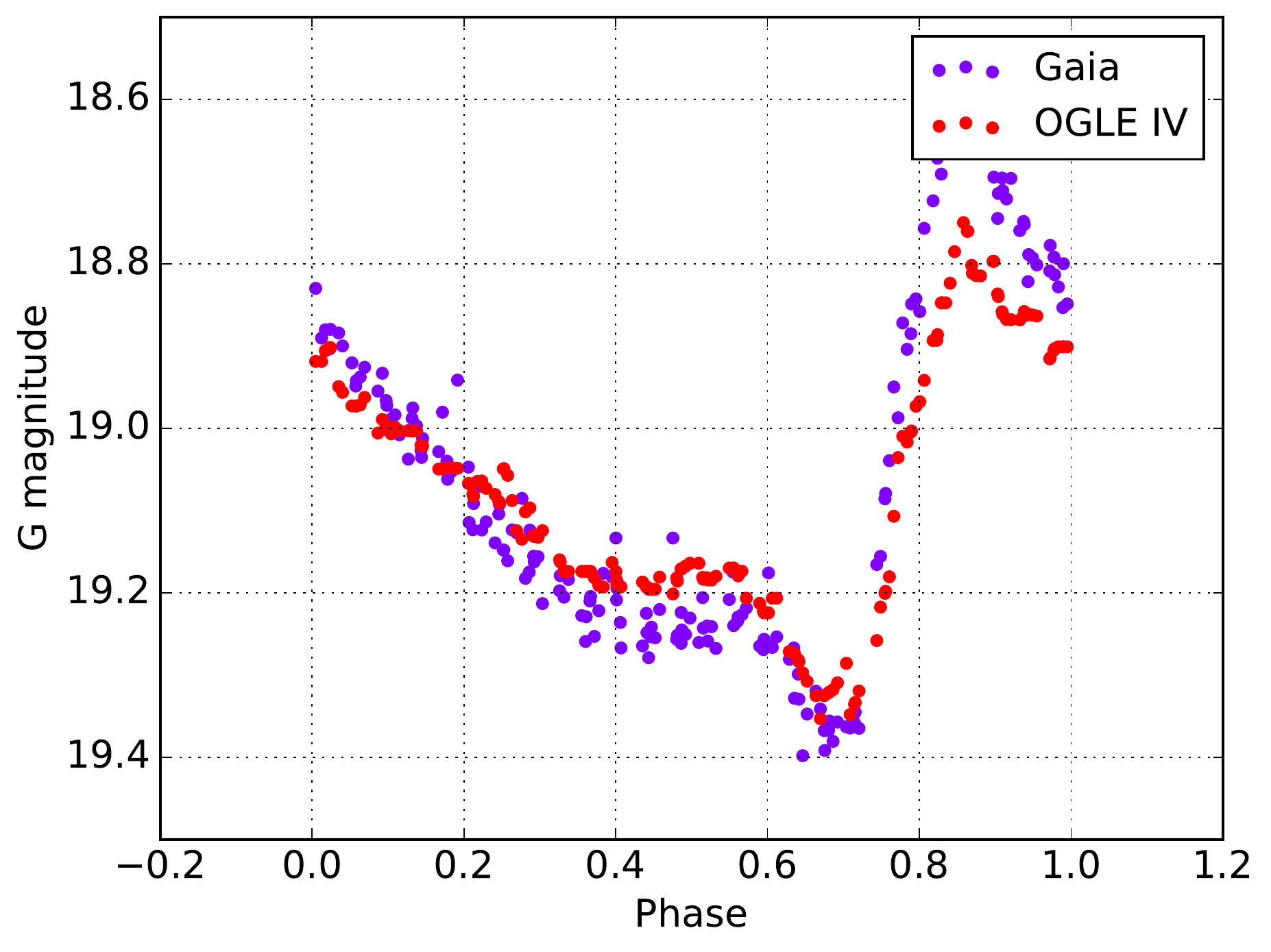}
\caption{Example of a folded light curve corresponding to a Gaia RR~Lyrae star compared to a magnitude-converted and interpolated OGLE counterpart. The interpolation process hides the real dispersion present in OGLE, which is generally greater than in Gaia.}\label{fig:946_folded_cepheid}
\end{figure}

Light curves included in {\GDR1} were also compared to OGLE IV SEP catalogue. Since OGLE uses $V$ and $I$ filters, it was necessary to transform them into $G$ magnitudes, which was possible thanks to the internal work done by the DPAC variability group. Additionally, to ease the comparison task, OGLE light curves were linearly interpolated to match the data points present in the folded Gaia light curves as shown in Fig.~\ref{fig:946_folded_cepheid}. This is a simple approach, the magnitude transformation is not perfect and the interpolation is more difficult in regions with fewer measurements, but it has been shown to be good enough to discard the presence of extreme outliers. 

Considering the whole sample, we found an average RMS of $0.04\pm0.02$ (the average {\gmag} magnitude is $\sim18.99$~mag). After a visual inspection of transformed OGLE and Gaia folded light curves with larger RMS, we did not identify any significant outlier.

The determination of the light curves of variable stars is not limited to the presence of accurate photometry, but also it is fundamental to have reliable registered times for each measurement. To validate this aspect, we computed and compared the time separations between the moment of maximum and minimum magnitude in the Gaia and OGLE light curves. As a complementary test, we also computed $v = \frac{\left(t^{\mathrm{max}}_{\mathrm{OGLE}} - t^{\mathrm{max}}_{\mathrm{Gaia}} \right)}{p}$, where $t^{\mathrm{max}}$ are the times of maximum magnitude and $p$ the period, and we considered the decimal part of $v$, which should be close to 0.000 or close to 0.999 if the variable has gone through the full variability cycle an integer number of times. Both validations were executed considering the whole group of variables together, since it is expected that in individual cases there can be variations due to sources not pulsating completely regularly. Based on statistical tests, we did not find any significant discrepancy in the reported times between catalogues.

%
%
%

\subsubsection{Comparing distributions of variable stars to constant stars.}\label{chap:cu9val_var_dist}

The Hipparcos catalogue and its variability classification was used as main reference for creating two different subsets of Gaia sources with constant and variable stars. Then, these groups were compared to check whether:

    \begin {itemize}
    \item Parallaxes are not affected by variability
    \item No correlation exists between parallaxes or parallax uncertainties and periods, amplitudes, mean $G$ magnitude or colours
    \item Mean $G$ magnitude are within known min/max magnitudes for variable stars.
    \end {itemize}

The cross-matched group formed by constant stars contained 36\,661 sources with a mean {\gmag} magnitude of $8.27\pm1.11$, while the variable stars group was composed by 1\,820 sources with a mean {\gmag} magnitude of $8.26\pm1.10$. Based on statistical tests, we found that the normalized parallax difference distributions between these two groups were consistent and, for periodic stars, that no correlation were identified with periods or amplitudes. Hence, stellar variability does not seem to have a major effect in the reported {\GDR1} parallaxes.

\subsection{Summary of the photometric validation}\label{sec:summary-photo}

With very precise photometry for (much) more than one billion stars, 
the {\gaia} photometry is on the verge of becoming a standard for
several decades. It is thus extremely important to understand the
properties and limitations of $G$ photometry for {\GDR1}.

It appears that systematics are present at the 10 mmag 
level with a strong variation with magnitude. This
is well above the standard uncertainties for bright stars and
could originate from saturation and gate configuration changes.
These points will be solved for the DR2. 

As for the photometric precision, the standard uncertainties may be 
underestimated for the most precise, \beforeReferee{though by a factor smaller than 1.7, but is}
\afterReferee{but they are} probably correctly estimated for most of the other stars.

\section{Conclusions and recommendations for data usage}\label{sec:conclusions}

This paper summarizes the results of the validation tests applied to the
first Gaia data release as a final 
quality control before its publication. These tests have both 
confirmed the global quality of the data and shown several shortcomings
due to the preliminary nature of the release, based on a 
limited set of observations and processed using initial 
versions of the processing pipelines, see 
\citet{DPACP-14,DPACP-12,DPACP-8}
for a more detailed discussion on these issues.

We advise the users of {\GDR1} to keep these shortcomings
in mind for its scientific exploitation since they may
have relevant effects on the final results extracted from
them. In the next sections we discuss some of the main
limitations arising from them, but the limitations
for the use of the Gaia data in any specific case
should be carefully assessed as a part of the data analysis.

\subsection{Effect of correlations}\label{sec:effect-correlations}

The astrometric data in DR1 is provided with formal uncertainties for each one of
the parameters (five in the case of TGAS and two in the case of the main
catalogue). Although these standard uncertainties are enough when using each
of the parameters in isolation, they do not contain the complete information 
about the error distribution of the astrometric data. Indeed, the astrometry
of a star in the Gaia catalogue is the result of the Astrometric Global 
Iterative Solution -- \cite{DPACP-14} -- and therefore its parameters
(whether two or five) are obtained from a joint fitting during the Source
Update stage. Thus, strictly speaking, the error distributions of these 
parameters can only be described by a joint distribution of all of them. 

For this reason DR1 provides, in addition to the standard uncertainties, a correlation
matrix for the astrometric parameters: a correlation value is given in
dimensionless units (values in the range $[-1,1]$) for each pair of 
parameters. This matrix should be used for the error analysis when the
astrometric parameters are jointly used. For instance, the calculation of the
transverse spatial velocity of a star requires the use of its parallax (for
the distance) and the proper motions in right ascension and declination; 
therefore the three correlations between them will be needed for the
error analysis, since if the correlations are high the three uncertainties cannot 
be treated as being fully independent. If the correlations are not
included, the dispersion of velocities could be underestimated, for 
instance.

\begin{figure}
 	\centering
   \includegraphics[width=0.49\columnwidth,height=0.4\columnwidth]{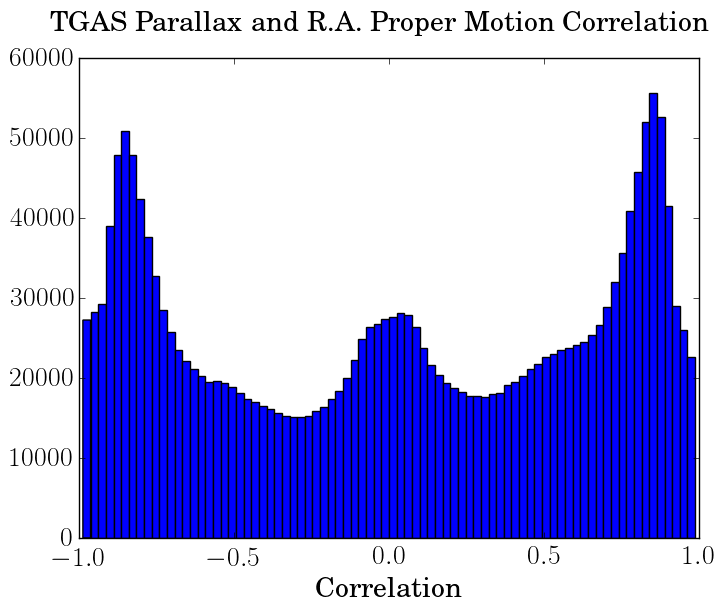}
   \includegraphics[width=0.49\columnwidth,height=0.4\columnwidth]{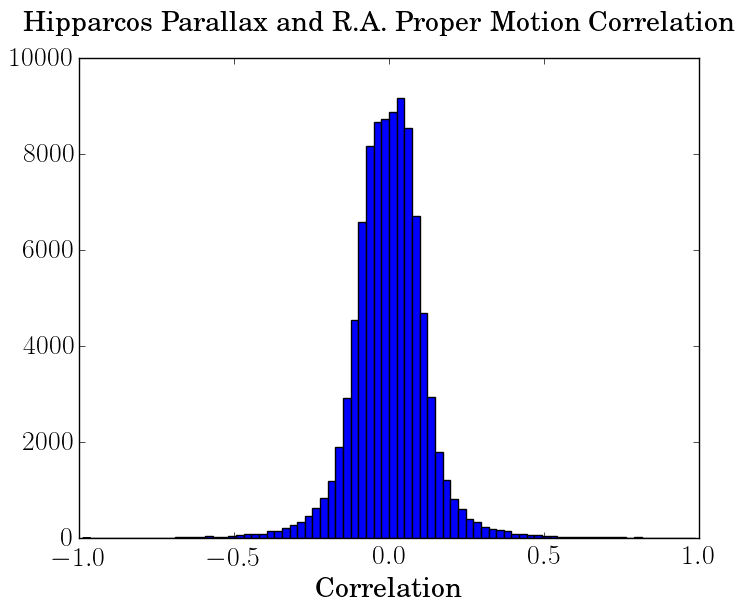}
   \caption{Histogram of the correlations between proper motion in right ascension and parallax 
   for the whole TGAS dataset (left)
   and for its subset of Hipparcos stars alone (right).\label{fig:pmra-pi-correlation}}
 \end{figure}

It is also important to note that in Gaia DR1, due to 
the limited timespan and number of observations, the values of these correlations
can be large. For instance, \figref{fig:pmra-pi-correlation}a shows the
histogram of the $\mu_{\alpha *}$ and $\varpi$ correlations in the TGAS dataset. It is
clear that the fraction of stars with high correlations is large. However, although 
this applies to most TGAS stars, the Hipparcos subset is strikingly different, 
\figref{fig:pmra-pi-correlation}b, as the precise first epoch Hipparcos 
positions allowed to better decouple the proper motion from the parallax.

The usage of the {\GDR1} covariance matrix between parameters should however be done with some caution. All the tests done against external catalogues using the covariance matrix to compute the residuals $R_\chi$ indicate a much larger number of outliers than when using only each astrometric parameter normalised residuals independently. The abnormally high values of $R_\chi$ can be seen in  \figref{fig:cu9val_wp944_hipPMchi2}a for the Hipparcos catalogue and they \beforeReferee{are most probably explaining} \afterReferee{most probably explain} the bright Gaia sources mis-match with UCAC4 (\figref{fig:wp944_ucac4}) as well as the high number of LMC member stars removed by a $\chi^2$ test. 
Moreover, a strong increase of the $R_\chi$ residual for bright sources has been seen on the Hipparcos proper motions (\figref{fig:cu9val_wp944_hipPMchi2}b) as well as on the ICRF2 QSO positions. This indicates that a censorship using the covariance matrix will induce a censorship on the magnitude too.

And again, beside the correlations between astrometric parameters, there are also
correlations between stars which produce systematics at small 
scales (\secref{sec:summary-astro}).

\subsection{Censorships and truncations, completeness}

As discussed in \secref{sec:completeness}, {\GDR1} is incomplete in several ways. There are global effects, small scale effects and effects related to crowding, angular separation, brightness, colour and position that make the incompleteness of the catalogue very difficult to describe. For this reason the use of {\GDR1} for star count analysis, although not impossible, should be done with great care. Specially in small fields the complex features of the completeness caused by under-scanning and lack of on-board resources, as depicted in \figref{fig:942_holes} should be taken into account.

\subsection{Data transformation and error distributions}

Besides the above described limitations due to the characteristics
of {\GDR1}, related to its preliminary nature, we want to conclude 
this paper with a warning to the user about potential biases
introduced by the use of transformed quantities. We will not discuss
this issue in full since it is not the goal of this text, and 
we rather refer the reader to other texts.

First of all, the TGAS dataset in {\GDR1} provides an unprecedented
set of stellar parallaxes, more than two million. But most frequently 
the users of these data will rather be interested in obtaining 
stellar distances from the parallaxes, and the first obvious idea
will be just to apply the well known relation 

$d = \frac{1}{\varpi}$

\noindent where $d$ is the distance in parsecs and $\varpi$ is the parallax
in arcseconds. Although this relation is formally true, the presence
of observational errors complicates its use for the {\it estimation}
of distances from parallaxes. Notice that we use on purpose the word {\it
estimation} because in practice this is the most we can do to obtain a distance
from a parallax: build an estimator. Due to the observational error
the observed parallax will be a value {\it around} the true parallax,
determined by some statistical distribution describing the error. In
the case of Gaia this distribution is almost gaussian, its width given
by the standard uncertainties in the catalogue and centered (unbiased) in the
true value within the limits of the systematics described in previous
sections. 

A discussion on how to use the observed parallaxes, understood as
these realisations of the error distributions was already presented
at the time of the release of the Hipparcos catalogue in \citet{1997IAUJD..14E...5B}
and a further discussion can be found in \citet{1999ASPC..167...13A}. 
We refer the reader to these papers, which warn about
the truncation of samples based on the relative parallax error and 
the bias in the estimated distances if one just naively inverts the
observed parallaxes.

Solving these problems is not obvious. Simple procedures
can help to some extent, for instance never average distances
obtained from inverting observed parallaxes, but rather first average
the parallaxes and then invert the result -- see \citet{1999ASPC..167...13A} --.
But a proper solution would require a careful analysis of the 
problem in hand to define an unbiased estimator of the distances
needed, for instance using a Bayesian estimator. We refer the reader
to \citet{2015PASP..127..994B} for a discussion of this kind of
methods. Beside distances, another application of parallaxes is
the computation of an absolute magnitude; here again, the formal
expression $M_G = m_G - 10 + 5 \log(\varpi) - A_G$ has to face
the non-linear use of parallaxes having an observational error.

\begin{figure}
 \centering
   \includegraphics[width=0.49\columnwidth]{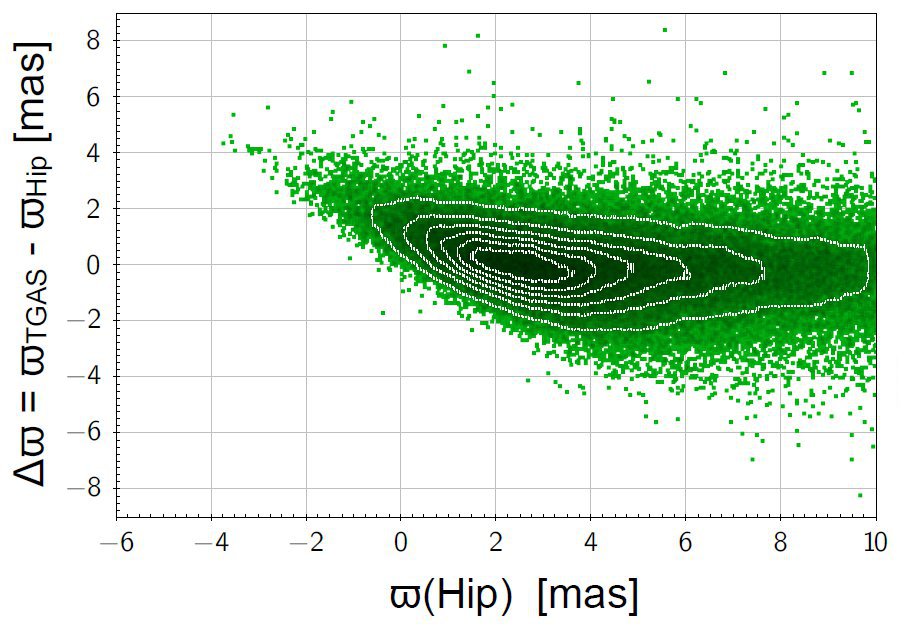}
   \includegraphics[width=0.49\columnwidth]{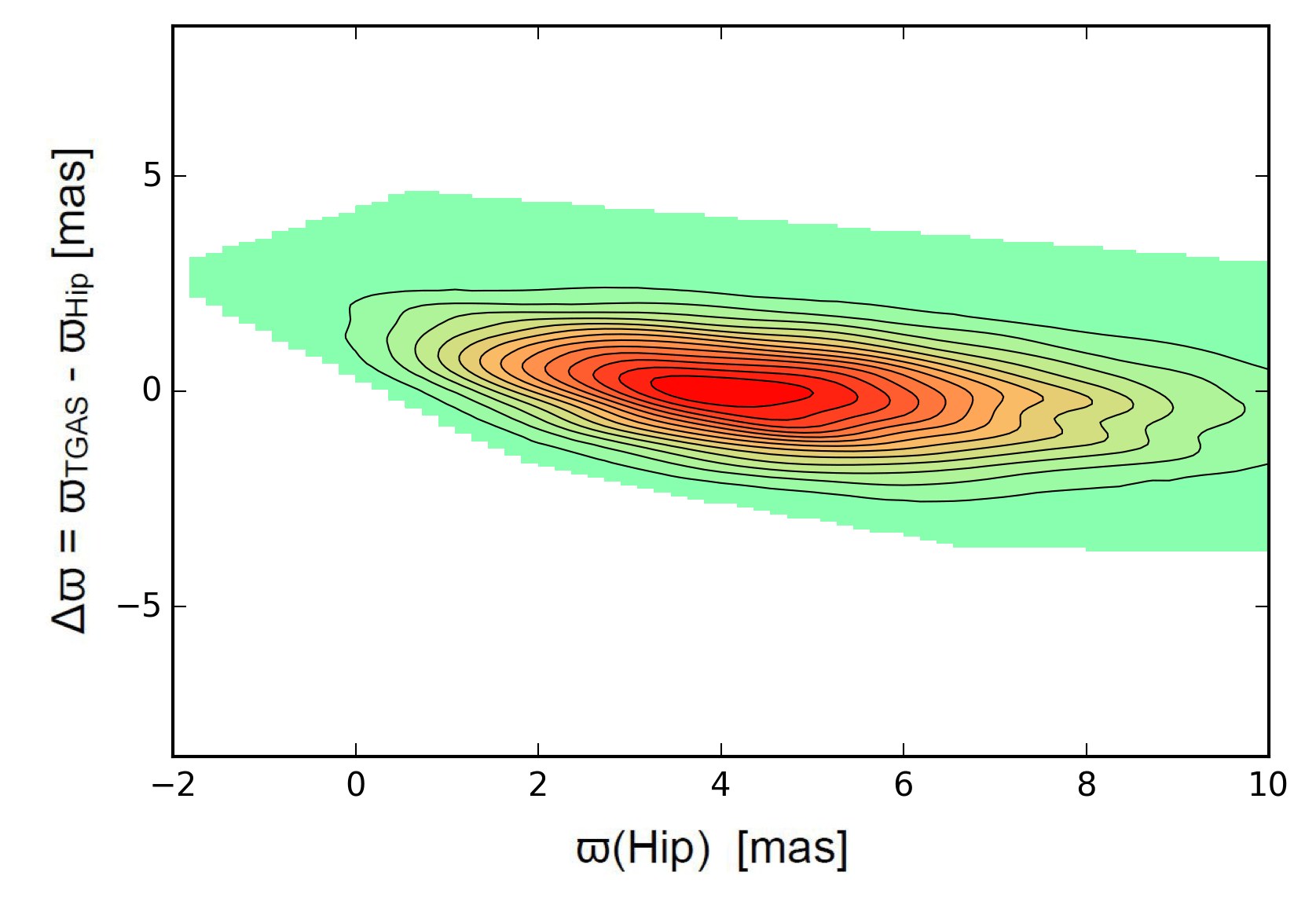}
   \caption{TGAS minus Hipparcos parallaxes vs Hipparcos parallaxes, source: L. Lindegren (left).
   Simulation based on completely unbiased sets of Hipparcos-like and TGAS-like 
   parallaxes (right). \label{fig:pi-Hip-Tgas}}
 \end{figure}

Beyond the problems with the use of trigonometric parallaxes discussed
in the papers cited above we also want to add a word of warning 
about the comparison of the Gaia DR1 parallaxes with parallaxes from
other sources. In this case the properties of the error distribution
in {\em each} catalogue, and their combined effect, should be properly taken
into account when drawing conclusions about the comparison. We will illustrate
this with a couple of examples. First, to compare the Hipparcos and
TGAS parallaxes one can draw a plot of the differences between them versus
the Hipparcos parallaxes. The result can be seen in \figref{fig:pi-Hip-Tgas}a,
and to the unaware reader this figure can suggest a strong systematic difference
between the two sets for small values of the parallax $\varpi<2$ mas.
However, such a behaviour is just what one can expect when drawing this
figure when the two sets of parallaxes have significantly different values
of the uncertainties. Figure \ref{fig:pi-Hip-Tgas}b shows this using simulated data. 
Starting from a set of error-free (simulated) parallaxes imitating
the distribution of the dataset used in the previous figure, two sets of 
parallaxes were generated: one with uncertainties around 1~mas (Hipparcos-like) 
and another one with uncertainties around 0.3~mas (TGAS-like). As can be seen
in the figure, in spite of the simulation being completely bias-free
and therefore without any systematic difference between the two sets
of parallaxes, the figure is similar to the one from real Hipparcos data
and could (wrongly) suggest the presence of systematic effects in one
or another catalogue. In fact, the asymmetric top-tail in these figures
is just an effect of the longer tail of negative parallaxes in the 
Hipparcos data when compared with the TGAS data.

\begin{figure}
 \centering
   \includegraphics[width=0.8\columnwidth]{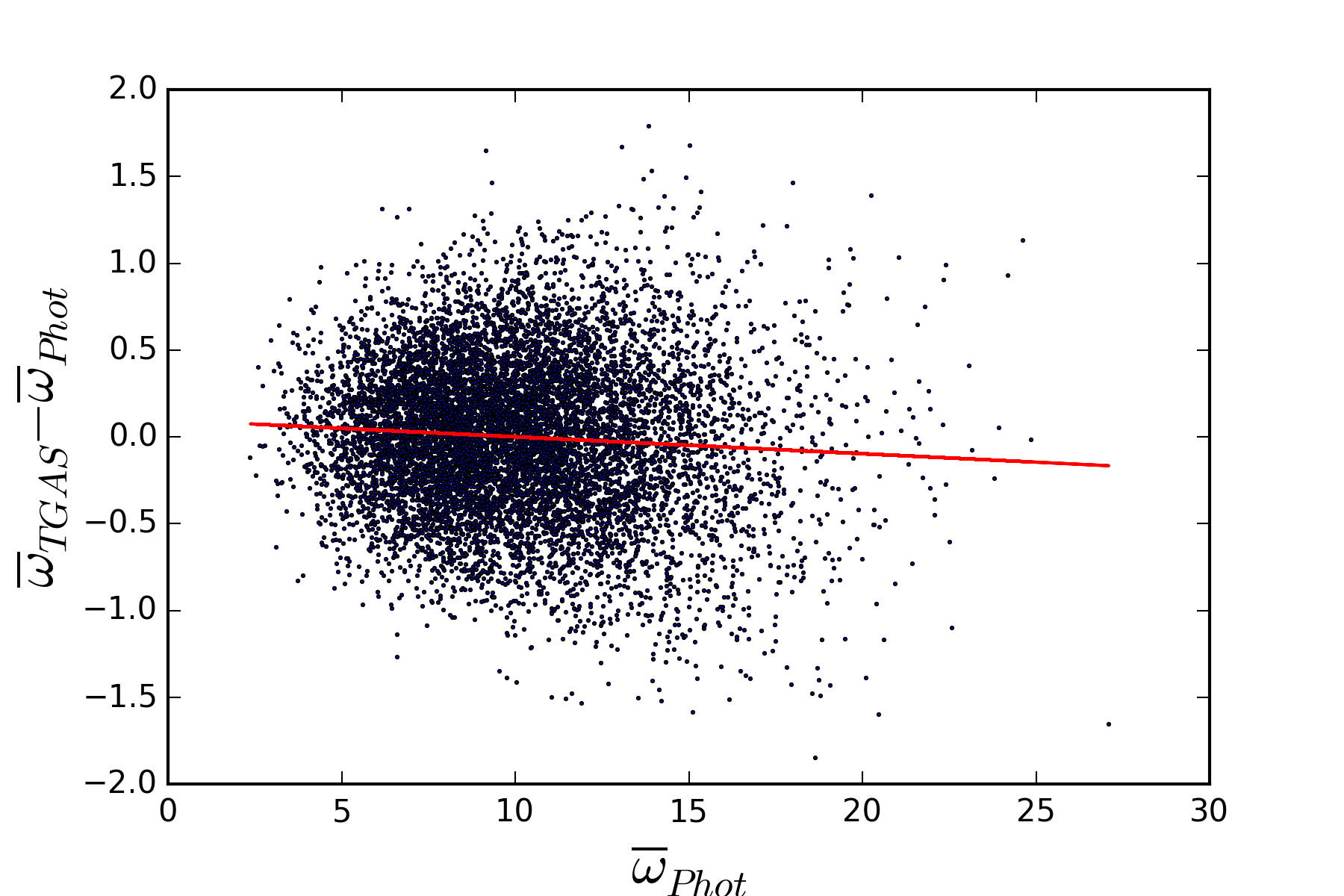}
   \caption{Simulation comparing photometric parallaxes with TGAS-like parallaxes. Notice that
            in spite of the complete absence of biases and therefore any
            systematic difference, there is an apparent systematic difference
            between the two datasets, specially for large parallaxes. \label{fig:DifPi-Phot}}
 \end{figure}

A second example of such effects deriving from the error distributions in the 
parallaxes is present when comparing trigonometric parallaxes versus photometric
or spectroscopic parallaxes. In this case the effect does not come from the
different magnitudes of the errors but from their different distributions, the first
ones being gaussian and the second one (derived from magnitudes or spectra)
being log-normally distributed. Figure \ref{fig:DifPi-Phot} shows another simulation
illustrating this effect. Starting from a set of error-free (simulated) parallaxes 
two sets of parallaxes were generated: one with log-normal errors (photometric-like) 
and another one with normal errors, in both cases with a standard deviation of 0.3~mas. 
Again, the figure could suggest to the unaware reader a systematic effect, making
the TGAS parallaxes smaller than the photometric ones, specially for large parallaxes
(short distances). The linear fit (red line)
added to the figure stresses this effect. However, as stated, the simulation is
completely bias-free and therefore this effect comes purely from the properties
of the error distributions of the two datasets
and the complete (anti)correlation between abscissa and ordinate
\citep[see also][Fig. 4]{1999ASPC..167...13A}.

The discussions presented above about the proper use of the parallaxes
also extend to the case of the $G$ magnitude contained
in Gaia DR1. The archive does not contain, on purpose, standard uncertainties for
these magnitudes. Instead, errors are given for the fluxes from which
these magnitudes are obtained, along with the fluxes themselves. The problem
in this case is again that the obtention of the desired quantity, 
the magnitude $m$, from the observed quantity, the observed flux
$F$ is non-linear:

$m = -2.5 \log(F) + C_0$

\noindent where $C_0$ is the zero point of the photometric band.
As in the case of the parallax this non-linearity will introduce
biases if not properly taken into account, although in this case
the effect is less severe because the relative errors are smaller.

We recall here however that the flux uncertainty provided in {\GDR1} corresponds to the observed scatter which can be much lower than the systematics and \beforeReferee{is therefore not}\afterReferee{may therefore not be} fully representative of the actual uncertainties, especially for bright stars.

\subsection{Conclusion}

At the end of this paper, it is needed to recall that the validation, 
by its very nature, has insisted more on the various problems found rather 
than on the intrinsic quality of the Catalogue. The summary about 
the Catalogue completeness can be found in \secref{sec:summary-content},
what was found about astrometry in \secref{sec:summary-astro}, and conclusions 
about photometry are given in \secref{sec:summary-photo}.

It must nevertheless be underlined that the {\GDR1} represents 
a major breakthrough since the Hipparcos Catalogue
on the direct measurement of the solar neighbourhood. With $20\times$ 
more stars than Hipparcos, and a median precision $3\times$ better,
it will provide new basis for studies on stellar physics and galactic
structure, provided the limitations shown above are accounted for.

With the promise of soon being superseded by the {\gaia} DR2 data,
{\GDR1} proves the ESA cornerstone mission concept, the good health of the 
instruments, the capabilities of the on-ground reconstruction, 
and the strong dedication of the community members involved in the project.

\begin{acknowledgements}

Funding for the DPAC has been provided by national institutions, in
particular the institutions participating in the Gaia Multilateral Agreement: 
the Centre National d'Etudes Spatiales (CNES), 
the European Space Agency in the framework of the Gaia project.

This research has made an extensive use of Aladin and the SIMBAD, VizieR databases 
operated at the Centre de Donn\'ees 
Astronomiques (Strasbourg) in France 
and of the software TOPCAT \citep{2005ASPC..347...29T}.
This work was supported by the MINECO (Spanish Ministry of Economy) - FEDER through grant ESP2014-55996-C2-1-R,  MDM-2014-0369 of ICCUB (Unidad de Excelencia `Mar\'ia de Maeztu') and the European Community's Seventh Framework Programme (FP7/2007-2013) under grant agreement GENIUS FP7 - 606740.
We acknowledge the computer resources, technical expertise and assistance provided by the Red Espa\~nola de Supercomputaci\'on and specially the MareNostrum supercomputer at the Barcelona Supercomputing Center.
AH, MB and JV acknowledge financial support from NOVA (Netherlands Research School for Astronomy), and from NWO in the form of a Vici grant. R.L. acknowledges support from the French
National Research Agency (ANR) through the STILISM project.
This work has been possible thanks to the support and efficiency 
of A. Brown, G. Gracia and J. Hern\'andez, to cite a few only. In particular,
we thank F. van Leeuwen for drawing our attention towards the property of the 
Hipparcos subset with respect to \figref{fig:pmra-pi-correlation},
and L. Lindegren for many inputs. 
\end{acknowledgements}

\bibliographystyle{aa} 
\bibliography{dpac,biblio} 

\appendix

\section{Gaia archive interface validation}\label{sec:betatest}

\subsection{Testing Methodology}\label{sec:wp910:method}

This section discusses the validation procedures employed in testing the design and interfaces of the archive systems delivering the {\GDR1}  data to the end user community. 

The design of the Gaia Archive was such as to fulfil the set of data access requirements gathered through a community scoping exercise. The Gaia user community were asked to suggest a number of ``Gaia data access scenarios'' and
enter them on the Gaia Data Access wiki pages at {\small\url{http://great.ast.cam.ac.uk/
Greatwiki/GaiaDataAccess}}. All scenarios received to March 2012 were considered and analysed, and presented in the DPAC Gaia data access scenarios scoping document GAIA-C9-TN-LEI-AB-026\footnote{{\scriptsize\url{http://www.rssd.esa.int/doc_fetch.php?id=3125400}}}. The Gaia ESA Archive~\citep{DPACP-19} was designed to take into account these user requirements. 

Within the Catalogue validation exercise a ``Gaia Beta Test Group'' (BTG) was constituted with a remit to perform a range of usage tests on the Gaia Data Archive and associated access clients and interfaces. The BTG is composed of members from across the DPAC, with expertise in all areas of Gaia. In addition the BTG includes members from the astronomical data centres associated with DPAC. 

The BTG generated a range of archive tests, documented the results of these tests, and raised fault reports in cases where the tests failed. These issues were reported through the DPAC ticketing system, with each being assigned to the relevant members of the Gaia Archive team. 

A range of the test queries generated have subsequently been re-used as part of the user documentation associated with the {\GDR1} release, in particular many queries have entered the  {\GDR1} Cookbook\footnote{{\scriptsize\url{https://gaia.ac.uk/science/gaia-data-release-1/adql-cookbook}}}. 

\subsection{Testing the Main {\GDR1}  Archive}\label{sec:wp910:main}

The main website access to the {\GDR1}  data is accessible at {\small\url{http://archives.esac.esa.int/gaia/}}. This was made available to the BTG at an early stage, initially populated with simulation data. Testing commenced early 2016, with an initial focus on the web interfaces to the archive. This included queries constructed via the simple form based archive pages, or more complex queries using ADQL (Astronomical Data Query Language\footnote{Documentation for ADQL available at {\scriptsize\url{http://www.ivoa.net/documents/REC/ADQL/ADQL-20081030.pdf}.}}, a IVOA\footnote{International Virtual Observatory Alliance: {\scriptsize\url{http://www.ivoa.net}}} standard. 

Later testing exercised remote programmatic access utilising the IVOA Table Access Protocol\footnote{see the IVOA Standard definition at {\scriptsize\url{http://www.ivoa.net/documents/TAP/20100327/}}} interface. 

Issues raised included those related to the user interface issues and also to the archive documentation. Functionality issues covered topics such as simplifying bulk data download, to use of server side storage, to inconsistencies in data table schemas. 

At the time of {\GDR1} release to the community, all raised issues classified as high priority have been fixed or resolved. Some lower priority issues will be addressed in upcoming maintenance releases, these being documented at the time of public data release. 

\subsection{Testing the {\GDR1}  Partner Archives}\label{sec:wp910:partner}

The {\GDR1} will also be released through a number or `partner' data centres. These provide alternative access points to the Gaia data, and additionally each provides some specific functionalities not available through the main ESA Gaia archive. 

The Gaia partner archives publishing {\GDR1} data are available at the following access points:

\begin{itemize}
\item Centre de Donn\'{e}es astronomiques de Strasbourg (CDS): {\small\url{http://cdsweb.u-strasbg.fr/gaia#gdr1}}
\item Leibniz-Institute for Astrophysics Potsdam (AIP): {\small\url{https://gaia.aip.de/}}
\item Astronomisches Rechen-Institut (ARI),  Zentrum f\"{u}r Astronomie der Universit\"{a}t Heidelberg: {\small\url{http://gaia.ari.uni-heidelberg.de/}}
\item ASI Science Data Center, Italian Space Agency  (ASDC): {\small\url{http://gaiaportal.asdc.asi.it/}}
\end{itemize}

Each partner data centre was provided with the {\GDR1}  data in early August 2016 in advance of the {\GDR1}  data release. This enabled a range of tests of the interfaces to be carried out by the BTG. All issues found were reported to the operators of these partner data centres. 

%

\section{Acronyms}\label{sec:acronyms}
{\small
\begin{tabular}{ll}
\hline\hline
\textbf{Acronym}  &  \textbf{Description}  \\ \hline
2MASS & Two-Micron All Sky Survey \\
AC & Across scan (direction) \\
ACS & Advanced Camera for Surveys (HST) \\
AGIS & Astrometric Global Iterative Solution \\
AL & ALong scan (direction) \\
BGM & Besan{\c{c}}on Galaxy Model \\
BP & Gaia Blue Photometer \\
CCD & Charge-Coupled Device \\
CFHT & Canada-France Hawaii Telescope \\
DPAC & Data Processing and Analysis Consortium \\
EPSL & Ecliptic Pole Scanning Law \\
GC & Globular cluster \\
HIP & Hipparcos catalogue \\
HPM & High Proper Motion \\
HST & Hubble Space Telescope \\
HealPix & Hierarchical Equal Area isoLatitude Pixelisation \\
IGSL & Initial Gaia Source List \\
LMC & Large Magellanic Cloud \\
MAD & Median Absolute Deviation \\
NSL & Nominal Scanning Law \\
OGLE & Optical Gravitational Lensing Experiment \\
PSF & Point Spread Function \\
RAVE & RAdial Velocity Experiment \\
RECONS & REsearch Consortium On Nearby Stars \\
RP & Gaia Red Photometer \\
SDSS & Sloan Digital Sky Survey \\
SED & Spectral Energy Distribution \\
SMC & Small Magellanic Cloud \\
TDSC & Tycho Double Star Catalogue \\
TGAS & Tycho-Gaia Astrometric Solution \\
URAT & USNO star catalogue \\
WFPC2 & Wide-Field and Planetary Camera 2 (HST) \\
\hline
\end{tabular} 
}

\section{Statistics}\label{sec:statistics}
	  \begin{figure*}[!t]
	  \centering
	    \includegraphics[width=0.33\textwidth]{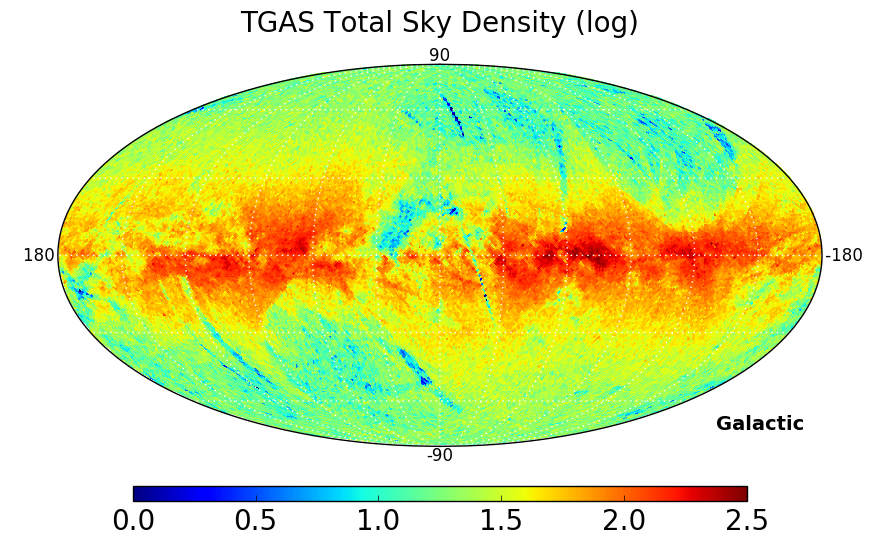}
        \includegraphics[width=0.33\textwidth]{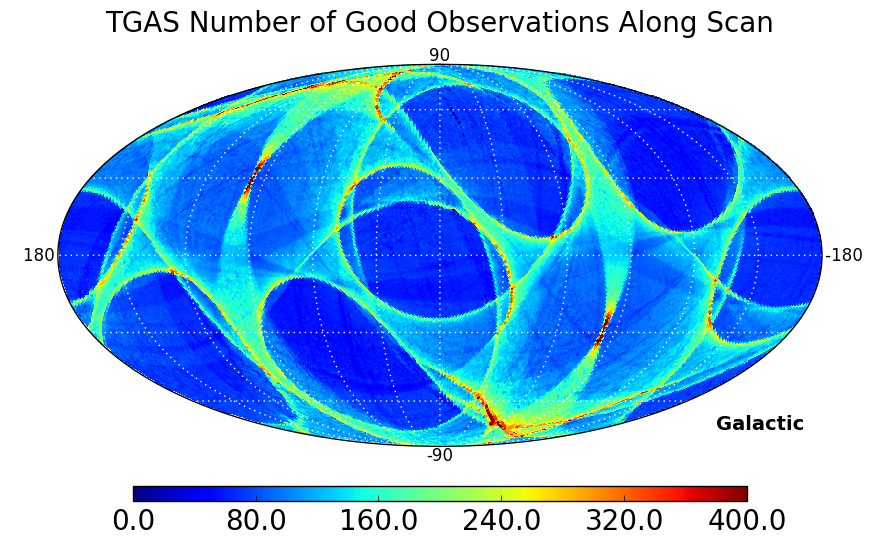}
        \includegraphics[width=0.33\textwidth]{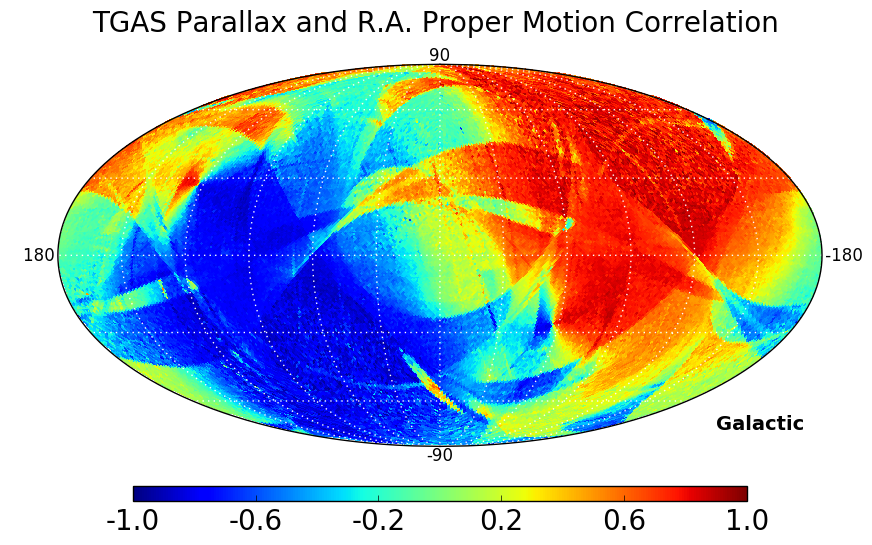}
    	      \caption{Sky map in galactic coordinates of TGAS: logarithm of star density (left),
	      number of good observation AL (center),
	       of the correlation between parallax and proper motion in right ascension (right).}
            \label{fig:TGAShealpix_stat_TotalSkyDensityLogCount-GAL_HealpixMapSAM}
    \end{figure*}
    
     \begin{figure*}[!h]
	  \centering
	    \includegraphics[width=0.33\textwidth]{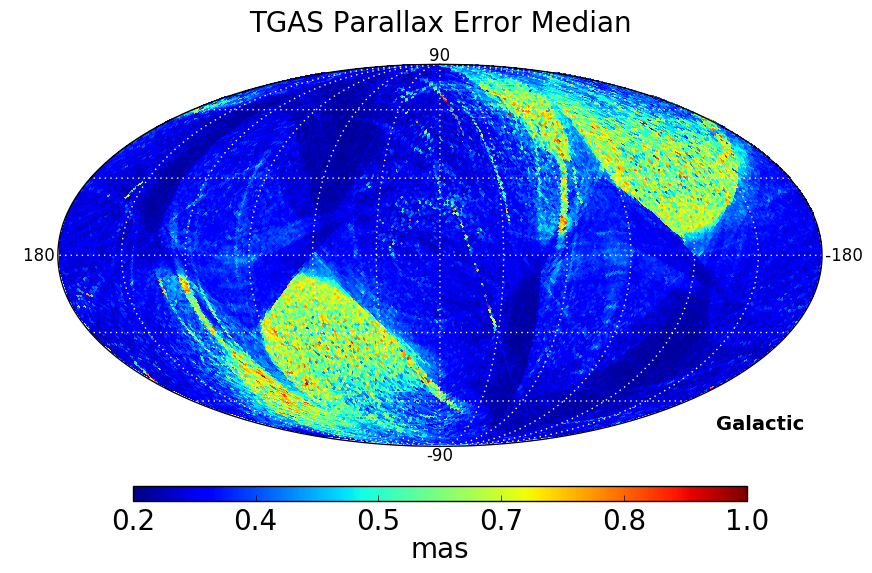}
	    \includegraphics[width=0.33\textwidth]{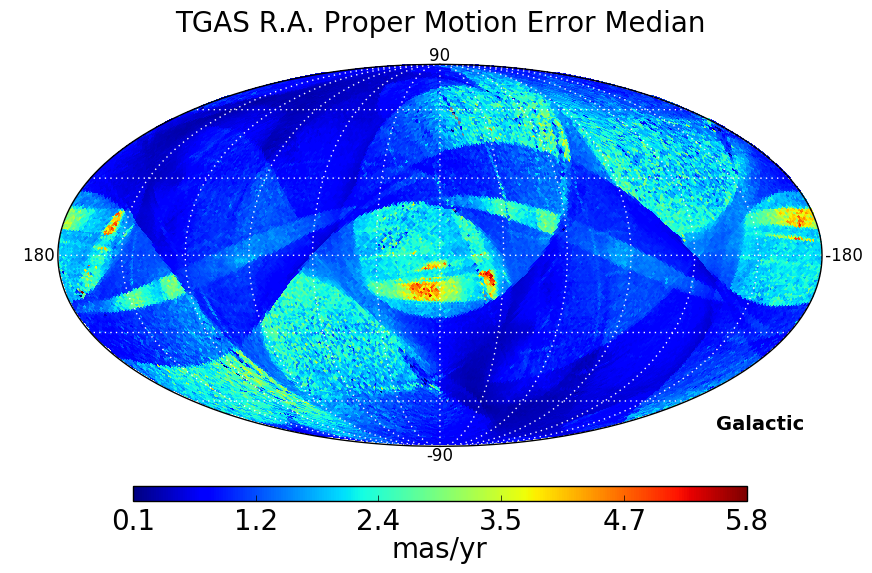}
        \includegraphics[width=0.33\textwidth]{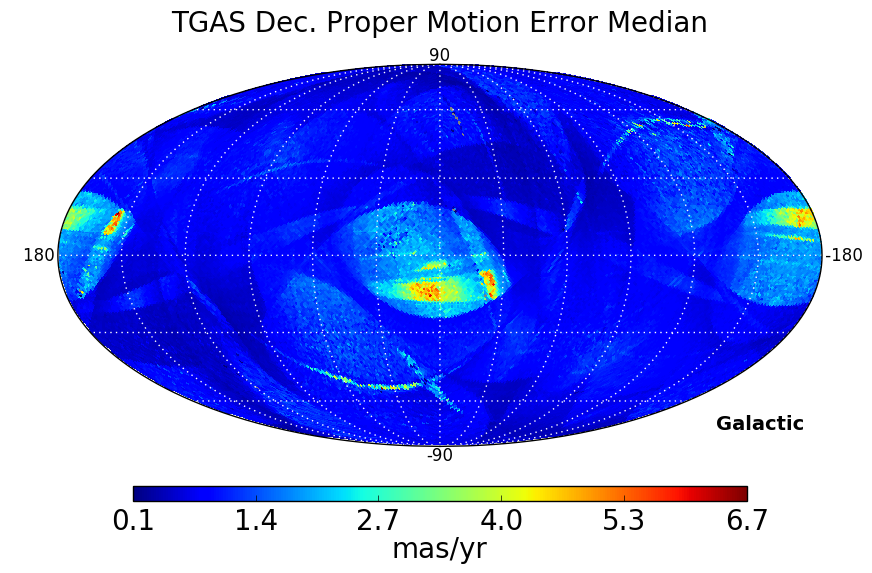}
  	      \caption{Sky map in galactic coordinates of the standard uncertainties of TGAS: parallaxes (mas, left),
	      proper motions in right ascension (\masyr, center) and
	      proper motions in declination (right). 
	      Note that the \beforeReferee{amplitude is however much reduced}\afterReferee{precision is however much better} for the subset of Hipparcos stars.} 
              \label{fig:Stats_HealpixMedianError-GAL_HealpixMapMedianSAM}
     \end{figure*}

    \begin{figure*}[!h]
	  \centering
	    \includegraphics[width=0.33\textwidth,height=0.25\textwidth]{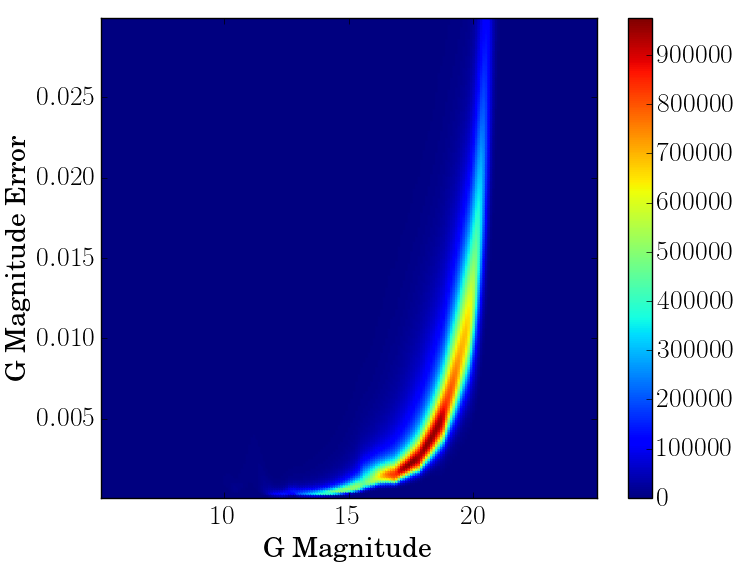}
      \includegraphics[width=0.33\textwidth,height=0.25\textwidth]{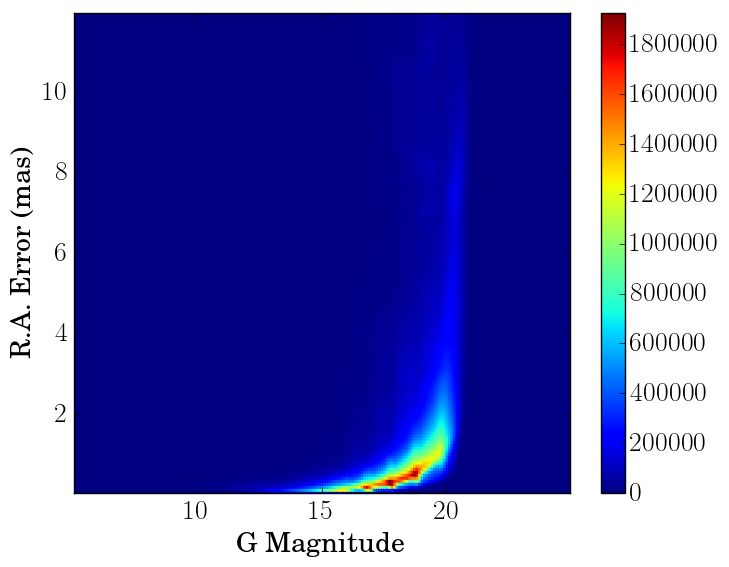}
	  \includegraphics[width=0.33\textwidth,height=0.25\textwidth]{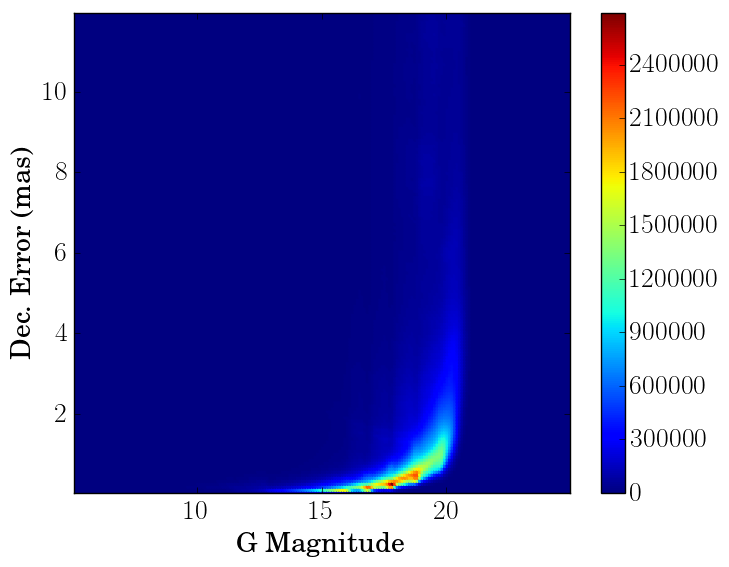}
      \caption{Distribution of the standard uncertainty of $G$ magnitude (left),
      of right ascension (center) and 
      of declination (right) as a function of $G$.}\label{fig:Histogram2DDecErrorAgainstMagG_Histogram2DSAM}
    \end{figure*}

     \begin{figure*}[!h]
	  \centering
	    \includegraphics[width=0.33\textwidth]{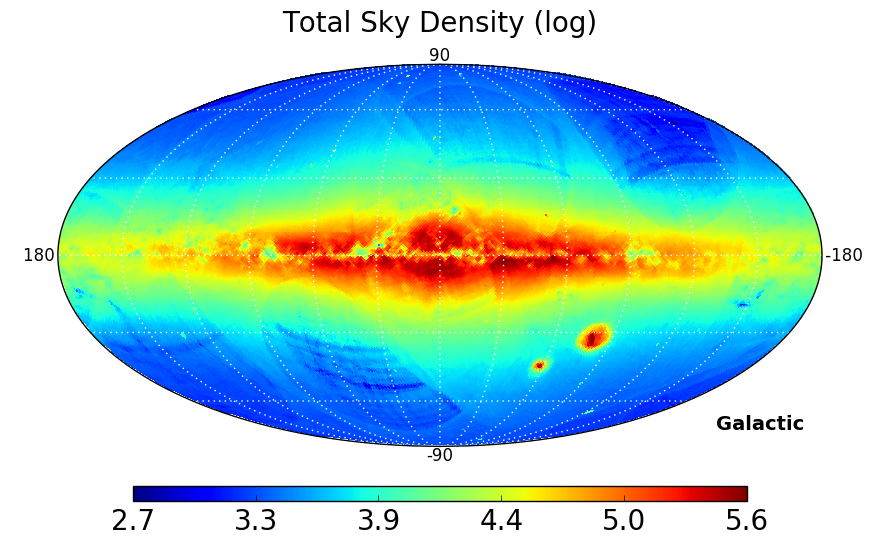}
        \includegraphics[width=0.33\textwidth]{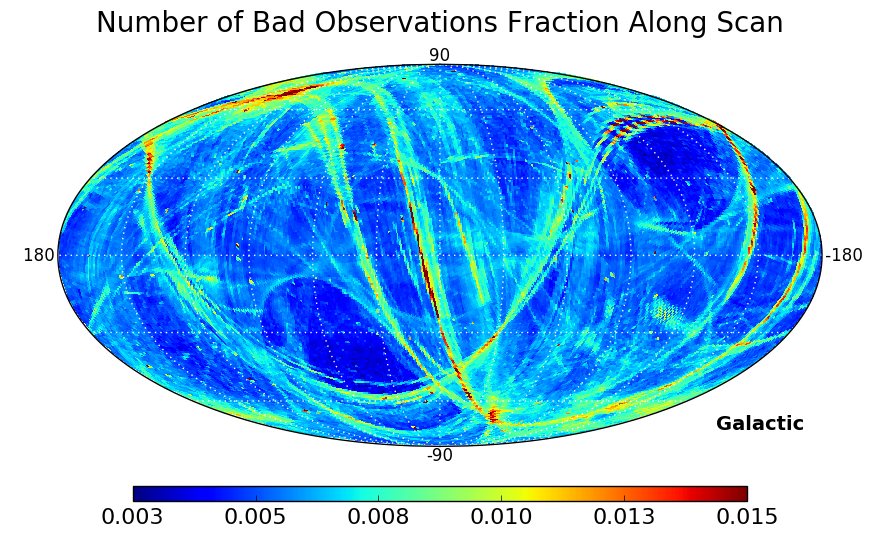}
        \includegraphics[width=0.33\textwidth]{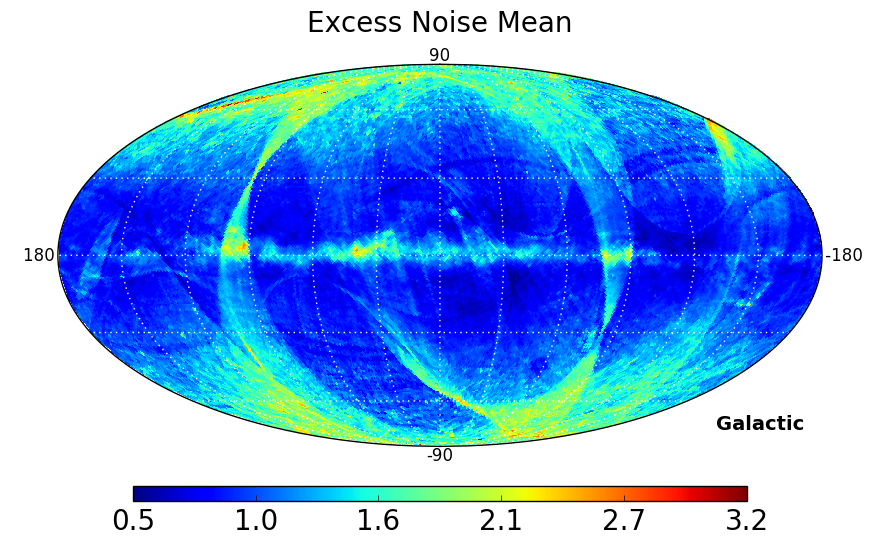}
  	      \caption{Sky map in galactic coordinates of the whole catalogue: logarithm of star density (left), 
	      fraction of bad observations (center) and
	      excess noise (right) showing the areas with potential problems, e.g. due to the ecliptic pole scanning law.} 
              \label{fig:GDR1healpix_stat_TotalSkyDensityLogCount-GAL_HealpixMapSAM}
     \end{figure*}

An overview and discussion of the contents of {\GDR1} can be found in \cite{DPACP-8} and full details are available in the archive documentation\footnote{\scriptsize\url{http://gaia.esac.esa.int/documentation/GDR1/Catalogue_consolidation/sec_cu1cva/sec_cu9gat.html}}. 


\newcommand\gdrtotnum{\ensuremath{1\,142\,679\,769}}
\newcommand\gdrsecnum{\ensuremath{1\,140\,622\,719}}
\newcommand\tgasnum{\ensuremath{2\,057\,050}}
\newcommand\tycnum{\ensuremath{1\,963\,415}}
\newcommand\hipnum{\ensuremath{93\,635}}
\newcommand\varnum{\ensuremath{3194}}
\newcommand\cepnum{\ensuremath{599}}
\newcommand\cepnumnew{\ensuremath{43}}
\newcommand\rrlnum{\ensuremath{2595}}
\newcommand\rrlnumnew{\ensuremath{343}}

\subsection{Selected TGAS statistics}

Figure~\ref{fig:TGAShealpix_stat_TotalSkyDensityLogCount-GAL_HealpixMapSAM}a shows the star density of TGAS in galactic coordinates. Besides the physical features like the galactic disk this figure also shows clearly the traces of the incompleteness discussed in previous sections; artefacts in the shape of the Gaia scanning law show regions of under-densities arising from the removal of stars with low number of observations in under-scanned regions. We remind again the reader of the incompleteness of this release discussed in \secref{sec:completeness}.

Figure \ref{fig:Stats_HealpixMedianError-GAL_HealpixMapMedianSAM} shows the distribution of the errors in TGAS astrometry over the sky. As can be seen the distribution of these errors is quite inhomogeneous around the sky, with large regions with small errors and some regions with high errors. These features are also present in the distributions of the errors of other parameters, like the magnitudes. Therefore we advise the reader to always use the errors given in the catalogue for the analysis of the data and never rely on an average error. Also, as discussed in 
\secref{sec:effect-correlations}, the correlations between the astrometric parameters should be taken into account for the error analysis. These correlations can be significant in {\GDR1} and its sky distribution is very inhomogeneous, as illustrated in \figref{fig:TGAShealpix_stat_TotalSkyDensityLogCount-GAL_HealpixMapSAM}c. Notice the large areas with significant positive or negative correlations. 

Although these uncertainties and correlations represent the behaviour of most TGAS stars, 
it is important to note that the corresponding figures with the Hipparcos subset 
alone are very different, due to much smaller uncertainties and correlations, 
see e.g. \figref{fig:pmra-pi-correlation}b.

\subsection{Selected global statistics}

Figure~\ref{fig:GDR1healpix_stat_TotalSkyDensityLogCount-GAL_HealpixMapSAM}a shows the star density in galactic coordinates of the global {\GDR1} dataset. Although less prominent than in \figref{fig:TGAShealpix_stat_TotalSkyDensityLogCount-GAL_HealpixMapSAM}a the artefacts in the shape of the scanning law due to the incompleteness caused by the selection applied is still present, and should be taken into account for star count analysis, as already discussed. 

On the other hand, \figref{fig:Histogram2DDecErrorAgainstMagG_Histogram2DSAM}
illustrate the distribution of the errors in magnitude and position as a function of the $G$ magnitude. As illustrated by these figures the behaviour of the errors approximately follow the mean dependence on $G$ expected for the mission Science Performance estimations\footnote{\scriptsize\url{http://www.cosmos.esa.int/web/gaia/science-performance}}, but also show features due to the effects of on-board priorization of the Calibration Faint Stars at every magnitude (vertical lines), some jumps due to the effects of the CCD gates (at the bright end) and a wide dispersion around these mean relations due to the varying number of observations and star colours. Again, we advise the reader to always use the errors given in the catalogue for the analysis of the data and never rely on average errors or error relations.

\end{document}